\documentclass[10pt, pre, amsmath, onecolumn, showpacs, superscriptaddress, nofootinbib]{revtex4-1}
\usepackage{graphicx,color}
 \graphicspath{{./}{./figures/}}
\usepackage{ams math}
\usepackage{enumerate}
\usepackage{mathbbol}
\usepackage{mathtools}
\usepackage{amsfonts}
\usepackage{natbib}

\usepackage{bm}
\usepackage{hyperref}
\hypersetup{
    colorlinks=true,       
    linkcolor=red,          
    citecolor=blue,        
    filecolor=magenta,      
    urlcolor=cyan           
}

\usepackage{color}
\usepackage[dvipsnames]{xcolor}

\usepackage{cancel}

\usepackage[caption=false]{subfig}
\usepackage{bbm}

\usepackage{soul}
\setstcolor{red}
\usepackage{multirow}

\newcommand{\nn}{\nonumber}

\newcommand{\sgn}{{\mathrm{sgn}}}

\newcommand{\be}{\begin{equation}}
\newcommand{\ee}{\end{equation}}
\newcommand{\bea}{\begin{eqnarray}}
\newcommand{\eea}{\end{eqnarray}}
\newcommand{\beq}{\begin{eqnarray}}
\newcommand{\eeq}{\end{eqnarray}}

\newcommand{\Ai}{{\rm Ai}}

\newlength{\bilderlength}

\usepackage{subfig}

\begin{document}

\title{Spatio-temporal fluctuations in the passive and active Riesz gas on the circle}

\author{L\'eo \surname{Touzo}}
\affiliation{Laboratoire de Physique de l'Ecole Normale Sup\'erieure, CNRS, ENS and PSL Universit\'e, Sorbonne Universit\'e, Universit\'e Paris Cit\'e,
24 rue Lhomond, 75005 Paris, France}
\author{Pierre Le Doussal}
\affiliation{Laboratoire de Physique de l'Ecole Normale Sup\'erieure, CNRS, ENS and PSL Universit\'e, Sorbonne Universit\'e, Universit\'e Paris Cit\'e,
24 rue Lhomond, 75005 Paris, France}
\author{Gr\'egory \surname{Schehr}}
\affiliation{Sorbonne Universit\'e, Laboratoire de Physique Th\'eorique et Hautes Energies, CNRS UMR 7589, 4 Place Jussieu, 75252 Paris Cedex 05, France}

\date{\today}

\begin{abstract} 
We consider a periodic version of the Riesz gas consisting of $N$ classical particles on a circle,
interacting via a two-body repulsive potential which behaves locally as a power law of the distance, $\sim g/|x|^s$
for $s>-1$. Long range (LR) interactions correspond to $s<1$, short range (SR) interactions to $s>1$, 
while the cases $s=0$ and $s=2$ describe the well-known log-gas and the Calogero-Moser (CM) model respectively.
We study the fluctuations of the positions around the equally spaced crystal configuration,
both for Brownian particles -- passive noise -- 
and for run-and-tumble particles (RTP) -- active noise. 
We focus on the weak noise regime where the equations of motion can be linearized,
and the fluctuations can be computed using the Hessian matrix.
We obtain exact expressions for the space-time correlations,
both at the macroscopic and microscopic scale, for $N \gg 1$ and 
at fixed mean density $\rho$. They are characterized by
a dynamical exponent $z_s=\min(1+s,2)$. We also obtain the gap
statistics, described by a roughness exponent $\zeta_s=\frac{1}{2} \min(s,1)$.
For $s>0$ in the Brownian case, we find that in a broad window of time, i.e. 
for $\tau=1/(g \rho^{s+2}) \ll t \ll N^{z_s} \tau $,
the root mean square displacement of a particle exhibits sub-diffusion
as $t^{1/4}$ for SR as in single-file diffusion, and $t^{\frac{s}{2(1+s)}}$ for LR interactions. Remarkably, this
coincides, including the amplitude, with a recent prediction obtained using macroscopic
fluctuation theory. 
These results also apply to RTPs beyond a characteristic 
time-scale $1/\gamma${, with $\gamma$ the tumbling rate,} and a length-scale $\hat g^{1/z_s}/\rho$ with $\hat g=1/(2\gamma \tau)$. 
Instead, for either shorter times or shorter distances, the active noise leads to 
a rich variety of static and dynamical regimes, with distinct exponents, for which we obtain
detailed analytical results. For $-1<s<0$, the displacements are bounded,
leading to true crystalline order at weak noise. The melting transition,
recently observed numerically, 
is discussed in light of our calculation. 
Finally, we extend our method to the active Dyson Brownian motion 
and to the active Calogero-Moser model in a harmonic trap, 
generalizing to finite $\gamma$ the results of our earlier work. 
Our results are compared with the mathematics literature whenever possible. 
\end{abstract}

\maketitle

\tableofcontents

\section{Introduction}

\subsection{Overview}






Characterizing the collective behavior of many interacting active particles is a significant challenge that has attracted considerable attention in recent years, with important implications for biological and artificial systems, ranging from bacterial colonies to robotic swarms~\cite{Marchetti,Bechinger,Schranz,Ramaswamy2017,Berg2004}. Active particles, which consume energy from their environment to generate self-propulsion, exhibit rich and non-equilibrium behaviors distinct from the one of Brownian (passive) particles. One of the simplest yet much studied models for active matter is the run-and-tumble particle (RTP), where the dynamics is driven by a telegraphic noise~\cite{TailleurCates,W02,HJ95,ML17}.
In one dimension, which we focus on below, 
these particles switch, with rate~$\gamma$, between two discrete states of motion with opposite velocities $\pm v_0$, alternating between runs and tumbles. In the presence of mutual interactions, and even in the absence of alignment forces, these systems of RTPs exhibit non-trivial collective dynamics, particularly in confined geometries \cite{DKM19,slowman,slowman2,bodineau2018,us_bound_state,Maes_bound_state,nonexistence,MBE2019,KunduGap2020,LMS2019}. For short range interactions or in the presence of boundaries they often exhibit spontaneous phase separations and clustering \cite{BC2022,FHM2014,FM2012,Buttinoni2013,CT2015,Agranov2021,Agranov2022,Metson2022,MetsonLong,Dandekar2020,Thom2011}.

For long range interactions (besides the case of indirect hydrodynamic interactions~\cite{Janus,PireySphere2023,HydroConfined}) very few studies exist, while their Brownian counterparts have a long history \cite{DauxoisPhysRep2009,Dauxois_book,Lewin}. A particularly intriguing question is: How do spatio-temporal fluctuations in interacting systems differ between passive and active particles, especially when long-range interactions are present~? RTPs with long-range interactions provide a compelling framework for studying these properties. Recently, we introduced and studied the active Dyson Brownian motion (DBM) where the RTPs, on the real line, interact via repulsive logarithmic potential in the presence of an external quadratic well~\cite{ADBM1,ADBM2}. This generalizes the (passive) DBM well known in the context of random matrix theory where it describes the dynamics of the eigenvalues of a matrix Brownian motion \cite{Mehta_book,bouchaud_book,Spohn2}. In the passive case, the system reaches a Gibbs equilibrium state, which is often called the {\it log-gas}, characterized
by the Wigner semi-circle density and local correlations, for which a number of analytical predictions are available \cite{Mehta_book,Forrester_book}. 
By contrast, in the active case, we showed that the system reaches a non-Gibbsian stationary state,
for which only a few analytical results were obtained~\cite{ADBM1,ADBM2}. In the weak noise regime (i.e., for small enough velocity $v_0$) the global density retains a semi-circular shape, while for stronger noise it takes a bell-shape form. 
Correspondingly, the system is close to a crystal for weak noise, while the formation of clusters is observed at stronger noise. 
These differences are apparent in the spatial correlations between the particles. 
In the weak noise limit (together with the limit $\gamma \to 0^+$) we performed an expansion around the crystalline ground state and 
obtained a detailed description of the equal time covariance of the particle displacements~\cite{ADBM2}. This allowed for instance
to compute the variance of the gaps between particles. However our method did not allow us
to study the dynamics of the active system for a finite value of $\gamma$.


In the present paper, we consider a natural generalization of the active DBM, the {\it active Riesz gas}, for which
we study the statics and dynamics. It consists of a gas of $N$ identical RTPs interacting via a repulsive power-law potential of the form $1/|x_i - x_j|^s$ with $s>-1$. In the limit $s \to 0$, the active Riesz gas corresponds to the active DBM. Its passive version, the Riesz gas model for Brownian particles, has attracted considerable interest across both the physics and mathematics literature, 
encompassing its equilibrium \cite{Riesz,leble2017,leble2018,riesz3,Agarwal2019,jit2021,leble_loggas,BoursierCLT,BoursierCorrelations,jit2022,santra2022,dereudre2023number,Lelotte2023,Beenakker_riesz,Serfaty_lecturenotes,Riesz_FCS,UsRieszCumulants} and, more recently, its 
dynamical \cite{Huse_riesz,DynamicsRankPLD,DynamicsRankFlack,SerfatyDynamics2022,DFRiesz23,SerfatyDynamics2023} behavior. Depending on the value of $s$, the interaction is either long range (LR) for $-1<s<1$, or short range (SR) for $s > 1$ with markedly different behavior in both cases. 
For instance, the displacement in time of a tracer grows as $t^{1/4}$ for SR interaction (as in single-file diffusion
\cite{SingleFileMajumdar1991,SingleFileKrapivsky2015,passiveEW})\footnote{Note that in
\cite{UsFluctuationsArxiv} we have also obtained $t^{1/4}$ for the Calogero-Moser model.} ,
while it is logarithmic for $s=0$ \cite{Spohn1} and anomalous with an exponent that depends continuously on $s$ for $0<s<1$, as found recently using
an hydrodynamic approach \cite{DFRiesz23}. 
Apart from the log-gas ($s=0$), the class of Riesz gases includes two other important well known models: namely the one-dimensional Coulomb gas $s=-1$ \cite{Lenard,Baxter1963,Kunz,AM,dhar2017exact} and the Calogero-Moser (CM) model ($s=2$)
\cite{Calogero75,Moser76,Agarwal2019}. Interestingly, even in the case of the Riesz gas with Brownian particles, there are still some open questions. For instance a recent work points out to a possible phase transition for $-1 < s \leq 0$ from a crystalline phase to a fluid \cite{Lelotte2023}. 
Some observables such as the variance of the gaps or the linear statistics have only been studied very recently in the statics \cite{BoursierCorrelations} in the LR case, but
the dynamical correlations have not been investigated. It would thus be of great interest to address these questions both in the passive and in the
active Riesz gas.

\begin{figure}
    \centering
    \includegraphics[width=0.4\linewidth]{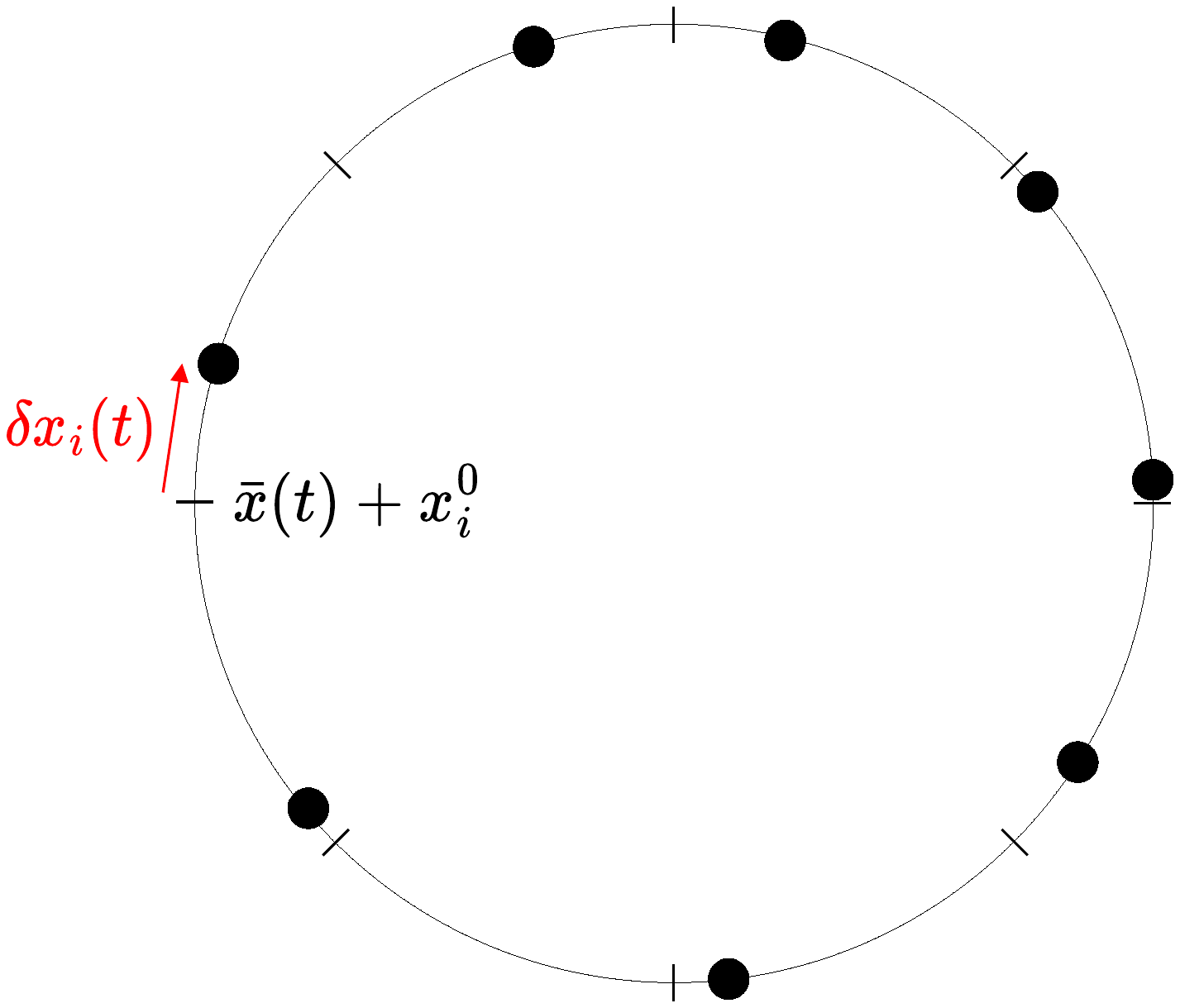}
    \caption{Schematic representation of the particle system in the periodic geometry. In the ground state the particles are equally spaced. We consider the evolution of the displacements $\delta x_i(t)$ of each particles with respect to its position in the ground state due to the presence of passive or active noise. As defined in \eqref{def_delta_x_intro}, $\bar x(t)$ denotes the position of the center of mass which decouples from the $\delta x_i(t)$.}
    \label{fig:sketch_circle}
\end{figure}

Here we investigate these questions in the weak noise regime, where the positions of the particles are close to a perfect crystal up to small displacements (phonon degrees of freedom). Upon linearizing the equations of motion, we can compute the spatio-temporal correlations of the displacements for various types of noise. We will
focus here both on the white noise (Brownian particle) and on the telegraphic noise (RTP). 
We will first study the Riesz gas on a circle, see Fig. \ref{fig:sketch_circle}, and in a second part we will consider the active DBM and active CM models in a harmonic well, generalizing to finite $\gamma$ the results of \cite{ADBM2}. 

Note that the active Riesz gas has been investigated very recently in the case $s=-1$ \cite{activeRankDiff}, and that in a companion paper the
active CM model is also studied \cite{activeCM}. 
We also mention that the active harmonic chain was studied in \cite{HarmonicChainRevABP,SinghChain2020,PutBerxVanderzande2019,HarmonicChainRTPDhar,HarmonicBasu}.
In that case the linear approximation is exact. As discussed below, this case shows similarities with the SR regime of the Riesz gas.



The paper is organized as follows. In the next subsection \ref{subsec:goal} we present a short summary of the main goal and results of the paper.
A more detailed account of the results is given in Section \ref{sec:detailedsum}. The results for the Riesz gas on a circle are derived in Section \ref{sec:Brownian} for the Brownian particles,
and in Section \ref{sec:RTP} for the RTPs. The results for the active DBM and active CM in presence of an harmonic well are given in Section \ref{sec:harmonic_trap}. 
A conclusion and discussion are given in Section \ref{sec:conclusion}. The appendices contain some additional technical details.

\subsection{Goal and summary of main results} \label{subsec:goal}

In the first part of this paper we consider $N$ classical particles on a circle of perimeter $L$, interacting via a two-body interaction potential such that
in the ground state the particles are equally spaced, i.e. they form a perfect lattice. In addition these particles experience either passive or active independent noises,
which produce displacements $\delta x_i$ for each particle away from their position in the ground state (see Fig. \ref{fig:sketch_circle}).
We consider two models: (i) particles submitted to Brownian noise, in which case the system reaches a Gibbs equilibrium at temperature $T$, 
(ii) run-and-tumble particles (RTPs) with velocity $\pm v_0$ and tumbling rate $\gamma$, in which case the system reaches a stationary state at long time. 
While a single, free RTP behaves as a Brownian particle at large time with effective temperature { (or diffusion constant)} $T_{\rm eff}= v_0^2/(2 \gamma)$,
the behavior of interacting RTPs is known to exhibit much richer behavior. 
We consider here the situation of weak noise (low temperature or weak active noise) such that
the relative displacements of neighboring particles remain small (being linear in temperature). 
Although we start by studying a general interaction potential, we will mainly focus in this paper on the case of the Riesz gas, i.e., with periodic power law interaction potential $\propto g |x|^{-s}$, where $s>-1$ and $g$ is the interaction strength. In each case we obtain a quantitative criterion
for the weak noise approximation to be valid.

Our aim is to compute the space-time correlation functions of the 
displacements of the particles in the limit of large $N \gg 1$. We 
consider the natural scaling where $L \gg 1$ 
with a fixed mean density $\rho=N/L$.
In that limit one can study observables either at microscopic scales,
i.e., for distances of the order of the lattice spacing, or observables at macroscopic scales,
i.e., for distances of the order of the size $L$ of the ring. Before providing a detailed exposition of our main findings in the next section, it is useful to start with a brief overview.

We first study the stationary correlations,
i.e., at equilibrium for the Brownian case, and in the stationary state for RTPs (annealed initial conditions). 
For $s>0$ they are characterized by a ``roughness'' exponent $\zeta_s = \frac{1}{2} \min(s,1)$ and a dynamical exponent $z_s= \min(1+s,2)$.
We study the variance of the displacement $\langle \delta x_i^2\rangle$ of a given particle. We find that in both cases (Brownian and RTP) 
for $s<0$ this variance 
reaches a finite limit as $N$ increases, while for $s > 0$ it grows with $N$ as $N^{2 \zeta_s}$. 
In the first case $s<0$ we thus obtain that there is translational order at sufficiently low temperature, and we obtain an estimate for a possible melting transition temperature 
for Brownian particles ($T=T_M$) and RTPs ($T_{\rm eff}= T_{\rm eff,M}$).
We also compute the variance of the ``gaps'' (i.e. of the relative distance) between two particles separated (in the ground state) by $k$ lattice spacings (see Fig.~\ref{fig:sketch_circle}).
We find that for $s>0$ it grows as $\sim k^{2 \zeta_s}$ at large $k$ on microscopic scales, and reaches $\sim N^{2 \zeta_s}$ at macroscopic scales. 
We also compute the correlations between the gaps associated to two
different intervals and find a rich behavior depending on the overlaps between these two intervals. 
We find generally that the correlations for the RTPs become similar to the Brownian ones at length scales larger than
a crossover distance $k \gg \hat g^{1/z_s}$ where $\hat g=g \rho^{s+2}/(2 \gamma)$.

Next we study the dynamical correlations, starting with the annealed initial condition. 
Within the linear expansion around the ground state, the eigenmodes of 
the Hessian matrix are simply plane waves, which allows to obtain the exact spectrum of relaxation times. We show
that there are two important time scales: $\tau= 1/(g \rho^{s+2})$ which sets the time scale of
local equilibration, and $N^{z_s} \tau$ which sets a priori the time scale of equilibration
for the full circle. In the case of the RTPs there is an additional characteristic time $1/\gamma$ (the persistence time).
The dimensionless parameter $\hat g=g \rho^{s+2}/(2 \gamma)$, plays an important role, being the ratio of the persistence time
to the local equilibration time set by the interactions.

The simplest dynamical observable is the variance of the displacement of a given particle during time $t$. For the Brownian
particles and for $s>0$, we find that there are three regimes in time. At very short time $t \ll \tau$, the particle has not yet
experienced the interaction and undergoes free diffusion. In a very broad intermediate time window $\tau \ll t \ll N^{z_s} \tau$, 
we find that $\langle (\delta x_i(t) - \delta x_i(0))^2\rangle$ grows subdiffusively as $\sim t^{1/2}$ in the case of short-range
interactions, similar to the behavior of single-file systems, as mentioned above. For long-range interactions, $0<s<1$, the growth becomes even slower, as $\sim t^{s/(1+s)}$.
Remarkably, this result agrees (including numerical prefactors) with the one obtained in a recent work on the Riesz gas, which uses a completely different, hydrodynamic method \cite{DFRiesz23}. 
Finally for $t \gg N^{z_s} \tau$, the displacement saturates to its equilibrium value with variance $\sim N^{2 \zeta_s}$. 
In the case of the RTPs there is an additional ballistic regime at short time $t \ll 1/\gamma$, with however a 
velocity $v_R=v_R(\hat g)$ which is renormalized by the interaction, and is a decreasing function of the parameter $\hat g$. 
In the case of very long range interactions $s<0$, the intermediate regime is absent, as equilibration of two-point correlations
occurs on a time scale $\tau$. 

To quantify further the time evolution of the motion of a particle, we compute the covariance of its displacements
during two different times $t$ and $t'$. We find that at large times it takes the same form as the two-time correlation
of a fractional Brownian motion with Hurst index $H = 1/4$ for $s>1$ (i.e., for short range interactions) and $\frac{s}{2(1+s)}$ for $0<s<1$ (i.e., for long range interactions). 
Again this agrees with the result obtained in the recent work 
\cite{DFRiesz23}.

To characterize the spatial dependence of the collective dynamics we compute the covariance between the displacements during time $t$ of two particles 
separated by $k$ lattice spacings. We find that this covariance at the microscopic level takes a dynamical scaling form 
as a function of the scaling variable $k/(t/\tau)^{1/z_s}$ and, in the case of the RTPs, of the additional variable $\gamma t$. We identify different dynamical regimes, which
in the case of the RTPs are quite rich, each of them being characterized by scaling functions that we compute explicitly. We perform a similar study
for the time correlation of the gaps. 

We also study the dynamics starting from a perfect lattice (quenched initial conditions) and obtain the two-point two-time 
correlation of the displacements. In the case of the Brownian particles, the variance of the time displacement of a given particle 
$\langle (\delta x_i(t) - \delta x_i(0))^2\rangle$ exhibits similar regimes as for the annealed initial condition, 
with however different amplitudes. In the large time anomalous diffusion regime our results again agree 
with those of \cite{DFRiesz23}. In the case of the RTPs we find new dynamical regimes as compared to the annealed initial
condition. In particular for $s>0$ we show the existence of a super-diffusive regime which generalizes the one found recently for the active harmonic chain \cite{SinghChain2020}. 

In the last part of the paper we return to the problem of the active DBM ($s=0$) in a harmonic well, studied in \cite{ADBM2} for $\gamma=0^+$
and extend our study to the active CM model ($s=2$) also in a harmonic well. In the weak noise regime we obtain the scaling forms
of the space-time covariance of the particle displacements for arbitrary $\gamma$. 
Both models exhibit a bulk regime of correlations inside the support of the semi-circle density. In addition the
active DBM also exhibits a distinct edge regime, which is absent for the CM model. 
We find that in the active CM the variance of the gaps grows quadratically with $k$ on microscopic scales. 
In the case of the active DBM we find good agreement with numerical simulations displayed here.
The comparison with numerics for the active CM model is performed in a companion paper \cite{activeCM}. Finally, we discuss the extension of 
the present results to more general models of active particles, such as active Ornstein-Uhlenbeck processes or active Brownian particles with Riesz interactions.






\section{Details of main results} \label{sec:detailedsum}

Our model is described by the following equation of motion for $N$ particles at positions $x_i(t)$ on a ring of perimeter~$L$ (see Fig.~\ref{fig:sketch_circle}),
\be \label{Eq_def_intro}
\dot x_i(t) = -\sum_{j(\neq i)} W'(x_i(t)-x_j(t)) + \zeta_i(t) \quad , \quad \zeta_i(t) = \begin{cases} \sqrt{2T} \xi_i(t) \quad \text{(Brownian)} \\ v_0 \sigma_i(t) \quad \ \ \ \text{(RTP)}
\end{cases}
\ee
where $W(x)$ is the interaction potential with the periodicity of the ring.
{As announced above}, we consider two cases: (i) the Brownian particles, where the $\xi_i(t)$ are independent and identically distributed (i.i.d.) standard Gaussian white noises
and (ii) the RTPs, in which case $\sigma_i(t)=\pm 1$
are i.i.d telegraphic noises with rate $\gamma$. Although we will derive formulas for more general interaction
potential, our main focus here is the repulsive Riesz gas. On the real axis it corresponds to
\be 
W'(x)=-g \, \frac{\sgn(x)}{|x|^{s+1}}
\ee
with $g>0$, and on the ring we use the properly periodized version, see Eq.~\eqref{defRiesz} below. We restrict here to $s>-1$.
We parametrize the positions of the particles in terms of their displacements $\delta x_i(t)$ from the equally spaced ground state
configuration 
\be \label{def_delta_x_intro}
x_i(t) = \bar x(t) + x_i^0  + \delta x_i(t)  \quad , \quad x_{i+1}^0-x_i^0= \frac{L}{N} \quad , \quad \bar x(t) = \frac{1}{N} \sum_i x_i(t) 
\ee
where the center of mass $\bar x(t)$ diffuses at large time with coefficient $T/N$ {for Brownian particles}, respectively $T_{\rm eff}/N$ {\rm for RTPs}, and does not play a role in the following.

We focus here on the weak noise regime where one can linearize the equation of motion \eqref{Eq_def_intro} around the
ground state positions $x_i^0$, i.e. we assume the relative displacements $\delta x_j - \delta x_i $ to be small.
The domain of validity of this approximation is discussed right below. The inverse relaxation times of the system are thus equal to the eigenvalues $\mu_q$ of
the Hessian matrix. We have computed them for a general interaction, see Eq. \eqref{eigenvals}. In the case of the Riesz gas they
are given for any $N$ as 
\be \label{mu_Riesz_intro}
\mu_q = g\rho^{s+2} f_s\left( \frac{q}{N} \right) \quad \text{with} \quad f_s(u)= 4(s+1) \sum_{\ell=1}^{\infty} \frac{\sin^2(\pi \ell u)}{\ell^{s+2}} \quad , \quad q=1,...,N-1 \;,
\ee
with $\mu_q=\mu_{N-q}$. The function $f_s(u)=f_s(1-u)$ is increasing for $u \in [0,1/2]$ and vanishes at the edges as a power law
\be \label{fasympt_intro}
f_s(u) \underset{u\to 0}{\sim} a_s u^{z_s} \quad \text{with} \quad z_s = \min(1+s,2) \;,
\quad \text{and} \quad a_s = \begin{cases} 2\pi^{s+\frac{3}{2}} \frac{\Gamma(\frac{1-s}{2})}{\Gamma(1+\frac{s}{2})} \quad \ \; \text{for } -1<s<1 \\
4\pi^2 (s+1) \zeta(s) \quad \text{for } s>1 \end{cases} \;.
\ee
We have obtained the general formula for the covariance for arbitrary $N$, both for Brownian particles and 
for RTPs, see Eqs. \eqref{cov_brownian} and \eqref{cov_rtp}. Using these formulas we have studied various observables
in the thermodynamic limit $N \gg 1$, $L \gg 1 $ with a fixed density $\rho=N/L$.

\begin{figure}
    \includegraphics[width=0.32\linewidth]{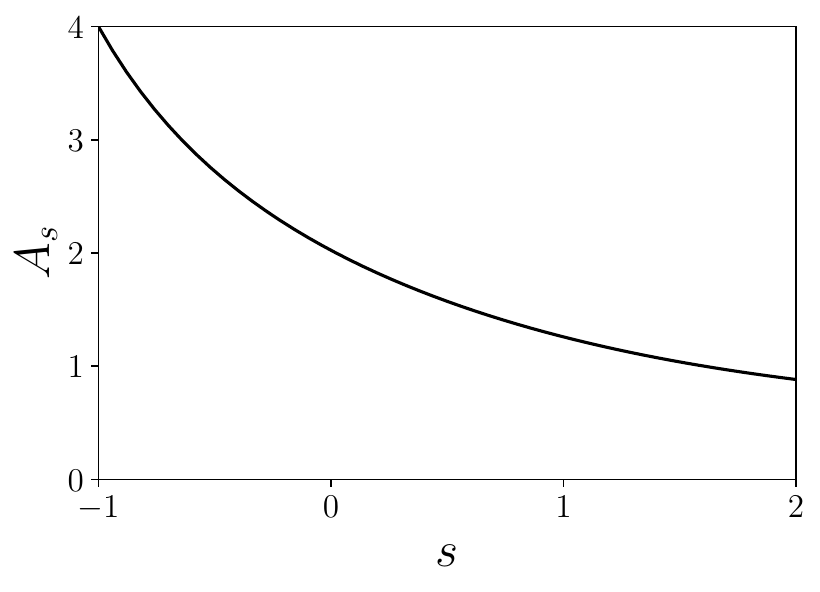}
    \includegraphics[width=0.32\linewidth]{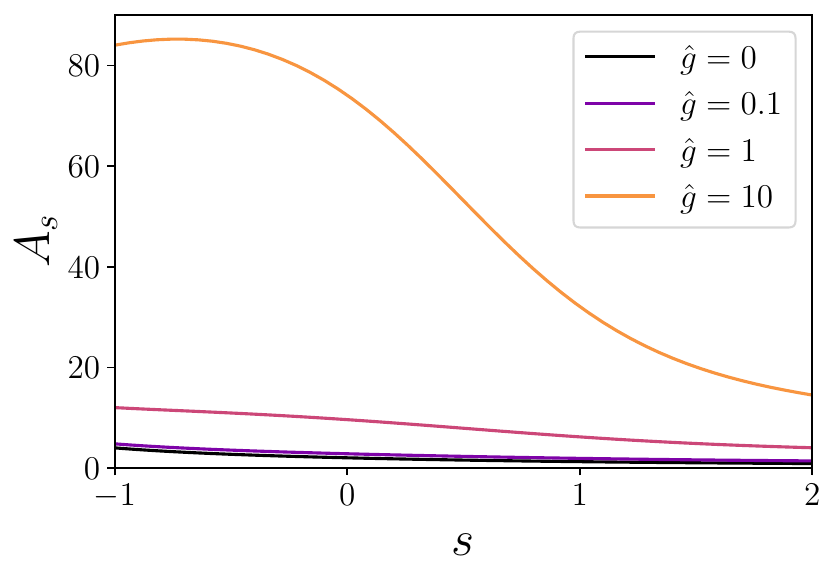}
    \caption{{Plot of $A_s$ vs $s$, as defined in Eq. (\ref{TG_intro})), which gives the temperature
    $T_G=A_s g \rho^s$ below which the present method should be valid. Left: Brownian case (i.e., $\hat g = 0$). Right: RTP case, and for different values of $\hat g > 0$.}}
    \label{FigAs} 
\end{figure}

We start by giving an estimate of the domain of validity of the linear approximation used in this paper. 
We find that this approximation is expected to be accurate whenever $T \ll T_G$ (Brownian) or $T_{\rm eff} \ll T_G$ (RTP)
where 
\be \label{TG_intro}
T_G = A_s  g\rho^s   \quad , \quad A_s^{-1} = 2(s+2)^2 \int_0^{1/2} du  \frac{\sin^2\left( \pi u  \right)}{f_s(u)\left(1+\hat g f_s(u) \right)} \quad , \quad \hat g = \begin{cases} 0 \quad \ \ \, \text{ for the Brownian} \\
\frac{g \rho^{s+2}}{2\gamma} \text{ for the RTP} \end{cases} \;.
\ee  
In the low temperature regime, the displacements are Gaussian for Brownian particles, while it is not necessarily the case for the RTPs. For both models we have determined the covariance of the displacements
in that regime. For the Brownian particles, the amplitude $A_s$ is a decreasing function of $s$, with $A_0=2.02441$ and $A_2=0.883237$, 
and it converges to $4$ as $s \to -1$. The amplitude $A_s$ is plotted in Fig. \ref{FigAs} for various values of $\hat g$. 

Note that for $s<0$, and for $s=0$ with $\beta = g/T<1$, in the Brownian case the particles can cross, while otherwise they never cross. 
Since we are studying here the low temperature regime these crossings are negligible as long as one remains in the domain
of validity discussed above. 

We now detail our results,
first for the Brownian particles and then for the RTPs.

\subsection{Brownian particles} \label{main_results_brownian}

\subsubsection{Static correlations} 

We first computed the equal time correlations at equilibrium. For $s \geq 0$
we find that the variance of the displacement of a given particle grows unboundedly with $N$ and behaves as 
\be \label{var_Riesz_liquid_intro}
\langle \delta x_i^2 \rangle \simeq \begin{dcases} \frac{T}{\pi^2 g\rho^2} (\ln N + \gamma_E) \;, \quad \quad \ \ \text{for } s=0\;, \\ 
\frac{N^{s} T \zeta(s+1) \Gamma(1+\frac{s}{2})}{\pi^{s+\frac{3}{2}} \Gamma(\frac{1-s}{2}) g\rho^{s+2}} \;, \quad \text{for } 0<s<1\;, \\ 
\frac{N}{\ln N} \frac{T}{24 g\rho^3} \;, \quad \quad \quad \quad \quad \ \  \text{for } s=1\;, \\ 
\frac{NT}{12 (s+1) \zeta(s) g\rho^{s+2}} \;, \quad \quad \, \text{for } s>1 
\end{dcases}
\ee
($\gamma_E$ is Euler's constant).
There is thus no long range translational order for $s \geq 0$. The absence of a phase transition at finite temperature
was proved recently in the case of the log-gas $s=0$ in \cite{leble_loggas}. 

\begin{figure}
    \includegraphics[width=0.32\linewidth]{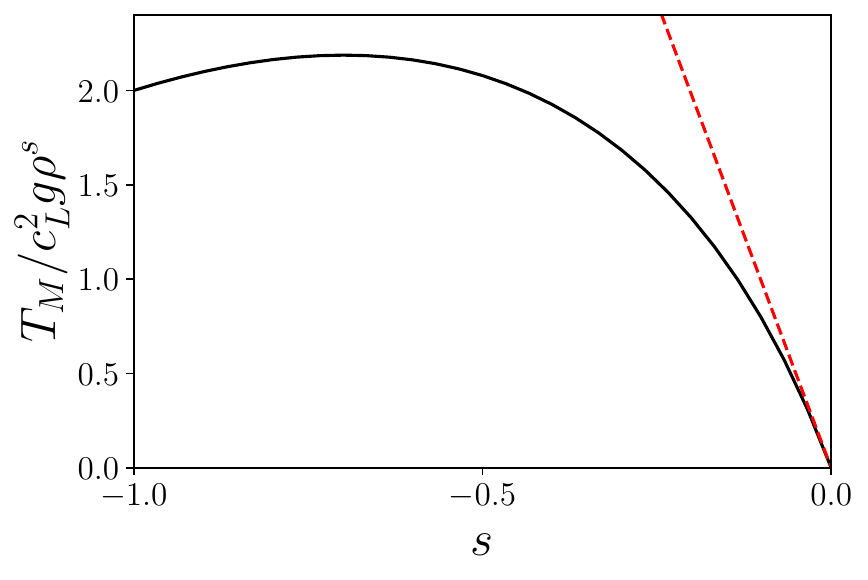}
    \includegraphics[width=0.32\linewidth]{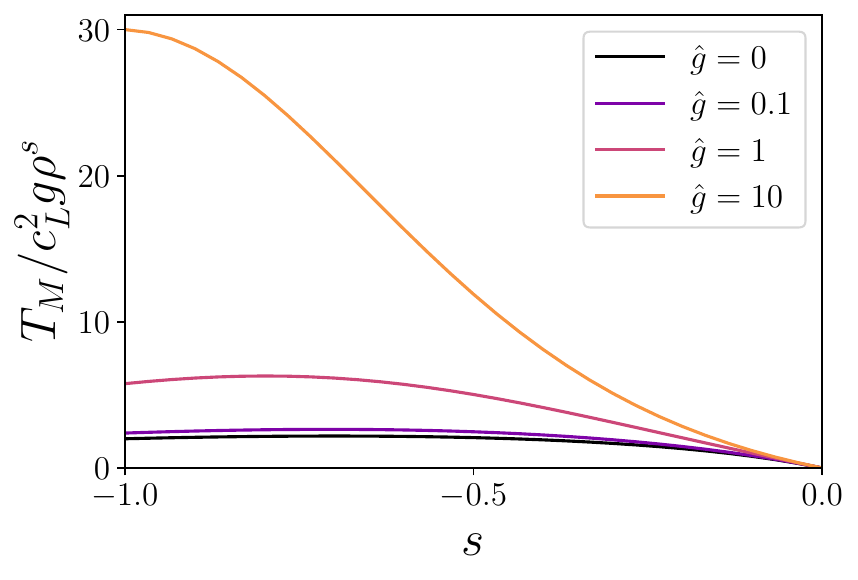}
    \caption{{Left: Plot of the ratio of the melting temperature over the Lindemann coefficient, $T_M/c_L^2$, versus $s$ for the Brownian case.
    The dashed red line shows the linear approximation in the limit $s\to 0^-$, given in (\ref{TM_small_s}}). Right: Plot of $T_M/c_L^2$ versus $s$ for different values of $\hat g$.}
    \label{Fig_TM}
\end{figure}

For $-1<s<0$ however the variance of
the displacements is bounded within our linear theory, and reads
\be \label{var_Riesz_solid_intro}
\langle \delta x_i^2 \rangle \simeq \frac{2T}{g\rho^{s+2}} \int_0^{1/2} \frac{du}{f_s(u)} \;.
\ee 
This result suggests that at low enough temperature the system exhibits long range translational order,
as was discussed recently in \cite{Lelotte2023}. In Ref.~\cite{Lelotte2023} numerical evidence was
found for a melting transition to a fluid phase at high temperature. 
We can estimate very roughly the melting transition temperature from
the crystal to the disordered state (assuming that it exists) via a Lindemann argument $\rho^2 \langle \delta x_i^2 \rangle = c_L^2$,
where $c_L$ is the phenomenological Lindemann coefficient (of order $c_L \simeq 0.05-0.2$ for 3D crystals) \cite{lindemann,navarro}. This gives
the following estimate for the melting temperature
\be \label{T_melting_intro}
T_M = \frac{1}{2} g\rho^{s} \frac{c_L^2}{\int_0^{1/2} \frac{du}{f_s(u)}} \;.
\ee 
The ratio $T_M/c_L^2$ is plotted in Fig. \ref{Fig_TM}. In the limit of the log-gas $s\to 0^-$ it vanishes as 
$T_M \sim \pi^2 c_L^2 g |s|$. More generally, one can check that $T_M$ is
below $T_G$. Especially for small $s$ the linear approximation should work 
up to the melting temperature. Note that for $s=-1$ (i.e., the $1d$ Coulomb gas) it was shown rigorously that the system is a crystal at any temperature \cite{Kunz,AM,Lewin}. 
In the numerical study of \cite{Lelotte2023} the melting temperature was indeed found to increase as $s$ decreases. 
The behavior shown in Fig. \ref{Fig_TM} for the ratio $T_M/c_L^2$ shows that for $s \lesssim -1/2$, the Lindemann coefficient $c_L$ cannot
be independent of $s$, which would contradict the aforementioned results \cite{Kunz, Lelotte2023}. 
Hence a more predictive theory of the possible melting transition remains a challenge for the future. 

%

Next we computed the variance of the gaps between two particles separated {(in the ground state)} by $k$ lattice spacings, with $k \ll N$. 
In the remainder of the paper the variable $k$ will be assumed a positive integer. 
We find
that for $s>0$ it grows at large $k$ as a power law
\be \label{gapscases_intro}
D_k(0) = \langle (\delta x_i-\delta x_{i+k})^2 \rangle \simeq \begin{dcases} \frac{T k^s}{\pi \tan \left( \frac{\pi s}{2} \right) g \rho^{s+2}} \;, \quad \text{for } 0<s<1 \;, \\
\frac{T k}{(s+1) \zeta(s) g\rho^{s+2}} \;, \quad \text{for } s>1 \;. \end{dcases}
\ee
The field of displacements thus exhibits a roughness exponent $\zeta=s/2$ for the long-range case and $\zeta=1/2$ in the short-range case. 
These gaps were also studied recently in the mathematics literature for $0<s<1$ \cite{BoursierCLT,BoursierCorrelations}, where it was proved 
that their scaled distribution converges to a Gaussian. The same power law $k^s$ behavior was found for $0<s<1$ (see however \cite{footnoteBoursier}).
In the case of the log-gas $s=0$ we find
\bea \label{gaps_s0}
D_k(0) \simeq  \frac{2 T}{\pi^2 g \rho^2} \ln(k) \;.
\eea
This formula can be compared to the well known result for the variance of the number of particles inside a fixed interval~\cite{Mehta_book,Forrester_book,Bourgade21},
with agreement in the leading logarithmic behavior, as discussed in Section \ref{sec:gapstat}.
In the case $s<0$ the variance of the gaps saturates at large $k$, as the displacements of the two particles become uncorrelated.
Note that the calculation of the variance of the gaps also allows to determine the range of validity of the linear
approximation given in \eqref{TG_intro}, see the derivation in Section \ref{sec:gapstat}. It also allows 
to determine the translation order correlation function within our linear Gaussian approximation.
In the case of the log-gas one finds that it decays to zero at large $k$ as 
\be 
S(k) =  \langle e^{ 2 i \pi \rho (\delta x_{j+k}-\delta x_j) } \rangle \underset{k \gg 1}{\propto} \,    k^{-\frac{4 T}{g}}  \;,
\ee
indicating the absence of translational order (in agreement with \cite{leble_loggas}). The decay exponent is in agreement with predictions from
bosonization \cite{Haldane,GiamarchiBook,Forrester1984} and with exact results for special values 
of $\beta$ \cite{Forrester_book,Forrester93,Forrester95,Forrester1998} (these results are obtained from the connection
of the circular log-gas to the circular $\beta$ ensemble (C$\beta$E) in random matrix theory and to interacting fermions on the circle,
see \cite{Smith2021}). It was also discussed recently in \cite{Lelotte2023} (see Eq. (14) there), and 
in \cite{NelsonZhang2020} in the context of grain boundaries in 2D crystals. 
For $0<s<1$ the translational order is destroyed faster, with a 
decay to zero at large $k$ as a stretched exponential, 
which becomes exponential for $s>1$. Finally for $s<0$ there is true translational order at low temperature,
as discussed above, and the translational order correlation 
saturates at large $k$ to a non-zero value.
\\

\begin{figure}
    \centering
    \includegraphics[width=0.55\linewidth,trim={0 7cm 0 7cm},clip]{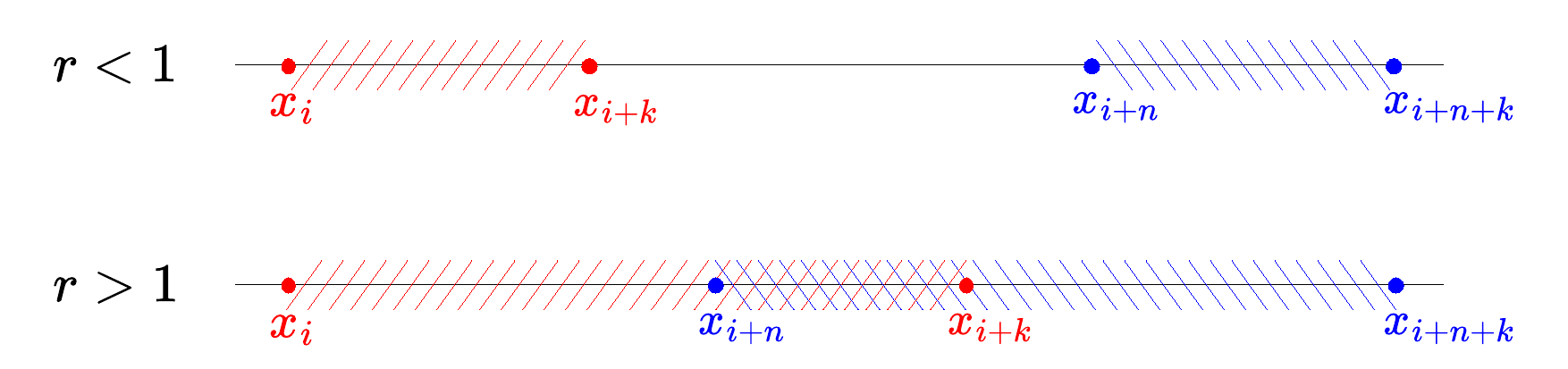}
    \includegraphics[width=0.3\linewidth,trim={0 0.3cm 0 0},clip]{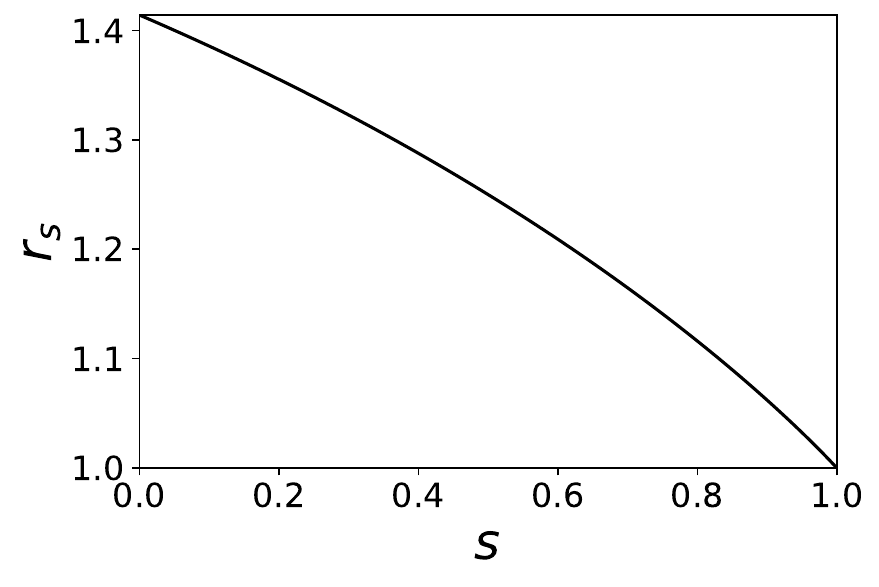}
    \caption{Left: $D_{k,n}(0)$ (see e.g. Eq. \eqref{LRgapcorr_intro}) measures the covariance between the red gap and the blue gap, where the $x_i$ are the positions of the particles in the Riesz gas. For $r=k/n<1$ the two intervals are non overlapping, while for $r>1$ the two interval overlap.
    Right: Plot of the value $r_s$ for which $D_{k,n}(0)$ changes sign, versus the Riesz interaction exponent $s$.}
    \label{gaps_sketch}
\end{figure}

We also studied the covariance of two gaps of sizes $k$, shifted by $n$, see Fig.~\ref{gaps_sketch}. Both variables $k$ and $n$ are assumed
to be positive integers.
In the long-range case $0<s<1$ we find that for
$k,n,|n-k| \gg 1$ this covariance behaves as
\be \label{LRgapcorr_intro}
D_{k,n}(0) = \langle (\delta x_{i}-\delta x_{i+k}) (\delta x_{i+n}-\delta x_{i+n+k}) \rangle \simeq \frac{T}{g\rho^{s+2}a_s} \frac{\pi^{s+1}}{2^{s}} \frac{|n-k|^s + (n+k)^s -2n^s}{\sin(\frac{\pi s}{2})\Gamma(1+s)} \;,
\ee
where $a_s$ is given in \eqref{fasympt_intro}.
It thus exhibits an algebraic roughness at large separations. 
Interestingly, for $k \leq n$
this correlation is negative, 
i.e. two gaps which are disjoint are always anti-correlated. This can be expected since 
if a given gap expands, the surrounding gaps are more likely to be reduced. 
The formula \eqref{LRgapcorr_intro} exhibits however a more subtle feature.
Let us denote $r=k/n$, such that the "overlap ratio" is $\frac{k-n}{k} = (r-1)/r$. We find that 
there is a transition to positive correlation when the two gaps overlap sufficiently. It occurs at a 
critical value of $k/n=r_s>1$, which is defined as the root of the equation $(r-1)^s+(r+1)^s=2$
and is plotted in Fig.~\ref{gaps_sketch}. 

To compare with the rigorous bounds for the gap correlations obtained in \cite{BoursierCorrelations}, let us 
now consider the regime where $k$ remains fixed and $n$ becomes large. In that regime we find
\be \label{Dkn_large_n}
D_{k,n}(0) \simeq -\frac{T}{g\rho^{s+2}a_s} \frac{\pi^{s+1}}{2^{s}} \frac{s(1-s) \, k^2}{n^{2-s}\sin(\frac{\pi s}{2})\Gamma(1+s)} \;.
\ee 
Note that the dependence $\sim n^{2-s}$ is consistent with the bounds obtained in \cite{BoursierCorrelations},
however the prefactor and its $k$ dependence were not obtained there. 
In the case of the 1d log-gas $s=0$, the $1/n^2$ decay of these correlations was rigorously proved in 
\cite{ErdosCorrelations}. Here by taking the limit $s \to 0$ in \eqref{Dkn_large_n}, one obtains a decay 
$D_{k,n}(0) \simeq -\frac{T k^2}{\pi^2 g\rho^2 n^2}$ in agreement 
with the $1/n^2$ decay, and with a negative amplitude. In addition, again for $s=0$, we find for $k,n,|n-k| \gg 1$, all three of the same order,
\be \label{gapslog}
D_{k,n}(0) \simeq \frac{T}{\pi^2 g\rho^2} \ln |1-\frac{k^2}{n^2}| \;.
 \ee
which again gives $D_{k,n}(0) \simeq -\frac{T k^2}{\pi^2 g\rho^2 n^2}$ for $n/k \gg 1$.
Note that 
the covariance is negative for $r=\frac{k}{n} <\sqrt{2}$ and positive for $r= \frac{k}{n} >\sqrt{2}$
 in agreement with the limit $r_{s=0^+}=\sqrt{2}$. The divergence in \eqref{gapslog} for $k/n=1$
 is only apparent, since for $|n-k| = O(1)$ the formula does not hold and there is another regime 
 which is detailed in Section \ref{sec:gapstat}.

In the short-range case $s>1$, the correlations are important only when the intervals overlap and one finds for $k,n$ 
large with $r=k/n$ fixed
\be
\frac{1}{n} D_{k,n}(0) 
\simeq  \begin{cases} 0 \hspace*{1.95cm} \quad \text{if } r\leq 1 \;, \\ \frac{4 \pi^2 T}{g\rho^{s+2}a_s} (r-1)  \quad \text{if } r>1 \;. \end{cases}
\ee
A more precise estimate of the residual correlations for $r<1$ is obtained below in \eqref{gapsSR}.
\\

\subsubsection{Dynamical correlations} 

Next we have computed dynamical correlations. First we have studied the time evolution of the displacement of a given particle,
more precisely we have obtained the variance of this quantity. Let us consider the case $s>0$. At short time
$t \ll \tau=g \rho^{s+2}$, the displacement is simply free diffusion with diffusion coefficient $T$. 
At very large time $t \gg N^{z_s} \tau$, it converges to its equilibrium limit $2\langle \delta x_i^2 \rangle$, given 
in \eqref{var_Riesz_liquid_intro} which behaves as $\propto N^{z_s-1}$. For large $N$ there is thus
a very broad intermediate time regime $\tau \ll t \ll N^{z_s} \tau$. In this regime we obtain
\be \label{displacement_Riesz_intro}
C_0(t) = \langle (\delta x_i(t) - \delta x_i(0))^2 \rangle \simeq \begin{dcases}
\frac{2T}{\pi^2 g\rho^2} \left( \ln \big(2\pi^2 g\rho^2 t\big) + \gamma_E - \frac{1}{\pi^2g\rho^2 t} \right) + O\left(\frac{1}{t^2}\right)\;, \quad \text{for } s=0 \;, \\
U_s \frac{T \, t^{\frac{s}{s+1}}}{g^{\frac{1}{s+1}}\rho^{\frac{s+2}{s+1}}} \;, \hspace*{5.65cm}
\quad \text{for } 0<s<1 \;, \\ 
2T \sqrt{\frac{t}{\pi g\rho^3 \ln(g\rho^3 t)}} \;, \hspace*{4.65cm} \quad \text{for } s=1 \;, \\ 
\frac{2T}{\sqrt{\pi (s+1) \zeta(s)}} \sqrt{\frac{t}{g\rho^{s+2}}}\;,  \hspace*{3.9cm}
\quad \text{for } s>1 \;,\end{dcases}
\ee
with
\be \label{Us_val_intro}
U_s = \frac{ 4\Gamma\left(\frac{1}{s+1}\right)}{\pi s} \left[ \frac{\Gamma\left(1+\frac{s}{2}\right)}{2\sqrt{\pi} \, \Gamma\left( \frac{1-s}{2} \right)} \right]^{\frac{1}{s+1}} \;.
\ee
This corresponds to an "annealed" initial condition, i.e., at $t=0$ the system is at Gibbs equilibrium. 
Note that in the short-range case $s>1$, we find the $\sqrt{t}$ behaviour of single-file diffusion,
while in the long range case $0<s<1$ one finds a continuously varying exponent.
Another setting which is often also 
considered is the "quenched" initial condition, which in the present case is realized by choosing $\delta x_i(0)=0$. 
In the quenched case we find that all formulas above for $s>0$ also hold, up to an additional multiplicative factor $2^{-\frac{1}{s+1}}$ for $0<s<1$, and $2^{-\frac{1}{2}}$ for $s\geq 1$. In the case $s=0$ the above formula holds upon an additional multiplicative factor $1/2$ together with the replacement $t \to 2 t$. 

Remarkably, our results for both annealed and quenched 
exactly coincide including the exact prefactors (at any $T$ for $0<s<1$ and at $T\ll g\rho^s$ for $s>1$) with formulas (2) and (3) obtained in the recent work \cite{DFRiesz23}.
In that work the authors used a completely different method based on macroscopic fluctuation theory (MFT), in terms
of the density field. It is quite intriguing that the present approach in terms of displacements, and which is a priori restricted to small temperatures, allows to obtain the same asymptotics (in particular in the long-range case). 
We want to stress that in all cases our method allows to obtain predictions beyond the asymptotics (such as the whole
crossover regimes at small and large times) at sufficiently low temperature. Note that the cases $s=0$ and $s=1$ were not treated in details in \cite{DFRiesz23}.
In the case of the log-gas we can compare our results with previous works \cite{Spohn1,Spohn2}. We find
that our leading order agrees with the exact result, including the exact prefactor, see for instance Eq. (1.4) in \cite{Spohn1} which corresponds to $g=1$ and $T=1/2$. Note that the corrections are power law in time. The hypergeometric function in \eqref{crossoverhypergeo} describes the full crossover between the
short time regime ($t \ll \tau$) and the large time regime ($t \gg \tau$). 

Finally let us mention the case $s<0$. In that case the variance of the displacement of a particle during time $t$ converges to 
a finite value on a timescale $\tau$, given by $2 \langle \delta x_i^2\rangle $ see Eq. \eqref{var_Riesz_solid_intro}. 
We show below that it reaches this asymptotic value with power law corrections in time
$O((t/\tau)^{-\frac{|s|}{s+1}})$, see \eqref{decroissance} for details. 
\\


Because of the long range interactions in space, the motion of a particle also exhibits long range correlations in time.
To characterize these correlations we have computed the two-time correlations of the displacement of a particle.
In the annealed case, i.e. at equilibrium, one can show the relation
\be \label{relC0twotime_intro}
C_0(t_1,t_2) = \langle (\delta x_i(t_1) - \delta x_i(0)) (\delta x_i(t_2) - \delta x_i(0)) \rangle =  \frac{1}{2} [ C_0(t_1) + C_0(t_2) - C_0(|t_1-t_2|) ] \;,
\ee 
where $C_0(t)$ is given in Eq. (\ref{displacement_Riesz_intro}). In particular, it implies that at small times $t_1,t_2,|t_1-t_2|\ll \tau = 1/(g \rho^{s+2})$ we find the free diffusion result $C_0(t_1,t_2)=2 T \min(t_1,t_2)$, while at large times $t_1,t_2,|t_1-t_2|\gg \tau$, one has
\be \label{multitime_corr_Riesz}
C_0(t_1,t_2) \simeq \begin{dcases} \frac{U_s T}{2g^{\frac{1}{s+1}}\rho^{\frac{s+2}{s+1}}} (t_1^{\frac{s}{s+1}} + t_2^{\frac{s}{s+1}} - |t_1-t_2|^{\frac{s}{s+1}})\;,
\quad \hspace{1.2cm} \text{for } 0<s<1 \;, \\ 
\frac{T}{\sqrt{\pi (s+1) \zeta(s)g\rho^{s+2}}} (\sqrt{t_1} + \sqrt{t_2} - \sqrt{|t_1-t_2|})\;,
\quad \text{for } s>1 \;. \end{dcases}
\ee
This correlation function coincides with the one of a fractional Brownian motion (fBm) of Hurst index $H=s/(2 (s+1))$ in the long-range case and $H=1/4$ in the short-range case. It also coincides with the recent result of Ref. \cite{DFRiesz23}, obtained using MFT in the annealed case and an additional approximation
which relates the two time current fluctuation to the tracer position fluctuations. For $s=1$ there are additional
logarithmic corrections, see Section \ref{sec:multitime}. 
In the case of the log-gas $s=0$, i.e. the Dyson Brownian motion on the circle, we find
\be  \label{res_C0}
C_0(t_1,t_2) \simeq \frac{T}{\pi^2 g \rho^2} \left( \log \big( \frac{2 \pi^2 g \rho^2 t_1 t_2}{|t_1-t_2|} \big) +\gamma_E \right) \;,
\ee 
which is valid for all three times $t_1,t_2,|t_1-t_2|$ large and of the same order, while 
the regime of closer times $|t_1-t_2|=O(1)$ is discussed in Section \ref{sec:multitime}.
\\

We have also characterized the correlations (with annealed initial conditions) between the total displacement during time $t$ of two particles separated by $k$ lattice spacings. 
We recall that here and below the variable $k$ is assumed a positive integer.
We have shown that for $s>0$ their correlation function $C_k(t)$ takes the following scaling form at large $t$ and $k$, with $k \sim (t/\tau)^{1/z_s}$,
where $\tau=1/(g \rho^{s+2})$,
\be \label{Ck_scaling_intro}
C_k(t) = \langle (\delta x_i(t) - \delta x_i(0)) (\delta x_{i+k}(t) - \delta x_{i+k}(0))   \rangle \simeq \begin{dcases} T \left(\frac{t^s}{g\rho^{s+2}}\right)^\frac{1}{s+1} F_s\left( \frac{k}{(g\rho^{s+2}t)^{\frac{1}{s+1}}} \right)\;, \quad \text{for } 0<s<1 \;, \\
T \sqrt{\frac{t}{g\rho^{s+2}}} \, F_s\left( \frac{k}{\sqrt{g\rho^{s+2}t}} \right)\;, \quad \hspace{1.2cm} \text{for } s>1 \;. \end{dcases}
\ee 
The scaling function $F_s(x)$ for $0<s<1$ is given in \eqref{Fs_def1}. For small argument $x \sim k/(t/\tau)^{1/z_s} \ll 1 $ it 
behaves as $F_s(x) \simeq  U_s - \frac{1}{\pi \tan( \frac{\pi s}{2}) } |x|^s + \dots$
and matches the
result for $k=0$ given above in \eqref{displacement_Riesz_intro}. In addition this implies 
$C_0(t) - C_k(t) \simeq D_k(0) \sim k^s$ where we recall that $D_k(0)$ is the variance of the gaps given in
\eqref{gapscases_intro} (see below for the relation between these observables). For large argument $x \gg 1$
it decays as a power law $F_s(x) \sim 1/|x|^{2+s}$. Hence we find that the
correlation function decays as $C_k(t) \sim t^2/k^{2+s}$ in the regime of large separations $k/(t/\tau)^{1/z_s} \gg 1$.
This is quite remarkable since it correspond to a ballistic behavior in time. The rationale
may be that at large distance only fast propagating excitations (i.e. ballistic) survive. 

In the short range case $s>1$,
the scaling functions $F_s(x)$ for different $s$ are identical, up to an $s$ dependent rescaling, i.e.,  
\be \label{Fs_intro} 
F_s(x) = \frac{1}{\sqrt{(s+1)\zeta(s)}} F_1\left( \frac{x}{\sqrt{(s+1) \zeta(s)} } \right) \quad \text{with} \quad
F_1(x) = \frac{2}{\sqrt{\pi}} e^{-\frac{x^2}{4}} + |x| \left({\rm erf}\big(\frac{|x|}{2}\big)-1\right) \;,
\ee
where $F_1(0)=2/\sqrt{\pi}$ and $F_1(x)$ decays exponentially fast as 
$F_1(x) \simeq \frac{4}{\sqrt{\pi}} \frac{e^{-\frac{x^2}{4}}}{x^2}$ at large argument. 
For $s \to 1$ the scaling factor diverges and the correct scaling variable is $k/ \sqrt{t \log t}$,
see Section \ref{sec:spacetime_brownian} for details. 

In the case of the log-gas $s=0$ one finds 
\be \label{Ck_s0_intro}
C_k(t) \simeq 
\frac{T}{g\rho^2} F_0\left(\frac{k}{g\rho^2 t}\right) \; \quad , \quad F_0(x) = \frac{1}{\pi^2} \ln\left(1 + \frac{\pi^2}{x^2}\right) \;,
\ee
where $F_0(x)$ coincides with the limit $s \to 0$ of the scaling function $F_s(x)$ given in \eqref{Fs_def1}.
This formula is valid in the regime of large $k$ and $t$ with $k/(g\rho^2 t)$ fixed. The matching
to the above result for $k=0$ given in \eqref{displacement_Riesz_intro} in the limit $k/(g\rho^2 t) \ll 1$ requires a separate discussion 
which is detailed in
Section \ref{sec:spacetime_brownian}. Finally the case $s<0$ is also discussed in Section \ref{sec:spacetime_brownian}. 
\\

As we have discussed above the variance of the gaps grows as a power law as a function of the separation $k$ [see Eq.~(\ref{gapscases_intro})], which allows to define
an equilibrium roughness exponent. It is also interesting to investigate how the gaps are correlated in time (we consider again the annealed initial condition).
We show that their correlation function $D_k(t)$ 
takes the following scaling form at large $t$ and $k$, with $k \sim (t/\tau)^{1/z_s}$,
\be \label{Dk_scaling_intro}
D_k(t) = \langle (\delta x_i(t) - \delta x_{i+k}(t)) (\delta x_{i}(0) - \delta x_{i+k}(0))   \rangle 
\simeq \begin{dcases} T \left(\frac{t^s}{g\rho^{s+2}}\right)^\frac{1}{s+1} G_s\left( \frac{k}{(g\rho^{s+2}t)^{\frac{1}{s+1}}} \right)\;, \quad \text{for } 0<s<1 \;, \\
T \sqrt{\frac{t}{g\rho^{s+2}}} \, G_s\left( \frac{k}{\sqrt{g\rho^{s+2}t}} \right)\;, \quad \hspace{1.2cm} \text{for } s>1 \;, \end{dcases}
\ee 
where the scaling function $G_s(x)$ is given in \eqref{Gs_def1}.
At small time these correlations should yield back the variance of the gaps $D_k(0)$ given in \eqref{gapscases_intro}. Indeed, we find that, at large argument, the scaling function $G_s(x)$ behaves as 
\be 
G_s(x) \simeq \begin{dcases} \frac{4}{sa_s} (2 \pi)^s \cos(\frac{\pi s}{2}) \Gamma(1-s) x^s \;, \quad \text{for } -1<s<1 \;, \\ \frac{x}{(s+1)\zeta(s)}\;, \quad \hspace{2.55cm} \text{for } s>1 \;, \end{dcases}
\ee 
where $a_s$ is given in \eqref{fasympt_intro}.
In the small argument limit $x\to 0$, one finds that the scaling function behaves as $G_s(x) \sim x^2$ for all $s>0$.
Hence at large time, of the order or larger than $k^{z_s}$, the gap correlation $D_k(t)$ decays in time as
\be \label{Dk_decay_intro}
D_k(t) \sim k^2 t^{-\frac{2-s}{1+s}} \quad \text{for} \ s<1 \quad , \quad D_k(t) \sim k^2 t^{-1/2} \quad \text{for} \ s>1 \;.
\ee 

In the case of the log-gas $s=0$ we find 
\be
D_k(t) \simeq \frac{T}{g\rho^2} G_0 \left( \frac{k}{g\rho^2 t} \right) \quad , \quad G_0(x) = \frac{1}{\pi^2} \ln(1+\frac{x^2}{\pi^2}) \;.
\ee
Interestingly this scaling function is related to the one for the time displacements via $G_0(x)=F_0(1/x)$.
This may be a consequence of the "relativistic" invariance of the log-gas with dynamical exponent $z_0=1$. 
\\

Finally, note that there is an exact relation between the two observables $C_k(t)$ and $D_k(t)$, which holds for annealed initial condition and reads (see Section \ref{sec:observables}) 
\be \label{relationCkDk} 
C_0(t) - C_k(t) = D_k(0) - D_k(t) \;.
\ee 
This implies in particular that at large time and fixed separation $C_0(t) - C_k(t) \simeq D_k(0)$,
and that at large separation and fixed time $D_k(0) - D_k(t) \simeq C_0(t)$. This also implies
relations between the scaling functions $F_s(x)$ and $G_s(x)$.

\subsubsection{Macroscopic observables}

Until now we have focused on observables at microscopic scales. It is also interesting to study the static and dynamic fluctuations at 
the scale of the full circle. 
\\

Let us start with the covariance at equilibrium of the displacements of two particles separated
by $k = \kappa N$ with $\kappa = O(1)$. We find that for $0<s<1$ it takes the scaling form, for $0<\kappa < 1$,
\be \label{cov_kappa_longrange}
\langle \delta x_i \delta x_{i+k}\rangle \simeq \frac{\Gamma(1+\frac{s}{2})}{\pi^{s+\frac{3}{2}} \Gamma(\frac{1-s}{2})} \frac{T N^s}{g\rho^{s+2}}  \sum_{q=1}^{\infty} \frac{\cos(2\pi \kappa q )}{q^{s+1}} \quad , \quad 
\kappa=\frac{k}{N} \;.
\ee
It has a finite limit for $\kappa \to 0$ which matches the result for the on-site variance $k=0$ given in the second line of \eqref{var_Riesz_liquid_intro}. 
In the short-range case $s>1$ one finds
\be \label{cov_shortrange_intro}
\langle \delta x_i \delta x_{i+k}\rangle \simeq \frac{NT}{12(s+1)\zeta(s) g\rho^{s+2}} (1 - 6\kappa(1-\kappa)) \quad , \quad 
\kappa=\frac{k}{N} \;,
\ee
which again matches for $\kappa \to 0$ the result in the fourth line of \eqref{var_Riesz_liquid_intro}. 
Note that the covariance decreases with $\kappa$ until it reaches a negative minimum at $\kappa=1/2$.
For $s=1$ the result is given in \eqref{covar_s1}. The behavior near $\kappa=0$, which is non-analytic,
i.e. $O(\kappa^s)$ [see Eq. \eqref{matchvariance}] in the long range case, and $O(\kappa)$ in the short range case, can
be shown to match the large $k$ behavior of the gaps, see Section \ref{sec:var_brownian}. 

In the case of the log-gas $s=0$ we find for $0<\kappa<1$
\be
\langle \delta x_i \delta x_{i+k}\rangle \simeq \frac{T}{\pi^2 g\rho^2}  \sum_{q=1}^{\infty} \frac{\cos(2\pi \kappa q )}{q} = - \frac{T}{\pi^2 g\rho^2} \ln (2\sin(\pi \kappa)) \;.
\ee
This formula can be compared to the variance of the number of eigenvalues for CUE($\beta$) in a mesoscopic interval 
on the circle \cite{HughesCircle2001,NajnudelCircle2018}.  In the limit $\kappa \to 0$ one recovers 
the $\log N$ divergence of the variance in the first line of \eqref{var_Riesz_liquid_intro}. The precise matching is discussed in 
Section \ref{sec:var_brownian}. 
\\

Another way to characterize the space time correlations is to study the linear statistics. Consider a function $f(x)= \sum_{n \in \mathbb{Z}} \hat f_n e^{-2 i \pi \frac{n}{L} x}$ on the circle, periodic of period $L$,
with Fourier coefficients $\hat f_n$. These coefficients are assumed to decay sufficiently fast which corresponds
to $f(x)$ varying at the scale of the circle.
One defines the linear statistics
${\cal L}_N(t) = \sum_{i=1}^N f(x_i(t))$. We find that in the large $N$ limit, and for annealed initial condition, the covariance takes the form
\be
\langle {\cal L}_N(t) {\cal L}_N(t') \rangle_c \simeq \begin{dcases}  4\pi^{\frac{1}{2}-s} \frac{\Gamma(1+\frac{s}{2})}{\Gamma(\frac{1-s}{2})} \frac{ T N^s}{g\rho^s} \sum_{q=1}^{\infty} e^{-a_s q^{1+s} |\tilde t - \tilde t'|}  q^{1-s} | \hat f_q |^2 \;, \quad \text{for } -1<s<1 \;, \\
\frac{2 T N}{(s+1) \zeta(s) g\rho^s} \sum_{q=1}^{\infty} e^{- a_s q^2 |\tilde t - \tilde t'|} | \hat f_q |^2\;, \quad \hspace{1.95cm} \text{for } s>1 \;, \end{dcases}
\ee
where the times are expressed in units of the relaxation time at the scale of the full circle, i.e. one defines $t = N^{z_s} \tau \tilde t$
with $\tau= 1/g \rho^{s+2}$. In the case of the log-gas $s=0$, using the correspondence 
$\beta=g/T$ this formula recovers exactly the one proved in \cite{Spohn3} for the Dyson Brownian
motion on the circle. Here we obtain the case of general $s$. For $0<s<1$ and at equal time $t=t'$ it agrees 
with a recent result in the math literature \cite{BoursierCLT} (see Section \ref{sec:linear_stat} for details). 

\subsection{Run-and-tumble particles} \label{subsec:MainResultsRTP}

For the run-and-tumble particles, described by the equation of motion \eqref{Eq_def_intro}, 
there are two important 
parameters. The first is the "effective temperature" $T_{\rm eff}$ which measures the amplitude of
the active noise, and the second is the ratio of the persistence time $1/\gamma$ to the local interaction time $\tau=1/(g \rho^{s+2})$,
\be 
T_{\rm eff}= \frac{v_0^2}{2 \gamma} \quad , \quad \hat{g}= \frac{g\rho^{s+2}}{2\gamma} = \frac{1}{2 \gamma \tau} \;.
\ee 
There are two special limiting cases, for large and small $\gamma$ respectively
: (i) the diffusive limit which corresponds to $\hat g \ll 1$ at fixed $T_{\rm eff}$, 
(ii) the  limit $\gamma \to 0$ with all the other parameters fixed, 
which corresponds to the simultaneous limit $T_{\rm eff} \sim \hat g \to +\infty$. This limit belongs to the family
of models which include the Jepsen gas \cite{jepsen,satya}. 

Under the time evolution \eqref{Eq_def_intro} the RTP system reaches a stationary state which is non-Gibbsian, 
and where the displacements are a priori non-Gaussian. We have obtained a formula for the two-point two-time correlations 
of the displacements of the particles, in this stationary state, for any $N$, see \eqref{cov_rtp}. 
We now present the results that we have obtained in the limit of large $N$ and large $L$ with fixed average density $\rho=N/L$.

All the results presented in this paper for run-and-tumble particle only rely on the fact that the two-time correlation of the driving noise takes an exponential form. They are thus also valid for other models of active particles, in particular active Ornstein Uhlenbeck particles (AOUPs) and active Brownian particles (ABPs), up to some
simple correspondence in the model parameters (see Appendix~\ref{app:other_active_models} for a mapping to these models).

\subsubsection{Equal time correlations in the stationary state}

We start with equal time correlations. For $s>0$, we find that the variance of the displacement of a given particle
is given by 
\be \label{var_RTP2_intro}
\langle \delta x_i^2 \rangle \simeq \frac{2 T_{\rm eff} N^{z_s-1}}{g\rho^{s+2} a_s} \sum_{q=1}^{+\infty} \frac{1}{q^{z_s} (1+ \hat g (\frac{q}{N})^{z_s})} \;
\quad , \quad z_s= \min(1+s,2) \;.
\ee
If $\hat g \ll N^{z_s}$, the last term in the denominator can be neglected, so that at leading order we recover that $\langle \delta x_i^2 \rangle$ is given by the Brownian result \eqref{var_Riesz_liquid_intro} with an effective temperature $T \to T_{\rm eff}=\frac{v_0^2}{2\gamma}$. In 
the case $s=0$ describing the active DBM on the circle, the large $N$ behavior of the variance has again the same
logarithmic behavior as for the Brownian particles, up to an additive constant $c(\hat g)$
characterizing the active noise, which is given in \eqref{var_log_rtp}, i.e. 
one finds $\langle \delta x_i^2 \rangle \simeq \frac{T_{\rm eff}}{\pi^2 g\rho^2} (\ln N + c(\hat g) )$.

For $s<0$, as in the Brownian case, we find that there is translational order at low enough
effective temperature, and expect a melting 
transition when 
\be  \label{TmeltingRTP_intro}
T_{\rm eff} = T_{\rm eff,M} = \frac{1}{2} g\rho^{s} \frac{c_L^2}{\int_0^{1/2} \frac{du}{f_s(u)\left( 1 + \hat g f_s(u) \right)}} \;,
\ee 
where $c_L$ is a phenomenological Lindemann constant. 
In the limit $s \to 0^-$ one finds that $T_{\rm eff,M}$ vanishes linearly in $s$, as in the Brownian case, i.e. $ T_{\rm eff,M} \sim \pi^2 c_L^2 g |s|$.
\\

Next we have computed the variance of the gaps between two particles separated by $k$ lattice spacings, with $k \ll N$. 
The most interesting situation occurs when $\hat g \gg 1 $. In that case a change of behavior occurs at a characteristic scale for $k \sim \hat g^{1/z_s} \gg 1 $, and the variance takes the scaling form for $s>0$
\be \label{scalinggaprtp}
D_k(0) = \langle (\delta x_{i+k}-\delta x_i)^2 \rangle \simeq \frac{ T_{\rm eff} k^{z_s-1}}{g \rho^{s+2}}   \, {\sf G}_s( k/\hat g^{\frac{1}{z_s}}) \;,
\ee 
where the scaling function ${\sf G}_s({\sf x})$ is given in \eqref{eqFgap}. 
At large distances ${\sf x} = k/\hat g^{1/z_s} \gg 1$, ${\sf G}_s({\sf x})$ converges to a constant ${\sf G}_s(+\infty)$, 
such that one recovers the Brownian result \eqref{gapscases_intro}
with an effective temperature $T_{\rm eff}$. 
However, on smaller scales, i.e. for ${\sf x} = k / \hat g^{1/z_s} \ll 1$, the activity plays an important role. 
Depending on $s$, there are three cases
to distinguish, namely
\bea
{\sf G}_s({\sf x}) \underset{{\sf x}\ll 1}{\propto} \begin{cases} {\sf x}^{1+s} \quad \text{for } 0 \leq s < 1/2 \\  {\sf x}^{2-s} \quad \text{for } 1/2 \leq s < 1 \\  {\sf x} \quad \quad \ \, \text{for } s>1 \;. \end{cases}
\eea

This leads to various regimes for the dependence in $k$ of the variance, for $1 \ll k  \ll \hat g^{1/z_s}$, summarized in Table \ref{table:gap_var_rtp}. Note that such an intermediate regime also exists for $-1/2<s<0$. The gaps in the case $s=0$ (log-gas also called 
active DBM) are discussed in detail in Section \ref{sec:Dk0_rtp}.

\begin{table}
\begin{center}
\begin{tabular}{|c|c|c|c|c|c|c|}
\hline
 & $-1<s<-1/2$ & $-1/2<s<0$ & $s=0$ & $0<s<1/2$ & $1/2<s<1$ & $s>1$ \\
\hline
$k \ll \hat g^{\frac{1}{z_s}}$ & $\sim cst$ & $\propto k^{1+2s}$ & $\propto k$ & $\propto k^{1+2s}$ & $\propto k^2$ & $\propto k^2$ \\  
\hline
$k \gg \hat g^{\frac{1}{z_s}}$ & $\sim cst$ & $\sim cst$ & $\propto \ln k$ & $\propto k^s$ & $\propto k^s$ & $\propto k$ \\
\hline
\end{tabular}
\end{center}
\caption{Different regimes for the variance of the gaps $D_k(0)$ for the RTP Riesz gas, as a function of the interparticle distance $k$ and the parameter $s$ of the interaction.}
\label{table:gap_var_rtp}
\end{table}


\subsubsection{Dynamical correlations: annealed initial conditions} 

\begin{figure}
    \centering
    \includegraphics[width=0.5\linewidth,trim={0 2cm 0 2cm},clip]{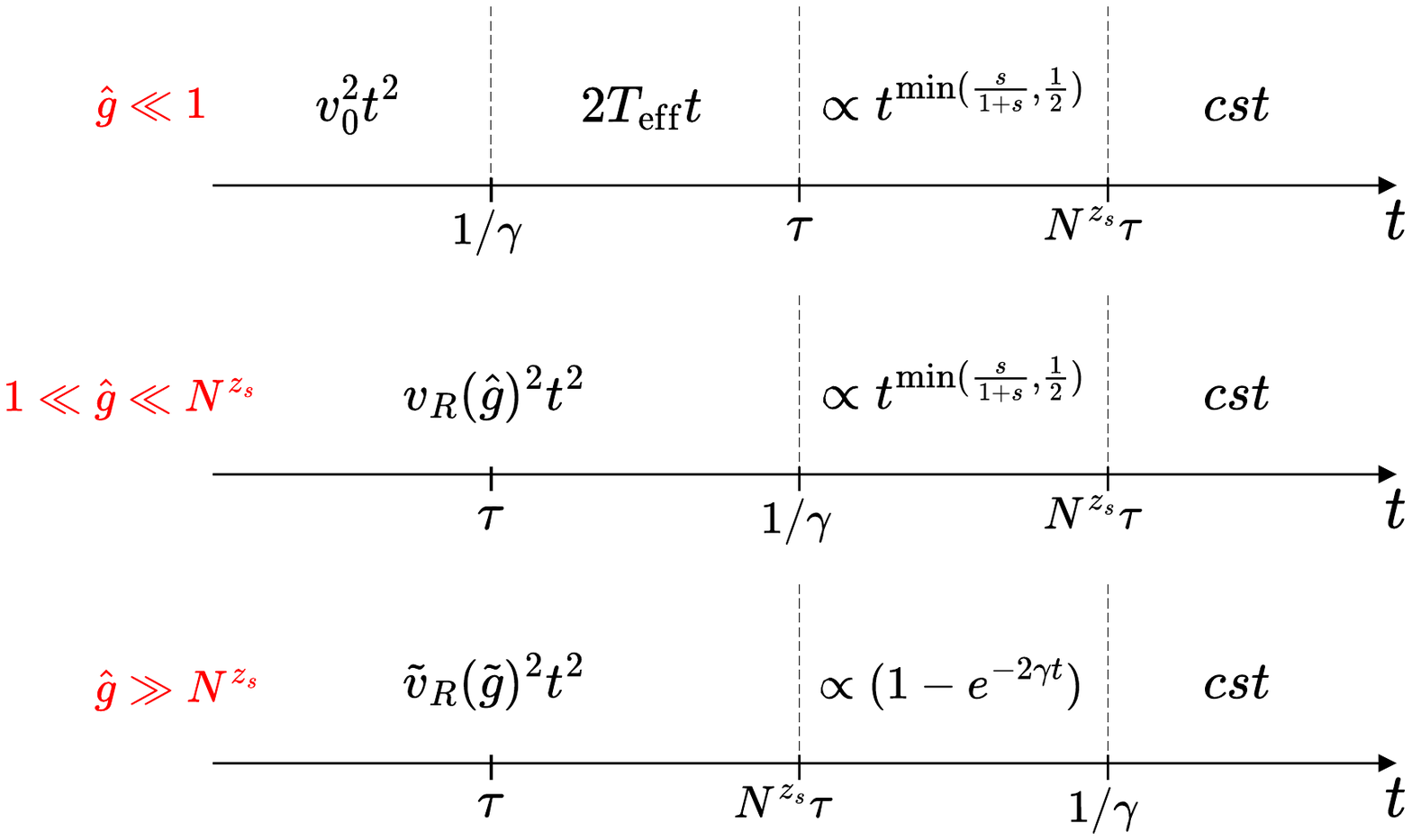}
    \caption{Sketch of the three different cases for the time evolution of the variance $C_0(t)$ of the displacement during time $t$ of a given particle for the RTP Riesz gas, depending on the ordering of the three important relaxation time scales $1/\gamma$, $\mu_{(N-1)/2} \sim \tau=1/(g \rho^{s+2})$ and $\mu_{1} \sim \tau N^{z_s}$ as the parameter $\hat g=1/(2\gamma \tau)$ increases. In the first case $\hat g \ll 1$ there is an intermediate free diffusion regime between the ballistic and the anomalous diffusion regime as time increases. In the second case $1 \ll \hat g \ll N^{z_s}$ there is a direct crossover from the ballistic to the anomalous diffusion regime.
The third case corresponds to very small $\gamma$ ($1/\gamma \gg N^{z_s} \tau $) and is discussed later in Section \ref{mainres_smallgamma}. 
In all three cases "{\it cst}" denotes the large time saturation value
    $2 \langle \delta x_i^2 \rangle$. }
    \label{fig:time_regimes_rtp}
\end{figure}

Next we have studied the variance $C_0(t)=\langle (\delta x_i(t) - \delta x_i(0))^2 \rangle$ 
of the displacement of a given particle
during time $t$, starting in the stationary state. 
There are three main situations depending on the relative values of the 
characteristic time scales $1/\gamma$, $\tau$ and $N^{z_s}\tau$, as 
illustrated in Fig. \ref{fig:time_regimes_rtp}. We focus 
on the first two situations where $\gamma$ is independent of $N$. 
In that case we have obtained a formula for the variance of the displacement during time $t$ in the
large $N$ limit, given in \eqref{disp_RTP_largeN}. From this formula we
have obtained the different time regimes. Here we summarize the results, more details
are given in Section \ref{sec:C0(t)}, see also Fig. \ref{fig:diagrameC0} for a diagram of the different regimes.

\begin{figure}[t]
    \centering
    \includegraphics[width=0.32\linewidth]{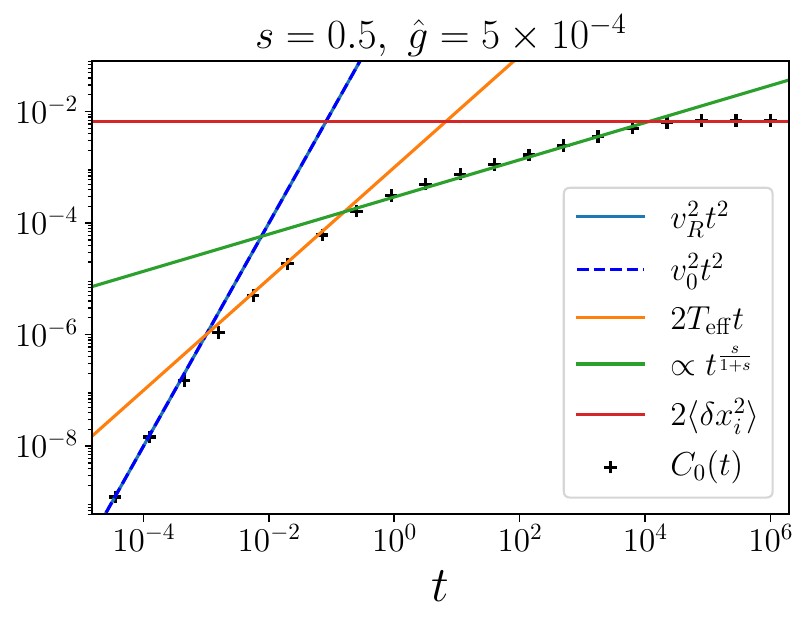}
    \includegraphics[width=0.32\linewidth]{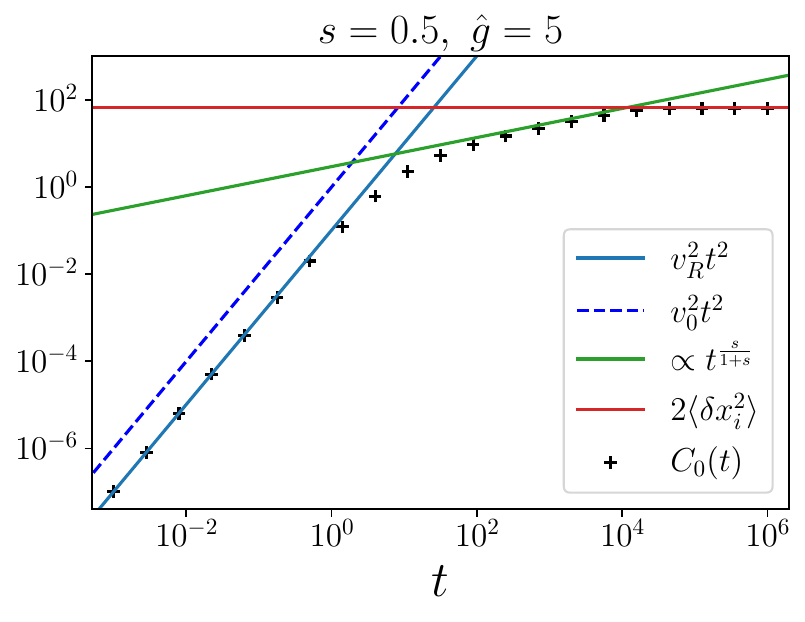}
    \includegraphics[width=0.32\linewidth]{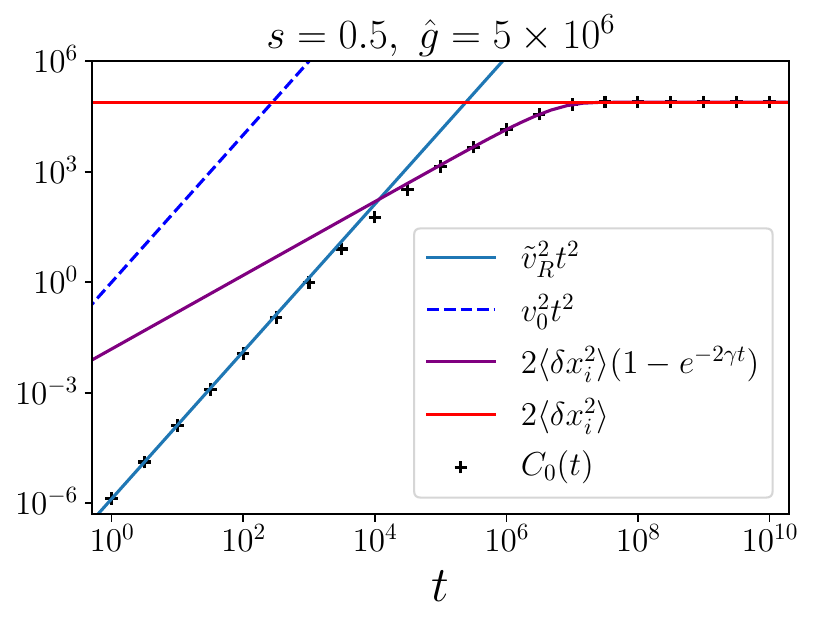}
    \caption{Left: Variance of the displacement $C_0(t)$ as a function of time, obtained by numerical computation of the sum \eqref{disp_RTP} for a RTP Riesz gas with $s=0.5$, $N=10001$, $g=1$, $\rho=1$, $v_0=1$ and $\gamma=1000$. The 4 time regimes are clearly visible. In the ballistic regime, the renormalized velocity $v_R(\hat g)$ is indistinguishable from the non-renormalized one $v_0$. Center: Same plot for $\gamma=0.1$. There is no free diffusion regime in this case. In the ballistic regime, there is a clear difference between $v_R(\hat g)$ and $v_0$. 
    Right: Same plot for $\gamma = N^{-z_s} \tilde \gamma = 10^{-7}$. In this case, both the free diffusion regime and the anomalous diffusion regime are absent. Instead, $C_0(t)$ converges exponentially to its large time limit after the ballistic regime.}
    \label{fig:disp_rtp}
\end{figure}

The first situation is when $1/\gamma \ll \tau= 1/(g \rho^{s+2})$, i.e. $\hat g \ll 1$ (first line in Fig. \ref{fig:time_regimes_rtp}). As time increases there is first 
a ballistic regime for $t \ll 1/\gamma$, with
$C_0(t) \simeq v_0^2 t^2$. At larger time $1/\gamma \ll t \ll \tau$ there is a crossover 
to a free diffusion regime $C_0(t) \simeq 2T_{\rm eff} t$. Finally, as time increases
further $\tau \ll t \ll N^{z_s} \tau$, the interactions become important 
and for $s>0$, $C_0(t)$ crosses over to the same anomalous single-file diffusion behavior as in the Brownian case.
The formulas are exactly the same as \eqref{displacement_Riesz_intro}, upon the replacement $T \to T_{\rm eff}$. 
This behavior is plotted in Fig. \ref{fig:disp_rtp} (left panel)
where the various crossovers are visible. 

When $\hat g$ becomes of order unity, i.e when the two time scales $1/\gamma$ and $\tau$ are
comparable, there is still a ballistic
regime at short time $t \ll 1/\gamma$, with however a velocity $v_R=v_R(\hat g)<v_0$ which is renormalized by the interactions
\be \label{ballistic_intro}
C_0(t) \simeq v_R(\hat g)^2 t^2 \quad , \quad v_R(\hat g)^2 = v_0^2 \int_0^{1/2}
 \frac{2 \, du}{1 + \hat g f_s(u)}  \;.
\ee 
The dependence of the renormalized velocity $v_R(\hat g)$ in $\hat g$ is as follows. As $\hat g \to 0$ it recovers $v_0$. In the opposite limit $\hat g \gg 1$ it behaves as
\bea
\frac{v_R(\hat g)^2}{v_0^2} \underset{\hat g \gg 1}{\propto} \begin{cases} \hat g^{-1} \quad \quad \text{for } -1 < s < 0 \;, \\  \hat g^{-\frac{1}{s+1}} \quad \text{for } 0 < s < 1 \;, \\  \hat g^{-1/2} \quad \; \text{for } s>1 \;. \end{cases}
\eea

When $\hat g$ becomes large, i.e. $\tau \gg 1/\gamma $, this is the second situation (second line in Fig. \ref{fig:time_regimes_rtp})
where the activity plays an important role. In that
case, for $s>0$, there is a direct crossover between the ballistic regime \eqref{ballistic_intro} for $t \ll 1/\gamma$
and the same anomalous single-file diffusion behavior as in the Brownian case for $1/\gamma \ll t \ll N^{z_s} \tau$.
This crossover is described
by a non-trivial scaling function 
\be \label{C0scaling-intro}
C_0(t) = T_{\rm eff} \tau \hat g^{1-\frac{1}{z_s}}  {\sf C}_s( \gamma t) \quad , \quad  {\sf C}_s(y) \propto \begin{cases} y^2 \quad \quad \,\,\, \text{ for } y \ll 1 \;, \\ y^{1-\frac{1}{z_s}} \quad \text{ for } y \gg 1 \;,  \end{cases}
\ee
where the scaling function ${\sf C}_s(y)$ is given in \eqref{C0scaling}. 
At large time
it recovers the anomalous diffusion behavior \eqref{displacement_Riesz_intro} (with $T$ replaced by $T_{\rm eff}$).
This crossover is apparent in Fig. \ref{fig:disp_rtp} (central panel). 

Finally, for $s>0$, in all cases shown in Fig. \ref{fig:time_regimes_rtp}, for much larger time $t = O(N^{z_s} \tau)$ (and $t\gg 1/\gamma$), 
$C_0(t)$ converges to its stationary value equal to $2\langle \delta x_i^2 \rangle$, which is itself equal to
its value for the Brownian case \eqref{var_Riesz_liquid_intro} with the replacement $T \to T_{\rm eff}$. 
For $s<0$ this convergence occurs on time scales of the order $\max(\tau, 1/\gamma)$. 

In summary, in all cases, $C_0(t)$ only differs from its Brownian counterpart (i) in the ballistic regime $t\ll 1/\gamma$ and (ii) within the crossover to free diffusion ($\hat g \ll 1$) or to anomalous diffusion ($\hat g \gg 1$) that follows, for $t \sim 1/\gamma$. 
The various behaviors of $C_0(t)$ obtained here are summarized in Table \ref{table:disp_rtp1}.



\begin{table}
\begin{center}
\begin{tabular}{|c|c|c|c|c|}
\hline
 & $t\ll 1/\gamma$ & $1/\gamma \ll t \ll \tau$ & $\tau \ll t \ll N^{z_s} \tau$ & $N^{z_s} \tau \ll t$ \\
\hline
$-1<s<1$ & \multirow{5}{*}{$\propto t^2$} & \multirow{5}{*}{$\propto t$} & $\sim cst$ & $\sim cst$ \\ 
\cline{1-1} \cline{4-5}
$s=0$ & & & $\propto \ln t$ & $\sim cst \times \ln N$ \\  
\cline{1-1} \cline{4-5}
$0<s<1$ & & & $\propto t^{\frac{s}{s+1}}$ & $\sim cst \times N^s$ \\
\cline{1-1} \cline{4-5}
$s=1$ & & & $\propto \sqrt{t/\ln t}$ & $\sim cst \times N/\ln N$ \\  
\cline{1-1} \cline{4-5}
$s>1$ & & & $\propto \sqrt{t}$ & $\sim cst \times N$ \\
\hline
\end{tabular}
\end{center}
\caption{Different time regimes for the variance $C_0(t)$ of the displacement of a given particle
during time $t$ for the RTP Riesz gas, as a function of the parameter $s$ of the interaction, when $1/\gamma \ll \tau$
($\hat g \ll 1$, first line in Fig.~\ref{fig:time_regimes_rtp}). If $1/\gamma \gg \tau$ ($\hat g \gg 1$, second line in Fig.~\ref{fig:time_regimes_rtp}), 
the displacement remains ballistic 
until $t\sim 1/\gamma$ (with however a renormalized velocity $v_R(\hat g)$), and the free diffusion regime is absent.
The last regime corresponds to a saturation to a stationary limit, which depends on $N$ for $s \geq 0$.
Apart from the ballistic regime, and the crossover away from the ballistic regime discussed in the text, 
see Eq.~\eqref{C0scaling-intro}, 
$C_0(t)$ behaves as for the Brownian particles with the replacement $T \to T_{\rm eff}$.}
\label{table:disp_rtp1}
\end{table}

\bigskip

We have also considered the two time correlation of the displacement $C_0(t_1,t_2)$. It can be simply obtained from $C_0(t)$ using 
the relation \eqref{relC0twotime_intro} 
which, one can show, is also valid for the RTP in the stationary state (it only requires time translation invariance).
From the above results for $C_0(t)$, when $t_1,t_2,|t_1-t_2|$ are of the same order, we obtain (i) a free diffusion regime 
for $1/\gamma \ll t_1,t_2,|t_1-t_2| \ll \tau$ (which exists for $\hat g \ll 1$)
with $C_0(t_1,t_2) \simeq 2 T_{\rm eff} \min(t_1,t_2)$ (ii) a large time regime for $t_1,t_2,|t_1-t_2| \gg \max(1/\gamma,\tau)$
where $C_0(t_1,t_2)$ has the same form as for a fractional Brownian motion (recalling however that in the RTP case
the displacements are not Gaussian). Finally in the ballistic regime, we obtain for $t_1,t_2,|t_1-t_2| \ll 1/\gamma$,
\be
C_0(t_1,t_2) \simeq v_R(\hat g)^2  t_1 t_2 \;,
\ee
with the same renormalized velocity $v_R$, {as given in Eq. (\ref{ballistic_intro})}. 
\\


Next, we have computed both the covariance $C_k(t)$ between the total displacement during time $t$ 
of two particles separated by $k$ lattice spacings [defined in Eq. (\ref{Ck_scaling_intro})], and the time correlation of the gaps $D_k(t)$ [defined in Eq. (\ref{Dk_scaling_intro})]. 
As in the Brownian case, see \eqref{relationCkDk}, these two quantities are connected through the 
relation $C_0(t)-C_k(t)=D_k(0)-D_k(t)$.
We find that when either $t/\tau \gg 1$, or $\hat g \gg 1$, or $k \gg 1$, 
$C_k(t)$ and $D_k(t)$ can each be written in two equivalent scaling forms
\be \label{Ck_scalingintro}
C_k(t) \simeq T_{\rm eff} \tau \hat g^{1-\frac{1}{z_s}}  \mathcal{\tilde F}_s( k/\hat g^{1/z_s} , \gamma t) 
= T_{\rm eff} \tau (t/\tau)^{1-\frac{1}{z_s}} \mathcal{F}_s \left(\frac{k}{(t/\tau)^{\frac{1}{z_s}}},\gamma t \right) \;,
\ee
where the scaling functions are given in \eqref{Ck_scaling1} and \eqref{Ck_scaling2}, while 
\be \label{Dk_scalingintro}
D_k(t) \simeq T_{\rm eff} \tau \hat g^{1-\frac{1}{z_s}}  \mathcal{\tilde G}_s \left( \frac{k}{\hat g^{1/z_s}} , \gamma t \right) = T_{\rm eff} \tau (t/\tau)^{1-\frac{1}{z_s}} \mathcal{G}_s \left(\frac{k}{(t/\tau)^{\frac{1}{z_s}}},\gamma t \right) \;,
\ee
where the scaling functions are given in \eqref{Dk_scaling1} and \eqref{Dk_scaling2}. We recall that the dynamical exponent is $z_s=1+s$ for $s<1$ and $z_s=2$ for $s>1$.
For each observable the two forms are equivalent, each being useful in a different regime. There are in total three scaling
variables denoted by $x=k/(t/\tau)^{1/z_s}$, ${\sf x} = k/\hat g^{1/z_s}$
and $y=\gamma t$. 

\begin{figure}
\hspace{-4.2cm}
\begin{minipage}[c]{.36\linewidth}
    \centering
    \raisebox{-\height}{\includegraphics[width=\linewidth,trim={0 0cm 2cm 0.5cm},clip]{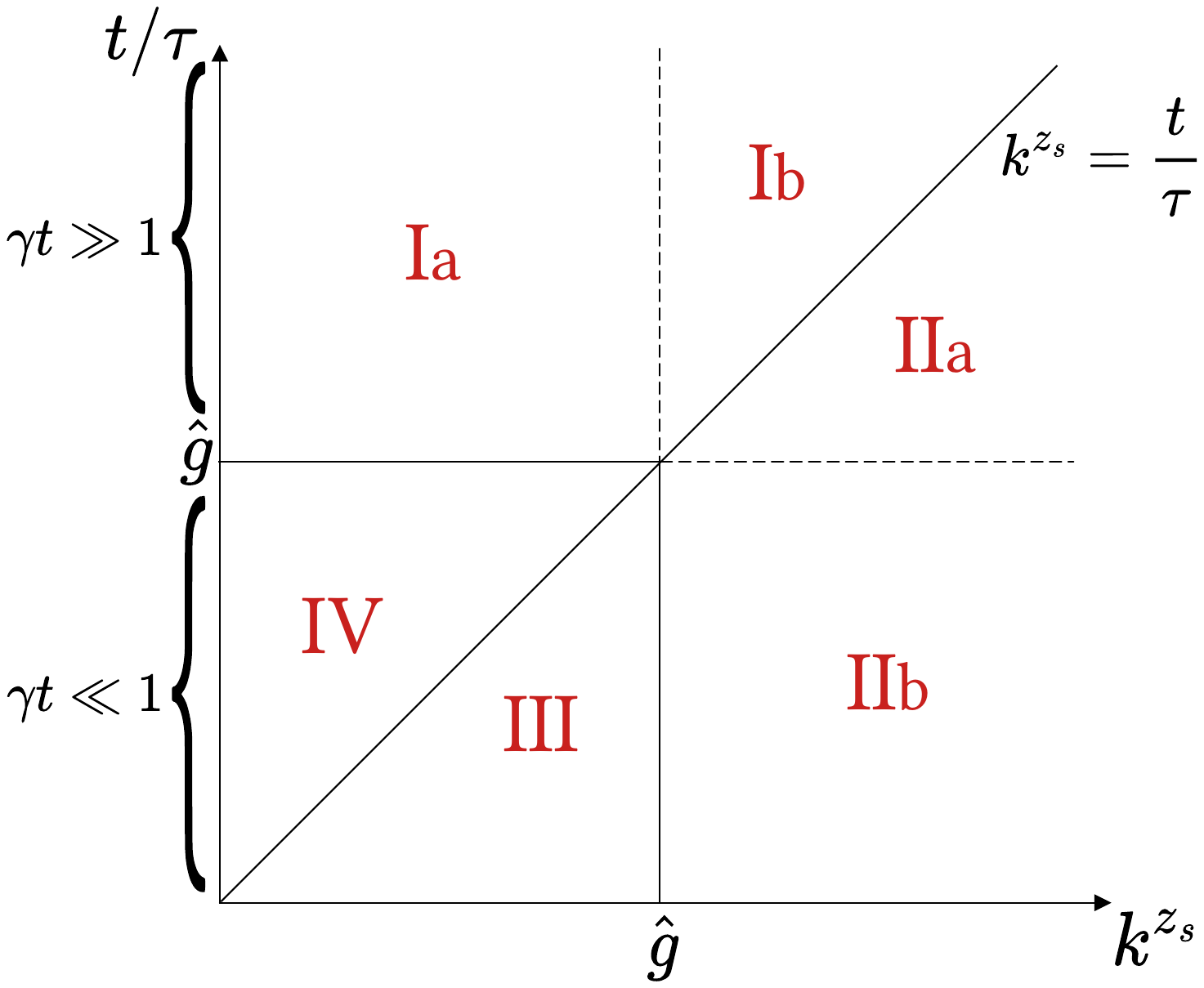}}
  \end{minipage}
  \begin{minipage}[c]{.45\linewidth}
    \centering
\raisebox{-\height}{\begin{tabular}{|c|c|c|}
\hline
 & $C_k(t)$ & $D_k(t)$  \\
\hline
Ia & $C_0(t)-C_k(t) \propto \begin{cases} k^2 \ , \ s>1/2 \\ k^{1+2s} \ , \ s<1/2  \\ \end{cases}$ & \multirow{2}{*}{$D_k(t) \propto \frac{k^2}{t^{\frac{2-s}{1+s}}}$} \\ 
\cline{1-2} 
Ib & $C_0(t)-C_k(t) \propto k^s$ &  \\  
\hline
IIa & \multirow{2}{*}{$C_k(t) \propto \frac{t^2}{k^{2+s}}$} & $D_k(0)-D_k(t) \propto t^{\frac{s}{1+s}}$  \\
\cline{1-1} \cline{3-3}
IIb & & $D_k(0)-D_k(t) =v_R(\hat g)^2 t^2$ \\  
\hline
III & $C_0(t)-C_k(t) \propto t^2 k^s$ & $D_k(0)-D_k(t) \propto t^2 k^s$ \\
\hline
IV & $C_0(t)-C_k(t) \propto \begin{cases} t^{\frac{3s}{1+s}} k^2 \ , \ s>1/2 \\ t k^{1+2s} \ , \ s<1/2  \\ \end{cases}$ & $D_k(0)-D_k(t) \propto \begin{cases} t^{\frac{3s}{1+s}} k^2 \ , \ s>1/2 \\ t k^{1+2s} \ , \ s<1/2  \\ \end{cases}$   \\
\hline
    \end{tabular}}
  \end{minipage}
    \caption{Left panel: the six asymptotic regimes of the space-time correlation function $C_k(t)$ 
    and of the time correlation of the gaps $D_k(t)$, 
    defined in \eqref{Ck_scaling_intro} and \eqref{Dk_scaling_intro}, for the RTP Riesz gas, represented in the plane $(k^{z_s}, t/\tau)$, where $\hat g=1/(2 \gamma \tau)$, in the
    limit where $k,t/\tau,\hat g \gg 1$. Right panel: Table of the leading behavior of $C_k(t)$ and $D_k(t)$ for $0<s<1$
    in the six regions of the left panel, as a function of time $t$ and separation $k$.
    }
    \label{fig:diagrameCkDk_intro}
\end{figure}

In the limit where {time, separation and activity are large, i.e. $t/\tau \gg 1$, $\hat g \gg 1$, and $k \gg 1$},
this leads to six different regimes, depending on whether each of the three scaling variables ${\sf x}, \, x, \, y$
is large or small. These six regimes are represented in Fig.~\ref{fig:diagrameCkDk_intro}
and correspond to the six regions delimited in the left panel. Here we discuss only $0<s<1$,
the other cases are discussed in Sections \ref{sec:Ck_rtp} and \ref{sec:Dk_rtp}. 

Let us describe these regimes qualitatively, first for $C_k(t)$ and then for $D_k(t)$, 
see the table in Fig.~\ref{fig:diagrameCkDk_intro} for more
quantitative details and Section \ref{sec:Ck_rtp} and Section \ref{sec:Dk_rtp} for an extensive study. In regions I, III and IV $C_k(t)$ is approximately
equal to $C_0(t)$, which was already discussed above, but the difference $C_0(t)-C_k(t)$ exhibits distinct
behaviors in these regions. In regions Ib and IIa the covariance $C_k(t)$ behaves as for the Brownian particles
(with an effective temperature $T_{\rm eff}$), where region Ib corresponds to large time while region IIa
corresponds to large separations, see discussion below \eqref{Ck_scaling_intro}. The other regions are specific to the RTP gas.
In the region Ia, although $\gamma t \gg 1$ -- hence $C_0(t) \sim t^{\frac{s}{1+s}}$
behaves as for Brownian particles -- the difference $C_0(t)-C_k(t) \simeq D_k(0)$ 
is given by the variance of the gap, see the first line of Table \ref{table:gap_var_rtp}. The 
regions II, III and IV exhibit ballistic behavior to leading order. The regions IIa and IIb are equivalent to leading order for $C_k(t)$,
and correspond to large separation, where $C_k(t)$ is ballistic and decays
to zero. Note that the ballistic behavior in region IIa was already noted above in the case of the Brownian particles.
In regions III and IV, $C_k(t)$ behaves as $C_0(t) \simeq v_R(\hat g)^2 t^2$, but
while the difference $C_0(t)-C_k(t)$ is also ballistic in region III,
it has a non trivial power law dependence in time in region IV. 

The regimes for $D_k(t)$ are closely related and are represented in the same figure, {see the right panel of Fig.~\ref{fig:diagrameCkDk_intro}}. 
The region I corresponds to the large time behavior with a decay of $D_k(t)$ which is a power law in time,
identical to the decay of $D_k(t)$ at large time in the case of Brownian particles, see \eqref{Dk_decay_intro}. In regions II, III and IV, the leading behavior of $D_k(t)$ is given by $D_k(0)$ (see Table \eqref{table:gap_var_rtp}), but the behavior of 
$D_k(0)-D_k(t)$ differs between these regions: Region IIa corresponds to the large gap limit of the Brownian regime, where $D_k(0)-D_k(t) \simeq C_0(t) \sim t^{\frac{s}{1+s}}$. 
In region IIb the same relation holds but now $C_0(t)$ is ballistic. The behavior of 
$D_k(0)-D_k(t)$ in regions III and IV can be obtained from $C_k(t)$ using the relation 
\eqref{relationCkDk} and is specific to the RTP gas. 

In summary the six regions in Fig.~\ref{fig:diagrameCkDk_intro} correspond to different behaviors for
either $C_k(t)$ or $D_k(t)$ or both. Across each of the lines which separate the neighboring regions 
there is a crossover in the behavior of $C_k(t)$ and $D_k(t)$
which is described by a distinct scaling function 
of one of the arguments $x=k/(t/\tau)^{1/z_s}$, ${\sf x} = k/\hat g^{1/z_s}$
or $y=\gamma t$. These nine crossover functions are shown in the more detailed 
figures, Fig. \ref{fig:diagrameCk} and Fig. \ref{fig:diagrameDk}, 
and their analytic expressions are obtained in Sections \ref{sec:Ck_rtp} and \ref{sec:Dk_rtp}. 

\bigskip

\subsubsection{Dynamical correlations: quenched initial conditions} 

\begin{figure}
    \begin{minipage}[c]{.55\linewidth}
    \centering
    \raisebox{-\height}{\includegraphics[width=\linewidth,trim={0 8cm 0 8cm}, clip]{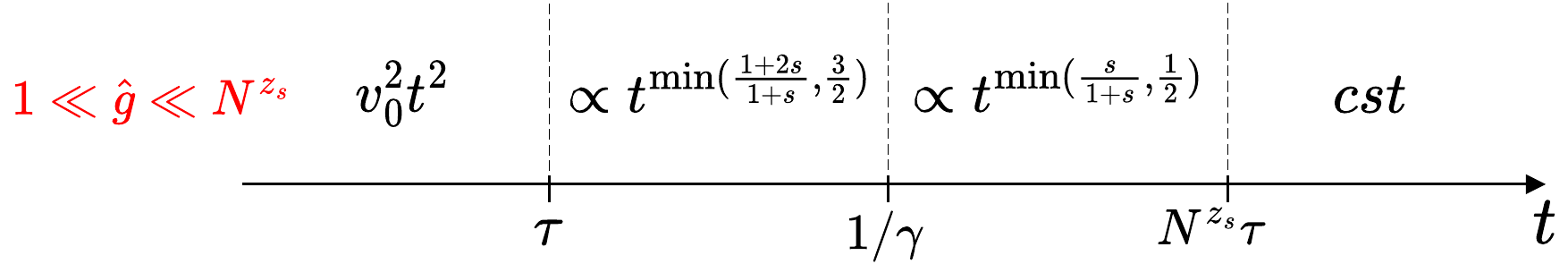}}
    \end{minipage}
    \hspace{0.8cm}
    \begin{minipage}[c]{.35\linewidth}
    \centering    
    \raisebox{-\height}{\includegraphics[width=\linewidth]{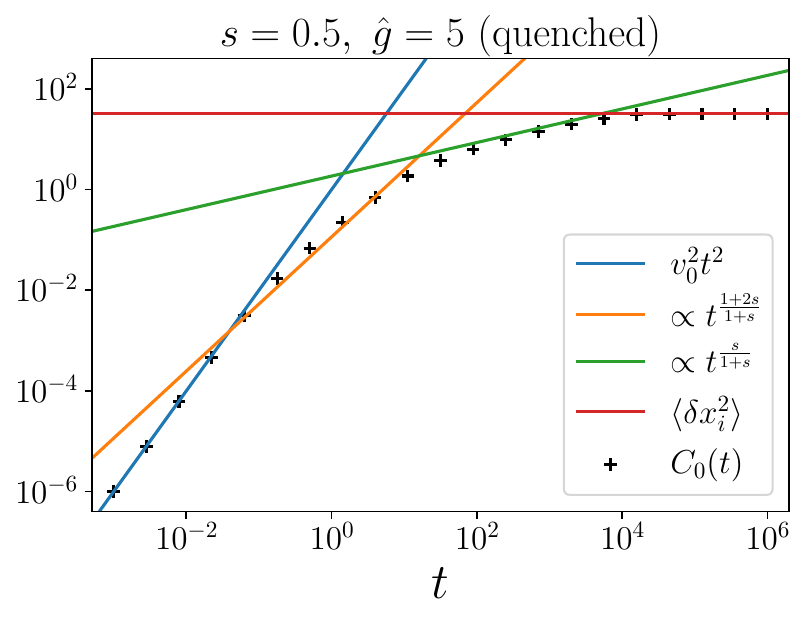}}
    \end{minipage}
    \caption{Left: Regimes for the variance of the displacement of a particle $C_0^{\rm qu}(t)$
    as a function of time for quenched initial conditions for $s>0$. Only the case $1 \ll \hat g \ll N^{z_s}$ 
    (equivalent of the central line in Fig.~\ref{fig:time_regimes_rtp}) is represented. 
The time evolution is first ballistic, then anomalous superdiffusive, then anomalous subdiffusive, and finally saturates. The anomalous superdiffusive regime has no analog in the annealed case. The case $\hat g \ll 1$ is identical to the annealed case up to numerical factors. The case $g \gg N^{z_s}$ was not studied in this paper for a quenched initial condition. Right: Plot of $C_0(t)$ as a function of time, obtained by numerical computation of the sum \eqref{C0_quenched_RTP} for a RTP Riesz gas with $s=0.5$, $N=10001$, $g=1$, $\rho=1$, $v_0=1$ and $\gamma=0.1$. The 4 time regimes are clearly visible.}
    \label{fig:diagram_quenched_C0}
\end{figure}

Next we consider the case of the quenched initial condition, i.e. $\delta x_i(0)=0$ for all $i$. Contrarily to the Brownian case, where
the quenched covariance can be obtained from the annealed one by a simple rescaling, the dynamical behavior is much richer for the RTPs. 
For instance, as shown in Fig. \ref{fig:diagram_quenched_C0}  the variance of the displacement of a particle $C_0^{\rm qu}(t)$ exhibits 
an additional time regime when $\tau \ll 1/\gamma$, i.e. $\hat g \gg 1 $, which corresponds to $\tau \ll t \ll 1/\gamma$, 
and is intermediate between the ballistic and the anomalous subdiffusion. In that regime, 
the dynamics for $s>0$ is super-diffusive with $C_0^{\rm qu}(t) \sim t^{(1+2s)/(1+s)}$
for $0<s<1$ while $C_0^{\rm qu}(t) \sim t^{3/2}$ for $s>1$. Interestingly, 
the $t^{3/2}$ growth was also obtained for harmonic chains of RTPs, which 
corresponds to our short-range case \cite{PutBerxVanderzande2019,SinghChain2020,HarmonicChainRevABP}.  Another noticeable difference with the annealed setting concerns the ballistic regime: since the particles are initialized with a uniform density, there is no renormalization of the velocity, i.e. one finds $v_R=v_0$ for any $\hat g$.

Correspondingly the space time correlation $C_k^{\rm qu}(t)$ exhibits a variety of behaviors 
as shown in Fig.~\ref{fig:diagram_quenched}. Comparing with Fig.~\ref{fig:diagrameCk} for the annealed
case one sees that some of the dynamical regimes are common to both, while others 
are new. We obtain exact formulas for the two-parameter dynamical scaling functions, see 
\eqref{scaling1} in terms of the variables 
$x=k/(2t/\tau)^{1/z_s}$ and $y=\gamma t$
and
\eqref{scalingquenched}
in terms of the variables 
${\sf x}=k/(\hat g)^{1/z_s}$ and $y=\gamma t$. The analysis of these formulas leads to
the dynamical regimes in Fig. 
\ref{fig:diagram_quenched} with five one-parameter scaling functions 
which describe the crossovers between the different dynamical regions. 
We refer the reader to Section \ref{sec:rtp_quenched} for the detailed discussion and calculations.
The scaling function $\Omega_s$ in Fig.~\ref{fig:diagram_quenched} (see Eq. \eqref{omega_s}) generalizes
for $0<s<1$ the one obtained for the harmonic chain, which 
corresponds to our short-range case \cite{SinghChain2020}.

\subsubsection{Extremely slow tumbling} \label{mainres_smallgamma}

As indicated in Fig. \ref{fig:time_regimes_rtp} there is a third case (lowest panel) when $\gamma$ is very small, i.e. either
of order $N^{-z_s}/\tau$ (the smallest relaxation rate in the system) or even smaller (as in Fig.~\ref{fig:time_regimes_rtp}). Once again we
restrict ourselves to the stationary (annealed) case. 

Let us first note that there is a well defined limit $\gamma=0^+$. It was studied previously in the case of the active DBM $s=0$ in \cite{ADBM2}.
In this limit the observables are obtained by averaging over the fixed points of the dynamics at fixed $\sigma_i$. 
Whenever these fixed points can be studied within the linear approximation, one recovers the
same formulas for fixed $N$ as above for the various observables by simply setting $\gamma=0$, see e.g.
\eqref{cov_gamma0} for the static covariance.
Their large $N$ behavior however may differ from the finite $\gamma$ case. In particular the variance of the gaps $D_k(0)$ for $\gamma=0^+$ 
diverges (for fixed $k$) as $N^{2s-1}$ for $1/2<s<1$ (resp. $N$ for $s>1$), while it remains finite for $s<1/2$ [see Eqs. \eqref{Dk0_smallgammahalf} and \eqref{Dk0_gamma0}].  
By contrast the gap variance is always finite for $\gamma>0$. Having estimated the gap variance we can obtain the range of validity of the linear approximation 
for $\gamma=0^+$. For $s<1/2$ the condition is $\rho^2 v_0^2 \tau^2 \ll B_s$ 
where $B_s$ is given in \eqref{validity_rtp}, while for $s>1/2$ it is much more restrictive and it reads
$\rho v_0 \tau \ll N^{-(2z_s-3)}$. 

Between this $\gamma=0^+$ limit and the finite $\gamma$ regime discussed in the rest of the paper, there is a crossover for $\gamma = \tilde \gamma N^{-z_s}$ with
fixed $\tilde \gamma$ of order unity. For instance, for the variance we obtain
\be
\langle \delta x_i^2 \rangle \simeq 
N^{2z_s-1} \frac{2 v_0^2 \tau^2}{a_s} \sum_{q=1}^{+\infty} \frac{1}{q^{z_s} (a_s q^{z_s} + \tilde g^{-1})} \;,
\ee
where we denote $\tilde g = 1/(2\tilde \gamma \tau) = N^{-z_s}\hat g$ assumed here to be of order unity. A similar result for the gaps is given in \eqref{crossoverDk0smallgamma}.This crossover is also valid for dynamical variables such as $C_0(t)$, if we also scale the time as $t\sim N^{z_s}$, see Eq. \eqref{C0_smallgamma}. 
This crossover is illustrated in the bottom panel of Fig.~\ref{fig:time_regimes_rtp} (represented there in the limit $\tilde g \gg 1$ where the 
time scales $1/\gamma$ and $N^{z_s} \tau$ are well separated). In the limit where $t \ll 1/ \gamma$ and $t \ll N^{z_s}\tau$, $C_0(t)$ remains ballistic with a renormalized velocity given by
\be
C_0(t) \simeq \tilde v_R(\tilde g)^2 t^2 \quad , \quad \tilde v_R(\tilde g)^2 = \frac{2v_0^2}{N} \sum_{q=1}^{\infty} \frac{1}{1+a_s\tilde g q^{z_s}} \;,
\ee
as indicated in the bottom panel of Fig.~\ref{fig:time_regimes_rtp}. For $\tilde g \gg 1$ and $t \gg N^{z_s} \tau$, 
one finds that $C_0(t) \simeq 2 \langle \delta x_i^2 \rangle (1-e^{- 2 \gamma t})$ as also indicated in Fig.~\ref{fig:time_regimes_rtp}.
Hence, in this regime the interactions do not seem to play a role in the relaxation dynamics.

\subsection{Active Dyson Brownian motion and active Calogero-Moser model in a harmonic trap} 

We now abandon the circle geometry and study the active Riesz gas on the real line in presence of an harmonic trap. 
We focus on the stationary state reached at large time.
The problem is quite difficult in general due to the finite support of the density, but there are two
simpler cases on which we will focus here, namely $s=0$, the active Dyson Brownian motion, and $s=2$ the
active Calogero-Moser model. 

\subsubsection{Active Dyson Brownian motion}

The active Dyson Brownian motion (active DBM) was introduced in \cite{ADBM1} as a generalization of the standard DBM in the presence of
active run-and-tumble noise. It is defined by the dynamical equation
\bea \label{ADBM_def} 
 \dot x_i(t) &=& - \lambda x_i(t) +  \frac{2\,g }{N} \sum_{j \neq i} 
\frac{1}{x_i(t)-x_j(t)} +  v_0 \sigma_i(t) \quad {\rm for} \ i=1,2, \cdots, N,
\eea
where the $\sigma_i(t)$ are again independent telegraphic noises with rate $\gamma$. 
The $N$ particles are confined in a harmonic potential $V(x)=\lambda x^2/2$. The scaling 
of the interaction with $N$ is chosen such that in the large $N$ limit the support of the density 
remains finite and converges to $[-2\sqrt{g/\lambda},2\sqrt{g/\lambda}]$. 
It is thus quite different from what was done for the Riesz gas on the circle. 
Let us recall that due to the diverging repulsion, the particles cannot cross and thus remain ordered.
In the absence of noise, $v_0=0$, they converge to equilibrium positions $x_{\rm eq,i} = \sqrt{\frac{2g}{\lambda N}} y_i$ 
where the $y_i$'s are the zeroes of the Hermite polynomial of degree $N$, i.e. $H_N(y_i) = 0$.
This leads to a semi-circle density in the limit of vanishing noise.
In \cite{ADBM1} we have discussed the various regimes of the model as the amplitude of the noise is varied. 
This is illustrated in Fig. 1 (bottom panel) of \cite{ADBM1} (equivalently in Fig. 1 of \cite{ADBM2}).
The important parameter is $v_0^2/(g\lambda)$. When it is of order unity or smaller, the mean particle density
retains a semi-circular shape. If it is very small, of order $1/N$, the particles become localized and form
a perfect crystal. Finally, when it is large, of order $N$, clusters of particles with effective contact interactions form and 
the density takes a bell shape. 


We study here the stationary state of the model in the weak noise limit $v_0^2/(g \lambda) \ll 1$ 
for arbitrary value of the tumbling rate $\gamma$. This generalizes our previous study \cite{ADBM2} 
where we restricted to the large persistence limit $\gamma=0^+$. Let us write the particle positions as
$x_i = x_{\rm eq,i} + \delta x_i$, where the $\delta x_i$'s are the small deviations 
from equilibrium which vanish as $v_0 \to 0$. To discuss the results one must distinguish two regions of the gas, which lead to a different scaling with $N$: the bulk region, where $i, N-i \gg 1$, and the edge region where  $i=O(1)$ or $N-i = O(1)$. 

In the bulk region we obtain the following formula for the equal time correlations
\begin{equation}
\langle \delta x_i \delta x_j \rangle \simeq \frac{v_0^2}{\lambda^2 N} \mathcal{C}_b^{\gamma/\lambda}\left( \frac{x_{{\rm eq},i}}{2\sqrt{g/\lambda}}, \frac{x_{{\rm eq},j}}{2\sqrt{g/\lambda}} \right) \ , \ 
\mathcal{C}_b^{\tilde \gamma}(x,y) = \sum_{k=1}^\infty \frac{U_{k-1}(x) U_{k-1}(y)}{k(k+2\tilde \gamma)} \ , \ U_{k-1}(x) = \frac{\sin(k \arccos(x))}{\sqrt{1-x^2}} \;.
\label{covADBM_largeN_intro}
\end{equation}
Note that this result, as well as all the other bulk results presented in Section~\ref{sec:harmonic_trap}, can be rewritten using that at large $N$, $\frac{x_{{\rm eq},i}}{2\sqrt{g/\lambda}} = \frac{y_{{\rm eq},i}}{\sqrt{2N}} \simeq G^{-1}(i/N)$ where $G(x)=\int_{-1}^x du \frac{2\sqrt{1-u^2}}{\pi}$ is the cumulative of the semi-circle density. Eq. \eqref{covADBM_largeN_intro} generalizes Eq. (12) in \cite{ADBM2}. 
We have tested numerically this prediction by comparing with numerical simulations in Fig.~\ref{figADBM1} below. For small enough ratio $v_0^2/(g\lambda)$ we find a perfect agreement for any value of $\gamma$. We have also obtained the 
two-time correlations for the displacements of bulk particles in the stationary state
given in \eqref{covADBM_largeN_time}, which has no equivalent in \cite{ADBM2}. Finally we have computed the variance of the gaps on macroscopic scales $k=\kappa N$
in \eqref{var_chn}. 

In the edge region, i.e. for $i$ and $j$ of order 1, we find
\begin{eqnarray}
\langle \delta x_i \delta x_j \rangle \simeq  \frac{v_0^2}{\lambda^2 N^{2/3}} \frac{1}{\Ai'(a_i) \Ai'(a_j)} \int_0^{+\infty} dx \ \frac{\Ai(a_i + x)\Ai(a_j + x)}{x(x+2 \hat \gamma)} \quad , \quad \hat \gamma=\frac{\gamma}{\lambda}N^{-1/3} \;,
\label{covADBM_edge_intro}
\end{eqnarray}
where $\Ai(x)$ is the Airy function and $a_i$ denotes its $i^{\rm th}$ zero.
Eq. \eqref{covADBM_edge_intro} generalizes Eq. (17) in \cite{ADBM2}.
Again, we have also obtained the 
two-time correlations for the displacements of the edge particles in the stationary state
given in \eqref{covADBM_largeN_time}, which has no equivalent in \cite{ADBM2}. 
Note that the overall size of fluctuations are $N^{-2/3}$, larger than the $1/N$ scaling in the bulk, and that the 
relaxation time scales at the edge scale as $N^{-1/3}$, and are much faster than in the bulk
(hence the dependence in $\hat \gamma$ in \eqref{covADBM_edge_intro}).

\subsubsection{Active Calogero-Moser model}

We have also studied the active Calogero-Moser model described by the equations of motion \eqref{Calogero} below.
This model is introduced and studied numerically in a companion paper \cite{activeCM} where the
phase diagram is discussed in details. Here we
give a calculation of the correlation functions in the stationary state for this model in the weak noise regime, for any $\gamma>0$.
The equilibrium positions of the particles are again given by the rescaled zeros of the Hermite polynomial $H_N(y)$
(leading to a semi-circle equilibrium density).
Due to the relation that exists between the Hessian matrices of the two systems, the covariance of the particle displacements in the weak noise limit has an expression very similar to \eqref{covADBM_largeN_intro}, but with $k$ replaced by $k^2$ in the denominator, displayed in \eqref{covCM_largeN} below. 
There is however an important difference with the active DBM: the absence of a distinct edge regime. This is due to 
the short range nature of the interactions.
Thus the bulk result \eqref{covCM_largeN} and its $1/N$ scaling remain valid even for edge particles.
An important result is the behavior of the variance of the gaps $\langle (\delta x_i - \delta x_{i+n})^2 \rangle \propto n^2$
for any $1 \leq n \ll N$. 
This implies in particular giant number fluctuations \cite{DasGiant2012}, i.e. that the variance of the number of particles in
an interval scales as the square of its length.



\section{Derivation of the results: Brownian gas}\label{sec:Brownian}

Consider $N$ Brownian particles on a ring of perimeter $L$, with positions $x_i(t)$ on the real axis modulo $L$, interacting through a potential $W(x)$ (of periodicity $L$, $W(x+L)=W(x)$), and 
described by the equations of motion
\be \label{Eq_def_brownian}
\dot x_i(t) = -\sum_{j(\neq i)} W'(x_i(t)-x_j(t)) + \sqrt{2T} \xi_i(t) \quad i=1,...,N \;,
\ee
where $\xi_i(t)$ are i.i.d. Gaussian white noises with unit variance. At equilibrium, the probability of observing a set of positions $\{x_i\}$ ($1\leq i \leq N$) is 
\be \label{Gibbs1}
P(\{x_i\}) \propto e^{-\frac{E(\{x_i\})}{T}} \quad , \quad E(\{x_i\}) = \sum_{i<j} W(x_i-x_j) \;.
\ee
We assume that the interaction potential $W(x)$ is even and monotonically decreasing around $x=0$, and such that the total energy $E(\{x_i\})$
is minimized when the particles are equally spaced. To study the fluctuations around the ground state, i.e. the equally spaced configuration, we
decompose $x_i(t)$ as
\be \label{def_delta_x}
x_i(t) = \bar x(t) + x_i^0  + \delta x_i(t)  \quad , \quad \bar x(t) = \frac{1}{N} \sum_i x_i(t) \;,
\ee
where $\bar x(t)$ is the position of the center of mass and the $x_i^0$ can be chosen as $x_i^0= (i - \frac{N+1}{2}) \frac{L}{N}$, $i=1,...,N$. By consistency
one must have 
\be  \label{constraint}
\sum_i \delta x_i(t)=0 \;.
\ee
Summing over $i$ \eqref{Eq_def_brownian} we see that the center of mass freely diffuses with a diffusion coefficient $\sqrt{\frac{2T}{N}}$, 
i.e. $\dot{ \bar x}(t) = \frac{\sqrt{2T}}{N} \sum_j \xi_j(t) = \sqrt{\frac{2T}{N} } \xi(t)$ (where $\xi(t)$ has the same law as the $\xi_i(t)$). 

\subsection{Dynamics for small deformations} \label{sec:Brownian_derivation}

In this paper we focus on the regime of small relative displacements which leads to small Gaussian fluctuations around the ground state. We Taylor expand
the r.h.s of \eqref{Eq_def_brownian} in powers of the $\delta x_i$'s around the ground state configuration. We 
assume that 
\be \label{cond_approx}
\forall j\neq i, \ \delta x_i - \delta x_j \ll 2\frac{W''(x_i^0-x_j^0)}{W'''(x_i^0-x_j^0)} \;,
\ee
which allows us to 
retain only the linear order in the expansion (see Appendix \ref{app:validity_condition}).
This leads to 
\be \label{Eq_delta_x}
\delta \dot{x}_i(t) = - \sum_{j=1}^N H_{ij} \delta x_j(t) + \sqrt{2T} \xi_i(t)  - \frac{\sqrt{2T}}{N} \sum_{j=1}^N \xi_j(t) 
\ee 
where
\be \label{defHessian}
H_{ij} = \frac{\partial^2 E}{\partial x_i \partial x_j}(\{ x_i^0 \}) = \begin{cases} \sum_{k(\neq i)} W''(\frac{L}{N}(i-k)) \ {\rm for} \ i=j \\ -W''(\frac{L}{N}(i-j)) \hspace{1cm} {\rm for} \ i\neq j \end{cases} \;.
\ee
This $N \times N$ Toeplitz matrix can be diagonalized with the eigenvectors $v^q_k=\frac{1}{\sqrt{N}} e^{2\pi i \frac{q}{N} k}$ 
(which form an orthonormal basis) and eigenvalues
\be \label{eigenvals}
\mu_q = \mu_{N-q}= \sum_{\ell=1}^{N-1} W''(\frac{L}{N} \ell) \left(1-\cos(\frac{2\pi q\ell}{N}) \right) = 2\sum_{\ell=1}^{N-1} W''(\frac{L}{N} \ell) \sin^2(\frac{\pi q\ell}{N}) \quad , \quad q=0,1,...,N-1 \;.
\ee

We now study the equilibrium dynamics (preparing the system at equilibrium at time $-\infty$).
This corresponds to the "annealed initial condition", the quenched case will be studied below.
Taking the Fourier transform of \eqref{Eq_delta_x} with respect to time one thus obtains by inversion in the frequency domain
\be \label{eqfourier} 
\delta \hat x_j(\omega) = \sqrt{2T} \sum_{k=1}^N  [i \omega \mathbb{1}_N + H]^{-1}_{jk} \hat \xi_k(\omega) - \frac{\sqrt{2T}}{N} \frac{1}{i\omega} \sum_{k=1}^N \hat \xi_k(\omega) \;,
\ee 
where $\mathbb{1}_N$ is the $N \times N$ identity matrix, $\delta \hat x_i(\omega) = \int_{-\infty}^{\infty} e^{-i \omega t} \delta x_i(t)\,dt$ and $\hat \xi_i(\omega)$ is a Gaussian white noise with correlations $\langle \hat \xi_i(\omega) \hat \xi_j(\omega') \rangle= 2 \pi \delta_{ij}\,\delta(\omega+\omega')$. 
Here and below the brackets denote averages over the Brownian noises.
We have used that
\be \label{identity_hessian_fourier}
\sum_{l=1}^N [i \omega \mathbb{1}_N + H]^{-1}_{jl} = \frac{1}{N} \sum_{q=0}^{N-1} \sum_{l=1}^N \frac{e^{2\pi i \frac{q}{N} (j-l)}}{i\omega + \mu_q} = \frac{1}{i\omega} \;,
\ee
since only the term $q=0$ remains. {The two-point, two-time correlation function can be obtained from \eqref{eqfourier} upon Fourier inversion, leading to}
\be
\langle \delta x_j(t) \delta x_k(t') \rangle = 2 T \int_{-\infty}^{+\infty} \frac{d\omega}{2 \pi} e^{i \omega (t-t')} \left( [\omega^2 \mathbb{1}_N + H^2]^{-1}_{jk} - \frac{1}{N\omega^2} \right) \;.
\ee
Using the eigensystem of $H$ given in \eqref{eigenvals}, this reads
\bea \label{cov_brownian}
\langle  \delta x_j(t)  \delta x_k(t') \rangle &=& \frac{2 T}{N}  \sum_{q=1}^{N-1} \int_{-\infty}^{+\infty}  \frac{d\omega}{2 \pi} \frac{e^{i \omega (t-t')}}{\omega^2 + \mu_q^2 }
e^{2\pi i \frac{q}{N} (j-k)} \nonumber \\
&=& \frac{2T}{N}  \sum_{q=1}^{(N-1)/2} \frac{e^{- \mu_q |t-t'|}}{\mu_q}  \cos\left({2\pi \frac{q}{N} (j-k)}\right) \;,
\eea
where we have made use of the symmetry $\mu_q=\mu_{N-q}$ in the last step. The last expression is exact only for odd values of $N$ (otherwise one simply needs to take the sum from $1$ to $N-1$ and remove the factor $2$), but since throughout the paper we will be focusing on the large $N$ limit this is irrelevant. Note that one can also compute the average $\langle \delta x_i(t) \rangle$ to leading order by keeping the next order term $\propto W''' \delta x^2$ in the Taylor expansion of \eqref{Eq_def_brownian}. This is performed in Appendix \ref{app:avg_nextorder} and one finds that $\langle \delta x_i(t) \rangle \propto T$.
Hence Eq. \eqref{cov_brownian} also gives the covariance to the same leading order in $T$.

\subsection{Observables, parameters and scaling regimes} \label{sec:observables}

\subsubsection{Observables}

Several observables of interest can be deduced from \eqref{cov_brownian}. We now introduce a few of them, which will be the focus for the rest of this section. The definitions of these observables is valid both for the Brownian and the RTP particles. The formulas 
given here using the eigenvalues $\mu_q$ are however only valid for the Brownian particles in their equilibrium state. 

We start with "static" quantities in the stationary state. 
The first quantity we consider is the variance of the displacement for a single particle, given by
\be \label{var_brownian}
\langle \delta x_i^2 \rangle = \frac{2T}{N} \sum_{q=1}^{(N-1)/2} \frac{1}{\mu_q} \;,
\ee
where here and below we omit the argument $t$ for time-independent quantities. We will also consider the static covariance
\be \label{covstat_brownian}
\langle \delta x_i \delta x_{i+k}\rangle = \frac{2T}{N}  \sum_{q=1}^{(N-1)/2} \frac{\cos\left(2\pi \frac{q}{N} k\right)}{\mu_q} \;.
\ee

Another quantity of interest is the variance of the gap between particle $i$ and $i+k$ in the stationary state defined as
\be \label{gap_brownian}
D_k(0) = \langle (\delta x_{i}-\delta x_{i+k})^2 \rangle = \frac{4T}{N} \sum_{q=1}^{(N-1)/2} \frac{1-\cos\left( \frac{2\pi kq}{N} \right)}{\mu_q} = \frac{8T}{N} \sum_{q=1}^{(N-1)/2} \frac{\sin^2\left( \frac{\pi kq}{N} \right)}{\mu_q} \;,
\ee
as well as the gap correlations,
\be \label{gapcor_brownian}
D_{k,n}(0) = \langle (\delta x_{i}-\delta x_{i+k}) (\delta x_{i+n}-\delta x_{i+n+k}) \rangle 
= \frac{8T}{N} \sum_{q=1}^{(N-1)/2} \frac{\sin^2\left( \frac{\pi kq}{N} \right)}{\mu_q}  \cos\left(2 \pi \frac{q}{N} n\right) \;.
\ee

We would also like to have some information on the equilibrium dynamics. We start with the time evolution of the displacement of a single particle, which we can obtain through the quantity
\be \label{disp_brownian}
C_0(t) = \langle (\delta x_i(t) - \delta x_i(0))^2 \rangle 
= \frac{4T}{N} \sum_{q=1}^{(N-1)/2} \frac{1 - e^{- \mu_q t}}{\mu_q} \;.
\ee
One may also consider the two-time correlations of these displacements,
\be \label{multitime_brownian}
C_0(t,t') = \langle (\delta x_i(t) - \delta x_i(0)) (\delta x_{i}(t') - \delta x_{i}(0))   \rangle 
= \frac{2T}{N} \sum_{q=1}^{(N-1)/2}  \frac{1 - e^{- \mu_q t} -  e^{- \mu_q t'} + e^{- \mu_q |t-t'|} }{\mu_q} \;.
\ee
Next, we will consider the correlations between the time displacements of different particles, 
which describes space-time correlations 
\be \label{spacetime_brownian}
C_k(t) = \langle (\delta x_i(t) - \delta x_i(0)) (\delta x_{i+k}(t) - \delta x_{i+k}(0))   \rangle 
= \frac{4T}{N} \sum_{q=1}^{(N-1)/2}  \frac{1 - e^{- \mu_q t}}{\mu_q} \cos\left( 2 \pi \frac{q k}{N} \right) \;.
\ee 
Finally, we can also study the correlations of the gaps at two different times 
\be \label{gaptime_brownian}
D_k(t) = \langle (\delta x_i(t) - \delta x_{i+k}(t)) (\delta x_{i}(0) - \delta x_{i+k}(0))   \rangle 
= \frac{8T}{N} \sum_{q=1}^{(N-1)/2}  \frac{e^{- \mu_q t}}{\mu_q} \sin^2\left(\pi \frac{q k}{N} \right) \;.
\ee 
For convenience, and without loss of generality, everywhere below we will consider the variables $k$ and $n$ to be positive integers.

Note that there are some relations between these different observables. For instance one can check using the above formulas that
\be \label{remarkrelation1} 
C_0(t) - C_k(t) = D_k(0) - D_k(t) \;.
\ee 
In particular this implies, since $D_k(t) \to 0$ as $t\to +\infty$,
\be \label{remarkrelation} 
\lim_{t \to +\infty} C_0(t) - C_k(t) = D_k(0) = \langle (\delta x_{i}-\delta x_{i+k})^2 \rangle \;,
\ee 
as well as, since $C_k(t) \to 0$ as $k\to +\infty$,
\be \label{remarkrelation2} 
\lim_{k \to +\infty} D_k(0) - D_k(t) = C_0(t) = \langle (\delta x_{i}(t) -\delta x_{i}(0))^2 \rangle \;.
\ee 
All of these results will be useful below. Another important relation is
\be \label{relation12times}
C_0(t,t') = \frac{1}{2} [C_0(t)+C_0(t') - C_0(|t-t'|)] \;.
\ee

These relations \eqref{remarkrelation1}, \eqref{remarkrelation}, \eqref{remarkrelation2} and \eqref{relation12times}
are also valid for the RTP gas in the stationary state (see below) since they
only require time-translational invariance. They are however not valid for quenched
initial conditions for which time-translational invariance is broken. 
\\

\subsubsection{Time-scales and large $N$ limit} 

Before taking the large $N$ limit, let us discuss the relevant time-scales of the problem. Each inverse eigenvalue $1/\mu_q$ gives the time-scale of the equilibration at a given length-scale, namely $L/(2q)$. Let us assume that $\mu_q$ is an increasing function of $q$ for $1\leq q\leq \frac{N-1}{2}$ (which will generally be the case if the strength of the interaction decreases with the distance). A priori, we expect that two of these time-scales will play a particular role. The smallest time-scale, $1/\mu_{(N-1)/2}$, is the local relaxation time due to the interactions at the scale of the lattice spacing $L/N$. Below this time-scale, the behaviour of the system is universal as the interactions do not have enough time to play a role. Consider for instance the displacement of a single particle \eqref{disp_brownian}. At small time $t\ll 1/\mu_{(N-1)/2}$, one can expand all the exponentials to first order, leading to
\be \label{C0_freediff}
C_0(t) \simeq \frac{2T}{N} (N-1) t \;,
\ee 
to which one can add the center of mass diffusion $\langle (\bar x(t) - \bar x(0))^2\rangle = \frac{2T}{N} t $. This corresponds to free diffusion. 
Similarly for the space-time correlations \eqref{spacetime_brownian}, we find for any $k\neq 0$,
\be \label{Ck_freediffusion}
C_k(t) \simeq \frac{4Tt}{N} \sum_{q=1}^{(N-1)/2} \cos\left( 2 \pi \frac{q k}{N} \right) = - \frac{2Tt}{N} \;.
\ee 
Taking additionally the center of mass into account, the particle motions are uncorrelated at these time scale\footnote{One indeed has $\langle (x_i(t) - x_i(0)) (x_{i+k}(t) - x_{i+k}(0)) \rangle = C_k(t) + \langle (\bar x(t) - \bar x(0))^2\rangle = 0$.}. 

The other time-scale which may play an important role is the largest one, $1/\mu_1$. It typically diverges with the system size (again for interactions which decrease with the distance). In theory it should correspond to the relaxation at the scale $L$ of the full-circle (or rather $L/2$, which is the maximum possible distance between particles due to the periodicity). 
However, depending on the interaction, it is possible that for some observables, the equilibration occurs earlier
in the large $N$ limit as we will discuss now.
 

From now on, we will focus on the limit of large $N$. 
One may imagine different ways to take this limit. In this paper we will mostly consider a fixed density scaling: we take the limits $N\to+\infty$ and $L\to+\infty$ simultaneously, with $\rho=N/L$ fixed. In this limit,
the inverse relaxation time spectrum $\mu_q$ defined in \eqref{eigenvals} becomes a function of the scaling variable $u=q/N$ (which is proportional to
the corresponding wavevector $p= 2 \pi q/L=2 \pi \rho u$) defined as (by symmetry we can restrict to $0<u<1/2$) 
\be \label{tilde_mu}
\tilde \mu(u)=\lim_{N\to+\infty} \mu_{q=uN} = 2 \sum_{\ell=1}^{+\infty} W''(\rho \ell) \sin^2(\pi u \ell) \;.
\ee
The above sum is convergent if the potential decays with the distance, or grows slower than linearly (corresponding to $s>-1$ in
the Riesz gas, see below). We restrict to this case in this paper. The large $N$ behavior of the observables defined above can then
be studied using this function. 
For instance, \eqref{disp_brownian} becomes, for any $t\ll 1/\mu_1$,
\be \label{disp_brownian_largeN}
C_0(t) = \langle (\delta x_i(t) - \delta x_i(0))^2 \rangle  \simeq 4T \int_0^{1/2} du \frac{1 - e^{- \tilde \mu(u) t}}{\tilde \mu(u)} \;.
\ee
This integral is well-defined for any function $\tilde \mu(u)$ of interest here. 

Let us now discuss
the time scale for equilibration. Independently of the formula \eqref{disp_brownian_largeN}, where the $N \to +\infty$ limit
has already be taken, 
we know from 
Eq. \eqref{disp_brownian} that for any $N$ and $t \to +\infty$, $C_0(t)$ converges to the limit
\be \label{disp_brownian_lim}
C_0(t) \to \frac{4T}{N} \sum_{q=1}^{(N-1)/2} \frac{1}{\mu_q} = 2 \langle \delta x_i^2 \rangle \;,
\ee 
and similarly $C_k(t) \to 2 \langle \delta x_i \delta x_{i+k} \rangle$. This happens once the particle displacement has completely equilibrated and no trace of the initial condition $\delta x_i(0)$ remains. 
The time-scale at which this happens strongly depends on the type of interaction. 
Indeed, the relevance of the largest time-scale $1/\mu_1$ depends on the behavior of $\tilde \mu(u)$ near $u=0$. If the integral $\int_0^{1/2} \frac{du}{\tilde \mu(u)}$ is convergent at $u=0$, 
then at large $N$ one has $\langle \delta x_i^2 \rangle \simeq 2T \int_0^{1/2} \frac{du}{\tilde \mu(u)}$, which is independent of $N$. 
In this case the expression \eqref{disp_brownian_largeN} remains valid at all times, and \eqref{disp_brownian_lim} in the large $N$ limit, coincides with the limit $t\to +\infty$ of \eqref{disp_brownian_largeN}. Then we see that the large time limit \eqref{disp_brownian_lim} is reached at finite time (of order $N^0$), and the only relevant time-scale is $1/\mu_{(N-1)/2} \simeq \tilde \mu(1/2)$.
The time-scale $1/\mu_1$ does not play any particular role because the sum in \eqref{disp_brownian} is never dominated by the large length-scales (small values of $q$). This happens if the interaction is very long range (for the Riesz gas case which we will consider in the rest of the paper it corresponds to $s<0$). On the other hand, if the integral $\int_0^{1/2} \frac{du}{\tilde \mu(u)}$
diverges at $u=0$, this generally means that $\langle \delta x_i^2 \rangle$ is an increasing function of $N$. Thus the time required to reach the limit \eqref{disp_brownian_lim} diverges when $N\to +\infty$. Indeed in this case, the large length-scales play a particular role, and the convergence to the large time limit requires convergence of the large length-scales, i.e. on a time-scale $1/\mu_1$. In this case, one observes a very broad third regime of intermediate times, $1/\mu_{(N-1)/2} \ll t \ll 1/\mu_1$, which strongly depends on the interaction.
\\

{\bf Remark:} Coming back to the scaling in the large $N$ limit, 
another possibility would be to fix the system size $L$ and to only take $N\to+\infty$, thus sending the density to infinity. In this case the interaction should be scaled appropriately so that the local equilibration time-scale $1/\mu_{(N-1)/2}$ remains finite. This will be discussed further below (see Appendix~\ref{app:other_scaling}).
\\

\subsection{Specialization to the Riesz gas}

\subsubsection{The periodic Riesz interaction}

From now on, we will specialize to the periodic Riesz gas. We choose the interacting potential (see below for its detailed form)
so that its derivative is given by 
\be \label{defRiesz}
W'(x) = \begin{dcases}  
-g \lim_{n\to\infty} \left(\sum_{m=-n}^n \frac{{\rm sgn}(x+m L)}{|x+mL|^{s+1}}  \right) \hspace{2.95cm} \text{for } -1<s < 0 \;, \\
-g \lim_{n\to\infty} \left(\sum_{m=-n}^n \frac{{\rm sgn}(x+m L)}{|x+mL|}  \right) = - g \frac{\pi}{L} \cot\big( \frac{\pi x}{L} \big)  \quad \text{for } s=0 \;, \\
- g \sum_{m=-\infty}^\infty \frac{{\rm sgn}(x+m L)}{|x+mL|^{s+1}} \hspace{4.25cm} \text{for } s>0 \;. \\
\end{dcases}
\ee 
These sums can also be expressed using Hurwitz's zeta function, which for $r>1$ is defined by $\zeta(r,a)=\sum_{k=0}^{\infty} (k+a)^{-r}$.
One has, for $0< x< 1$ and for $s>-1$,
\be 
W'(x) = - \frac{g}{L^{s+1}} \left(\zeta(1+s,\frac{x}{L} ) - \zeta(1+s,1-\frac{x}{L}) \right) \;.
\ee 
Note that each term as a pole $1/s$ at $s=0$, which however cancels in the difference so this function
is well defined for $s>-1$ (and equal to the limits indicated in \eqref{defRiesz}). 
In the case $s=0$ this interaction corresponds to the log-gas on a circle, and the equation of motion \eqref{Eq_def_brownian} identifies 
with the one of the Dyson Brownian motion on the circle with parameter $\beta=g/T$, see e.g. \cite{Spohn3,ForresterCircularBM}.
Its equilibrium Gibbs measure is $\propto \prod_{i<j} |\sin( \frac{\pi}{L} (x_i-x_j)|^\beta$, and it enjoys connections 
with the CUE$(\beta)$ random matrix ensemble and the Calogero-Sutherland model of interacting fermions~\cite{Spohn3}.
\\

{\bf Remarks:} 
\vspace*{0.5cm}

\noindent $\bullet$ An equivalent set of formulas for $s\notin \{ 1, 3,...\}$ for both the potential and its derivative are as follows
\bea \label{defRieszFourier}
&& W(x) = \frac{1}{2} \pi^{s-\frac{1}{2}} \frac{\Gamma(\frac{1-s}{2})}{\Gamma(1+\frac{s}{2})} \frac{g}{L^{s}} \sum_{p \neq 0} |p|^{s-1} \, e^{2 i \pi p \frac{x}{L}} = \pi^{s-\frac{1}{2}} \frac{\Gamma(\frac{1-s}{2})}{\Gamma(1+\frac{s}{2})} \frac{g}{L^{s}} {\rm Re} \, {\rm Li}_{1-s}(e^{2 i \pi \frac{x}{L}}) \;, \\
&& W'(x) = i \pi^{s+\frac{1}{2}} \frac{\Gamma(\frac{1-s}{2})}{\Gamma(1+\frac{s}{2})} \frac{g}{L^{s+1}} \sum_{p \neq 0} |p|^{s} \, {\rm sgn} (p) \, e^{2 i \pi p \frac{x}{L}} =  -2 \pi^{s+\frac{1}{2}} \frac{\Gamma(\frac{1-s}{2})}{\Gamma(1+\frac{s}{2})} \frac{g}{L^{s+1}} \, {\rm Im} \, {\rm Li}_{-s}(e^{2 i \pi \frac{x}{L}}) \;.
\eea 
The equivalence between these formulas and the sum over images in \eqref{defRiesz} is shown in Appendix~\ref{app:RieszCorresp}. 
\\

\noindent $\bullet$ In the case $s=-1$ the formula \eqref{defRieszFourier} leads to a well defined limit for the periodic potential $W(x)$, for $0<x<L$,
\be 
W(x) = g L \, \left( \frac{1}{6} - \frac{x}{L}\left(1-\frac{x}{L} \right)\right) \;,
\ee 
where we have used that ${\rm Re} {\rm Li}_{2}(e^{2 i \pi x}) = - \pi^2 x(1-x) + \frac{\pi^2}{6}$. 
At short distance it recovers the linear 1D Coulomb potential. This model was studied in \cite{Kunz} 
where it was shown that the system is a crystal at all temperature.


\subsubsection{The eigenvalues of the Hessian matrix}

Our starting point is the expression in Eq. \eqref{defRiesz} which, after differentiation, leads, for any $s>-1$, to
\be \label{W2_Riesz}
W''(x) = (s+1)g \sum_{m=-\infty}^\infty \frac{1}{|x+mL|^{s+2}} \;.
\ee
For this interaction potential, the eigenvalues of the Hessian given in \eqref{eigenvals} take the form (for any $N$) 
\be \label{eigRiesz0}
\mu_q = \frac{2(s+1)g}{L^{s+2}} \sum_{m=-\infty}^\infty \sum_{\ell=1}^{N-1} \frac{ \sin^2(\frac{\pi q\ell}{N})}{|\frac{\ell}{N}+m|^{s+2}} = \frac{4(s+1)gN^{s+2}}{L^{s+2}} \sum_{\ell=1}^\infty \frac{ \sin^2(\frac{\pi q\ell}{N})}{\ell^{s+2}} \;,
\ee
where the second equality is obtained by noting that we can combine both sums by extending the summation over $\ell$ to all integers, and using the symmetry $\ell \to - \ell$.
We can now write this expression as 
\be \label{mu_Riesz}
\mu_q = g\rho^{s+2} f_s\big( \frac{q}{N} \big) \quad \text{with} \quad f_s(u)= 4(s+1) \sum_{\ell=1}^{\infty} \frac{\sin^2(\pi \ell u)}{\ell^{s+2}} \;,
\ee
which is valid for any $N$ (hence the function $\tilde \mu$ defined above is equal to $\tilde \mu(u)=g\rho^{s+2} f_s(u)$). The function $f_s(u)$ can be written as 
\be  
f_s(u)= 
2 (s+1) \left( \zeta(2+s) -  {\rm Re} \, {\rm Li}_{2+s}(e^{2 i \pi u}) \right) \underset{s=2n}{=}  2 (s+1)  \zeta(2+s) (1 - \frac{B_{2+s}(u)}{B_{2+s}(0)} ) \;,
\ee 
where the $B_{2+s}(u)$ are Bernoulli polynomials and the last identity holds for $s=2n$ even positive integer.
In particular one has
\be \label{fs_even}
f_0(u) = 2\pi^2 u(1-u) \quad , \quad f_2(u) = 2 \pi^4 u^2 (1-u)^2 \quad , \quad f_4(u)= \frac{2}{9} \pi^6 u^2(1-u)^2 (1+2u(1-u)) \;.
\ee

For any $s>-1$, the function $f_s(u)$ is increasing on the interval $[0,\frac{1}{2}]$, from $f_s(0)=0$ to $f_s\big(\frac{1}{2}\big)=\sum_{\ell=1}^{\infty} \frac{4(s+1)}{(2\ell+1)^{s+2}}=4(1-2^{-(s+2)})(s+1)\zeta(s+2)$. In the limit $u\ll 1$, for $-1<s<1$, the sum over $\ell$ in \eqref{eigRiesz0}
can be replaced by an integral and we obtain the asymptotic behaviour
\be
f_s(u) \simeq 4(s+1) u^{s+1} \int_0^{+\infty} d\lambda \frac{\sin^2(\pi \lambda)}{\lambda^{s+2}} = 2\pi^{s+\frac{3}{2}} \frac{\Gamma(\frac{1-s}{2})}{\Gamma(1+\frac{s}{2})} u^{s+1}
\ee
(useful alternative expressions for this integral obtained via identities for the $\Gamma$-function are given in App.~\ref{app:integrals}, along with some other integrals used in the rest of the paper). This corresponds to the long-range (LR) case. For $s>1$, i.e. in the short-range (SR) case, the sum over $\ell$ in \eqref{eigRiesz0} is dominated by the first terms, so that one can expand the sine function to obtain $f_s(u) \simeq 4\pi^2 (s+1) \zeta(s) u^2$ for $u \ll 1$. Finally, in the marginal case $s=1$, one has
\be \label{fasympt_s1}
f_1(u) = 4\pi^2 (3-2\ln (2\pi u))u^2 + O(u^3) \;.
\ee
Leaving aside the case $s=1$, this can be summarized as
\be \label{fasympt}
f_s(u) \underset{u\to 0}{\sim} a_s u^{z_s} \quad \text{with} \quad z_s = \min(s+1,2) \quad \text{and} \quad a_s = \begin{cases} 2\pi^{s+\frac{3}{2}} \frac{\Gamma(\frac{1-s}{2})}{\Gamma(1+\frac{s}{2})} = \frac{2^{1+s} \pi^{2+s}}{\cos(\frac{\pi s}{2})\Gamma(1+s)} \quad \text{for } -1<s<1 \;, \\
4\pi^2 (s+1) \zeta(s) \quad \quad \quad \quad \quad \quad \quad \, \text{for } s>1 \;. \end{cases}
\ee
More precisely, for the SR case the series is $f_s(u)=a_s u^2 + \tilde a_s u^{1+s} + O(u^4)$ for $1<s<3$ where $\tilde a_s$ is the analytic
continuation of $a_s$ to $s>1$ (note that $\tilde a_s<0$ for any $1<s<3$).
As we will see below, $z_s$ gives the dynamical exponent. Finally, note that $f_s(u)$ has a finite limit as $s\to -1^+$, which is independent of $u$ at leading order,
\be
f_s(u) \underset{s\to-1^+}{=} 2 + 2\left(\gamma_E + \ln \left(2\sin(\pi u)\right)\right)(1+s) + O\left((1+s)^2\right) \;,
\ee
where $\gamma_E$ is Euler's constant. 

These results enable us to evaluate the relevant time-scales in the case of the Riesz gas: the local equilibration time is given by $1/\mu_{(N-1)/2} \sim 1/(g\rho^{s+2}) \equiv \tau $, while $1/\mu_1 \sim N^{z_s}/(g\rho^{s+2}) = N^{z_s} \tau$ corresponds to the global equilibration time (if different from the local one, i.e. for $s \geq 0$). Finally, as mentioned above we could have kept $L$ fixed when taking the limit $N\to +\infty$, and instead scaled $g$ as $N^{-(s+2)}$ in order to keep the time-scale $\tau = 1/(g\rho^{s+2})$ independent of $N$. All the results presented here would remain valid with this scaling. See Appendix \ref{app:other_scaling} for more details.
\\

{\bf Comparison with the harmonic chain.} One may also consider a harmonic chain of particles with interaction strength $K$, as was done in \cite{HarmonicChainRevABP,SinghChain2020,PutBerxVanderzande2019,HarmonicChainRTPDhar}. In this case the linear approximation is exact and the Hessian matrix reads (with the notations of \cite{SinghChain2020}) $H_{ij}=K(2\delta_{i,j}-\delta_{i,j+1}-\delta_{i,j-1})$ (with periodic conditions). 
It is diagonalized by the same eigenvectors $v^q_k=\frac{1}{\sqrt{N}} e^{2\pi i \frac{q}{N} k}$, with eigenvalues given by
\be \label{eigvals_harmonic_chain}
\mu_q = 4K \sin^2 \left(\frac{\pi q}{N}\right) \quad , \quad q=1,...,N-1 \;.
\ee
This means that the results of this paper can be applied to the harmonic chain, both for the Brownian and the RTPs,
by replacing $\tau=1/(g\rho^{s+2}) \to \tau_K = 1/K$ and $f_s(u) \to f_{\rm harmo}(u)=4\sin^2(\pi u)$.
In particular, one has for $u \to 0$, $f_{\rm harmo}(u) \simeq 4\pi^2 u^2$. This means that for the asymptotic regimes which are dominated by the smallest eigenvalues, i.e. for most results of this paper, the harmonic chain coincides with a short-range Riesz gas with the formal replacement $(s+1)\zeta(s) \to 1$.


\subsection{Variance, covariance and melting transition} \label{sec:var_brownian}

\subsubsection{Variance}

Let us start by examining the variance of the displacement of a given particle $\langle \delta x_i^2 \rangle$ at equilibrium. For $s>0$, the sum in \eqref{var_brownian} is dominated by small values of $q$ and we can write for large $N$
\be \label{var_Riesz_liquid}
\langle \delta x_i^2 \rangle \simeq \frac{2T}{g\rho^{s+2}} \frac{N^{z_s-1}}{a_s} \sum_{q=1}^{\infty} \frac{1}{q^{z_s}} = \begin{dcases} 
\frac{N^{s} T \zeta(s+1) \Gamma(1+\frac{s}{2})}{\pi^{s+\frac{3}{2}} \Gamma(\frac{1-s}{2}) g\rho^{s+2}} \quad \text{for } 0<s<1\;, \\ 
\frac{NT}{12 (s+1) \zeta(s) g\rho^{s+2}} \quad \quad \, \text{for } s>1\;.
\end{dcases}
\ee
Thus the variance diverges with the system size, suggesting a liquid phase. For $-1<s<0$, the sum in \eqref{var_Riesz_liquid} diverges. In this case, all the values of $q$ are relevant and we should instead replace the sum in \eqref{var_brownian} by an integral in the large $N$ limit,
\be \label{var_Riesz_solid}
\langle \delta x_i^2 \rangle \simeq \frac{2T}{g\rho^{s+2}} \int_0^{1/2} \frac{du}{f_s(u)} \;.
\ee 
Since $f_s(u) \sim a_s u^{s+1}$ the integral is well-defined. Thus, in this case the variance of the displacement is finite, which suggests that a solid phase exists at low temperature, as observed in \cite{Lelotte2023}. Hence we can estimate the melting transition temperature $T_M$ by writing that, at this temperature,
\be \label{melting_crit}
\rho^2 \langle \delta x_i^2 \rangle = c_L^2
\ee 
where $c_L$ is the Lindemann constant. This gives
\be \label{T_melting}
T_M = \frac{1}{2} g\rho^{s} \frac{c_L^2}{\int_0^{1/2} \frac{du}{f_s(u)}} \;.
\ee 
The melting temperature $T_M$ converges to $0$ for $s\to 0^-$, compatible with the numerical observations of \cite{Lelotte2023}. More precisely, for $s\to 0^-$ the integral in the denominator is dominated by the edge behavior and one has
\be
\int_0^{1/2} \frac{du}{f_s(u)} \simeq \frac{1}{a_s} \int_0^{1/2} \frac{du}{u^{s+1}} \sim \frac{1}{2\pi^2|s|} \;.
\ee
To leading order we thus have that $T_M$ vanishes as $s \to 0$ as
\be \label{TM_small_s}
T_M \underset{s\to 0^-}{\sim} \pi^2 c_L^2 g |s| \;.
\ee
The expression \eqref{T_melting} should thus be accurate for $s$ near zero (since the approximation method used in this paper is valid at small temperatures). However, for $s\to -1^+$, we find that the ratio $T_M/c_L^2$ has a finite limit $2 g \rho^{-1}$. In addition, it is concave and has a local maximum on $[-1,0]$, see Fig. \ref{Fig_TM}. 
As discussed in the introduction, because of the rigorous result $T_M=+\infty$ for $s=-1$ \cite{Kunz}, and the numerical results of \cite{Lelotte2023}, these arguments are incompatible with a Lindemann coefficient which would be constant over the whole interval $s \in [-1,0]$. 




The limiting cases $s=0$ and $s=1$ have to be treated separately. The case $s=1$ belongs to the liquid phase. In this case the sum is still dominated by small $q$, and we can use \eqref{fasympt_s1} to obtain
\be \label{var_Riesz_s1}
\langle \delta x_i^2 \rangle \simeq \frac{N}{\ln N} \frac{T}{24 \, g\rho^3} \;.
\ee
The log-gas case $s=0$ marks the frontier between the liquid and solid regimes. 
For the log-gas it was shown that there is no crystallisation transition \cite{leble_loggas},
and that the system is in a liquid regime with translational quasi-order (see below). 
In this case, we can use the exact expression for $f_0(x)$ given in \eqref{fs_even}, which leads to
\be \label{varloggas}
\langle \delta x_i^2 \rangle = \frac{T}{2\pi^2 g\rho^2} \sum_{q=1}^{N-1} \frac{N}{q(N-q)} =\frac{T}{\pi^2 g\rho^2} \sum_{q=1}^{N-1} \frac{1}{q} = \frac{T}{\pi^2 g\rho^2} (\ln N + \gamma_E + O(N^{-1})) \;,
\ee
where $\gamma_E$ is Euler's constant. Thus at large $N$ the variance of the displacement diverges, as in the liquid regime.
\\

\subsubsection{ Covariance (macroscopic regime)} 

Let us now consider the covariance
\be \label{covstat_brownian}
\langle \delta x_i \delta x_{i+k}\rangle = \frac{2T}{N}  \sum_{q=1}^{(N-1)/2} \frac{\cos\left(2\pi \frac{q}{N} k\right)}{\mu_q} = \frac{2T}{Ng\rho^{s+2}}  \sum_{q=1}^{(N-1)/2} \frac{\cos\left(2\pi \frac{q}{N} k\right)}{f_s(\frac{q}{N})} \;.
\ee
In the large $N$ limit, if $k\ll N$, for any $s\geq 0$ one has $\langle \delta x_i \delta x_{i+k}\rangle \simeq \langle \delta x_i^2 \rangle$ at leading order in $N$, and one needs to scale $k\sim N$ to get a non-trivial result. Hence in this subsection we are studying the covariance in the macroscopic regime,
at the scale of the full circle.
In this case, the sum is dominated by small values of $q$. Writing $k=\kappa N$ with $0\leq\kappa\leq 1$, the expression \eqref{covstat_brownian} becomes at large $N$ (for $s\neq 1$, $s\geq 0$)
\be \label{cov_brownian_general}
\langle \delta x_i \delta x_{i+k}\rangle \simeq \frac{2TN^{z_s-1}}{g\rho^{s+2}a_s}  \sum_{q=1}^{\infty} \frac{\cos(2\pi \kappa q )}{q^{z_s}} \;.
\ee
In the short-range case $s>1$, this can be computed explicitly using that $\sum_{q=1}^\infty \frac{\cos(q\theta)}{q^2} = \frac{\pi^2}{6} + \frac{|\theta|}{2}\left(\frac{|\theta|}{2}-\pi\right)$ for $\theta \in [-2\pi,2\pi]$, and one obtains
\be \label{cov_shortrange}
\langle \delta x_i \delta x_{i+k}\rangle \simeq \frac{NT}{12(s+1)\zeta(s) g\rho^{s+2}} (1 - 6\kappa(1-\kappa)) \quad , \quad \kappa=\frac{k}{N} \;.
\ee
For $\kappa=0$ we recover the variance as given in \eqref{var_Riesz_liquid}. The covariance then decreases as a function of $\kappa$ until it reaches a negative minimum at $\kappa=1/2$,
\be
\langle \delta x_i \delta x_{i+N/2}\rangle \simeq -\frac{NT}{24(s+1)\zeta(s) g\rho^{s+2}} = -\frac{1}{2} \langle \delta x_i^2 \rangle \;.
\ee
Note that the integral of this covariance over the whole system vanishes as it should since $\sum_i \delta x_i=0$. 
As a function of $\kappa$, Eq. \eqref{cov_shortrange} identifies with
the covariance of a Brownian bridge conditioned to have a zero total integral. 

For $s=1$ we have, using \eqref{fasympt_s1},
\be \label{covar_s1}
\langle \delta x_i \delta x_{i+k}\rangle \simeq \frac{N T}{4\pi^2 g\rho^3 \ln N}  \sum_{q=1}^{\infty} \frac{\cos(2\pi \kappa q )}{q^2} = \frac{N}{\ln N} \frac{T}{24 g\rho^3} (1 - 6\kappa(1-\kappa)) \;.
\ee

For $0<s<1$, Eq.~\eqref{cov_brownian_general} reads
\be \label{cov_kappa_longrange}
\langle \delta x_i \delta x_{i+k}\rangle \simeq \frac{\Gamma(1+\frac{s}{2})}{\pi^{s+\frac{3}{2}} \Gamma(\frac{1-s}{2})} \frac{TN^s}{g\rho^{s+2}}  \sum_{q=1}^{\infty} \frac{\cos(2\pi \kappa q )}{q^{s+1}} \;,
\ee
which can be also expressed using polylogarithm functions. We can obtain the asymptotics for $\kappa \ll 1$, i.e. for $1\ll k\ll N$. In this limit
the dependence in $\kappa$ of the covariance is proportional to $1 + O(\kappa^s)$ and one obtains
\be \label{matchvariance}
\langle \delta x_i^2 \rangle - \langle \delta x_i \delta x_{i+k}\rangle = \frac{1}{2} D_k(0) \simeq \frac{T k^s}{2\pi \tan \left( \frac{\pi s}{2} \right) g \rho^{s+2}} \;,
\ee
which matches the large $k$ limit of the fixed $k$ expression in \eqref{gapss} below. 
The minimal value of the covariance, obtained for $\kappa=1/2$, is
\be
\langle \delta x_i \delta x_{i+N/2}\rangle \simeq \frac{\Gamma(1+\frac{s}{2})}{\pi^{s+\frac{3}{2}} \Gamma(\frac{1-s}{2})} \frac{TN^s}{g\rho^{s+2}} \sum_{q=1}^{\infty} \frac{(-1)^q}{q^{s+1}} = - (1-2^{-s}) \frac{N^{s} T \zeta(s+1) \Gamma(1+\frac{s}{2})}{\pi^{s+\frac{3}{2}} \Gamma(\frac{1-s}{2}) g\rho^{s+2}} = - (1-2^{-s}) \langle \delta x_i^2 \rangle \;,
\ee
which is again negative for any $0<s<1$.

For the log-gas $s=0$, we have for $0<\kappa<1$
\be \label{covarianceloggas}
\langle \delta x_i \delta x_{i+k}\rangle \simeq \frac{T}{\pi^2 g\rho^2}  \sum_{q=1}^{\infty} \frac{\cos(2\pi \kappa q )}{q} = - \frac{T}{\pi^2 g\rho^2} \ln (2\sin(\pi \kappa)) \;.
\ee
Once again this is negative for $\kappa=1/2$,
\be
\langle \delta x_i \delta x_{i+N/2}\rangle \simeq - \frac{T \ln 2}{\pi^2 g\rho^2} \;.
\ee
Note that \eqref{covarianceloggas} diverges for $\kappa \to 0$. Indeed we see that the covariance between particles separated by $k \sim N$ is of order $N^0$, while the variance \eqref{varloggas} is of order $\ln N$. However, taking $\kappa=k/N$ with $k$ of order 1 correctly recovers the leading term in $\ln N$ of the variance. 
Note that formulas similar to \eqref{covarianceloggas} appear in the context of counting statistics of eigenvalues of unitary random matrices \cite{Keating,NajnudelCircle2018,Smith2021}.
\\


For $s<0$, the situation is radically different since the covariance remains finite at large $N$. Indeed one has for $N \gg 1$ with $k\ll N$
\be \label{cov_sneg}
\langle \delta x_i \delta x_{i+k}\rangle \simeq \frac{2T}{g\rho^{s+2}}  \int_0^{1/2} du \frac{\cos(2\pi k u)}{f_s(u)} \;.
\ee
For large $k$ (but still with $k\ll N$) we thus have
\be \label{cov_sneg_largek}
\langle \delta x_i \delta x_{i+k}\rangle \simeq \frac{2T k^s}{g\rho^{s+2}a_s}  \int_0^{+\infty} dv \frac{\cos(2\pi v)}{v^{s+1}} \simeq \frac{2T k^{-|s|}}{g\rho^{s+2}a_s} (2\pi)^{-|s|} \cos(\frac{\pi |s|}{2}) \Gamma(|s|) \;.
\ee
Thus in this case the covariance decays 
on a scale of $k \sim N^0$. For $k$ of order $N$, the covariance is still given by \eqref{cov_kappa_longrange} and is very small, of order $N^{-|s|}$.

\subsection{Gap statistics and validity of the approximation} \label{sec:gapstat}

\subsubsection{Variance of the gaps}

We will now look at some statistics of the gaps, i.e. the distance between particles, at equilibrium. We first consider the variance of the distance between particles $i$ and $i+k$, given by \eqref{gap_brownian}. At large $N$, for any $k\ll N$, it reads, for any $s>-1$,
\be \label{gaps_step1}
D_k(0) = \langle (\delta x_{i}-\delta x_{i+k})^2 \rangle \simeq \frac{8T}{g\rho^{s+2}} \int_0^{1/2} du \frac{\sin^2(\pi k u)}{f_s(u)}  \;.
\ee
This integral is always well-defined since $f_s(u)=O(u^2)$ for $u\to 0$. For large $k$ (but still of order $N^0$), we must distinguish $s>0$
and $s<0$. In the case $s>0$ ($s\neq 1$), Eq. \eqref{gaps_step1} behaves as
\be \label{gapss}
D_k(0) \simeq \frac{8T}{g\rho^{s+2}k} \int_0^{k/2} dv \frac{\sin^2(\pi v)}{f_s\big(\frac{v}{k}\big)} \simeq \frac{8Tk^{z_s-1}}{g\rho^{s+2}a_s} \int_0^{+\infty} dv \frac{\sin^2(\pi v)}{v^{z_s}} \simeq \frac{4T}{g\rho^{s+2}a_s} \frac{\pi^{z_s-\frac{1}{2}}}{z_s-1} \frac{\Gamma(\frac{3-z_s}{2})}{\Gamma(\frac{z_s}{2})}  k^{z_s-1} \;,
\ee
where we recall that $z_s=\min(s+1,2)$ and $a_s$ is given in \eqref{fasympt}. In summary one finds
\be \label{gapscases}
D_k(0) \simeq \begin{dcases} \frac{T k^s}{\pi \tan \left( \frac{\pi s}{2} \right) g \rho^{s+2}} \quad \, \text{for } 0<s<1 \;, \\
\frac{T k}{(s+1) \zeta(s) g\rho^{s+2}} \quad \text{for } s>1 \;. \end{dcases}
\ee
The field of displacements thus exhibits a roughness exponent $\zeta=s/2$ in the LR case and $\zeta=1/2$ in the SR case. This result can be compared with mathematical works (see discussion in Section \ref{main_results_brownian}). 

For $s=1$, using \eqref{fasympt_s1} we obtain for $1 \ll k \ll N$
\be
D_k(0) \simeq \frac{2Tk}{\pi^2g\rho^3} \int_0^{k/2} dv \frac{\sin^2(\pi v)}{(3-2\ln (2\pi v)+2\ln k)v^2} \simeq \frac{Tk}{\pi^2g\rho^3 \ln k} \int_0^{+\infty} dv \frac{\sin^2(\pi v)}{v^2} = \frac{Tk}{2g\rho^3 \ln (k)} \;.
\ee

For the case $s=0$ we can use $f_0(u)=2\pi^2 u(1-u)$ and we find for $k \ll N$
\bea \label{gaps_s0}
D_k(0) \simeq \frac{4 T}{\pi^2 g \rho^2} \int_0^{1/2} du \frac{\sin^2(\pi k u)}{u(1-u)} 
= \frac{2 T}{\pi^2 g \rho^2} (\ln (2 \pi k)-\text{Ci}(2 \pi k)+\gamma_E) \underset{k\to+\infty}{\simeq} \frac{2 T}{\pi^2 g \rho^2} \ln(k) \;.
\eea

Since there is a relation between the gaps and the counting statistics (see below), we can now compare this formula to the known result for the variance of the number of particles inside a fixed interval $[a,b]$ for the log-gas, ${\cal N}_{[a,b]}$. Let us recall that for a given interval $[a,b]$ 
of microscopic size the variance behaves, for $\rho (b-a) \gg 1$, as
\be \label{varinterval}
{\rm Var} \, {\cal N}_{[a,b]} \simeq \frac{2}{\pi^2 \beta} (\ln (\pi \rho (b-a)) + c_\beta) \;,
\ee 
where $\beta=g/T$ and $c_\beta$ is a constant. For $\beta=1,2,4$ this is the Dyson-Mehta formula \cite{Mehta_book,Forrester_book}.
For general $\beta$, the leading term was proved in \cite{Bourgade21} and 
the constant $c_\beta$ was computed for any $\beta$ in 
\cite{Smith2021}. We can check that the leading logarithmic behavior in  \eqref{varinterval}
is in agreement with our result in \eqref{gaps_s0}
if we use the relation $\beta=g/T$ and
the estimate, valid for sufficiently large intervals,
\be 
{\rm Var} \, {\cal N}_{[a,b]} \simeq \rho^2 \langle (\delta x_i-\delta x_{i+k})^2 \rangle \;,
\ee 
where $a=x_i^0$, $b=x_{i+k}^0$, hence $\rho(b-a)=k$.

For $s<0$ the variance of the gaps saturates to a constant at large $k \gg 1$ (replacing $\sin^2(\pi k u) \to 1/2$ in \eqref{gaps_step1})
\be \label{gaps_sneg}
D_k(0) \simeq \frac{4T}{g\rho^{s+2}} \int_0^{1/2}  \frac{du}{f_s(u)}  = 2 \langle \delta x_i^2 \rangle \;,
\ee
as the displacements become independent at large distance each with variance given in \eqref{var_Riesz_solid}.
\\

Let us now discuss the fluctuations of the gaps in the macroscopic regime $k \sim N$ and for $s>0$.
As we have already noted in the previous subsection, the variance of the gaps is related to the covariance 
which we have computed in Section~\ref{sec:var_brownian}. For $k$ of order $N$, i.e. for $k=\kappa N$ with $0\leq \kappa \leq 1$, we thus have for $s>0$ (see \eqref{cov_brownian_general})
\be 
D_k(0) = \langle (\delta x_i - \delta x_{i+k})^2 \rangle = 2(\langle \delta x_i^2 \rangle - \langle \delta x_i \delta x_{i+k} \rangle) \simeq \frac{4TN^{z_s-1}}{g\rho^{s+2}a_s}  \sum_{q=1}^{\infty} \frac{1-\cos(2\pi \kappa q )}{q^{z_s}} \;.
\ee 
In particular, in the short range case $s > 1$ this tells us how the gap variance behaves at large $k$,
\be
D_k(0) \simeq \frac{NT}{(s+1)\zeta(s) g\rho^{s+2}} \kappa(1-\kappa) \;.
\ee


\subsubsection{Validity of the approximation} 

At the beginning of this paper we made the assumption that the system of Brownian particles only undergoes small relative Gaussian displacements around its ground state. For the Riesz gas, we can make this hypothesis more precise using the results of the present section. In this case, the assumption \eqref{cond_approx} becomes, using \eqref{W2_Riesz},
\be \label{cond_approx_Riesz1}
\forall 1\leq k\leq N-1, \ \sqrt{\langle (\delta x_i-\delta x_{i+k})^2\rangle} \ll 2\left|\frac{W''(x_i^0-x_{i+k}^0)}{W'''(x_i^0-x_{i+k}^0)}\right| = \frac{2L}{s+2} \left| \frac{\zeta(2+s,\frac{k}{N})+\zeta(2+s,1-\frac{k}{N})}{\zeta(3+s,\frac{k}{N})-\zeta(3+s,1-\frac{k}{N})} \right| \;,
\ee
where here $\zeta(a,x)$ denotes the Hurwitz zeta function. The r.h.s of \eqref{cond_approx_Riesz1} diverges for $k/N \to 1/2$, and increases faster than linearly as a function of $k/N$ up to $1/2$. Hence a sufficient condition for \eqref{cond_approx_Riesz1} is 
\be
\forall 1\leq k\leq (N-1)/2, \ 
\sqrt{D_k(0)} \ll \frac{2k}{(s+2)\rho} \;.
\ee
Thus, since we have shown that $\sqrt{D_k(0)}$ grows slower than linearly in $k$
, one only needs to check the condition for $k=1$~\footnote{We have also checked numerically that $\frac{D_k(0)}{k^2} \propto \frac{1}{k^2} \int_0^{1/2} du \frac{\sin^2(\pi k u)}{f_s(u)}$ is a decreasing function of $k$.} ,
\be \label{cond_approx_Riesz2}
\rho \sqrt{D_1(0)} 
\ll \frac{2}{s+2} \;.
\ee
Using \eqref{gaps_step1} we find that the condition for the linear approximation to be accurate reads
\be \label{validity_brownian}
T \ll T_G = A_s  g\rho^s   \quad , \quad A_s = \frac{1}{2(s+2)^2\int_0^{1/2} du \frac{\sin^2(\pi u)}{f_s(u)}} \;,
\ee  
and one has for instance $A_0=2.02441$ and $A_2=0.883237$, as well as $A_s \to 4$ as $s\to -1$.

In the case $s<0$ it is interesting to compare $T_G$ and the melting temperature $T_M$. Using \eqref{T_melting}
one finds the ratio
\be 
\frac{T_M}{T_G} =  c_L^2 (s+2)^2 \frac{\int_0^1 du \frac{\sin^2(\pi u)}{f_s(u)}}{\int_0^1 du \frac{1}{f_s(u)}} < 1 \;,
\ee 
since we can assume that $c_L<1/2$. As $s \to 0^-$ one finds $T_M/T_G \simeq c_L^2 \pi^2 |s|/A_0 \approx 4.8753 c_L^2 |s|$.
This confirms that for small $s$ the linear approximation should work very well up to the estimated melting temperature.

\subsubsection{Translational order correlation function}

The translation order correlation function
is usually defined as
\be 
S(k) = \langle e^{ \frac{2 i \pi}{a} (\delta x_{j+k}-\delta x_j) } \rangle = \langle e^{ 2 i \pi \rho (\delta x_{j+k}-\delta x_j) } \rangle \;,  
\ee 
where the lattice spacing is here $a=1/\rho$. 
True translational order is present when $S(k)$ decays to a non-zero constant at large $k$, and absent if it
decays to zero. If it decays to zero as a power law in $k$, it is usually called quasi-ordered.
For the Brownian particles, we can evaluate this correlation function 
within our linear Gaussian approximation. It is then related to the variance of the gap
as follows 
\be \label{gaussiandecay} 
S(k) 
\simeq e^{ - 2 \pi^2 \rho^2 \langle (\delta x_{j+k}-\delta x_j)^2 \rangle } \;.
\ee 

We can use our results above to determine the behavior of $S(k)$ at large $k$. 
The case of the log-gas $s=0$ is marginal, and one obtains from \eqref{gaps_s0} the power law decay
\be
S(k)  \underset{k \gg 1}{\simeq} \,    (e^{\gamma_E} 2 \pi k)^{-\frac{4 T}{g}} \;.
\ee 
One recovers the quasi-order of the log-gas on the circle. The exponent $4 T/g=4/\beta$ is in agreement
with the known results for the decay of the oscillating part of the density correlation \cite{Haldane,Forrester1984}, 
see Ref. \cite{Lelotte2023} for a recent 
discussion. 

For $0<s<1$ we see, from \eqref{gaussiandecay} and \eqref{gapscases}, that the translational order correlation 
decays to zero at large $k$ as a stretched exponential,
$\log S(k) \propto - k^s$, while the decay is exponential for $s>1$. 

Finally for $s<0$, we find true translational order at low temperature within our linear Gaussian approximation, and the translational order correlation 
saturates at large $k$ to the non-zero value
\be
S(k) \underset{k \gg 1}{\simeq}
e^{ - 8 \pi^2 \frac{T}{g\rho^{s}} \int_0^{1/2} \frac{du}{f_s(u)}
} \;,
\ee 
which takes values $\exp( - 4 \pi^2 c_L^2)$ near melting. It remains a challenge to compute it in the high temperature phase.

\subsubsection{Spatial correlations of the gaps} 

Let us now consider the correlations of the gaps \eqref{gapcor_brownian}. In the large $N$ limit with $k$ and $n\ll N$, we have
\be \label{gap_spacecorr_largeN}
D_{k,n}(0) = \langle (\delta x_{i}-\delta x_{i+k}) (\delta x_{i+n}-\delta x_{i+n+k}) \rangle \simeq \frac{8T}{g\rho^{s+2}} \int_0^{1/2} du \frac{\sin^2(\pi k u)}{f_s(u)} \cos(2\pi n u) \;.
\ee
Let us first consider the regime where both $k$ and $n$ are large. In that regime this integral is dominated by small values of $u$, so that we can use the asymptotic expression of $f_s(u)$ \eqref{fasympt}. In the short-range case $s>1$, we find for $k,n \gg 1$ (recalling that $k$ and $n$ are positive integers)
\be \label{Dkn_step1}
D_{k,n}(0) \simeq \frac{8T}{g\rho^{s+2}a_s} \int_0^{+\infty} du \frac{\sin^2(\pi k u)}{u^2} \cos(2\pi n u) =  \begin{cases} 0 \hspace{2.37cm} \text{if } k\leq n \;, \\ \frac{4 \pi^2 T}{g\rho^{s+2}a_s} (k-n) \quad \text{if } k>n \;, \end{cases}
\ee
i.e. at this order of approximation there are no correlations between the gaps if the two intervals do not overlap. For $s=2$ we can use the exact expression of $f_s(u)$ given in \eqref{fs_even} to obtain a more precise result in the case $k\leq n$. In addition, we can subtract the integral above since we know that it evaluates to zero in this case, and use that $1/(u^2(1-u)^2) - 1/u^2 = \frac{2-u}{u(1-u)^2}\simeq 2/u$ for small $u$. This leads to, for $k,n,n-k \gg 1$, and fixed ratio 
$k/n < 1$,
\be \label{s2} 
D_{k,n}(0) \simeq \frac{8T}{\pi^4 g\rho^4} \int_0^{+\infty} du \frac{\sin^2(\pi k u)}{u} \cos(2\pi n u) = \frac{2T}{\pi^4 g\rho^4} \ln |1-\frac{k^2}{n^2}| \quad , \quad \frac{k}{n} < 1 \quad , \quad s=2 \;.
\ee
The divergence at $k/n \to 1$ is regularized by including the neglected terms, as is done below in details for the case $s=0$.

More generally, in the short-range case, for $1<s<3$, using the same method,
we can subtract the integral above since we know that it evaluates to zero in this case, and use that $1/(f_s(u)) - 1/(a_s u^2) = 
(a_s u^2 - f_s(u))/(a_s u^2 f_s(u) ) \simeq - \tilde a_s/(a_s^2 x^{3-s})$
for small $u$ which for $k,n \gg 1$ leads to, for $k/n < 1$,
\bea  \label{gapsSR}
D_{k,n}(0) &\simeq& -\frac{8T \tilde a_s}{g\rho^{s+2} a_s^2} \int_0^{+\infty} du \frac{\sin^2(\pi k u)}{u^{3-s}} \cos(2\pi n u) \nn \\ &=& \frac{8T |\tilde a_s|}{g\rho^{s+2} a_s^2} \frac{\pi^{2-s}}{2^s} \cos(\frac{\pi s}{2}) \Gamma(s-2) \left(|n-k|^{2-s} + (n+k)^{2-s} -2n^{2-s} \right) \;,
\eea
which reproduces \eqref{s2} for $s \to 2$ using $a_2=2 \pi^4$, $\tilde a_2=- 4 \pi^4$. Note that this correlation
is negative for $1<s \leq 2$ and positive for $2<s<3$. Similar estimates exist in each interval $2m-1 < s < 2m+1$, for $m \geq 2$.

In the marginal case $s=1$ we have
\bea
D_{k,n}(0) &\simeq& \frac{2T}{\pi^2 g\rho^3} \int_0^{+\infty} du \frac{\sin^2(\pi k u)}{(3-2\ln (2\pi u))u^2} \cos(2\pi n u)
\simeq \frac{2Tk}{\pi^2 g\rho^3 \ln k} \int_0^{+\infty} du \frac{\sin^2(\pi v)}{v^2} \cos(2\pi \frac{n}{k} v) \nn \\
&=& \begin{cases} 0 \hspace{2.28cm} \text{if } k\leq n \\ \frac{T}{g\rho^3 \ln k} (k-n) \quad \text{if } k>n \end{cases} \;. 
\eea
\\

In the long-range case $0<s<1$, one finds for $n,k \gg 1$
\be \label{LRgapcorr}
D_{k,n}(0) \simeq \frac{8T}{g\rho^{s+2}a_s} \int_0^{+\infty} du \frac{\sin^2(\pi k u)}{u^{s+1}} \cos(2\pi n u) \simeq \frac{T}{g\rho^{s+2}a_s} \frac{\pi^{s+1}}{2^{s}} \frac{|n-k|^s + (n+k)^s -2n^s}{\sin(\frac{\pi s}{2})\Gamma(s+1)} \;.
\ee

Finally, in the case of the log-gas, $s=0$, we find for $k,n, |k-n| \gg 1$ all three being of the same order,
\be \label{gap_spacecorr_s0_large}
D_{k,n}(0) \simeq \frac{4T}{\pi^2 g\rho^2} \int_0^{+\infty} du \frac{\sin^2(\pi k u)}{u} \cos(2\pi n u) = \frac{T}{\pi^2 g\rho^2} \ln |1-\frac{k^2}{n^2}| \;,
\ee
which changes sign for $k/n=\sqrt{2}$. This formula ceases to be valid when $|k-n|= O(1)$, the apparent divergence at $k=n$ being an effect of the approximation used. 
In that case, the integral in \eqref{gap_spacecorr_largeN} is no longer dominated by small values of $u$ and one has to use the exact expression $f_s(u)=2\pi^2 u(1-u)$. For arbitrary integers $k>0$ and $n>0$, we find for $k \neq n$ the exact expression
\bea \label{gap_spacecorr_s0_exact}
D_{k,n}(0) &\simeq& \frac{4T}{\pi^2 g\rho^2} \int_0^{1/2} du \frac{\sin^2(\pi k u)}{u(1-u)} \cos(2\pi n u) \nn \\
&=& \frac{T}{\pi^2 g\rho^2} \left( \ln |1-\frac{k^2}{n^2}| + 2 \text{Ci}(2\pi n) - \text{Ci}(2\pi (n+k)) - \text{Ci}(2\pi |n-k|) \right) \;, 
\eea
while for $k=n$, one finds
\be
D_{k,n}(0) \simeq \frac{T}{\pi^2 g\rho^2} (2 \text{Ci}(2 \pi k )-\text{Ci}(4 \pi k )-\ln (\pi  k)-\gamma_E )
\simeq -\frac{T}{\pi^2 g\rho^2} (\log (\pi  k)+\gamma_E ) \;,
\ee
which corresponds to the limit $k-n \to 0$ in \eqref{gap_spacecorr_s0_exact}. Finally, in the limit where $k$, $n$ and $|n-k|$ are large Eq. \eqref{gap_spacecorr_s0_exact} recovers \eqref{gap_spacecorr_s0_large}. 
\\

One can also study the regime where $k$ remains fixed and $n$ becomes large, as discussed
in Section \ref{main_results_brownian}. The result in the LR case can be correctly
recovered from \eqref{LRgapcorr} by considering $k \ll n$ and expanding the sinus square term.
It gives \eqref{Dkn_large_n}.


\subsection{Time evolution of the displacement} \label{sec:disp}

Let us now study the time evolution of the displacement of a particle. More precisely, we consider the variance of this quantity, given in \eqref{disp_brownian}. We have seen in section \ref{sec:observables} that at times $t\ll \tau$, the particle diffuses freely (see \eqref{C0_freediff}). For $s<0$, the displacement saturates at a finite time $t\sim \tau$ to reach its stationary value, $2\langle \delta x_i^2 \rangle \sim T \tau$ -- see Eq. (\ref{var_Riesz_solid}). On the other hand, for $s>0$ it takes a much larger time $\sim N^{z_s} \tau$ to reach this stationary value. In between these two times, i.e. for $\tau \ll t \ll N^{z_s} \tau$, there is a very broad intermediate time regime where the time evolution of the particles strongly depends on the interaction. In the large $N$ limit, this regime is well described by \eqref{disp_brownian_largeN}, which in the case of the Riesz gas reads
\be \label{disp_step1}
C_0(t) = \langle (\delta x_i(t) - \delta x_i(0))^2 \rangle 
\simeq \frac{4T}{g\rho^{s+2}} \int_0^{1/2} du \frac{1 - e^{- g\rho^{s+2} f_s(u) t}}{f_s(u)} \;.
\ee
This function describes the crossover between the short time regime and the large time regime. Let us discuss
them now separately.

At short time $t \ll \tau$, one can expand the exponential in \eqref{disp_step1} and one finds 
$C_0(t) \simeq 2 T t$ (as was found in \eqref{C0_freediff} for arbitrary $N$). 

At large time $t \gg \tau=1/(g\rho^{s+2})$, this integral is dominated by small values of $u$, and we can use \eqref{fasympt} to write (for $s\neq 1$)
\be
C_0(t) 
\simeq 4Tt \int_0^{1/2} du \frac{1 - e^{- g\rho^{s+2} a_s u^{z_s} t}}{g\rho^{s+2} a_s u^{z_s} t} \simeq \frac{4Tt}{(g\rho^{s+2} a_s t)^{1/z_s}} \int_0^{+\infty} dv \frac{1-e^{-v^{z_s}}}{v^{z_s}} = \frac{4Tt^{\frac{z_s-1}{z_s}}}{(g\rho^{s+2} a_s)^{1/z_s}} \frac{\Gamma(1/z_s)}{z_s-1}  \;,
\ee
which leads to
\be \label{displacement_Riesz}
C_0(t) \simeq \begin{dcases} U_s \frac{T \, t^{\frac{s}{s+1}}}{g^{\frac{1}{s+1}}\rho^{\frac{s+2}{s+1}}}
\hspace{2.08cm} \text{for } 0<s<1 \\ 
\frac{2T}{\sqrt{\pi (s+1) \zeta(s)}} \sqrt{\frac{t}{g\rho^{s+2}}}
\quad \text{for } s>1 \end{dcases} \quad 
\text{with} \quad
U_s = \frac{ 4\Gamma\left(\frac{1}{s+1}\right)}{\pi s} \left[ \frac{\Gamma\left(1+\frac{s}{2}\right)}{2\sqrt{\pi} \, \Gamma\left( \frac{1-s}{2} \right)} \right]^{\frac{1}{s+1}} \;.
\ee
As discussed in Section \ref{main_results_brownian}, although our method here is completely different, our result exactly coincides with formulas (2) and (3) in \cite{DFRiesz23} (at any $T$ for $0<s<1$ and at $T\ll g\rho^s$ for $s>1$). Note that here we have assumed that at $t=0$ the system is at equilibrium, which corresponds to the ``annealed'' initial condition. The ``quenched'' case can also be treated in this setting, which we do in Section~\ref{sec:quenched_brownian}.
In the short-range case, we find the $\sqrt{t}$ behaviour of single-file diffusion as expected.

Once again, we need to treat separately the marginal cases $s=1$ and $s=0$. For $s=1$, using the small $u$ asymptotics \eqref{fasympt_s1} in \eqref{disp_step1}, we obtain (again for $t \gg \tau = 1/(g\rho^3)$) 
\bea \label{disp_s1}
C_0(t) 
&\simeq& 4Tt \int_0^{1/2} du \frac{1 - e^{- 4\pi^2 g\rho^3 u^2 (3-2\ln(2\pi u)) t}}{4\pi^2 g\rho^3 u^2 (3-2\ln(2\pi u)) t} \simeq \frac{2T}{\pi} \sqrt{\frac{t}{g\rho^3}} \int_0^{+\infty} dv \frac{1-e^{-v^2(3-2\ln v+\ln(g\rho^3 t))}}{v^2(3-2\ln v+\ln(g\rho^3 t))} \nn \\
&\simeq& \frac{2T}{\pi} \sqrt{\frac{t}{g\rho^3}} \int_0^{+\infty} dv \frac{1-e^{-v^2 \ln(g\rho^3 t)}}{v^2\ln(g\rho^3 t)} 
= 2T \sqrt{\frac{t}{\pi g\rho^3 \ln(g\rho^3 t)}}
\;.
\eea
\\

In the special case of the log-gas one can even compute analytically the full crossover function
between short time and large time. Indeed, for $s=0$, \eqref{disp_step1} reads without further approximation (using \eqref{fs_even})
\be
C_0(t) \simeq 4T \int_0^{1/2} du \frac{1 - e^{- 2\pi^2 g\rho^2 u(1-u) t}}{2\pi^2 g\rho^2 u(1-u)} \;.
\ee
This can be computed exactly by noticing that
\be \label{integral_log_gas_disp}
\partial_t C_0(t) 
\simeq 4T \int_0^{1/2} du \, e^{- 2\pi^2 g\rho^2 u(1-u) t} = \frac{T \sqrt{2} \, e^{-\frac{\pi^2 g\rho^2 t}{2}} \text{erfi}\left(\sqrt{\frac{\pi^2 g\rho^2 t}{2}}\right)}{\pi^{3/2} \sqrt{g\rho^2 t}} \;.
\ee
Integrating over $t$ (taking into account that the displacement should obviously be zero at $t=0$) we find
\be \label{crossoverhypergeo}
C_0(t) \simeq 2T t  \times \, _2F_2\left(1,1;\frac{3}{2},2;-\frac{\pi^2}{2} g\rho^2 t\right) \;,
\ee
which holds at large $N$ for any $t\ll N\tau$. This hypergeometric function thus describes the crossover
from $t \ll \tau$ to $t \gg \tau$.
When $t\ll \tau=1/(g\rho^2)$ we recover the free diffusion regime, and obtain the additional corrections
\be
C_0(t) \simeq 2T t - \frac{\pi^2}{3} g\rho^2 t^2 +O(t^3) \;.
\ee
For $t\gg\tau$ we find instead
\be \label{disp_s0_large_t}
C_0(t) \simeq \frac{2T}{\pi^2 g\rho^2} \left( \ln \big(2\pi^2 g\rho^2 t\big) + \gamma_E - \frac{1}{\pi^2g\rho^2 t} \right) + O(\frac{1}{t^2}) \;.
\ee

Finally let us return to $s<0$. As mentioned above the variance of the displacement of a particle during time $t$ converges to 
a finite value, given by $2 \langle \delta x_i^2\rangle $. The crossover towards this limit can be obtained more precisely as
\be \label{decroissance}
2\langle \delta x_i^2\rangle - C_0(t) 
\simeq \frac{4T}{g\rho^{s+2}} \int_0^{1/2} du \frac{e^{- g\rho^{s+2} f_s(u) t}}{f_s(u)} \simeq \frac{4T}{(a_s g\rho^{s+2})^{\frac{1}{1+s}}} \frac{\Gamma(\frac{|s|}{1+s})}{1+s} t^{-\frac{|s|}{1+s}} \;,
\ee
which exhibits a power law decay in time.

\subsection{Multi-time correlations} \label{sec:multitime}

To compute $C_0(t_1,t_2)$, we only need to notice that (see \eqref{relation12times})
\be
C_0(t_1,t_2) = \frac{1}{2} [ C_0(t_1) + C_0(t_2) - C_0(|t_1-t_2|) ] 
\ee
and to use the results of Section~\ref{sec:disp} in the large $N$ limit. For $0<s<1$ and $s>1$ the results are presented in
Section \eqref{main_results_brownian}. 


For $s=1$ we find when all times are large, i.e. $t_1,t_2,|t_1-t_2|\gg \tau $
\be
C_0(t_1,t_2) \simeq \frac{T}{\sqrt{\pi g\rho^3}} \left( \sqrt{\frac{t_1}{\ln(g\rho^3 t_1)}} + \sqrt{\frac{t_2}{\ln(g\rho^3 t_2)}} - \sqrt{\frac{|t_1-t_2|}{\ln(g\rho^3 |t_1-t_2|)}} \, \right) \;.
\ee

In the case of the log-gas $s=0$, one can use the analytical result for the full crossover function in \eqref{crossoverhypergeo} and we find the exact result valid for all times $t_1,t_2 \ll N \tau$
\be
\begin{split} C_0(t_1,t_2) \simeq T t_1  \times \, _2F_2\left(1,1;\frac{3}{2},2;-\frac{\pi^2}{2} g\rho^2 t_1\right) \,+\, T t_2  \times \, _2F_2\left(1,1;\frac{3}{2},2;-\frac{\pi^2}{2} g\rho^2 t_2\right) \\  - T |t_1-t_2|  \times \, _2F_2\left(1,1;\frac{3}{2},2;-\frac{\pi^2}{2} g\rho^2 |t_1-t_2| \right) \;, \end{split}
\ee
which for short times $t_1$, $t_2 \ll \tau$ gives the first corrections to free diffusion as
\be
C_0(t_1,t_2) \simeq 2T \min(t_1,t_2) - \frac{\pi^2}{3} g\rho^2 t_1 t_2 +O(t^3) \;,
\ee
and for $t_1, t_2, |t_1-t_2| \gg\tau$,
and for large times gives 
\be 
C_0(t_1,t_2) \simeq \frac{T}{\pi^2 g\rho^2} \left( \ln \big(\frac{2\pi^2 g\rho^2 t_1 t_2}{|t_1-t_2|}\big) +\gamma_E - \frac{1}{\pi^2g\rho^2} \big(\frac{1}{t_1} - \frac{1}{t_2} + \frac{1}{|t_1-t_2|}\big) \right) + O\left(\frac{1}{t^2}\right) \;.
\ee
The first two leading terms yield the result announced the introduction in Eq. (\ref{res_C0}).

\subsection{Space-time correlations} \label{sec:spacetime_brownian}

We now consider the correlations $C_k(t)$ between the displacements of two particles  
separated by $k-1$ particles, as defined in \eqref{spacetime_brownian}. Let us discuss first the case $s>0$. In that case we know from Section \ref{sec:observables} -- see discussion around \eqref{Ck_freediffusion} -- that the correlations for $k \geq 1$ vanish for $t\ll \tau$, while for $t \gg N^{z_s} \tau$, $C_k(t)$ converges to its equilibrium value
$2\langle \delta x_i \delta x_{i+k} \rangle$. As for the single-particle variance, we want to know how this quantity behaves in the intermediate time regime $\tau\ll t \ll N^{z_s} \tau$. We consider two particles labelled $i$ and $i+k$, and we take the large $N$ limit with $k\ll N$. This gives
\be \label{Ck_step1}
C_k(t) \simeq \frac{4T}{g \rho^{s+2}} \int_0^{1/2} du \frac{1 - e^{-g\rho^{s+2} f_s(u) t}}{f_s(u)} \cos( 2 \pi k u ) \;.
\ee
At large times $t \gg \tau$, the integral is again dominated by small $u$, and we can thus write (for $s\neq 1$, $s>0$)
\bea \label{Ck_brownian1}
C_k(t) &\simeq& 4T \int_0^{1/2} du \frac{1 - e^{- g \rho^{s+2} a_s u^{z_s} t}}{g \rho^{s+2} a_s u^{z_s}} \cos( 2 \pi k u ) \nn \\
&\simeq& \frac{4T t^{\frac{z_s-1}{z_s}}}{(g \rho^{s+2} a_s)^{1/z_s} } \int_0^{+\infty} dv \frac{1 - e^{-v^{z_s}}}{v^{z_s}} \cos\left( \frac{2 \pi k v}{(g\rho^{s+2}a_s t)^{1/{z_s}} } \right) \;.
\eea
Thus the correlation function takes the the scaling form in the regime of both $t$ and $k$ large (recalling the definition $\tau=1/(g\rho^{s+2})$)
\be \label{Ck_scaling}
C_k(t) \simeq \begin{dcases} T \; \tau^\frac{1}{s+1} t^{\frac{s}{s+1}} F_s\left( \frac{k}{(t/\tau)^{\frac{1}{s+1}}} \right) \quad \text{for } 0<s<1 \;, \\
T \sqrt{\tau t} \, F_s\left( \frac{k}{\sqrt{t/\tau}} \right) \hspace{1.6cm} \text{for } s>1 \;. \end{dcases}
\ee 
For $0<s<1$, the scaling function reads
\be \label{Fs_def1}
F_s(x) = \frac{4}{a_s^{\frac{1}{s+1}}} \int_0^{+\infty} dv \frac{1-e^{-v^{s+1}}}{v^{s+1}} \cos \left( 2\pi a_s^{-\frac{1}{s+1}} x v \right)  \;.
\ee 
At small argument $x = k/(t/\tau)^{\frac{1}{s+1}} \to 0$ the function $F_s(x)$ has a finite limit, equal to $F_s(0)=U_s$, with $U_s$ given in \eqref{displacement_Riesz}, recovering the result \eqref{displacement_Riesz} for $C_0(t)$.
To study its behavior for $x \ll 1$ we change the integration variable to $w=2\pi a_s^{-\frac{1}{s+1}} x v$
and obtain
\be \label{smallx}
F_s(0) - F_s(x) \simeq  \frac{4}{a_s} ( 2\pi x)^s \int_0^{+\infty} dw \frac{1}{w^{s+1}} (1-\cos(w))
= \frac{x^s}{\pi \tan( \frac{\pi s}{2}) } \;,
\ee
where we neglected the stretched exponential term which is subleading
in that limit. We can now recall the relation~\eqref{remarkrelation} between the large time limit of the correlation $C_0(t)-C_k(t)$ and the
variance of the gaps $D_k(0)$. In view of this relation, it is reasonable to expect that the behavior \eqref{smallx} of the scaling function
in the limit 
 $x = k/(t/\tau)^{\frac{1}{s+1}} \ll 1$ connects to the behavior of the variance
 of the gaps $D_k(0)$ in the regime where $k = O(1)$ but large $k \gg 1$. Indeed one can check that 
 upon inserting \eqref{smallx} in \eqref{Ck_scaling} the time dependence cancels out and one obtains exactly the result obtained
 above in \eqref{gapscases}, showing perfect matching between the two regimes.

One can also study the behavior of the scaling function for large $x \gg 1$. After two integration by parts, one finds
\bea
F_s(x) &=& \frac{(s+1)a_s^{\frac{1}{s+1}}}{\pi^2 x^2} \int_0^{+\infty} dv \cos( 2\pi a_s^{-\frac{1}{s+1}} x v) \left( (s+1) v^{s-1} e^{-v^{s+1}} + \frac{(s+2)e^{-v^{s+1}}}{v^2} -\frac{(s+2)(1-e^{-v^{s+1}})}{v^{s+3}} \right) \nn \\
&\underset{x\gg 1}{\simeq}& \frac{s(s+1)a_s^{\frac{1}{s+1}}}{2\pi^2 x^2} \int_0^{+\infty} dv \frac{\cos(2\pi a_s^{-\frac{1}{s+1}} x v)}{v^{1-s}} = \frac{\Gamma(2+s) \cos(\frac{\pi s}{2}) a_s}{2^{1+s} \pi^{2+s} x^{2+s}} = \frac{s+1}{x^{2+s}} \;,
\eea
i.e., a power law decay, which implies that
\be \label{Ck_brownian_largek}
C_k(t) \simeq (s+1) T g\rho^{s+2} \frac{t^2}{k^{2+s}}
\ee
in the regime $k/(t/\tau)^{\frac{1}{1+s}} \gg 1$. The ballistic time dependence is an interesting result. A possible interpretation is that the ballistic component of the motion propagates faster than the anomalous diffusion component, and therefore dominates the large distance correlations.

In the short-range case $s>1$, the scaling function can be computed more explicitly,
\be \label{Fs} 
F_s(x) = \frac{2}{\pi\sqrt{(s+1)\zeta(s)}} \int_0^{+\infty} dv \frac{1-e^{-v^2}}{v^2} \cos\left( \frac{xv}{\sqrt{(s+1)\zeta(s)}} \right)
= \frac{1}{\sqrt{(s+1)\zeta(s)}} F_1\left( \frac{x}{\sqrt{(s+1) \zeta(s)} } \right)\;,
\ee
where
\be
F_1(x) = 
\frac{2}{\sqrt{\pi}} e^{-\frac{x^2}{4}} + |x| \left({\rm erf}\big(\frac{|x|}{2}\big)-1\right) \;.
\ee
For small $x>0$, one has $F_1(x)=\frac{2}{\sqrt{\pi}} - x + O(x^2)$. 
Computing $C_0(t)-C_k(t)$ for $x=k/\sqrt{t/\tau} \ll 1$ by 
inserting this behavior in \eqref{Fs} and then in \eqref{Ck_scaling} for $s>1$,
we check once again, in agreement with \eqref{remarkrelation}, that the result matches exactly the expression for $D_k(0)$ for $s>1$ in \eqref{gapscases}, which is linear in $k$.

At large $x$, one has $F_1(x) \simeq \frac{4}{\sqrt{\pi}} \frac{e^{-\frac{x^2}{4}}}{x^2}$, leading to an exponential decay
\be
C_k(t) \simeq 4 T \sqrt{\frac{(s+1)\zeta(s)g\rho^{s+2}}{\pi}} \; \frac{t^{3/2}}{k^2} \exp \left(-\frac{k^2}{4(s+1)\zeta(s) g\rho^{s+2} t} \right) \;,
\ee
at variance with the long range case.
\\

We note that $\zeta(s)$ diverges as $s \to 1^-$ which signals a change of behavior as a function of time in \eqref{Ck_scaling}.
Indeed, for $s=1$ we find (by a computation similar to \eqref{disp_s1}, with $\tau=1/(g\rho^3)$)
\be
C_k(t) \simeq T \sqrt{\frac{t}{\tau \ln(t/\tau)}} \, F_1 \left( \frac{k}{\sqrt{(t/\tau) \ln(t/\tau)}} \right) \;.
\ee

In the case of the log-gas, $s=0$, from \eqref{Ck_step1} and \eqref{fs_even} we obtain
\be
C_k(t) \simeq 4T \int_0^{1/2} du \frac{1 - e^{- 2\pi^2 g\rho^{2} u(1-u) t}}{2\pi^2 g \rho^{2} u(1-u)} \cos( 2 \pi k u ) \;.
\ee
Taking the derivative with respect to $t$ leads, at large $t$ with $k/t$ fixed, to
\be
\partial_t C_k(t) \simeq 4T \int_0^{1/2} du \, e^{- 2\pi^2 g\rho^{2} u(1-u) t}\cos( 2 \pi k u ) \simeq \frac{2T}{\pi^2 g\rho^2 t} \int_0^{+\infty} dv \, e^{-v}\cos\left( \frac{k v}{\pi g\rho^2 t} \right) = \frac{2T}{\pi^2 g\rho^2 t} \frac{1}{1+\big(\frac{k}{\pi g\rho^2 t}\big)^2} \;.
\ee
Integrating over $t$ one obtains the behaviour of the correlations in the scaling regime of large $t$ and large $k$ with $k \sim t/\tau$, $\tau=1/(g\rho^2)$,
\be \label{Ck_s0}
C_k(t) \simeq 
\frac{T}{\pi^2 g\rho^2} \ln \left( 1+ \big(\frac{\pi g\rho^2 t}{k}\big)^2 \right) 
= T \tau F_0\left(\frac{k \tau}{t}\right) \;,
\ee
where $F_0(x)$ coincides with the limit $s \to 0$ of the scaling function in \eqref{Fs_def1} (using $a_0=2\pi^2$),
\be 
F_0(x) = \frac{1}{\pi^2} \ln(1 + \frac{\pi^2}{x^2}) \;.
\ee 
We note that in the limit where the scaling variable $k \tau/t \ll 1$, the above result \eqref{Ck_s0} reproduces the
same leading $\ln t$ behavior at large time as in the calculation at $k=0$ presented in \eqref{disp_s0_large_t}, with however 
a diverging time independent term $\propto - \ln k$. More precisely \eqref{Ck_s0} leads to
$C_k(t) \simeq \frac{2 T}{\pi^2 g \rho^2} (\ln(\pi g \rho^2 t) - \ln k$) for $k \ll t/\tau$.
To understand the matching between this formula, which is valid in the regime of large $k,t$ with $k/(g\rho^2 t)$ fixed, in the limit $k/(g\rho^2 t) \ll 1$, 
and the regime $k=O(1)$ at large $t$, we note that the following difference converges to a finite, $k$-dependent value at large time,
\be \label{limitC}
\lim_{t \to +\infty}  C_0(t)-C_k(t) = \frac{4 T}{\pi^2 g \rho^2} \int_0^{1/2} du \frac{\sin^2(\pi k u) }{ u(1-u)} = \frac{2 T}{\pi^2 g \rho^2} \left(
\ln(2 \pi k) + \gamma_E -\text{Ci}(2 \pi k) \right) \;.
\ee 
This is identical to the calculation in \eqref{gaps_s0} for $D_k(0)$, which is expected because of the relation \eqref{remarkrelation}. Since $\text{Ci}(2 \pi k) \to 0$ at large $k$ we find a perfect matching between the two regimes including constant terms. For $k\gg t/\tau$, the correlations decay as $C_k(t)\sim \frac{T g\rho^2 t^2}{k^2}$.
\\

We now discuss the case $s<0$. In that case, as discussed in section \ref{sec:observables}, $C_k(t)$ converges for $t \gg \tau$ to the value (see \eqref{cov_sneg} and \eqref{cov_sneg_largek})
\be \label{Ck_stepinfty}
C_k(\infty) = 2 \langle \delta x_i \delta x_{i+k} \rangle  
 \simeq \frac{4T}{g \rho^{s+2}} \int_0^{1/2} du \frac{\cos( 2 \pi k u)}{f_s(u)} \simeq_{k \gg 1} \frac{4T k^{-|s|}}{g\rho^{s+2}a_s} (2\pi)^{-|s|} \cos(\frac{\pi |s|}{2}) \Gamma(|s|)
 \;,
\ee
which decays as $k^{- |s|}$ for large distance $k \gg 1$. To exhibit a dynamical scaling function we consider the
difference 
\be  \label{Ck_difference} 
C_k(\infty) - C_k(t) 
 \simeq \frac{4T}{g \rho^{s+2}} \int_0^{1/2} du \frac{e^{-g\rho^{s+2} f_s(u) t}}{f_s(u)} \cos( 2 \pi k u ) \;,
\ee 
which in the scaling regime of large $k, t$ with the ratio $x= k/(t/\tau)^{\frac{1}{s+1}}$ fixed, takes 
a scaling form similar to
\eqref{Ck_scaling}
\be \label{Ck_scaling_sneg}
C_k(\infty) - C_k(t) \simeq T \tau^\frac{1}{1+s} t^{-\frac{|s|}{1+s}} \tilde F_s\left( \frac{k}{(t/\tau)^{\frac{1}{1+s}}} \right)
\ee 
with 
\be \label{Fs_defnegative}
\tilde F_s(x) = \frac{4}{a_s^{\frac{1}{s+1}}} \int_0^{+\infty} dv \frac{e^{-v^{s+1}}}{v^{s+1}} \cos \left( 2\pi a_s^{\frac{1}{s+1}} x v \right) \;.
\ee
This function has a finite limit at small argument $k \ll (t/\tau)^{\frac{1}{s+1}}$, i.e  $\tilde F_s(0) = 4 a_s^{-\frac{1}{1+s}}\frac{\Gamma(\frac{|s|}{1+s})}{1+s}$.
One thus finds a perfect matching with the formula for $k=0$ \eqref{decroissance}.
At large argument $k \gg (t/\tau)^{\frac{1}{s+1}}$ we have
\be \tilde F_s(x) \simeq \frac{2^{2+s} \pi^s}{a_s} \cos( \frac{\pi s}{2}) \Gamma(|s|) x^{- |s|} \;,
\ee 
leading to $C_k(\infty)-C_k(0) \propto k^{-|s|}$ and independent of time.

\subsection{Time correlations of the gaps $D_k(t)$} \label{sec:Dk_brownian}

For any $s>-1$, the correlation between the gaps at time $0$ and time $t$, $D_k(t)$, is given by 
formula \eqref{gaptime_brownian}. In the large $N$ limit it becomes 
\be
D_k(t) \simeq \frac{8T}{g\rho^{s+2}} \int_0^{1/2} du \frac{e^{-g\rho^{s+2} f_s(u) t}}{f_s(u)} \sin^2(\pi k u) \;.
\ee
At short time $t \ll \tau=1/(g \rho^{s+2})$, one may expand the exponential, leading to
\be \label{Dk_freediffusion}
D_k(t) \simeq D_k(0) - 2Tt \;,
\ee
i.e. the free diffusion regime. We now discuss large time $t \gg \tau$, starting with $s>0$.

Consider first the regime of fixed $k$ and $t \gg \tau$. 
For any $s>-1$, the integral is dominated by small $u$ and one can expand the sinus and use \eqref{fasympt} to obtain
\be \label{Dk_larget}
D_k(t) \simeq \frac{8T \pi^2 k^2}{g\rho^{s+2} a_s}  \int_0^{+\infty} du \, u^{2-z_s} e^{-g\rho^{s+2} a_s u^{z_s} t} 
= \begin{dcases} \frac{8 \pi^2 T}{(a_s g\rho^{s+2})^{\frac{3}{1+s}}} \frac{\Gamma(\frac{2-s}{1+s}) }{1+s} \frac{k^2}{t^{\frac{2-s}{1+s}}} \hspace{0.93cm} \text{for } -1<s<1 \;, \\
\frac{T}{2 \sqrt{\pi} \, ((s+1)\zeta(s) g \rho^{s+2})^{3/2}} \frac{k^2}{\sqrt{t}} \quad \text{for } s>1 \;.\end{dcases}
\ee 
For $s=1$, one can use the asymptotics \eqref{fasympt_s1} to compute the logarithmic corrections as usual.
\\


{\bf Scaling function}. In the scaling regime where both $t$ and $k$ are large with $k \sim (t/\tau)^{1/z_s}$, where $z_s=\min(2,1+s)$, we obtain a scaling form analogous to the one for $C_k(t)$,
\be \label{Dk_scaling}
D_k(t) \simeq \begin{dcases} T \tau^{\frac{1}{s+1}} t^\frac{s}{s+1} G_s\left( \frac{k}{(t/\tau)^{\frac{1}{s+1}}} \right) \quad \text{for } 0<s<1 \;, \\
T \sqrt{\tau \, t} \, G_s\left( \frac{k}{\sqrt{t/\tau}} \right) \hspace{1.45cm} \text{for } s>1 \;, \end{dcases}
\ee 
with the scaling function
\be \label{Gs_def1}
G_s(x) = \frac{8}{a_s^{1/z_s}} \int_0^{+\infty} dv \frac{e^{-v^{z_s}}}{v^{z_s}} \sin^2 \left( \pi a_s^{-1/z_s} x v \right) \;.
\ee
For $x=k/(t/\tau)^{1/z_s} \ll 1$, one can expand the sine function to first order to obtain
\be 
G_s(x) \sim \begin{dcases} 
\frac{8\pi^2}{a_s^{\frac{3}{1+s}}} \frac{\Gamma(\frac{2-s}{1+s})}{1+s} x^2 \hspace{1.2cm} \text{for } 0<s<1 \;, \\ \frac{x^2}{2\sqrt{\pi}((s+1)\zeta(s))^{3/2}} \quad \text{for } s>1 \;. \end{dcases}
\ee
This matches exactly the fixed $k$ regime obtained above. 
Thus at large time, i.e. of the order or larger than $k^{s+1}$,
$D_k(t)$ decays as $k^2 t^{-\frac{2-s}{1+s}}$ for $s<1$ and as $k^2/\sqrt{t}$ for $s>1$. 
At large argument $x \gg 1$ one has (in this regime the integral \eqref{Dk_scaling} is dominated by small values of $v\sim 1/x$ and the exponential can be approximated to 1)
\be 
G_s(x) \simeq \begin{cases} \frac{4}{sa_s} (2 \pi)^s \cos(\frac{\pi s}{2}) \Gamma(1-s) x^s \quad \text{for } 0 <s<1 \;, \\ \frac{x}{(s+1)\zeta(s)} \hspace{3.2cm} \text{for } s>1 \;, \end{cases}
\ee 
which inserted in \eqref{Dk_scaling} reproduces exactly the equilibrium result for gap statistics $D_k(0)$ in \eqref{gapscases}.
This shows the matching to the small time regime. The next order at large $x$ can be obtained simply by using the identity \eqref{remarkrelation1} and the fact that $C_k(t)$ decays at large $k$,
\be
D_k(0) - D_k(t) = C_0(t) - C_k(t) \underset{k^{z_s} \gg t/\tau}{\simeq} C_0(t) \propto t^{1-\frac{1}{z_s}} \;,
\ee
where $C_0(t)$ is given in \eqref{displacement_Riesz}. This can also be shown without using the relation \eqref{remarkrelation1} by averaging the $\sin^2$ to $1/2$ in the difference $D_k(0)-D_k(t)$.

For $s=0$, the scaling function \eqref{Gs_def1} can be computed exactly
\be \label{G0def}
G_0(x) = \frac{1}{\pi^2} \ln(1+\frac{x^2}{\pi^2})\;.
\ee
As noted in Section \ref{main_results_brownian} this scaling function is related to the one for the time displacements via $G_0(x)=F_0(1/x)$,
some manifestation of "relativistic" invariance of the case with dynamical exponent $z_0=1$.
From \eqref{G0def} we correctly recover \eqref{Dk_larget} for $x\ll 1$ and \eqref{gaps_s0} for $x \gg 1$.

For $s<0$, we have a similar crossover between \eqref{Dk_larget} for $k\ll (t/\tau)^{\frac{1}{s+1}}$ and $D_k(0)\simeq 2\langle \delta x_i^2 \rangle$ for $k\gg (t/\tau)^{\frac{1}{s+1}}$, but we need to keep the exact expression of $f_s(u)$ to correctly recover the second limit.


\subsection{Quenched initial condition} \label{sec:quenched_brownian}

The results obtained above for the time-dependent quantities were obtained by drawing the initial condition from the equilibrium distribution, i.e. they correspond to an ``annealed'' initial condition. We will now see what happens if we consider instead a ``quenched'', i.e. deterministic, initial condition. Let us assume that the initial density is uniform, so that we have $\delta x_i(0)=0$ for all $i$. This corresponds to preparing the system in the ground state at $T=0$. Integrating \eqref{Eq_delta_x} with this initial condition leads to
\be
\delta x_i(t) = \sqrt{2T} \sum_{j=1}^{N} \int_0^t dt_1 [e^{(t_1-t)H}]_{ij} \left( \xi_j(t_1) - \frac{1}{N} \sum_{k=1}^N \xi_k(t_1) \right) \;,
\ee
where $H$ is the Hessian matrix defined in \eqref{defHessian}. Using that $\langle \xi_i(t) \xi_j(t')\rangle = \delta_{ij} \delta(t-t')$, we obtain
\be
\langle \delta x_j(t) \delta x_k(t') \rangle_{\rm qu} = 2T \int_0^{\min(t,t')} dt_1 \left([e^{(2t_1-t-t')H}]_{jk} -\frac{1}{N} \right) \;,
\ee
where $\langle \cdot \rangle_{\rm qu}$ denotes an average over the noise starting from a quenched initial condition, and where for the last term we have used that
\be \label{identity_expH}
\sum_{m,n=1}^N [e^{(t_1-t)H}]_{jm} [e^{(t_1-t')H}]_{nk} = \frac{1}{N^2} \sum_{q=0}^{N-1} \sum_{m,n=1}^N e^{\mu_q(2t_1-t-t')} e^{2\pi i \frac{q}{N}(j-m+k-n)} = 1 \;,
\ee
since only the term $q=0$ gives a non-zero contribution. Finally, decomposing the matrix $H$ in its eigenbasis and performing the integral we obtain the equivalent of \eqref{cov_brownian} for a quenched initial condition,
\be
\langle \delta x_j(t) \delta x_k(t') \rangle_{\rm qu} = \frac{T}{N} \sum_{q=1}^{N-1} \frac{e^{-\mu_q|t-t'|}-e^{-\mu_q(t+t')}}{\mu_q} e^{2\pi i \frac{q}{N}(j-k)} = \frac{2T}{N} \sum_{q=1}^{(N-1)/2} \frac{e^{-\mu_q|t-t'|}-e^{-\mu_q(t+t')}}{\mu_q} \cos(2\pi \frac{q}{N}(j-k)) \;.
\ee

Let us now compute again the dynamical quantities from this expression, starting with the variance of the displacement after time $t$ \eqref{disp_brownian}. Since $\delta x_i(0)=0$, we simply have in this case
\be \label{C0_quenched}
C_0^{\rm qu}(t) = \langle (\delta x_i(t) - \delta x_i(0))^2 \rangle_{\rm qu} = \langle \delta x_i(t)^2 \rangle_{\rm qu} = \frac{2T}{N} \sum_{q=1}^{(N-1)/2} \frac{1-e^{-2\mu_q t}}{\mu_q} \;.
\ee
Thus, the quenched result can be directly obtained from the annealed result through a transformation $t \to 2t$ and $T \to T/2$, i.e. one has
$C_0^{\rm qu}(t) = \frac{1}{2} C_0^{\rm ann}(2 t)$.
For $t\gg \tau$, this leads to
\be \label{displacement_Riesz_quenched}
C_0^{\rm qu}(t) \simeq \begin{dcases} U_s \frac{T \, t^{\frac{s}{s+1}}}{2^{\frac{1}{s+1}} g^{\frac{1}{s+1}}\rho^{\frac{s+2}{s+1}}}
\hspace{1.37cm} \text{for } 0<s<1 \;, \\
T \sqrt{\frac{2t}{\pi g\rho^3 \ln (g\rho^3 t)}} \hspace{1.26cm} \text{for } s=1 \;, \\
\frac{T}{\sqrt{\pi (s+1) \zeta(s)}} \sqrt{\frac{2t}{g\rho^{s+2}}}
\quad \text{for } s>1 \;, \end{dcases}
\ee
and for $s=0$,
\be \label{disp_s0_quenched}
C_0^{\rm qu}(t) \simeq \frac{T}{\pi^2 g\rho^2} \left( \ln \big(4\pi^2 g\rho^2 t\big) + \gamma_E - \frac{1}{2\pi^2g\rho^2 t} \right) + O(\frac{1}{t^2}) \;.
\ee
We thus simply have a factor $2^{-\frac{1}{s+1}}$ for $0<s<1$ and $2^{-\frac{1}{2}}$ for $s\geq 1$ compared to the annealed case. 
This holds until it reaches its asymptotic large time limit equal to the stationary variance, $ C_0^{\rm qu}(\infty) = \langle \delta x_i^2 \rangle$, half
of the annealed one. Once
again quite remarkably this coincides with the result obtained through MFT in \cite{DFRiesz23}.

For the time correlations we have
\be
C_0^{\rm qu}(t_1,t_2) = \langle \delta x_i(t_1) \delta x_i(t_2) \rangle_{\rm qu} = \frac{2T}{N} \sum_{q=1}^{(N-1)/2} \frac{e^{-\mu_q|t-t'|}-e^{-\mu_q(t+t')}}{\mu_q} \;.
\ee
This can be easily computed from our result for $C_0^{\rm qu}(t)$ by noticing that
\be
C_0^{\rm qu}(t_1,t_2) = C_0^{\rm qu} (\frac{t_1+t_2}{2}) - C_0^{\rm qu} (\frac{|t_1-t_2|}{2}) \;.
\ee
For $ \tau \ll t \ll N^{z_s}\tau$, and for $0<s<1$, we obtain for instance
\be
C_0^{\rm qu}(t_1,t_2) \simeq \frac{U_s T}{2} \left((t_1+t_2)^{\frac{s}{s+1}} - |t_1-t_2|^{\frac{s}{s+1}}\right) \;,
\ee
which again coincides with the result obtained in Ref. \cite{DFRiesz23}.

For the space time correlation $C_k(t)$ we obtain
\be \label{Ck_quenched_brownian}
C_k^{\rm qu}(t) = \langle \delta x_i(t) \delta x_{i+k}(t) \rangle_{\rm qu} = \frac{2T}{N} \sum_{q=1}^{(N-1)/2} \frac{1-e^{-2\mu_q t}}{\mu_q} \cos(2\pi \frac{qk}{N}) \;.
\ee
Once again we have the mapping $t \to 2t$ and $T \to T/2$ with the annealed case \eqref{spacetime_brownian}.
Finally, note that the gap correlations identically vanish, $D_k^{\rm qu}(t)=0$ in the quenched case.

\subsection{Linear statistics} \label{sec:linear_stat}

For this section we go back to the annealed case. Let us consider a function $f(x)= \sum_{n \in \mathbb{Z}} \hat f_n e^{-2 i \pi \frac{n}{L} x}$ on the circle, periodic of period $L$,
with Fourier coefficients $\hat f_n$. 
We define the linear statistics
\be 
{\cal L}_N(t) = \sum_{i=1}^N f(x_i(t)) \;.
\ee 
We can use the low temperature expansion, which reads to leading order $O(T)$,
\bea  
\langle {\cal L}_N(t) {\cal L}_N(t') \rangle_c &\simeq& 
\sum_{j, k=1}^N f'(x_j^0) f'(x_k^0) \langle  \delta x_j(t)  \delta x_k(t') \rangle =
\frac{2T}{N}  \sum_{q=1}^{(N-1)/2} \frac{e^{- \mu_q |t-t'|}}{\mu_q}  \big| \sum_{j=1}^N f'(x_j^0) e^{2 \pi i \frac{q}{N} j}\big|^2 \\
&=& \frac{8\pi^2 T}{N} \rho^2  \sum_{q=1}^{(N-1)/2} \frac{e^{- \mu_q |t-t'|}}{\mu_q}  q^2 \hat f_q \hat f_{-q} \;.
\eea  

Consider now the large $N$ limit, and let us first discuss a function $f(x)$ which varies at the scale of the full circle. In that
case the Fourier coefficient $\hat f_q$ are non-zero only for $q\ll N$. Since these modes are very slow
and equilibrate on time scale $N^{z_s} \tau$ with $\tau= 1/g \rho^{s+2}$, 
to obtain a limit we rescale the time difference as $|t-t'| = N^{z_s} \tau |\tilde t - \tilde t'|$.
 For the log-gas $s=0$ we obtain
 \bea  \label{linearstatlog}
 \langle {\cal L}_N(t) {\cal L}_N(t') \rangle_c &\simeq& \frac{4 T}{g}  \sum_{q=1}^{\infty} e^{-2\pi^2 q |\tilde t- \tilde t'|}  q |\hat f_q |^2 \;.
 \eea 
In this case, more precisely for the Dyson Brownian motion on the circle with parameter $\beta$,
these linear statistics were studied in \cite{Spohn3}. 
We can check that Eq. \eqref{linearstatlog} agrees with Eq. (2.9) in that paper (with the correspondence $\beta=g/T$,
a density $\rho=1/(2 \pi)$ there, and a rescaling of time by $N$).
Note that in \cite{Spohn3} the space time density-density correlations were also obtained (see Eqs. (2.10-2.15) there).

In a recent work the power spectrum statistics 
$S_N(\omega)= \frac{N}{(2 \pi)^2} \sum_{j,k} \langle \delta \theta_j \delta \theta_k \rangle e^{i \omega (j-k)}$
was studied for the circular $\beta$ random matrix ensemble (C$\beta$E) for $\beta=1,2,4$ \cite{ForresterWitte},
where $\delta \theta_j= \theta_j - \langle \theta_j \rangle$ and $\theta_j \in [0,2 \pi]$
are the eigenvalues. This corresponds to our model with $s=0$ and $\beta=g/T$, 
with $x_j= \frac{L}{2 \pi} \theta_j$. Our prediction for this quantity to first order in $1/\beta$ is thus
\bea
S_N(\omega)= \frac{1}{2\pi^2\beta} \frac{1}{\frac{\omega}{2\pi}(1-\frac{\omega}{2\pi})} \;,
\eea 
for $\omega=2 \pi q/N$, $q=1,\dots,N-1$ (and zero otherwise). Since
this result is valid for large $\beta$, it is not immediate to compare it with the formula given in \cite{ForresterWitte}.

Returning to linear statistics, for
generic $s>-1$ ($s\neq 1$) we obtain
\be
\langle {\cal L}_N(t) {\cal L}_N(t') \rangle_c \simeq \begin{dcases}  4\pi^{\frac{1}{2}-s} \frac{\Gamma(1+\frac{s}{2})}{\Gamma(\frac{1-s}{2})} \frac{ T N^s}{g\rho^s} \sum_{q=1}^{\infty} e^{-a_s q^{1+s} |\tilde t - \tilde t'|}  q^{1-s} | \hat f_q |^2 \quad \text{for } -1<s<1 \;, \\
\frac{2 T N}{(s+1) \zeta(s) g\rho^s} \sum_{q=1}^{\infty} e^{- a_s q^2 |\tilde t - \tilde t'|} | \hat f_q |^2 \hspace{2.25cm} \text{for } s>1 \;. \end{dcases}
\ee
For $0<s<1$ we can compare with the recent result of \cite{BoursierCLT}, Thm 2 (1.12-1.13) for the covariance at equal time $t=t'$. 
We find general agreement. To compare the prefactor 
we need to set $g=2 s$, $\rho=1$, $k \to 2 \pi q$ and $T=1/\beta$. We find that the amplitude there is $2^{-2 s}$ times our amplitude.
Note that the formula below (2.4) for the Fourier transform is inconsistent there. 
In the short range case $s>1$ we note that the equal time variance is proportional to $\frac{1}{L} \int_0^L dx f(x)^2$, which
shows the absence of correlations at the scale of the full circle. 

To explore microscopic scales we assume that $\hat f_q = \frac{1}{N} \tilde f(q/N)$ as $N \to +\infty$. This means that 
in that limit one gets $f(x) = \int_0^{1/2} du \tilde f(u) e^{-2 i \pi u \rho x}$.
At large $N$, assuming that the integral converges, the variance is then given in the microscopic scales by
\be 
\langle {\cal L}_N(t) {\cal L}_N(t') \rangle_c \simeq \frac{8\pi^2 T}{g\rho^s}  \int_0^{1/2} du \frac{e^{- g\rho^{s+2} f_s(u) |t-t'|}}{f_s(u)} u^2 | \tilde f(u) |^2 \;,
\ee
which is of order unity.


\section{Active gas: the RTP case} \label{sec:RTP}

We now consider the same model as in the previous section (for now we consider again a general interaction potential $W(x)$), but with run-and-tumble noise instead of Brownian noise. The equations of motion for the $N$ RTP particles at positions $x_i$ on the circle of perimeter $L$ read
\be \label{Eq_def_RTP}
\dot x_i(t) = -\sum_{j(\neq i)} W'(x_i(t)-x_j(t)) + v_0 \sigma_i(t) \;,
\ee
where the $\sigma_i(t)= \pm 1$ are independent telegraphic noises with rate $\gamma$. As in the Brownian case, we assume that we are in the regime of small relative displacements, i.e. we make again the assumption \eqref{cond_approx}. The regime of validity will be ascertained below. This enables us to approximate the dynamical equation \eqref{Eq_def_RTP} as
\be \label{Eq_delta_x_RTP}
\frac{d}{dt} \delta x_i(t) = - \sum_{j=1}^N H_{ij} \, \delta x_j(t) + v_0 \sigma_i(t) - \frac{v_0}{N} \sum_{j=1}^N \sigma_j(t)\;,
\ee 
where we recall that the $\delta x_i(t)$ are defined in \eqref{def_delta_x}, and the Hessian matrix $H$ is defined in \eqref{defHessian}, with spectrum given in \eqref{eigenvals}. We are now going to study the stationary state. We will compute the two-point two-time correlation function of the
displacements. Note that in the RTP case the distributions of the displacements, even in the linear model approximation,
are not Gaussian \cite{HarmonicChainRTPDhar}. However we can still compute exactly the two-point functions. 
Taking the Fourier transform with respect to time one thus obtains by inversion in the frequency domain
\be 
\delta \hat x_j(\omega) = v_0 \sum_{k=1}^N  [i \omega \mathbb{1}_N + H]^{-1}_{jk} \hat \sigma_k(\omega) - \frac{v_0}{N} \frac{1}{i\omega} \sum_{k=1}^N \hat \sigma_k(\omega) \;,
\ee 
where $\mathbb{1}_N$ is the $N \times N$ identity matrix, $\delta \hat x_i(\omega) = \int_{-\infty}^{\infty} dt \, e^{-i \omega t} \delta x_i(t)$ and $\hat \sigma_j(\omega)=\int_{-\infty}^{\infty} dt \, e^{-i \omega t} \sigma_i(t)$, and we have used \eqref{identity_hessian_fourier}. The correlations of $\sigma_i(t)$ are given by $\langle \sigma_i(t) \sigma_j(t') \rangle = e^{-2\gamma|t-t'|}\delta_{ij}$, from which we get in Fourier space
\be
\langle \hat \sigma_i(\omega) \hat \sigma_j(\omega') \rangle = \frac{4\gamma}{\omega^2+4\gamma^2} 2\pi \delta(\omega+\omega')
\ee
(in this section and the next we use brackets to denote averages over the RTP noise). Thus, the two point correlations of $\delta x_i(t)$ in the stationary state read
\be
\langle \delta x_j(t) \delta x_k(t') \rangle = v_0^2 \int_{-\infty}^{+\infty} \frac{d\omega}{2 \pi} \frac{4\gamma \, e^{i \omega (t-t')} }{\omega^2+4\gamma^2} \left( [\omega^2 \mathbb{1}_N + H^2]^{-1}_{jk} - \frac{1}{N\omega^2} \right) \;.
\label{rtp_corr_cm}
\ee
Using the eigensystem of $H$ given in \eqref{eigenvals}, this reads
\bea  \label{cov_rtp}
\langle  \delta x_j(t)  \delta x_k(t') \rangle &=& \frac{v_0^2}{N} \sum_{q=1}^{N-1} e^{2\pi i \frac{q}{N} (j-k)} \int_{-\infty}^{+\infty} \frac{d\omega}{2 \pi} \frac{e^{i \omega (t-t')}}{\omega^2 + \mu_q^2 } \frac{4\gamma}{\omega^2+4\gamma^2} \nn \\
&=& \frac{2v_0^2}{N} \sum_{q=1}^{(N-1)/2} \frac{\mu_q e^{-2\gamma|t-t'|} -2\gamma e^{- \mu_q |t-t'|}}{\mu_q(\mu_q^2-4\gamma^2)}  \cos\left(2\pi \frac{q}{N} (j-k)\right)  
\eea 
where we have made use of the symmetry $\mu_q=\mu_{N-q}$ in the last step. As before, the last expression is exact only for odd values of $N$.
\\

{\bf Riesz gas}. We now study the various observables defined in Section \ref{sec:observables}, and specialize the study to the periodic Riesz gas for $s>-1$,
i.e. with interaction defined in \eqref{defRiesz}. 
In that case the eigenvalues $\mu_q$ (inverse relaxation times) are given in Eq. \eqref{mu_Riesz} in terms of the function $f_s(u)$ defined there.
In addition to the local interaction time $\tau=1/(g \rho^{s+2})$, the new 
physically important parameters in the case of the RTP are the "effective temperature" $T_{\rm eff}$ and a dimensionless
interaction strength $\hat g$, which is also equal (up to a factor $1/2$) to the ratio of the persistence time $1/\gamma$ to the local interaction time $\tau$,
\be 
T_{\rm eff}= \frac{v_0^2}{2 \gamma} \quad , \quad \hat{g}= \frac{g\rho^{s+2}}{2\gamma} = \frac{1}{2\gamma \tau} \;.
\ee 
We will
express the results below using these parameters. The diffusive limit corresponds to $\hat g \ll 1$ at fixed $T_{\rm eff}$. 
The limit $\gamma \to 0$ (Jepsen gas) corresponds to the simultaneous limit $T_{\rm eff} \sim \hat g \to +\infty$.

\subsection{Variance and melting transition} \label{sec:var_rtp}

The variance of a particle position is given for any $N$ by
\be \label{var_RTP}
\langle \delta x_i^2 \rangle = \frac{2 v_0^2}{N} \sum_{q=1}^{(N-1)/2} \frac{1}{\mu_q (\mu_q +2\gamma)} = \frac{2T_{\rm eff}}{N g\rho^{s+2}} \sum_{q=1}^{(N-1)/2} \frac{1}{f_s(\frac{q}{N}) (1+ \hat g f_s(\frac{q}{N}))} \;,
\ee
where $\mu_q$ is defined in Eq. \eqref{mu_Riesz}. 
We will now consider separately the cases $s>0$ and $s<0$.

For $s>0$ the sum in \eqref{var_RTP} is dominated by the small values of $q$. In the large $N, L$ limit with fixed mean density $\rho=N/L$ we obtain, using that $f_s(u)\sim a_s u^{z_s}$ for $u \ll 1$,
\be \label{var_RTP2}
\langle \delta x_i^2 \rangle \simeq \frac{2 T_{\rm eff} N^{z_s-1}}{g\rho^{s+2} a_s} \sum_{q=1}^{\infty} \frac{1}{q^{z_s} (1+ \hat g a_s (\frac{q}{N})^{z_s})} \;.
\ee
If $\hat g \ll N^{z_s}$, the last term in the denominator can be neglected, so that at leading order we recover that $\langle \delta x_i^2 \rangle$ is given by the Brownian result \eqref{var_Riesz_liquid} with an effective temperature $T \to T_{\rm eff}=\frac{v_0^2}{2\gamma}$, which we
call $\langle \delta x_i^2 \rangle_{\rm Brownian}$. To be more precise, let us evaluate the difference with the Brownian case. It reads
\be
\langle \delta x_i^2 \rangle - \langle \delta x_i^2 \rangle_{\rm Brownian} = - \frac{2 T_{\rm eff} \hat g}{g\rho^{s+2} N} \sum_{q=1}^{(N-1)/2} \frac{1}{1+ \hat g f_s(\frac{q}{N})}
= - \frac{2 T_{\rm eff} \hat g}{g\rho^{s+2}} \int_{0}^{1/2} \frac{du}{1+ \hat g f_s(u)} + O(\frac{1}{N}) \;. 
\ee
This integral is always well-defined, thus the difference with the Brownian case is always of order $N^0$. In particular for $s=0$, using $f_s(u)=2\pi^2 u(1-u)$, this leads to
\be \label{var_log_rtp}
\langle \delta x_i^2 \rangle = \frac{T_{\rm eff}}{\pi^2 g\rho^2} \left(\ln N + \gamma_E - \pi \hat g \frac{\ln \left(1+ \pi^2 \hat g +\pi\sqrt{2\hat g (1+\frac{\pi^2}{2}\hat g})\right)}{\sqrt{2\hat g (1+\frac{\pi^2}{2} \hat g)}} + O(N^{-1})\right) \;.
\ee

For $-1<s<0$, similar to the Brownian case, the sum is dominated by $q$ of order $N$, and thus the variance has a finite limit when $N \to +\infty$,
\be \label{var_RTP_solid}
\langle \delta x_i^2 \rangle \simeq \frac{2 T_{\rm eff}}{g\rho^{s+2}} \int_0^{1/2} \frac{du}{f_s(u)\left( 1 + \hat g f_s(u) \right)} \;.
\ee
In this case, we recover the passive result \eqref{var_Riesz_solid} only in the limit $\hat g \ll 1$. Similar to the Brownian case, we find a melting transition, which in this case occurs when (using the same criterion \eqref{melting_crit})
\be  \label{TmeltingRTP}
T_{\rm eff} = T_{\rm eff,M} = \frac{1}{2} g\rho^{s} \frac{c_L^2}{\int_0^{1/2} \frac{du}{f_s(u)\left( 1 + \hat g f_s(u) \right)}} \;.
\ee 
In the limit $s \to 0^-$ one finds that $T_{\rm eff,M}$ vanishes linearly in $s$, as in the Brownian case,
\be 
T_{\rm eff,M} \sim \pi^2 c_L^2 g |s|.
\ee
Indeed using $f_s(u)\sim a_s u^{s+1}$ we obtain,
\be
\int_0^{1/2} \frac{du}{f_s(u)(\hat g f_s(u) + 1)} \simeq \int_0^\epsilon \frac{du}{a_s u^{s+1} (\hat g a_s u^{s+1} + 1)} = \frac{\hat g}{(\hat g a_s)^{\frac{1}{s+1}}} \int_0^{(\hat g a_s)^{\frac{1}{s+1}} \epsilon} \frac{dv}{v^{s+1} (v^{s+1} + 1)} \simeq \frac{\hat g }{|s|} (\hat g a_s)^{\frac{s-1}{s+1}} \epsilon^s \simeq \frac{1}{2\pi^2|s|} \;.
\ee

In the limit $\hat g \ll 1$, \eqref{var_RTP_solid} recovers the Brownian case and we have $T_{\rm eff,M} = T_M$. In the opposite limit $\hat g \gg 1$, one finds two regimes. If $-1<s<-1/2$ we find that \eqref{var_RTP_solid} becomes
\be \label{var_veryRTP1}
\langle \delta x_i^2 \rangle \simeq 2 \left(\frac{v_0}{g\rho^{s+2}}\right)^2 \int_0^{1/2} \frac{du}{f_s(u)^2} \quad , \quad T_{\rm eff,M} \simeq \frac{g \rho^s}{2} \frac{ \hat g \, c_L^2}{  \int_0^1 du  \frac{1}{f_s(u)^2} } \;.
\ee
Note that for these values of $s$ the variance has a finite limit as $\gamma \to 0$. 

If $-1/2<s<0$, the integral in \eqref{var_veryRTP1} diverges at $u=0$. In this case, the integral in \eqref{var_RTP_solid} is dominated by small values of $u$ and we get
\be \label{var_rtp_sneg_2}
\langle \delta x_i^2 \rangle \simeq \frac{v_0^2}{\gamma g \rho^{s+2} a_s}   \int_0^{1/2} \frac{du}{u^{s+1} \left(\hat g a_s u^{s+1} +1 \right)} \simeq \frac{v_0^2 (\hat g a_s)^{\frac{s}{s+1}}}{\gamma g \rho^{s+2} a_s} \int_0^{+\infty} \frac{dv}{v^{s+1} \left(v^{s+1} +1 \right)} = \frac{\pi v_0^2 \hat g^{\frac{s}{s+1}}}{(s+1) a_s^{\frac{1}{s+1}} \sin(\frac{2s+1}{s+1}\pi) \gamma  g \rho^{s+2}} \;,
\ee 
which leads to the expression for the melting temperature for $\hat g \gg 1$,
\be
T_{\rm eff,M} \simeq \frac{s+1}{2\pi} a_s^{\frac{1}{s+1}} \sin\big(\frac{2s+1}{s+1}\pi \big) g\rho^s \hat g^{\frac{|s|}{1+s}} c_L^2  \;,
\ee
where $a_s$ is given in \eqref{fasympt}.
The power of $\hat g$ is continuous at $s=-1/2$, but the amplitude vanishes as $s \to -1/2^+$ which suggests an additional dependence in 
$\log \hat g$ for $s=-1/2$.
\\

{\bf Covariance.} 
The covariance is given by
\be  \label{covar_RTP}
\langle \delta x_i \delta x_{i+k} \rangle =  \frac{2T_{\rm eff}}{N g\rho^{s+2}} \sum_{q=1}^{(N-1)/2} \frac{\cos( 2 \pi k \frac{q}{N})}{f_s(\frac{q}{N}) (1+ \hat g f_s(\frac{q}{N}))} \;.
\ee
For $s \geq 0$, similarly to the case of the variance studied above, we find that the last term in the denominator is irrelevant as long as $\hat g \ll N^{z_s}$, and thus we recover the Brownian results of Section~\ref{sec:var_brownian} with a temperature $T \to T_{\rm eff}$ for all range of $k=1,\dots,N$. 

For $s<0$, for $k\ll N$, we can replace the sum by an integral, which leads to
\be
\langle \delta x_i \delta x_{i+k}\rangle \simeq \frac{2T_{\rm eff}}{g\rho^{s+2}}  \int_0^{1/2} du \frac{\cos(2\pi k u)}{f_s(u)(1+\hat g f_s(u))} \;.
\ee
The asymptotic behavior at large $k$ then recovers the one in \eqref{cov_sneg_largek} with $T \to T_{\rm eff}$, provided $k \gg \hat g^{1/(1+s)}$. 


\subsection{Gap statistics and validity of the approximation} \label{sec:Dk0_rtp}

\subsubsection{Variance of the gaps}

Let us consider now the gap statistics, starting with the variance at fixed time. We have
\be \label{Dk0_rtp}
D_k(0) = \langle (\delta x_{i+k}-\delta x_i)^2 \rangle = \frac{4v_0^2}{N} \sum_{q=1}^{(N-1)/2} \frac{1-\cos\left( \frac{2\pi kq}{N} \right)}{\mu_q(\mu_q+2\gamma)} = \frac{8T_{\rm eff}}{N g\rho^{s+2}} \sum_{q=1}^{(N-1)/2} \frac{\sin^2\left( \frac{\pi kq}{N} \right)}{f_s(\frac{q}{N})(1+\hat g f_s(\frac{q}{N}))} \;.
\ee
At large $N$, with $k \ll N$, this becomes
\be
D_k(0)  \simeq \frac{8 T_{\rm eff}}{g \rho^{s+2}} \int_0^{1/2} du  \frac{\sin^2\left( \pi k u  \right)}{f_s(u)\left(1+\hat g f_s(u) \right)} \;. \label{gaprtp}
\ee
We now study the behavior for large $k$, i.e. $1 \ll k \ll N$. 

Let us start with $s>0$. One finds that there is a change of behavior at a characteristic scale for $k$,
given by $k \sim \hat g^{1/z_s}$. This scale is large when $\hat g \gg 1$, the regime of most interest, where the system is strongly active.
In general the variance of the gaps takes the following scaling form
\be \label{scalinggaprtp}
D_k(0)  \simeq \frac{ T_{\rm eff} k^{z_s-1}}{g \rho^{s+2}}   \, {\sf G}_s( k/\hat g^{\frac{1}{z_s}}) \;.
\ee 
For large $k$ the integral in \eqref{gaprtp} is dominated by small values of $u$, which leads to the following expression for 
the scaling function 
\be \label{eqFgap} 
{\sf G}_s({\sf x}) = \frac{8}{a_s} \int_0^{+\infty} dv  \frac{\sin^2(\pi v)}{v^{z_s}\left(1+\frac{a_s}{{\sf x}^{z_s}} v^{z_s} \right)} 
= \frac{8}{a_s} {\sf x}^{1-z_s} \int_0^{+\infty} dw  \frac{\sin^2(\pi {\sf x} w)}{w^{z_s}\left(1+ a_s w^{z_s} \right)}  \;,
\ee
where the first form is more convenient to study the small ${\sf x}$ asymptotics, and the second to study the 
large ${\sf x}$ regime for $s<1/2$. Note that since $z_s=\min(2,1+s)$ the integrals are convergent for $s>-1/2$.

At large distances ${\sf x} = k/\hat g^{\frac{1}{z_s}} \gg 1$, ${\sf G}_s({\sf x}) \to {\sf G}_s(+\infty) = \frac{4}{a_s} \frac{\pi^{z_s-\frac{1}{2}}}{z_s-1} \frac{\Gamma(\frac{3-z_s}{2})}{\Gamma(\frac{z_s}{2})}$ and \eqref{scalinggaprtp} recovers the Brownian result \eqref{gapscases} with an effective temperature $T_{\rm eff}$.
However, on smaller scales, i.e. for ${\sf x} = k / \hat g^{\frac{1}{z_s}} \ll 1$, the activity plays an important role. There are three cases
to distinguish. 

For $s>1/2$ the small ${\sf x}$ behavior of the integral in \eqref{eqFgap} is obtained by expanding the sinus. 
In the short-range case $s>1$ ($z_s=2$), one finds
\bea
{\sf G}_s({\sf x}) \simeq \frac{8 \pi^2}{a_s} {\sf x}
\int_0^{+\infty} \frac{dw}{1 + a_s w^{2} } 
= \frac{4 \pi^3}{a_s^{3/2}} {\sf x} \;,   \quad , \quad {\sf x} \ll 1 
\eea 
leading to
\be \label{Dk0_smallsep_SR}
D_k(0) 
\simeq 
 \frac{v_0^2 k^2}{(2 (s+1)\zeta(s) g \rho^{s+2})^{3/2} \gamma^{1/2} } \;.
\ee 
For $1>s>1/2$ one finds with a similar computation
\be \label{gap_rtp_ysmall} 
{\sf G}_s({\sf x}) \simeq  \frac{8 \pi^2}{a_s} {\sf x}^{2-s}
\int_0^{+\infty} dw \frac{w^{1-s}}{1 + a_s w^{s+1} } 
= \frac{8 \pi^3}{a_s} \frac{{\sf x}^{2-s}}{a_s^{\frac{2-s}{1+s}}(1+s) \sin( \frac{2-s}{1+s} \pi) } \quad , \quad {\sf x} \ll 1  \;,
\ee
which gives
\be \label{Dk0_smallsep_LR1}
D_k(0) \simeq
\frac{8\pi^3 v_0^2 k^2}{(s+1) \sin( \frac{2-s}{1+s} \pi) (a_s g\rho^{s+2})^{\frac{3}{1+s}} (2\gamma)^{\frac{2s-1}{1+s}}}
\;. 
\ee
In both cases, we obtain a $k^2$ behavior. 

For $0 < s < 1/2$, the integral in \eqref{gap_rtp_ysmall} diverges at infinity. One uses instead the
first integral in \eqref{eqFgap} which leads to the asymptotics
\be 
{\sf G}_s({\sf x}) \simeq \frac{8}{a_s^2} {\sf x}^{1+2s} \int_0^{+\infty} dv  \frac{\sin^2(\pi v)}{v^{2(1+s)}} = \frac{4}{a_s^2} \frac{\pi^{2s+\frac{3}{2}}}{2s+1} \frac{\Gamma(\frac{1}{2}-s)}{\Gamma(s+1)} {\sf x}^{1+2s} \quad , \quad {\sf x} \ll 1 \;.
\ee
Hence we find that for $s<1/2$ the variance scales as $k^{1+2 s}$
\be \label{gap_rtp_half}
D_k(0) \simeq \frac{4 v_0^2}{(a_s g \rho^{s +2})^2} \frac{\pi^{2s+\frac{3}{2}}}{2s+1} \frac{\Gamma(\frac{1}{2}-s)}{\Gamma(s+1)} k^{1+2s} \;.
\ee
It is important to note that this behavior is independent of $\gamma$, hence it holds in the limit $\gamma \to 0^+$. 
This is by contrast with the case $s>1/2$ where, within our linear approximation, the variance diverges as $\gamma \to 0$,
see discussion below. 
\\

Let us now discuss separately the case of the log-gas $s=0$. In that case the scaling form \eqref{scalinggaprtp}
is still valid when $k, \hat g \gg 1$ with ${\sf x}=k/\hat g$ fixed, and the scaling function 
is given explicitly by
\be 
{\sf G}_0({\sf x}) = \frac{2}{\pi^2} 
   \left(\log(\frac{{\sf x}}{\pi })+\gamma - 
   \text{Ci}\left(\frac{{\sf x}}{\pi}\right) \cos
   \left(\frac{{\sf x}}{\pi}\right) -\text{Si}\left(\frac{{\sf x}}{\pi }\right) \sin
   \left(\frac{{\sf x}}{\pi }\right) \right) +\frac{1}{\pi} \sin(\frac{{\sf x}}{\pi}) \;.
\ee 
Its limiting behaviors are as follows.
For ${\sf x} \ll 1$ we find ${\sf G}_0({\sf x}) \simeq {\sf x}/\pi^2$, which predicts 
the linear behavior for $k \ll \hat g$
\be \label{gap_rtp_log_lin}
D_k(0) \simeq \frac{v_0^2 k}{(\pi g \rho^2)^2} \;,
\ee
i.e. the formula \eqref{gap_rtp_half} still holds. 
In the opposite limit ${\sf x}=k/\hat g \gg 1$ we find ${\sf G}_0({\sf x})\simeq  \ln (\frac{{\sf x}}{\pi}) + \gamma_E $, i.e.
\be \label{gap_rtp_log_log2}
D_k(0) \simeq \frac{2 T_{\rm eff}}{\pi^2 g \rho^2} (\ln (\frac{k}{\pi \hat g}) + \gamma_E) \;.
\ee
One can also obtain the result in the case where $k \gg 1$ but with $\hat g$ fixed. 
 The 
leading behavior is correctly given by the Brownian result \eqref{gaps_s0} with $T \to T_{\rm eff}$, but the constant term is modified and now
depends explicitly on $\hat g$. Indeed, computing the difference with the Brownian result as in Section~\ref{sec:var_rtp}, one finds a constant term which should be added in the expansion, leading for $k \gg 1$ at fixed $\hat g$ to 
\be \label{gap_rtp_log_log}
D_k(0) \simeq \frac{2 T_{\rm eff}}{\pi^2 g \rho^2} \left(\ln (2 \pi k)+\gamma_E - \pi \hat g \frac{\ln \left(1+ \pi^2 \hat g +\pi\sqrt{2\hat g (1+\frac{\pi^2}{2}\hat g})\right)}{\sqrt{2\hat g (1+\frac{\pi^2}{2} \hat g)}} \right) \;.
\ee
One can check that in the limit of $\hat g \gg 1$ it matches correctly \eqref{gap_rtp_log_log2}.
\\

Let us now consider the case $-1<s<0$. In this case, the variance of the gaps converges to a constant at large $k$,
\be \label{gap_rtp_sneg}
D_k(0) \to \frac{4 T_{\rm eff}}{g \rho^{s+2}} \int_0^{1/2} \frac{du}{f_s(u)\left(1+\hat g f_s(u) \right)} = 2 \langle \delta x_i^2 \rangle \;.
\ee
However, for $s>-1/2$, the result \eqref{gap_rtp_half} is still valid for $k \ll \hat g^\frac{1}{1+s}$ when $\hat g \gg 1$. In this case, the integral in \eqref{gap_rtp_sneg} scales as $\hat g^{\frac{s}{s+1}}$ (see \eqref{var_rtp_sneg_2}), and one can check that the limit \eqref{gap_rtp_sneg} is indeed reached for $k \sim \hat g^\frac{1}{1+s}$. On the other hand, for $-1/2>s>-1$, the limit is reached much faster and the regime in $k^{1+2s}$ does not exist even when $\hat g \gg 1$. Table~\ref{table:gap_var_rtp} summarizes the different regimes.

\subsubsection{Validity of the approximation}

The criterion for the validity of the linear approximation remains exactly the same as in the Brownian case, namely \eqref{cond_approx_Riesz2}. In the case of the RTP this leads to
\be \label{validity_rtp}
T_{\rm eff} \ll T_G = A_s(\hat g) \, g\rho^s   \quad , \quad A_s(\hat g) = \frac{1}{2(s+2)^2\int_0^{1/2} du \frac{\sin^2(\pi u)}{f_s(u) (1+\hat g f_s(u))}} \;.
\ee  
For $\hat g \ll 1$ this is the same criterion as in the Brownian case. For $\hat g \gg 1$, with the same argument as in the previous subsection we have
\be
\int_0^{1/2} du \frac{\sin^2(\pi u)}{f_s(u) (1+\hat g f_s(u))} \simeq \begin{dcases} \frac{1}{\hat g} \int_0^{1/2} du \frac{\sin^2(\pi u)}{f_s(u)^2} \hspace{1.52cm} \text{for } s<\frac{1}{2} \;, \\ \frac{\pi^3}{(s+1) \sin(\frac{2-s}{1+s}\pi) a_s^{\frac{3}{1+s}} \hat g^{\frac{2-s}{1+s}}} \quad \text{for } \frac{1}{2}<s<1 \;, \\ \frac{\pi^3}{2a_s^{3/2} \hat g^{1/2}} \hspace{2.97cm} \text{for } s>1 \;, \end{dcases}
\ee
which implies, for $\hat g \gg 1$,
\be \label{asympt_Bs}
A_s(\hat g) \simeq B_s \, \hat g^{\nu_s} \quad , \quad \nu_s = \begin{cases} 1 \hspace{0.4cm} \text{ for } s<\frac{1}{2} \\
\frac{2-s}{1+s} \text{ for } \frac{1}{2}<s<1 \\
\frac{1}{2} \hspace{0.35cm} \text{ for } s>1 \end{cases} \;,
\ee
where $B_s$ is a constant which only depends on $s$. Recalling that $\hat g = g\rho^{s+2}/(2\gamma)$, this leads to the following validity criterion
\bea \label{validity_rtp}
v_0^2 \ll B_s g^2 \rho^{2s+2} \ &\text{for }& s<\frac{1}{2} \;, \nn \\
\frac{v_0^2}{(2\gamma)^{\frac{2s-1}{s+1}}} \ll B_s g^{\frac{3}{1+s}} \rho^{\frac{s+4}{s+1}} \ &\text{for }& \frac{1}{2}<s<1 \;, \\
\frac{v_0^2}{\sqrt{2\gamma}} \ll B_s g^{3/2} \rho^{\frac{3}{2}s+1} \ &\text{for }& s>1 \;. \nn
\eea
It is interesting to note that, while for $s>1/2$ this criterion becomes increasingly difficult to satisfy as $\gamma$ decreases (keeping all
the other parameters fixed), for $s<1/2$ it becomes completely independent of $\gamma$ when $\gamma$ is small. Thus our approximation should allow us to study the limit $\gamma \to 0$, but only for $s<1/2$.
\\


\subsection{Time evolution of the displacement} \label{sec:C0(t)}

\begin{figure}
    \centering
    \includegraphics[width=0.4\linewidth,trim={0 0.3cm 0 0.8cm},clip]{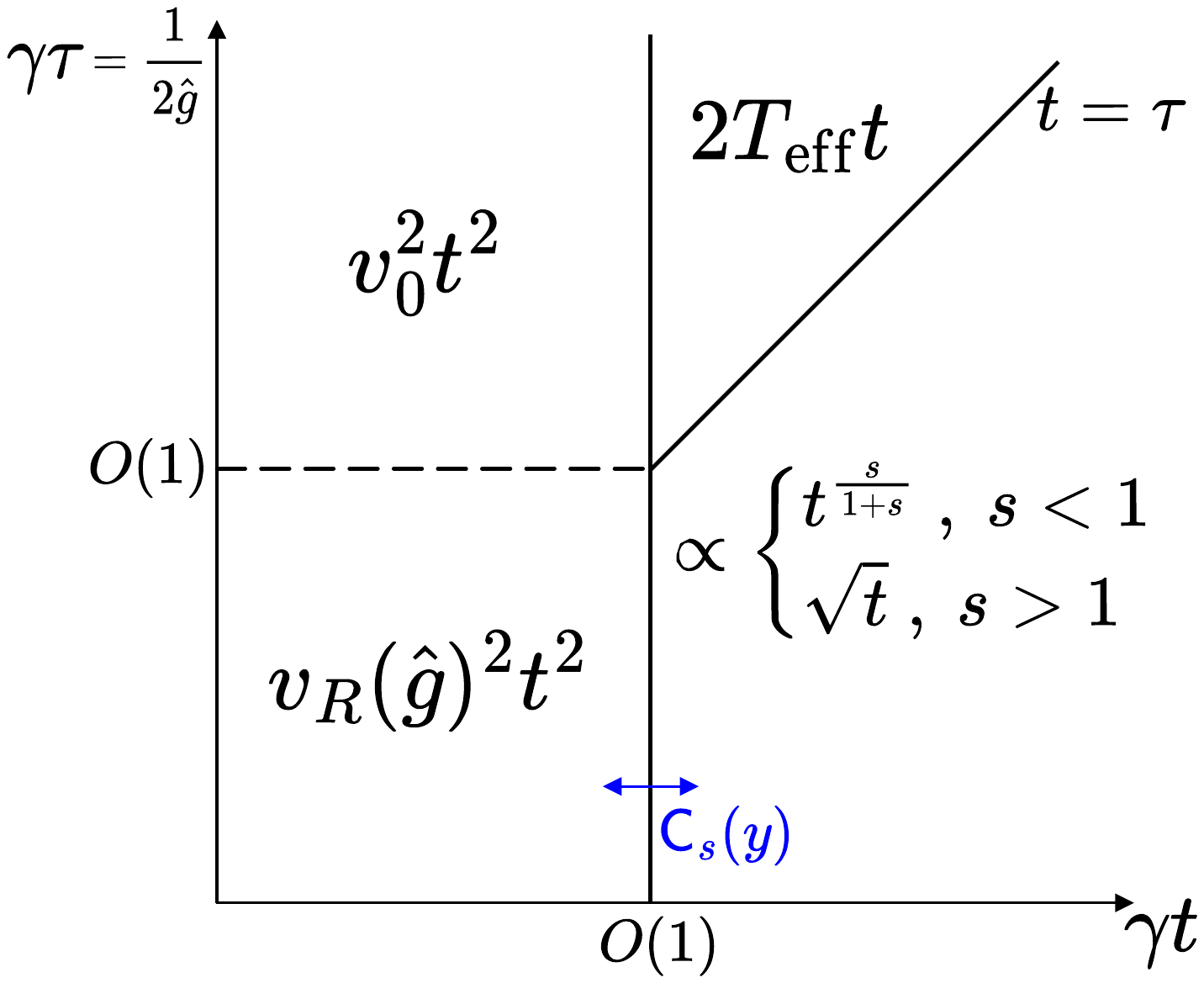}
    \caption{The different regimes for $C_0(t)$, the time evolution of the displacement of a given RTP,
    in the plane $\gamma t$, $\gamma \tau=1/(2 \hat g)$. For $\gamma t \ll 1$ there is a ballistic regime with
    a renormalized velocity $v_R=v_R(\hat g)$ which depends on the parameter $\hat g$. The upper triangle corresponds
    to free diffusion behavior, while the right portion corresponds to anomalous diffusion. The behavior becomes identical
    to the Brownian particles in the limit $\gamma t \gg 1$. For large $\hat g$ there is however a non-trivial
    crossover scaling function of $y=\gamma t$ which interpolates directly between ballistic behavior and anomalous diffusion. }
    \label{fig:diagrameC0}
\end{figure}

We now consider the time evolution of the displacement, in the stationary state, given by 
\be  \label{disp_RTP} 
C_0(t) = \langle (\delta x_i(t) - \delta x_i(0))^2 \rangle 
= \frac{4v_0^2}{N} \sum_{q=1}^{(N-1)/2} \frac{\mu_q (1-e^{-2\gamma t}) -2\gamma (1-e^{- \mu_q t})}{\mu_q(\mu_q^2-4\gamma^2)} \;.
\ee 

As long as $t\ll N^{z_s} \tau$, one can take the large $N$ limit and write
\be  \label{disp_RTP_largeN} 
C_0(t) \simeq
 4 T_{\rm eff} \tau \int_0^{1/2} du \frac{(1-e^{- f_s(u) t/\tau})-\hat g f_s(u) (1-e^{-2\gamma t})}{f_s(u)(1-\hat g^2 f_s(u)^2)} \;.
\ee

Let us analyze this formula in the different time regimes. Compared to the Brownian case, there is an additional time-scale $1/\gamma$.
One can distinguish the following regimes, summarized in Fig.~\ref{fig:diagrameC0}. 

\begin{itemize}

\item For $\gamma t \ll 1$, one can distinguish two cases. The first case is $t\ll \tau$, 
which corresponds to the region left of the line $\gamma t=1$
and above the diagonal in Fig.~\ref{fig:diagrameC0}. In that region one can clearly expand both exponentials in \eqref{disp_RTP_largeN} to second order, leading to
\be \label{ballistic}
C_0(t) \simeq v_R(\hat g)^2 t^2 \quad , \quad v_R(\hat g)^2 = v_0^2 \int_0^{1/2}
 \frac{2 \, du}{1 + \hat g f_s(u)}  \;.
\ee 
This correspond to a ballistic motion, with a velocity $v_R=v_R(\hat g) < v_0$ which is renormalized by the interactions.
This velocity depends on the ratio of time scales $\hat g=g \rho^{s+2}/(2\gamma)$ and tends to $v_0$ when $\hat g \to 0$ (upper left region
in Fig.~\ref{fig:diagrameC0}). 
It turns out that this result remains valid also below the diagonal for arbitrary $t/\tau$. In particular, we show in Appendix~\ref{app:Ckrtp_argument} that for $\hat g \gg 1$ (i.e. in the region near the $x$-axis in the figure) the main contribution to the integral in \eqref{disp_RTP_largeN} comes from $u\sim \hat g^{-1/z_s}$, and one can show that we can again expand the two exponentials.

Let us now study the behavior of the function $v_R(\hat g)$ in the limit of large $\hat g$,
i.e. in the region near the $x$-axis in the figure. For $-1<s<0$ it decays as $1/\hat g$ since one has
\be 
v_R(\hat g)^2 = v_0^2 \int_0^{1/2}
 \frac{2 \, du}{1 + \hat g f_s(u)} \simeq \frac{v_0^2}{\hat g}  \int_0^{1/2}
 \frac{2 \, du}{f_s(u)}  \;,
\ee 
where the integral in the prefactor of the $1/\hat g$ decay is the same as the one appearing in 
the expressions of the variance of the displacements and of the melting temperature for Brownian particles \eqref{T_melting}.

For $s>0$, the integral in \eqref{ballistic} is dominated by small $u\sim 1/\hat g^{1/z_s}$ and we may write
\be \label{vR_large_g}
v_R(\hat g)^2 \simeq v_0^2 \int_0^{1/2} \frac{2du}{1+\hat g a_s u^{z_s}} \simeq \frac{ v_0^2}{(\hat g a_s)^{1/z_s}} \int_0^{+\infty} \frac{2dv}{1+v^{z_s}} = \begin{cases} \frac{2\pi}{(s+1)\sin(\frac{\pi}{s+1})} \frac{v_0^2}{(a_s \hat g)^{\frac{1}{s+1}}} \text{ for } 0<s<1 \;, \\ \frac{\pi v_0^2}{\sqrt{a_s \hat g}} \hspace{2.42cm} \text{ for } s>1 \;, \end{cases}
\ee
i.e. a power law decay at large $\hat g$. 


\item If $\hat g \ll 1$, there exists a regime where $1/\gamma \ll t \ll \tau$. This corresponds to the upper right triangular region in Fig.~\ref{fig:diagrameC0}. In this case, the second term in the numerator and in the denominator of \eqref{disp_RTP_largeN} are both negligible, and one can expand the remaining exponential to first order, leading to
\be
C_0(t) \simeq \frac{v_0^2 t}{\gamma} = 2 T_{\rm eff} t \;,
\ee
which corresponds to free diffusion with an effective temperature $T_{\rm eff} = \frac{v_0^2}{2\gamma}$.

\item The last region corresponds to large times $t \gg 1/\gamma$ and $t\gg \tau$, right region in Fig.~\ref{fig:diagrameC0}. 
In that region
the integral in \eqref{disp_RTP_largeN}
is dominated by $u\sim (t/\tau)^{-1/z_s}$ (see Appendix~\ref{app:Ckrtp_argument}), and thus one can show that we recover the large time limit of the Brownian case \eqref{displacement_Riesz} 
with $T$ replaced by $T_{\rm eff}$, leading to anomalous diffusion, see  Fig.~\ref{fig:diagrameC0}.


\end{itemize} 

We can now discuss the crossover in $C_0(t)$ between ballistic and anomalous diffusion as $\gamma t$ increases (i.e. along horizontal lines
in Fig.~\ref{fig:diagrameC0}). At the top, for $\hat g \ll 1$, the system becomes equivalent to Brownian particles for $\gamma t \geq 1$
and the crossover coincides with the one for Brownian particles, with an intermediate free diffusion regime. This is shown in
Fig. \ref{fig:disp_rtp} and corresponds to the first line in Fig. \ref{fig:time_regimes_rtp} in Section \ref{subsec:MainResultsRTP}.
On the other hand for $\hat g \gg 1$ the activity plays an important role. The 
crossover occurs directly from the ballistic to the anomalous diffusion regions, see Fig.~\ref{fig:diagrameC0}.
It corresponds to the second line in Fig. \ref{fig:time_regimes_rtp}. It is described
by a non trivial scaling function, which is apparent in Fig. \ref{fig:disp_rtp} and 
its analytic form reads for $s>0$ 
\be \label{C0scaling}
C_0(t) = T_{\rm eff} \tau \hat g^{1-\frac{1}{z_s}}  {\sf C}_s( \gamma t) \quad , \quad
{\sf C}_s(y) = \frac{4}{a_s^{1/z_s}} \int_0^{+\infty} dv \frac{1-e^{-2 y v^{z_s}} - v^{z_s}(1-e^{-2 y})}{v^{z_s}(1 - v^{2 z_s})} \;.
\ee
For $y \ll 1$ we have ${\sf C}_s(y) \simeq \frac{8\pi}{a_s^{1/z_s} z_s \sin(\frac{\pi}{z_s})} y^2$, recovering \eqref{vR_large_g}, while for $y \gg 1$ we have ${\sf C}_s(y) \simeq 4 a_s^{-1/z_s} \frac{\Gamma(1/z_s)}{z_s-1} (2y)^{1-\frac{1}{z_s}}$, recovering \eqref{displacement_Riesz} (with $T$ replaced by $T_{\rm eff}$).
\\

\subsection{Multi-time correlations}

Let us now consider the covariance between the displacement at two different times $t_1$ and $t_2$. It reads
\bea \label{multitime_rtp}
C_0(t_1,t_2) &=& \langle (\delta x_i(t_1) - \delta x_i(0)) (\delta x_i(t_2) - \delta x_i(0)) \rangle \\
&=& \frac{2v_0^2}{N} \sum_{q=1}^{(N-1)/2}  \frac{\mu_q (1-e^{-2\gamma t_1} -e^{-2\gamma t_2} + e^{-2\gamma |t_1-t_2|}) -2\gamma (1-e^{- \mu_q t_1}-e^{- \mu_q t_2} + e^{- \mu_q |t_1-t_1|})}{\mu_q(\mu_q^2-4\gamma^2)} \nn \\
&=& \frac{1}{2} [ C_0(t_1) + C_0(t_2) - C_0(|t_1-t_2|) ] 
\nn \;,
\eea
i.e. it is connected to $C_0(t)$ through the same identity as in the Brownian case. We thus have a free diffusion regime for $1/\gamma \ll t_1,t_2,|t_1-t_2| \ll \tau$, and a fractional Brownian motion at large times, as in the Brownian case. In the ballistic regime, $t_1,t_2,|t_1-t_2| \ll 1/\gamma$, we obtain
\be
C_0(t_1,t_2) \simeq v_R^2  t_1 t_2 \;.
\ee

\subsection{Space-time correlations $C_k(t)$} \label{sec:Ck_rtp}




We now study the covariance between the displacements during time $t$ of two RTPs separated by $k$ lattice spacings, given for any $N$ by 
\be  \label{Ck_rtp_def} 
C_k(t) = \langle (\delta x_i(t) - \delta x_i(0)) (\delta x_{i+k}(t) - \delta x_{i+k}(0)) \rangle 
= \frac{4v_0^2}{N} \sum_{q=1}^{(N-1)/2} \frac{\mu_q (1-e^{-2\gamma t}) -2\gamma (1-e^{- \mu_q t})}{\mu_q(\mu_q^2-4\gamma^2)} \cos(2\pi k \frac{q}{N}) \;.
\ee 
We consider the large $N$ limit of this formula which reads, for $k \ll N$,
\be  \label{starting4} 
C_k(t) \simeq
 4 T_{\rm eff} \tau \int_0^{1/2} du \cos(2\pi k u) \frac{(1-e^{- f_s(u) t/\tau})-\hat g f_s(u) (1-e^{-2\gamma t})}{f_s(u)(1-\hat g^2 f_s(u)^2)} \;.
\ee

Apart from the adimensional effective temperature $T_{\rm eff}/(g\rho^s)$, the important dimensionless parameters are $t/\tau$, $\gamma t$, $\hat g$ and $k$ (not all of them are independent since $\hat g = 1/(2\gamma \tau)$). As usual, the limit $\hat g \ll 1$ corresponds to the diffusive case described in Section~\ref{sec:spacetime_brownian}. For $k=0$, we recover the variance of the displacement during time $t$, $C_0(t)$. When $1/\gamma \ll t \ll \tau$, we note that this implies $\hat g \ll 1$, which allows us to neglect the last term of both numerator and denominator. 
Expanding the first exponential, we recover the free diffusion regime,
where the particles diffuse independently and are uncorrelated with $C_{k}(t) \approx 0$ for any $k \neq 0$ (see \eqref{Ck_freediffusion}). In the ballistic regime $\gamma t \ll 1$,
we can expand both exponentials and write
\be \label{Ck_ballistic}
C_k(t) \simeq 2 v_0^2 t^2 \int_0^{1/2} du \frac{\cos( 2 \pi k u) }{1+\hat g f_s(u)} \;, 
\ee 
which defines again a "renormalized velocity" $v_R=v_R(\hat g,k)$, which now depends on $k$. 
\\

\begin{figure}
    \centering
    \includegraphics[width=0.7\linewidth,trim={0 0.5cm 0 0.5cm},clip]{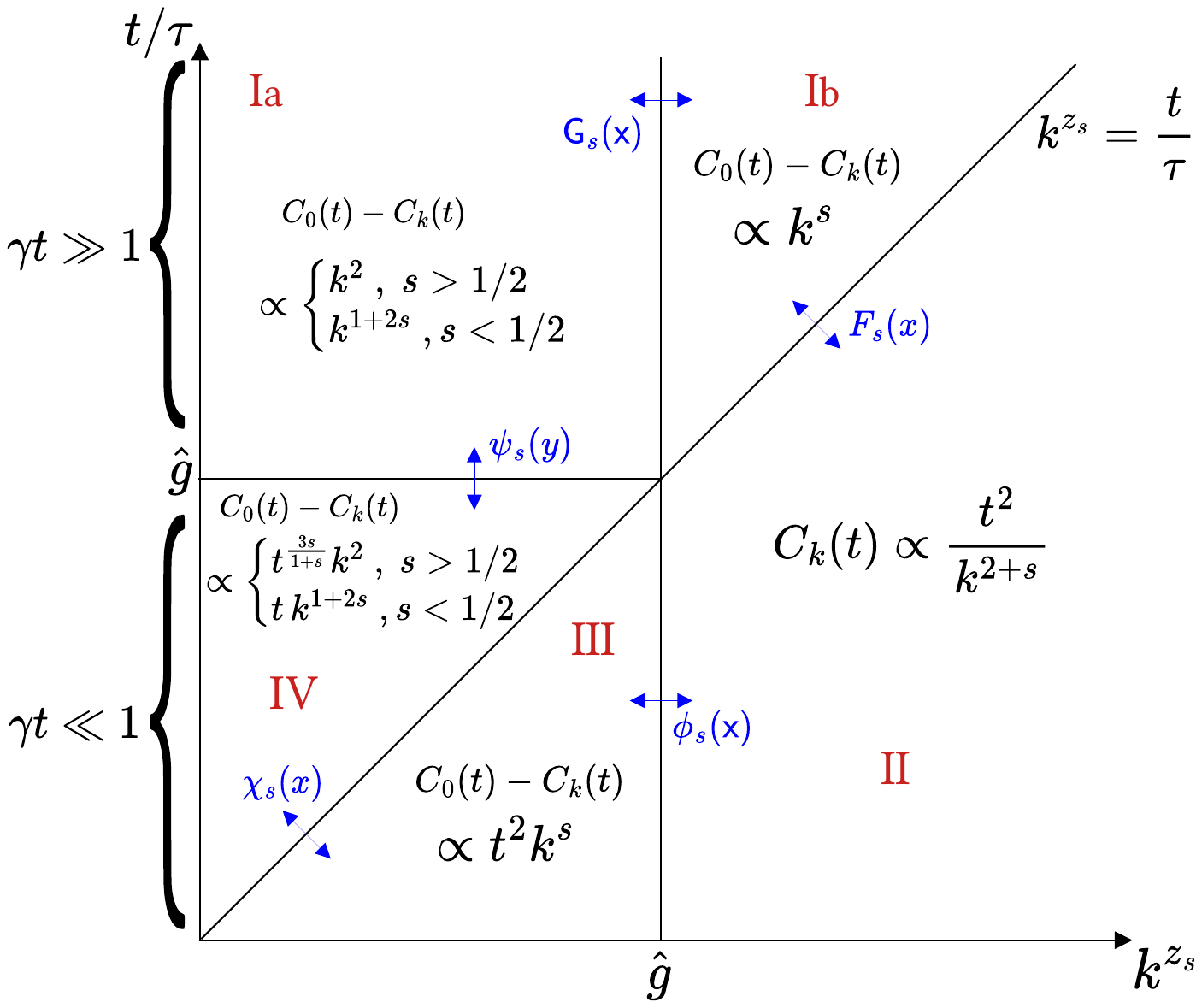}
    \caption{The five asymptotic regimes of the space-time correlation function $C_k(t)$ defined in \eqref{Ck_rtp_def},
    for $0<s<1$, represented in the plane $(k^{z_s}, t/\tau)$, where $\hat g=1/(2 \gamma \tau)$, in the
    limit where $k,t/\tau,\hat g \gg 1$. In that limit $C_k(t)$ admits two-argument scaling forms 
    \eqref{Ck_scaling1} and \eqref{Ck_scaling2}. The solid lines indicate the location of crossover regions
    between different asymptotic forms which are explicitly indicated. In the large time limit $C_0(t)-C_k(t) \to D_k(0)$
    the variance of the gaps defined in \eqref{Dk0_rtp} with a crossover between regions Ia and Ib described by the function ${\sf G}_s({\sf x})$, with ${\sf x}= k/\hat g^{1/z_s}$, in \eqref{scalinggaprtp}.
    The regions Ib and II correspond to an effective Brownian regime with the crossover function $F_s(x)$ of the scaling variable $x=k/(t/\tau)^{1/z_s})$ 
    as found for Brownian particles in \eqref{Ck_scaling}. The short time regime $y= \gamma t \ll 1$ is ballistic and corresponds to regions 
    II and III, where the renormalized velocity $v_R$ exhibits a crossover described by the function $\phi_s({\sf x})$ of the scaling variable ${\sf x}= k/\hat g^{1/z_s}$ given in \eqref{Ck_ballistic2}. The last region IV $y=\gamma t \ll 1, x = k/(t/\tau)^{1/z_s}) \ll 1$ is neither ballistic nor Brownian
    and the crossovers towards region III is described by the scaling function $\chi_s(x)$ of the scaling variable $x=k/(t/\tau)^{1/z_s}$ given in \eqref{Ck_smallsep_smalltime}
    and towards region Ia by the scaling function $\psi_s(y)$ of the scaling variable $y=\gamma t$ given in \eqref{Ck_smallsep}.}
    \label{fig:diagrameCk}
\end{figure}

{\bf Scaling form.} When at least one of the parameters $t/\tau$, $\hat g$ or $k$ is large, the integral is dominated by small values of $u$, and thus one can approximate $f_s(u)$ using \eqref{fasympt}, for any $s>0$. Let us clarify this in the three cases. If $k$ is large, then it is clear that the integral is dominated by $u\sim 1/k \ll 1$. 
If $\hat g \gg 1$, we show in Appendix~\ref{app:Ckrtp_argument} that the integral is dominated by $u\sim \hat g^{-1/z_s}$.
Finally, for $t/\tau \gg 1$, we show in Appendix~\ref{app:Ckrtp_argument} that the main contribution comes from $u\sim (t/\tau)^{-1/z_s}$.

In the three regimes where either $t/\tau \gg 1$, or $\hat g \gg 1$, or $k \gg 1$, we can thus write $C_k(t)$ under the scaling form (writing $v=(a_s \hat g)^{1/z_s} u$)
\be \label{Ck_scaling1}
C_k(t) \simeq T_{\rm eff} \tau \hat g^{1-\frac{1}{z_s}}  \mathcal{\tilde F}_s( k/\hat g^{1/z_s} , \gamma t) \quad , \quad
\mathcal{\tilde F}_s({\sf x},y) = \frac{4}{a_s^{1/z_s}} \int_0^{+\infty} dv \cos(2 \pi a_s^{-1/z_s} {\sf x} v)\frac{1-e^{-2 y v^{z_s}} - v^{z_s}(1-e^{-2 y})}{v^{z_s}(1 - v^{2 z_s})} \;.
\ee
The integral is absolutely convergent for $s>-1/2$. This scaling form is particularly interesting when $\hat g, k \gg 1$ with $k \sim \hat g^{1/z_s}$ and fixed $\gamma t$, i.e. 
a finite number of tumblings per particle. An alternative equivalent scaling form, which is more practical to study the Brownian limit, is (setting $v=(a_s t/\tau)^{1/z_s} u$)
\be \label{Ck_scaling2}
C_k(t) \simeq T_{\rm eff} \tau (t/\tau)^{1-\frac{1}{z_s}} \mathcal{F}_s \left(\frac{k}{(t/\tau)^{\frac{1}{z_s}}},\gamma t \right) \quad , \quad \mathcal{F}_s(x,y) = \frac{4}{a_s^{1/z_s}} \int_0^{+\infty} dv \cos(2 \pi a_s^{-\frac{1}{z_s}}x v)\frac{1-e^{- v^{z_s}} - v^{z_s} \frac{1-e^{-2 y}}{2y}}{v^{z_s}(1 - v^{2 z_s}/(2y)^2)} \;.
\ee
The two scaling forms are equivalent, and we can use one or the other depending on the limit that we want to study.

We now study $C_k(t)$ in the joint limit $\hat g \gg 1$, $k\gg 1$ and $t/\tau \gg 1$. We focus on $s>0$. There are five regions with different asymptotic behaviours which are displayed in Fig.~\ref{fig:diagrameCk}. We show how they can be recovered using the two scaling forms introduced above.
\\

\textit{Brownian limit}. The Brownian case is recovered in the limit $y=\gamma t \to +\infty$, keeping $x=k/(t/\tau)^{1/z_s}$ fixed. Indeed, when $y \to +\infty$, we see that $\mathcal{F}_s(x,y) \to F_s(x)$ and we recover the scaling form for the Brownian case. Note that $\gamma t \gg 1$ is equivalent to $t/\tau \gg \hat g$, thus taking $\gamma t \to +\infty$ with $k/(t/\tau)^{\frac{1}{z_s}}$ fixed implies $k \gg \hat g^{\frac{1}{z_s}}$. The scaling form \eqref{Ck_scaling2} is useful to compute corrections to the Brownian limit.

As a reminder, in the Brownian regime we have
\be \label{Ck_scaling_brownian}
C_k(t) \simeq \begin{dcases} T_{\rm eff} \tau (t/\tau)^{\frac{s}{s+1}} F_s\left( \frac{k}{(t/\tau)^{\frac{1}{s+1}}} \right) \quad \text{for } 0<s<1 \\
T_{\rm eff} \sqrt{\tau \, t} \, F_s\left( \frac{k}{\sqrt{t/\tau}} \right) \hspace{1.55cm} \text{for } s>1 \end{dcases} \quad , \quad F_s(x) = 4 \, a_s^{-\frac{1}{z_s}} \int_0^{+\infty} dv \frac{1-e^{-v^{z_s}}}{v^{z_s}} \cos \left( 2\pi a_s^{-\frac{1}{z_s}} x v \right)
\ee 
($F_s(x)$ can be computed explicitly for $s>1$, see \eqref{Fs}). For $x \ll 1$, the asymptotics of $F_s(x)$ are given by
\be
F_s(x) \simeq  \begin{cases} U_s - \frac{1}{\pi \tan( \frac{\pi s}{2}) } x^s \hspace{2.7cm} \text{for } 0<s<1 \;, \\ \frac{2}{\sqrt{\pi(s+1) \zeta(s)}} - \frac{x}{(s+1)\zeta(s)} + O(x^2) \quad \text{for } s>1 \;, \end{cases}
\ee 
which leads to, in the large time regime $k^{z_s} \ll t/\tau$,
\be \label{CkIb}
C_0(t)-C_k(t) \simeq \begin{cases}  \frac{T_{\rm eff} \tau}{\pi \tan( \frac{\pi s}{2}) } k^s \hspace{1.55cm} \text{for } 0<s<1 \;, \\ \frac{T_{\rm eff} \tau}{(s+1)\zeta(s)} k + O(x^2) \quad \text{for } s>1 \;, \end{cases}
\ee
which is compatible with $C_0(t)-C_k(t)=D_k(0)-D_k(t) \simeq D_k(0)$.
For $x \gg 1$ we find
\be 
F_s(x) \simeq  \begin{cases} \frac{s+1}{x^{2+s}} \hspace{3.2cm} \text{for } 0<s<1 \;, \\  \frac{4\sqrt{\pi(s+1)\zeta(s)}}{x^2} e^{-\frac{x^2}{2(s+1)\zeta(s)}} \quad \text{for } s>1 \;, \end{cases}
\ee 
which leads to, in the large distance regime $k^{z_s} \gg t/\tau$,
\be \label{CkII}
C_k(t) \simeq \begin{cases} (s+1)\frac{T_{\rm eff}}{\tau} \frac{t^2}{k^{2+s}}  \hspace{4.18cm} \text{for } 0<s<1 \;, \\ 4 T_{\rm eff} \sqrt{\frac{\pi(s+1)\zeta(s)}{\tau}} \, t^{3/2} \exp \left(-\frac{\tau k^2}{2(s+1)\zeta(s) t} \right) \quad \text{for } s>1 \;. \end{cases}
\ee
In summary, this Brownian behavior corresponds to the regions Ib and II on the right in Fig.~\ref{fig:diagrameCk}. 
The asymptotics \eqref{CkIb} corresponds to the regime Ib in Fig.~\ref{fig:diagrameCk}, i.e. above the diagonal,
while \eqref{CkII} corresponds to the top part of region II. The scaling function $F_s(x)$ describes the crossover
between these two regions as a function of the scaling variable $x=k/(t/\tau)^{1/z_s}$. 
\\

\textit{Large separation limit.} We now consider the limit of large distances, $k \gg \hat g^{1/z_s}$, keeping $\gamma t$ fixed. 
This means taking ${\sf x} \gg 1$ in \eqref{Ck_scaling1} while keeping $y$ fixed. This corresponds to the region II in Fig.~\ref{fig:diagrameCk}.
For $0<s<1$, we find for any $y>0$, $\mathcal{\tilde F}_s({\sf x},y) \simeq \frac{4(s+1)y^2}{{\sf x}^{2+s}}$ (see Appendix~\ref{app:computation_details}), i.e.
\be \label{Ck_ballistic3}
C_k(t) \simeq (s+1)\frac{T_{\rm eff}}{\tau} \frac{t^2}{k^{2+s}} \;.
\ee
Thus it remains exactly the same as the large distance limit of the the Brownian regime, with the same ballistic behaviour. 
\\

\textit{Ballistic limit.} Let us consider now the regime $y=\gamma t \ll 1$, with ${\sf x} = k/\hat g^{1/z_s}$ fixed (note that this implies $k \gg (t/\tau)^{\frac{1}{z_s}}$). This corresponds to the crossover between the regions II and III in Fig.~\ref{fig:diagrameCk}. In this case we can expand both exponentials in \eqref{Ck_scaling1} to obtain $\mathcal{\tilde F}_s({\sf x},y) \simeq 4 y^2 \phi_s({\sf x})$, which leads to the scaling form
\be \label{Ck_ballistic2}
C_k(t) \simeq v_0^2 t^2 \hat g^{-1/z_s} \phi_s(k/\hat g^{1/z_s}) \quad , \quad \phi_s({\sf x})= 
\frac{2}{a_s^{1/z_s}} \int_0^{+\infty} dv \frac{\cos (2 \pi a_s^{-1/z_s} {\sf x} v)}{1+v^{z_s}} \;.
\ee 
This is actually the limit $\hat g \gg 1$ of \eqref{Ck_ballistic}. For $s>1$, i.e. $z_s=2$, the integral can be evaluated, leading to an exponential decay in $k$,
\be \label{Ckexp}
C_k(t) \simeq \frac{v_0^2 t^2}{2\sqrt{(s+1)\zeta(s)\hat g}} \exp\left( -\frac{k}{\sqrt{(s+1)\zeta(s)\hat g}}\right) \;.
\ee
which describes the crossover between the regions II and III in the short-range case $s>1$. 
Note that contrarily to the long range case $s<1$, this result is different from the one 
obtained in \eqref{CkII} in the Brownian regime, although it it exponential in both cases. 
This means that for $s>1$ the region II actually splits into two different subregions.
For ${\sf x} = k/\sqrt{\hat g} \ll 1$ (region III) \eqref{Ckexp} leads to
\be
C_0(t) - C_k(t) \simeq v_0^2 t^2 \frac{k}{2(s+1)\zeta(s)\hat g} \;.
\ee

For $0<s<1$, there is no closed form expression for $\phi_s({\sf x})$ but the large $k$ asymptotics, i.e. for ${\sf x}=k/\hat g^{1/z_s} \gg 1$, can be obtained (see Appendix \ref{app:computation_details}). One finds that 
$\phi_s({\sf x}) \simeq (s+1) {\sf x}^{-(2+s)}$ for ${\sf x} \gg 1$, and thus we recover \eqref{Ck_ballistic3} (region II in Fig.~\ref{fig:diagrameCk}) 
\be \label{Ck_large_k_ballistic}
C_k(t) \simeq \frac{(s+1) \hat g}{k^{2+s}} v_0^2 t^2 = (s+1)\frac{T_{\rm eff}}{\tau} \frac{t^2}{k^{2+s}} \;.
\ee
In the opposite limit ${\sf x}=k/\hat g^{1/z_s} \ll 1$ (region III in Fig.~\ref{fig:diagrameCk}) we have
\be \label{phi_smallx}
\phi_s(0) - \phi_s({\sf x}) = \frac{4}{a_s^{1/z_s}} \int_0^{+\infty} dv \frac{\sin^2 (\pi a_s^{-1/z_s} {\sf x} v)}{1+v^{z_s}} \simeq \frac{4}{a_s} {\sf x}^{z_s-1} \int_0^{+\infty} dw \frac{\sin^2 (\pi w)}{w^{z_s}}
= \frac{{\sf x}^s}{2\pi \tan(\frac{\pi s}{2})} \;, 
\ee
which leads to
\be
C_0(t)- C_k(t) = \frac{k^s}{2\pi \tan(\frac{\pi s}{2})} \frac{v_0^2 t^2}{\hat g} \;,
\ee
where $C_0(t)=v_R^2 t^2$ and the renormalized velocity $v_R$ is given in \eqref{vR_large_g}. 

In the special case of the RTP log-gas, $s=0$, we can again obtain an explicit expression of $\phi_s({\sf x})$ for any ${\sf x}$ and one finds
\be
C_k(t)
\simeq \frac{v_0^2 t^2}{\hat g} \phi_0(\frac{k}{\hat g}) \quad , \quad \phi_0(x)= \frac{1}{2\pi^2} \left( (\pi -2 \text{Si}(\frac{x}{\pi})) \sin
   (\frac{x}{\pi})-2 \text{Ci}(\frac{x}{\pi}) \cos (\frac{x}{\pi}) \right) \simeq \begin{cases}  -\frac{1}{\pi^2} (\ln (\frac{x}{\pi}) + \gamma_E) 
   \quad , \ x \ll 1 \\
   \frac{1}{x^2} \hspace{2.6cm} , \ x \gg 1
   \end{cases} .
\ee 
\\

\textit{Large time limit.} Let us now consider the large time limit $y=\gamma t \gg 1$ in \eqref{Ck_scaling1}, with ${\sf x} = k/\hat g^{1/z_s}$ fixed (this implies that $k \ll (t/\tau)^{\frac{1}{z_s}}$). This describes the crossover between region Ia and Ib in Fig.~\ref{fig:diagrameCk}. 
In this large time regime, we can simply neglect the two exponentials, which implies that the following difference has reached
a time independent limit
\be \label{Ck_largetime_smallsep}
C_0(t) - C_k(t) \simeq T_{\rm eff} \tau \hat g^{1-\frac{1}{z_s}} \frac{8}{a_s^{1/z_s}} \int_0^{+\infty} dv \frac{\sin^2 (\pi a_s^{-1/z_s} {\sf x} v)}{v^{z_s}(1 + v^{z_s})} = D_k(0) \;,
\ee
which equals the variance of the gap as can be seen by 
comparing with the expression of $D_k(0)$ given in \eqref{scalinggaprtp}. This is compatible with the identity $C_0(t) - C_k(t) = D_k(0) - D_k(t)$ if $D_k(t) \ll D_k(0)$, which is indeed true in this regime, as we will see in Section~\ref{sec:Dk_rtp}. The asymptotics of $D_k(0)$ for $k \gg \hat g^{1/z_s}$ and $k \ll \hat g^{1/z_s}$ were derived in Section~\ref{sec:Dk0_rtp}. As a reminder, for $0<s<1$, we know that in this regime $C_0(t) \propto t^{\frac{s}{s+1}}$, and that $D_k(0) \propto k^s$ when $k \gg \hat g^{1/z_s}$ and $D_k(0) \propto k^2$ for $s>1/2$ and $\propto k^{1+2s}$ for $s<1/2$ when $k \ll \hat g^{1/z_s}$. The large ${\sf x}$ limit of $D_k(0)$ matches the small $x$ limit of the Brownian regime. The crossover between the two regions Ia and Ib is thus described by
the scaling form \eqref{scalinggaprtp} and the scaling function ${\sf G}_s({\sf x})$ with ${\sf x}=k/\hat g^{1/z_s}$.
\\

\textit{Small separation limit.} Let us now focus on the small separation regime. We saw that, at large times, i.e. in region Ia, the leading order of the difference $C_0(t)-C_k(t)$ is independent of time (since it is given by $D_k(0)$), while at small times it is ballistic in region III. In addition, the two limits also have a different dependence on $k$. One can ask how does on goes from regions Ia to III. As one can see in Fig.~\ref{fig:diagrameCk} there is an intermediate
region, region IV. 
To see this, we first study the limit  ${\sf x}=k/\hat g^{1/z_s} \ll 1$ at fixed $y=\gamma t$ of \eqref{Ck_scaling1}. 
This will describe the crossover from Ia to IV. One must distinguish two cases, $s<1/2$ and $s>1/2$.

For $s>1/2$, we can obtain the expansion $\mathcal{\tilde F}_s(0,y)-\mathcal{\tilde F}_s({\sf x},y) \simeq {\sf x}^2 \psi_s(y)$ from 
\eqref{Ck_scaling1} by expanding $1- \cos(\theta) \simeq \frac{\theta^2}{2}$ and one finds
\be \label{Ck_smallsep}
C_0(t) - C_k(t) = T_{\rm eff} \tau \hat g^{1-\frac{3}{z_s}} k^2 \psi_s(\gamma t) \quad , \quad \psi_s(y)= \frac{8\pi^2}{a_s^{3/z_s}}  \int_0^{+\infty} du \, u^2 \frac{1-e^{-2 y u^{z_s}} - u^{z_s}(1-e^{-2 y})}{u^{z_s}(1 - u^{2 z_s})} \;,
\ee
which implies that $C_0(t)-C_k(t) \propto k^2$ for any finite $y=\gamma t$. For $y=\gamma t \gg 1$, i.e. in region Ia, the crossover function $\psi_s(y)$ saturates to a finite limit
\be
\lim_{y \to \infty} \psi_s(y) = \frac{8\pi^2}{a_s^{3/z_s}} \int_0^{+\infty} du \frac{u^{2-z_s}}{1 + u^{z_s}} \;,
\ee
which, inserting back into \eqref{Ck_smallsep}, corresponds to $D_k(0)$ in the same limit, as expected [see \eqref{Dk0_smallsep_SR} and \eqref{Dk0_smallsep_LR1}]. In the opposite limit $y=\gamma t \ll 1$, i.e. within region IV, we find
\be \label{psi_smally}
\psi_s(y) \simeq \frac{8\pi^2}{a_s^{3/z_s}} (2y)^{\frac{3(z_s-1)}{z_s}} \int_0^{+\infty} dv \, \frac{1-\frac{1-e^{-v^{z_s}}}{v^{z_s}}}{v^{2(z_s-1)}} 
= \begin{cases} &\frac{8\sqrt{2\pi}}{3(s+1)\zeta(s)} y^{3/2} \quad \hspace*{1.24cm} \;\text{for } s>1 \;, \\ 
& \\
&\frac{8\pi^2}{3} \frac{(1+s) \Gamma(\frac{2-s}{1+s})}{s(2s-1)} (2y)^{\frac{3s}{1+s}} \quad \text{for } 1/2<s<1 \;. \end{cases}
\ee
We thus obtain $C_0(t)-C_k(t) \propto t^{3/2} k^2$ for $s>1$ and $C_0(t)-C_k(t) \propto t^{\frac{3s}{1+s}} k^2$ for $1/2<s<1$ 
within region IV in Fig.~\ref{fig:diagrameCk}. 

For $s<1/2$, the difference $\mathcal{\tilde F}_s(0,y)-\mathcal{\tilde F}_s({\sf x},y) \simeq {\sf x}^2 \psi_s(y)$
computed from \eqref{Ck_scaling1} 
is given by an integral which is 
dominated by large values of $v$ (of order $\sim 1/{\sf x}$). Performing a change of variable $w=a_s^{-1/z_s} {\sf x} v$, we get
\be
\mathcal{\tilde F}_s(0,y)-\mathcal{\tilde F}_s({\sf x},y) = 8 \int_0^{+\infty} \frac{dw}{{\sf x}} \sin^2(\pi w) \frac{1-e^{-2 a_s y (\frac{w}{{\sf x}})^{1+s}} - a_s(\frac{w}{{\sf x}})^{1+s}(1-e^{-2 y})}{a_s(\frac{w}{{\sf x}})^{1+s}(1 - a_s^2(\frac{w}{{\sf x}})^{2(1+s)})} \;.
\ee
For ${\sf x} \ll 1$ this becomes (we also assume ${\sf x}^{1+s}\ll y$, or $k^{1+s} \ll t/\tau$, i.e. we consider the limit ${\sf x} \ll 1$ with $y$ fixed)
\be \label{Ck_smallxhalf}
\mathcal{\tilde F}_s(0,y)-\mathcal{\tilde F}_s({\sf x},y) \simeq \frac{8}{a_s^2} {\sf x}^{1+2s} (1-e^{-2 y}) \int_0^{+\infty} dw \frac{\sin^2(\pi w)}{w^{2(1+s)}} = \frac{4}{a_s^2} \frac{\pi^{2s+\frac{3}{2}}}{2s+1} \frac{\Gamma(\frac{1}{2}-s)}{\Gamma(s+1)} (1-e^{-2y}) \hspace*{0.1cm} {\sf x}^{1+2s} \;,
\ee
which implies that the crossover form from region Ia to region IV takes the form for $s<1/2$
\be
C_0(t)-C_k(t) \simeq \frac{4}{a_s^2} \frac{\pi^{2s+\frac{3}{2}}}{1+2s} \frac{\Gamma(\frac{1}{2}-s)}{\Gamma(1+s)} \frac{T_{\rm eff} \tau}{\hat g} (1-e^{-2\gamma t}) k^{1+2s} \;.
\ee
Thus for $\gamma t\gg 1$ (region Ia) it recovers $D_k(0)$ as given by \eqref{gap_rtp_half}, while for $\gamma t \ll 1$ (region IV) one obtains $C_0(t)-C_k(t) \propto t k^{1+2s}$. 
\\

\textit{Small time and separation.} We now study the crossover between regions IV and III. This crossover occurs along the diagonal in Fig.~\ref{fig:diagrameCk}, 
for $k \sim (t/\tau)^{1/z_s}$. We thus return to the scaling form \eqref{Ck_scaling2} and take the limit $y=\gamma t \ll 1$, while keeping $x=k/(t/\tau)^{1/z_s}$ fixed. This leads to the following crossover scaling form (writing $\mathcal{F}_s(0,y) - \mathcal{F}_s(x,y) \simeq y^2 \chi_s(x)$)
\be \label{Ck_smallsep_smalltime}
C_0(t)-C_k(t) \simeq T_{\rm eff} \tau \left(\frac{t}{\tau} \right)^{1-\frac{1}{z_s}} (\gamma t)^2 \, \chi_s \left( \frac{k}{(t/\tau)^{1/z_s}} \right) \quad , \quad \chi_s(x) = \frac{32}{a_s^{1/z_s}} \int_0^{+\infty} dv \sin^2(\pi a_s^{-1/z_s}x v)\frac{1 - \frac{1-e^{- v^{z_s}}}{v^{z_s}}}{v^{2 z_s}} \;.
\ee
Note that the integral over $v$ in the crossover function $\chi_s(x)$ is convergent for any $s>0$. 

For $x\gg 1$, i.e. for $k^{z_s} \gg t/\tau$, this integral is dominated by small values of $v$ (of the order of $1/x$). Indeed, writing $w=a_s^{-1/z_s} x v$, we obtain
\be
\chi_s(x) = 32 x^{2z_s-1} \int_0^{+\infty} dw \sin^2(\pi w) \frac{1 - \frac{1-e^{- a_s (\frac{w}{x})^{z_s}}}{a_s(\frac{w}{x})^{z_s}}}{a_s^2 w^{2z_s}} \simeq \frac{16}{a_s} x^{z_s-1}\int_0^{+\infty} dw \frac{\sin^2(\pi w)}{ w^{z_s}} \;.
\ee
We thus recover the small ${\sf x}=k/\hat g^{1/z_s}$ limit of \eqref{Ck_ballistic2} (see Eq. \eqref{phi_smallx}), i.e. region III, with a $k^s$ dependence for $0<s<1$, and $k$ for $s>1$ of $C_0(t)-C_k(t)$. In the opposite limit $x\ll 1$, i.e. $k^{z_s} \ll t/\tau$ and region IV, for $s>1/2$ we can linearize the sine function in the definition of $\chi_s(x)$ to obtain
\be 
\chi_s(x) \simeq 32 \pi^2 a_s^{-3/z_s} x^2 \int_0^{+\infty} dv \frac{1-\frac{1-e^{-v^{z_s}}}{v^{z_s}}}{v^{2(z_s-1)}} \;, 
\ee
which matches the small $y=\gamma t$ limit of \eqref{Ck_smallsep} (see \eqref{psi_smally}). We thus recover the $\propto t^{\frac{3s}{1+s}} k^2$ behaviour (or $\propto t^{3/2} k^2$ for $s>1$) of the limit $k\ll \hat g$ in region IV. If $s<1/2$, the integral is dominated by large values of $v$ and we can instead write
\be
\chi_s(x) \simeq \frac{32}{a_s^2} x^{1+2s} \int_0^{+\infty} dw \frac{\sin^2(\pi w)}{w^{2(1+s)}} \;,
\ee
and we recover again the small $y=\gamma t$ limit of \eqref{Ck_smallsep} (see \eqref{Ck_smallxhalf}), region IV, where $C_0(t)-C_k(t) \propto t k^{1+2s}$.

\subsection{Time correlations of the gaps $D_k(t)$} \label{sec:Dk_rtp}

\begin{figure}
    \centering
    \includegraphics[width=0.7\linewidth,trim={0 0.5cm 0 0.5cm},clip]{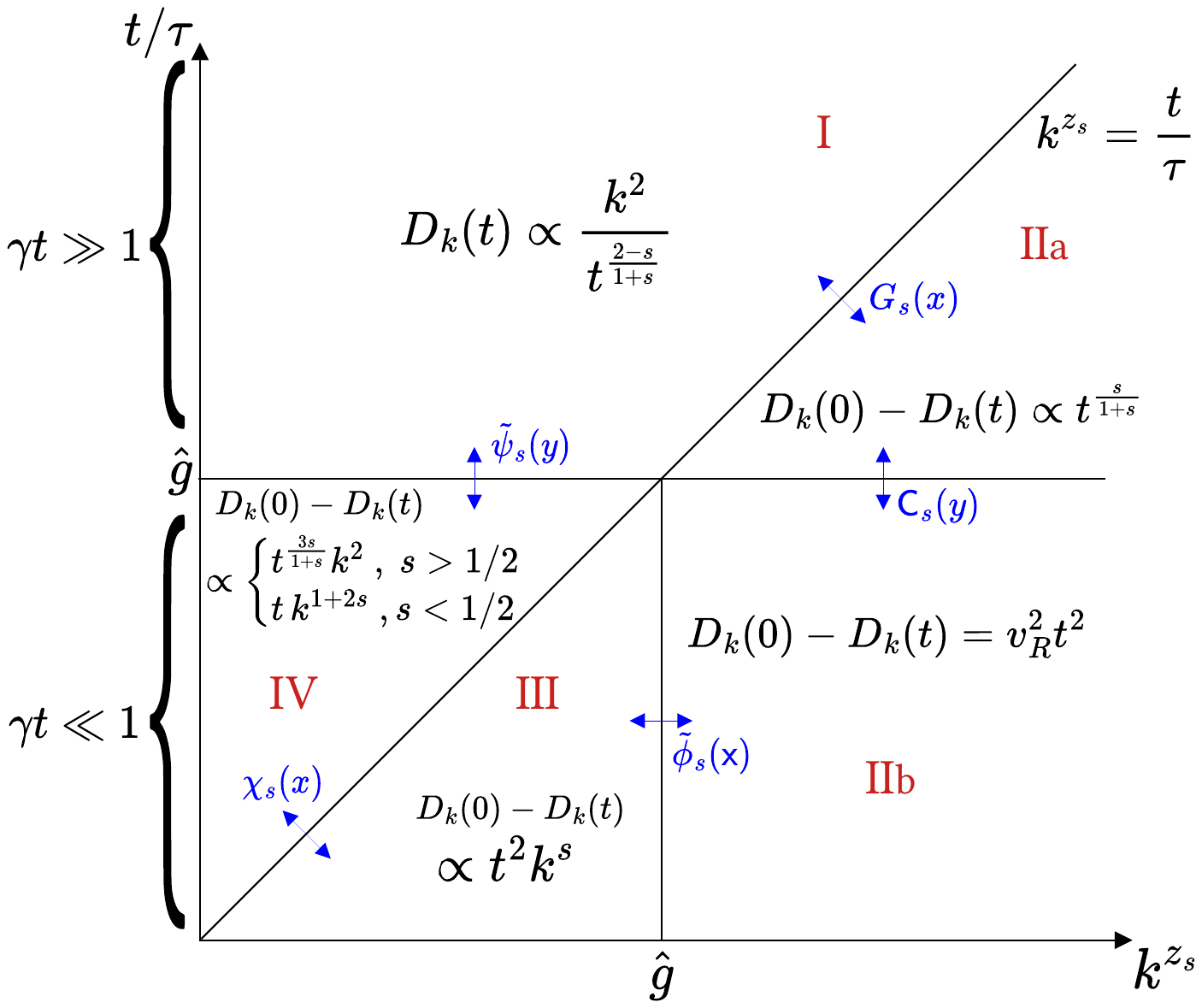}
    \caption{The five asymptotic regimes of the time correlation
    of the gaps $D_k(t)$ defined in \eqref{Dk_scaling_intro},
    for $0<s<1$, represented in the plane $(k^{z_s}, t/\tau)$, where $\hat g=1/(2 \gamma \tau)$, in the
    limit where $k,t/\tau,\hat g \gg 1$. In that limit $D_k(t)$ admits two-argument scaling forms 
    \eqref{Dk_scaling1} and \eqref{Dk_scaling2}. The solid lines indicate the location of crossover regions
    between different asymptotic forms which are explicitly indicated. In the large time limit, i.e. $t \gg 1/\gamma,\tau$, $D_k(t)$ decays to zero algebraically, corresponding
    to region I. In the large separation limit (regions IIa and IIb) one has $D_k(0)-D_k(t) \simeq C_0(t)$. This part of the diagram is divided into two regions IIa and IIb, depending on the behaviour of $C_0(t)$. The region IIa corresponds to the Brownian regime with anomalous diffusion, while IIb corresponds to the ballistic regime with a renormalized velocity $v_R=v_R(\hat g)$. The crossover between region I and IIa occurs along the diagonal and is described by 
    the scaling function ${\sf G}_s(x)$ where $x=k/(t/\tau)^{1/z_s}$, given in \eqref{Dk_scaling_brownian2}. The crossover between regions IIa and IIb 
    occurs along a horizontal line and is described by 
    the scaling function ${\sf C}_s(y)$ where $y=\gamma t$, given in \eqref{C0scaling}. 
    The bottom left square corresponds to short time and small gaps and is divided into the two regions III and IV. 
    The region III is ballistic but with a different dependence in $k$ from region IIb. The crossover from IIb to III is
    described by the scaling function $\tilde \phi_s({\sf x})$ where ${\sf x}=k/(\hat g)^{1/z_s}$, given in \eqref{Dk_scaling_ballistic}.
    The crossover from region III and IV occurs along the diagonal and is described by the scaling function $\chi_s(x)$ 
    with $x= k/(t/\tau)^{1/z_s}$, given in \eqref{Ck_smallsep_smalltime}. 
    Finally the crossover between region I and IV occurs along an horizontal line and is described by the scaling function 
    $\tilde \psi_s(y)$
    with $y=\gamma t$, given in \eqref{Dk_scaling_smallgaps}.}
    \label{fig:diagrameDk}
\end{figure}

In the RTP case, the time correlations for gaps of size $k \geq 1$ are given by
\be \label{Dk_rtp_def}
D_k(t) = \langle (\delta x_{i+k}(t)-\delta x_i(t))(\delta x_{i+k}(0)-\delta x_{i}(0)) \rangle = \frac{8v_0^2}{N} \sum_{q=1}^{(N-1)/2} \frac{\mu_q e^{-2\gamma t} -2\gamma e^{- \mu_q t}}{\mu_q(\mu_q^2-4\gamma^2)} \sin^2 \left( \frac{\pi k q}{N} \right) \;.
\ee
In the large $N$ limit with $k\ll N$ this becomes
\be  \label{Dk_RTP_largeN} 
D_k(t) 
\simeq \frac{8 T_{\rm eff}}{g\rho^{s+2}} \int_0^{1/2} du \frac{e^{-g\rho^{s+2} f_s(u) t}-\hat g f_s(u) e^{-2\gamma t}}{f_s(u)(1-\hat g^2 f_s(u)^2)} \sin^2(\pi k u) \;.
\ee
For $t=0$ we recover the gap variance at fixed time $D_k(0)$ given in \ref{sec:Dk0_rtp}. As usual the limit $\hat g \ll 1$ gives back the diffusive case studied in Section~\ref{sec:Dk_brownian}. In that case, i.e. $\hat g \ll 1$, there is a regime where $1/\gamma \ll t \ll \tau$. In this regime we recover the free diffusion result \eqref{Dk_freediffusion},
\be 
D_k(0) - D_k(t) \simeq 2 T_{\rm eff} t \;.
\ee
In the regime $\gamma t \ll 1$, we can expand both exponentials, leading to the ballistic growth
\be 
D_k(0) - D_k(t) \simeq 4 v_0^2 t^2 \int_0^{1/2} du \frac{\sin^2(\pi k u)}{1+\hat g f_s(u)} \;.
\ee
For $\hat g \ll 1$ this gives $D_k(0)-D_k(t) \simeq v_0^2 t^2$.
\\

{\bf Scaling form.} Due to the identity $D_k(0)-D_k(t) = C_0(t)-C_k(t)$, the function $D_k(t)$ has a lot of similarities with $C_k(t)$. As we did for $C_k(t)$, we will now focus on the limit $k,t/\tau,\hat g \gg 1$ (for $s>0$) and derive the different asymptotic regimes, which are represented in Fig.~\ref{fig:diagrameDk}. Some of the results can be directly derived from those obtained for $C_k(t)$ in the previous section. Using similar arguments to what was done for $C_k(t)$, in this limit we can write $D_k(t)$ under the scaling form
\be \label{Dk_scaling1}
D_k(t) \simeq T_{\rm eff} \tau \hat g^{1-\frac{1}{z_s}}  \mathcal{\tilde G}_s \left( \frac{k}{\hat g^{1/z_s}} , \gamma t \right) \quad , \quad
\mathcal{\tilde G}_s({\sf x},y) = \frac{8}{a_s^{1/z_s}} \int_0^{+\infty} dv \sin^2(\pi a_s^{-1/z_s} {\sf x} v)\frac{e^{-2 y v^{z_s}} - v^{z_s} e^{-2 y}}{v^{z_s}(1 - v^{2 z_s})} \;,
\ee
or alternatively
\be \label{Dk_scaling2}
D_k(t) \simeq T_{\rm eff} \tau (t/\tau)^{1-\frac{1}{z_s}} \mathcal{G}_s \left(\frac{k}{(t/\tau)^{\frac{1}{z_s}}},\gamma t \right) \quad , \quad \mathcal{G}_s(x,y) = \frac{8}{a_s^{1/z_s}} \int_0^{+\infty} dv \sin^2(\pi a_s^{-\frac{1}{z_s}}x v)\frac{e^{- v^{z_s}} - v^{z_s} \frac{e^{-2 y}}{2y}}{v^{z_s}(1 - v^{2 z_s}/(2y)^2)} \;.
\ee
\\

\textit{Brownian limit}. As for $C_k(t)$, the Brownian case is recovered in the limit $y=\gamma t \to +\infty$, keeping $x=k/(t/\tau)^{1/z_s}$ fixed in \eqref{Dk_scaling2} (indeed in this limit, $\mathcal{G}_s(x,y) \to G_s(x)$).
As a reminder, in the Brownian regime we found in Section \ref{sec:Dk_brownian},
\be \label{Dk_scaling_brownian2}
D_k(t) \simeq \begin{dcases} T_{\rm eff} \tau (t/\tau)^\frac{s}{s+1} G_s\left( \frac{k}{(t/\tau)^{\frac{1}{s+1}}} \right) \quad \text{for } -1<s<1 \\
T_{\rm eff} \sqrt{\tau \, t} \, G_s\left( \frac{k}{\sqrt{t/\tau}} \right) \hspace{1.55cm} \text{for } s>1 \end{dcases} \quad , \quad G_s(x) = 8 \, a_s^{-\frac{1}{z_s}} \int_0^{+\infty} dv \frac{e^{-v^{z_s}}}{v^{z_s}} \sin^2 \left(\pi a_s^{-\frac{1}{z_s}} x v \right) \;.
\ee 
The scaling function $G_s(x)$ describes the crossover between the two asymptotic regimes I (more precisely Ib) and IIa in Fig. \ref{fig:diagrameDk}. 
When $x=k/(t/\tau)^{1/z_s}\ll 1$, we find (regime I in Fig. \ref{fig:diagrameDk}) 
\be \label{Dk_Brownian_smallx}
D_k(t) \simeq \begin{dcases} 8\pi^2 \frac{\Gamma(\frac{2-s}{1+s})}{1+s} \left(\frac{\tau}{a_s}\right)^{\frac{3}{1+s}} T_{\rm eff} \frac{k^2}{t^{\frac{2-s}{1+s}}} \quad \text{for } 0<s<1  \;, \\ \frac{T_{\rm eff} \tau^{3/2}}{2\sqrt{\pi}((1+s)\zeta(s))^{3/2}} \frac{k^2}{\sqrt{t}} \hspace{1.25cm} \text{for } s>1  \;, \end{dcases}
\ee
while for $x=k/(t/\tau)^{1/z_s} \gg 1$ we have $D_k(0)-D_k(t) \simeq C_0(t)$, where $D_k(0)$ and $C_0(t)$ are given in \eqref{gapscases} and \eqref{displacement_Riesz} respectively with $T$ replaced by $T_{\rm eff}$. This corresponds to regime IIa in Fig. \ref{fig:diagrameDk}.
\\

\textit{Large time limit.} Let us now consider a different large time limit, namely $y=\gamma t \gg 1$ with ${\sf x} = k/\hat g^{1/z_s}$ fixed, using \eqref{Dk_scaling2}. Performing a change of variable $w=(2y)^{1/z_s}v$ we obtain
\bea
\mathcal{\tilde G}_s({\sf x},y) &=& \frac{8}{(2a_s y)^{1/z_s}} \int_0^{+\infty} dw \sin^2 \left(\frac{\pi {\sf x} w}{(2a_s y)^{1/z_s}} \right) \frac{e^{-w^{z_s}} - w^{z_s} \frac{e^{-2 y}}{2y}}{\frac{w^{z_s}}{2y}(1 - \frac{w^{2 z_s}}{4y^2})} \nn \\
&\simeq& \frac{8\pi^2}{a_s^{3/z_s}} {\sf x}^2 (2y)^{1-\frac{3}{z_s}} \int_0^{+\infty} dw w^{2-z_s} e^{-w^{z_s}} 
= \begin{cases} \frac{8\pi^2}{a_s^{\frac{3}{1+s}}} \frac{\Gamma(\frac{2-s}{1+s})}{1+s} {\sf x}^2 (2y)^{-\frac{2-s}{1+s}} \quad \text{for } 0<s<1 \\
\frac{1}{2\sqrt{\pi} ((1+s)\zeta(s))^{3/2}} \frac{{\sf x}^2}{\sqrt{2y}} \hspace{0.76cm} \text{for } s>1 \end{cases}
\;, \label{Dk_rtp_largetime}
\eea
which, inserting back into \eqref{Dk_scaling1}, recovers \eqref{Dk_Brownian_smallx}. This estimate \eqref{Dk_Brownian_smallx} is therefore valid beyond the Brownian regime, as long as $\gamma t \gg 1$ and $t/\tau \gg k^{z_s}$, i.e. within the regime I in Fig. \ref{fig:diagrameDk}. This is similar (upon exchanging time and space) to the large distance limit of $C_k(t)$, where the ballistic expression \eqref{Ck_ballistic3} also extends beyond the Brownian regime since it is also valid at short times.
\\

\textit{Large gap sizes.} For large gaps $k \gg \hat g^{1/z_s},(t/\tau)^{1/z_s}$, i.e. regimes IIa and IIb in Fig. \ref{fig:diagrameDk},
the asymptotic expression for $D_k(t)$ can be deduced directly from the relation $D_k(0)-D_k(t) = C_0(t)-C_k(t)$. Indeed in this regime, we know that $C_k(t) \ll C_0(t)$ (it decays as $k^{-(2+s)}$ for $0<s<1$ and exponentially in $k$ for $s>1$), and thus we have
\be
D_k(0)-D_k(t) \simeq C_0(t) \;.
\ee
In this regime, $D_k(0)$ is given by the Brownian expression \eqref{gapscases} with $T \to T_{\rm eff}$. For $\gamma t \gg 1$ (regime IIa), we are in the Brownian regime and $C_0(t)$ is given by \eqref{displacement_Riesz}. For $\gamma t \ll 1$ (regime IIb), one has instead $C_0(t) \simeq v_R^2 t^2$, where the renormalized velocity is given in \eqref{vR_large_g}.
\\

\textit{Small time and small gap.} In the remaining regimes III and IV, where $k\ll \hat g^{1/z_s}$ and $\gamma t \ll 1$, the asymptotic expression for $D_k(t)$ is again directly obtained from $D_k(0)-D_k(t) = C_0(t)-C_k(t)$. In particular, the crossover between these two regimes is again described by the function $\chi_s(x)$ defined in \eqref{Ck_smallsep_smalltime}. Note that here, $D_k(0)$ is given by its expression for $k \ll \hat g^{1/z_s}$, i.e. \eqref{Dk0_smallsep_SR}, \eqref{Dk0_smallsep_LR1} and \eqref{gap_rtp_half}, depending on the value of $s$. The crossover between regions IIb and III is described by the scaling form
\be \label{Dk_scaling_ballistic}
D_k(0)-D_k(t) \simeq v_0^2 t^2 \hat g^{-\frac{1}{z_s}} \tilde \phi_s ( k/\hat g^{1/z_s} ) \quad , \quad
\tilde \phi_s({\sf x}) = \frac{4}{a_s^{1/z_s}} \int_0^{+\infty} dv \frac{\sin^2(\pi a_s^{-1/z_s} {\sf x} v)}{1 + v^{z_s}} \;,
\ee
obtained by taking the limit $y \ll 1$ in \eqref{Dk_scaling1}. Note that for ${\sf x} \gg 1$
\be
\tilde \phi_s ({\sf x}) \simeq \frac{2}{a_s^{1/z_s}} \int_0^{+\infty} \frac{dv}{1+v^{z_s}} = \frac{2\pi}{(a_s)^{1/z_s} z_s \sin(\frac{\pi}{z_s})}  \;,
\ee
and for ${\sf x} \ll 1$
\be 
\tilde \phi_s({\sf x}) \simeq \frac{4}{a_s} {\sf x}^{z_s-1} \int_0^{+\infty} dw \frac{\sin^2(\pi w)}{w^{z_s}} = \frac{2\pi^{z_s-\frac{1}{2}}}{(z_s-1)} \frac{\Gamma(\frac{3-z_s}{2})}{\Gamma(z_s/2)} \frac{{\sf x}^{z_s-1}}{a_s} \;,
\ee
from which we recover regimes IIb and III respectively. To describe the crossover between I and IV, we must take the limit ${\sf x} \ll 1$ in \eqref{Dk_scaling1}. In the case $s>1/2$, we obtain the scaling form
\be \label{Dk_scaling_smallgaps}
D_k(t) \simeq T_{\rm eff} \tau \hat g^{1-\frac{3}{z_s}} k^2  \tilde \psi_s ( \gamma t ) \quad , \quad
\tilde \psi_s(y) 
= \frac{8 \pi^2}{a_s^{3/z_s}} (2y)^{\frac{z_s-3}{z_s}} \int_0^{+\infty} dw \, w^{2-z_s} \frac{e^{-w^{z_s}} - w^{z_s} \frac{e^{-2 y}}{2y}}{1 - \frac{w^{2 z_s}}{4y^2}} 
\;.
\ee
For $y \gg 1$ we have $\tilde \psi_s(y) \simeq 8\pi^2 a_s^{-3/z_s} \frac{\Gamma(\frac{3}{z_s}-1)}{z_s} (2y)^{\frac{z_s-3}{z_s}}$, which recovers the regime I as given by \eqref{Dk_Brownian_smallx} and \eqref{Dk_rtp_largetime}, while for $y \ll 1$ we have $\tilde \psi_s(0) - \tilde \psi_s(y) \simeq 8\pi^2 a_s^{-3/z_s} \frac{z_s \Gamma(\frac{3-z_s}{z_s})}{3(z_s-1)(2z_s-3)} (2y)^{\frac{3(z_s-1)}{z_s}} $, leading to the regime IV. For $0<s<1/2$, we could not write a scaling form simpler than \eqref{Dk_scaling1}. 

\subsection{Quenched initial condition} \label{sec:rtp_quenched}

As in the Brownian case, we have considered until now an ``annealed'' initial condition, i.e. the system was initialised in the stationary state. Let us now see how the results are affected if one considers instead a ``quenched'', i.e. deterministic, initial condition. We assume that the initial density is uniform, so that we have $\delta x_i(0)=0$ for all $i$. Integrating \eqref{Eq_delta_x_RTP} with this initial condition leads to
\be
\delta x_i(t) = v_0 \sum_{j=1}^{N} \int_0^t dt_1 [e^{(t_1-t)H}]_{ij} \left( \sigma_j(t_1) - \frac{1}{N} \sum_{k=1}^N \sigma_k(t_1) \right) \;,
\ee
where $H$ is the Hessian matrix defined in \eqref{defHessian}. We assume that the $\sigma_i$ are still initialized randomly to $\pm 1$ with equal probability. Using that $\langle \sigma_i(t) \sigma_j(t') \rangle = e^{-2\gamma|t-t'|}\delta_{ij}$, we obtain
\be
\langle \delta x_j(t) \delta x_k(t') \rangle_{\rm qu} = v_0^2 \int_0^{t} dt_1 \int_0^{t'} dt_2 \, e^{-2\gamma |t_1-t_2|} \left([e^{(t_1+t_2-t-t')H}]_{jk} -\frac{1}{N} \right) \;,
\ee
where we have used \eqref{identity_expH} to rewrite the last term. Finally, decomposing the matrix $H$ in its eigenbasis and performing the integral we obtain,
\be \label{covtwotime_quenched_RTP}
\langle \delta x_j(t) \delta x_k(t') \rangle_{\rm qu} 
= \frac{2v_0^2}{N} \sum_{q=1}^{(N-1)/2} \cos(2\pi \frac{q}{N}(j-k))\left[ \frac{e^{-2\gamma|t-t'|} - \frac{2\gamma}{\mu_q} e^{-\mu_q |t-t'|} - e^{-\mu_q t -2\gamma t'} - e^{-\mu_q t' -2\gamma t} }{\mu_q^2-4\gamma^2} + \frac{e^{-\mu_q(t+t')}}{\mu_q (\mu_q-2\gamma)} \right] 
\;.
\ee
We see that, contrary to the Brownian case, the quenched initial condition cannot be obtained from the annealed by some simple rescaling. We will see below that this leads to new dynamical regimes.

\subsubsection{Variance of the displacement $C_0^{\rm qu}(t)$} 

We now focus on the variance of the displacement at time $t$, $C_0^{\rm qu}(t)$. Equation \eqref{covtwotime_quenched_RTP} leads to
\be \label{C0_quenched_RTP}
C_0^{\rm qu} (t) = \langle (\delta x_i(t) - \delta x_i(0))^2 \rangle_{\rm qu} = \langle \delta x_i(t)^2 \rangle_{\rm qu} 
= \frac{2v_0^2}{N} \sum_{q=1}^{(N-1)/2} \frac{\mu_q \left( 1 + e^{-2\mu_q t} - 2e^{-(\mu_q + 2\gamma) t} \right) - 2\gamma \left( 1 - e^{-2\mu_q t} \right)}{\mu_q(\mu_q^2-4\gamma^2)}
\;.
\ee
In the large $N$ limit this can be written
\be \label{C0_quenched_RTP_largeN}
C_0^{\rm qu} (t) 
\simeq 2v_0^2 \int_0^{1/2} du \frac{\frac{f_s(u)}{\tau} \left( 1 + e^{-2f_s(u)t/\tau} -2e^{-(f_s(u)/\tau+2\gamma)t} \right) -2\gamma \left(1-e^{-2f_s(u)t/\tau} \right)}{\frac{f_s(u)}{\tau} \left( \frac{f_s(u)^2}{\tau^2} -4\gamma^2 \right)}
\;.
\ee
The different time regimes 
can be studied in the same way as was done in Section~\ref{sec:C0(t)} and App.~\ref{app:Ckrtp_argument} for the annealed case. 
Let us start with the ballistic regime $t \ll \tau, 1/\gamma$. 
In this case we can expand both exponentials to second order, leading to
\be
C_0^{\rm qu}(t) \simeq v_0^2 t^2 \;,
\ee
i.e. since the initial density is uniform we always have $v^{\rm qu}_R=v_0$.
Hence there is no renormalisation of the velocity because at $t=0$ the positions are $\delta x_i=0$, hence the force on each particle vanishes.
This is at variance with the annealed case, where the positions at $t=0$, hence the forces, are distributed according to the stationary measure.
One thus has $C_0^{\rm qu}(t) > C_0^{\rm ann}(t)$ in that regime. 

In the large time regime $t\gg \tau, 1/\gamma$, by a similar argument to App.~\ref{app:Ckrtp_argument}, the integral is dominated by $u\sim (t/\tau)^{-1/z_s} \ll 1$, which leads to 
\be
C_0^{\rm qu} (t) \simeq 2T_{\rm eff} \tau \int_0^{1/2} du \, \frac{1 - e^{-2 a_s u^{z_s} t/\tau}}{a_s u^{z_s}} \;.
\ee
This coincides with the limit $t\gg \tau$ of \eqref{C0_quenched} in the quenched Brownian case, and thus we recover
the anomalous diffusion \eqref{displacement_Riesz_quenched} in this regime, replacing $T \to T_{\rm eff}$. As for the Brownian case, the saturation value at infinite
time equals the stationary variance $C_0^{\rm qu} (\infty)= \langle \delta x_i^2 \rangle < C_0^{\rm ann} (\infty)= 2 \langle \delta x_i^2 \rangle $. 

To discuss the other regimes we need to distinguish the two cases of small and large $\hat g$ respectively.

In the case $\hat g \ll 1$, there is an intermediate time regime $1/\gamma \ll t \ll \tau$ 
where we find the free diffusion result $C_0^{\rm qu}(t) \simeq 2 T_{\rm eff} t$ (neglecting all the terms proportional to $\hat g$ and expanding the last exponential 
in \eqref{C0_quenched_RTP_largeN} 
to first order),
which is the same estimate as in the annealed case. Thus for $\hat g \ll 1$ the different time regimes are exactly the same as in the annealed case (up to some prefactors), see the top line of Fig.~\ref{fig:time_regimes_rtp}.


{ In the strongly active case $\hat g \gg 1$, we find however that there is an additional regime compared to the annealed case, when $\tau \ll t \ll 1/\gamma$.
This is represented in Fig. \ref{fig:diagram_quenched_C0} (left panel).
Indeed, in the annealed case, we argued (see App. \ref{app:Ckrtp_argument}) that the integral is dominated by $u \sim (t/\tau)^{-1/z_s}$, which leads to the same ballistic result $C_0(t)=v_R(\hat g) t^2$ as for $t\ll \tau,1/\gamma$.
Instead here, if we use a similar argument we find that the integral in \eqref{C0_quenched_RTP_largeN} is again dominated by $u \sim (t/\tau)^{-1/z_s}$, but this leads to a different time dependence. Comparing the different terms in the integral for $u \sim (t/\tau)^{-1/z_s}$, we see that for $\tau \ll t \ll 1/\gamma$ it can be approximated, for any $s>-1/2$, as (this formally corresponds to the limit $\gamma \to 0$)
\bea \label{C0_quenched_rtp_superdiff}
C_0^{\rm qu} (t) &\simeq& 2v_0^2 \tau^2 \int_0^{1/2} du \frac{ \left(1 - e^{-f_s(u)t/\tau} \right)^2}{f_s(u)^2} \simeq 2v_0^2 t^{2-\frac{1}{z_s}} (\frac{\tau}{a_s})^{1/z_s} \int_0^{+\infty} dv \left(\frac{1 - e^{-v^{z_s}}}{v^{z_s}}\right)^2 \nn \\
&=& \begin{dcases} \frac{4(2^{\frac{s}{1+s}}-1) \Gamma\left( -\frac{1+2s}{1+s} \right)}{(1+s) (a_s g\rho^{s+2})^{\frac{1}{1+s}}} v_0^2 \, t^{\frac{1+2s}{1+s}} \quad \text{for } -1/2<s<1 \;, \\ \frac{4(\sqrt{2}-1)v_0^2 \, t^{3/2}}{3\sqrt{\pi (s+1) \zeta(s) g\rho^{s+2}}} \hspace{1.7cm} \text{for } s>1 \;. \end{dcases}
\eea
This is a new superdiffusive (for $s>0$) anomalous regime
which was absent in the annealed setting. In the SR case, this result coincides with the one obtained in \cite{SinghChain2020} with the replacement $(s+1)\zeta(s) \to 1$. The scaling form (36) from this paper, which describes the crossover from this superdiffusive anomalous regime to the (subdiffusive) anomalous diffusion regime, can also be generalized to the Riesz gas. Assuming $\hat g \gg 1$ and $t\gg \tau$, the integral \eqref{C0_quenched_RTP_largeN} is dominated by small $u$ leading to (writing $w=(2a_s t/\tau)^{1/z_s} u$)
\bea
C_0^{\rm qu} (t) &\simeq& 2v_0^2 t^{2-\frac{1}{z_s}} \left(\frac{2\tau}{a_s}\right)^{1/z_s} {\sf C}_s^{\rm qu} (\gamma t) \;, \\
{\sf C}_s^{\rm qu} (y) &=& 2^{-2/z_s} \int_0^{+\infty} dw \frac{\frac{w^{z_s}}{2} \left( 1+e^{-w^{z_s}} -2e^{-(2y+\frac{w^{z_s}}{2})} \right) - 2y(1-e^{-w^{z_s}})}{\frac{w^{z_s}}{2} \left( \frac{w^{2z_s}}{4} -4y^2 \right)} \;. \nn
\eea
In the short range case $z_s=2$ this
coincides with equation (36) in \cite{SinghChain2020} setting $a_s=4\pi^2$. This scaling function describes the crossover from 
the superdiffusive regime for $t\ll 1/\gamma$ to the subdiffusive regime for $t\gg 1/\gamma$.

\subsubsection{Space time correlations $C_k^{\rm qu} (t)$} 

\begin{figure}
    \centering
    \includegraphics[width=0.7\linewidth]{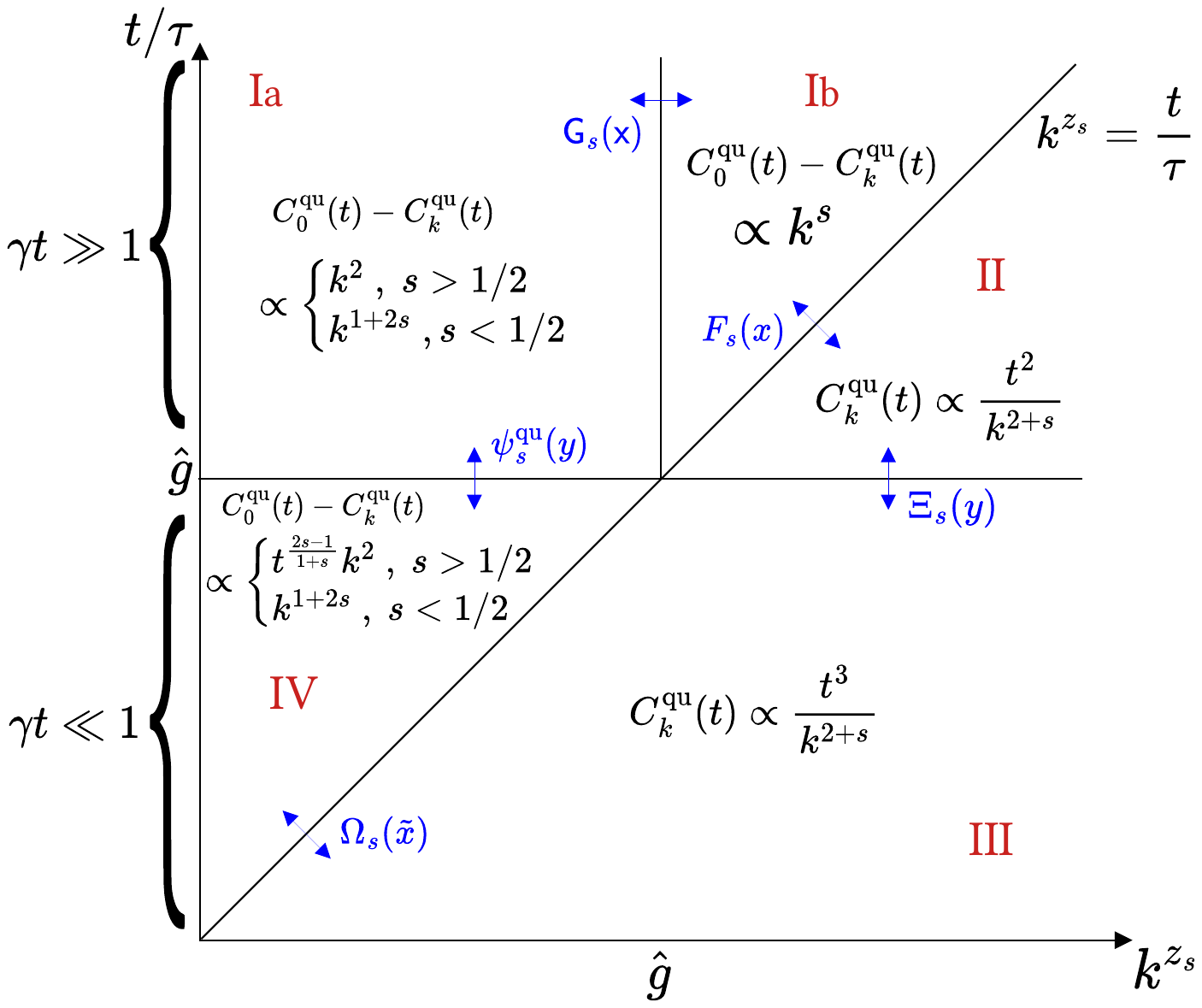}
    \caption{The five asymptotic regimes of the space-time correlation function $C^{\rm qu}_k(t)$ 
        with quenched initial conditions, 
    for $0<s<1$. They are represented in the plane $(k^{z_s}, t/\tau)$, where $\hat g=1/(2 \gamma \tau)$, in the
    limit where $k,t/\tau,\hat g \gg 1$. In that limit $C_k(t)$ admits two-argument scaling forms 
    \eqref{scaling1} and \eqref{scalingquenched}. The solid lines indicate the location of crossover regions
    between different asymptotic forms which are explicitly indicated. The diagram is qualitatively similar to
    the one for the annealed case shown in Fig. \ref{fig:diagrameCk}, with important differences. 
    The region III of Fig. \ref{fig:diagrameCk} has no analog here, instead $C^{\rm qu}_k(t)$
    decays as $k^{-(2+s)}$ as soon as one crosses the line $k \sim (t/\tau)^{1/z_s}$,
    i.e. $k$ larger than the dynamical scale. In addition, the region II of Fig. \ref{fig:diagrameCk} 
    is now split between regions II and III here, which differ by their asymptotic power of time.
    Furthermore, region IV exhibits the same dependence in $k$ as in the annealed case,
    but with a different time dependence. In particular for $s<1/2$ regions IV and Ia
    are identical in the quenched case. The behaviors in the three regions for $\gamma t \gg 1$
    are identical (up to numerical prefactors) to the annealed case. There are five
    scaling functions which describe the crossovers between these regions, which
    are explicitly given in the text. Three of them are specific to the quenched
    case, while the other two were already computed in the annealed case.
    The scaling function $\Omega_s(\tilde x)$ is the long range analog of the one
    obtained in \cite{SinghChain2020} in the short range case.}
    \label{fig:diagram_quenched}
\end{figure}

Similarly, the space-time correlations with quenched initial conditions read
\be
C_k^{\rm qu} (t) = \langle \delta x_i(t) \delta x_{i+k}(t) \rangle_{\rm qu} 
= \frac{2v_0^2}{N} \sum_{q=1}^{(N-1)/2} \frac{\mu_q \left( 1 + e^{-2\mu_q t} - 2e^{-(\mu_q + 2\gamma) t} \right) - 2\gamma \left( 1 - e^{-2\mu_q t} \right)}{\mu_q(\mu_q^2-4\gamma^2)} \cos(2\pi \frac{qk}{N})
\;,
\ee
or in the large $N$ limit,
\be \label{Ck_quenched_RTP_largeN}
C_k^{\rm qu} (t) 
\simeq 2v_0^2 \int_0^{1/2} du \frac{\frac{f_s(u)}{\tau} \left( 1 + e^{-2f_s(u)t/\tau} -2e^{-(f_s(u)/\tau+2\gamma)t} \right) -2\gamma \left(1-e^{-2f_s(u)t/\tau} \right)}{\frac{f_s(u)}{\tau} \left( \frac{f_s(u)^2}{\tau^2} -4\gamma^2 \right)} \cos(2\pi k u)
\;.
\ee
In the ballistic regime $t \ll \tau, 1/\gamma$, we find that the particles are uncorrelated, i.e. $C_k^{\rm qu} (t)=0$ for any $k\neq 0$. This is different from the annealed case. As in the annealed case, the correlations also vanish in the free diffusion regime $1/\gamma \ll t \ll \tau$ if it exists. 
We thus focus on $t \gg \tau$.

In the three regimes where either $t/\tau \gg 1$, or $\hat g \gg 1$, or $k \gg 1$, using the same arguments as in the annealed case (see Sec~\ref{sec:Ck_rtp} and Appendix~\ref{app:Ckrtp_argument}), $C_k^{\rm qu}(t)$ can be written under the scaling form (for $s>-1$) 
\bea \label{scaling1} 
&&C_k^{\rm qu} (t) \simeq 2v_0^2 t^{2-\frac{1}{z_s}} \left(\frac{2\tau}{a_s}\right)^{1/z_s} \mathcal{F}_s^{\rm qu} \left(\frac{k}{(2t/\tau)^{1/z_s}}, \gamma t \right) \;, \\
&&\mathcal{F}_s^{\rm qu} (x,y) = 2^{-2/z_s} \int_0^{+\infty} dw \frac{\frac{w^{z_s}}{2} \left( 1+e^{-w^{z_s}} -2e^{-(2y+\frac{w^{z_s}}{2})} \right) - 2y(1-e^{-w^{z_s}})}{\frac{w^{z_s}}{2} \left( \frac{w^{2z_s}}{4} -4y^2 \right)} \cos(2\pi a_s^{-1/z_s} x w) \;, \nn
\eea
which coincides with equation (45) of Ref.~\cite{SinghChain2020} for $z_s=2$ and $a_s=4\pi^2$. 
Alternatively, one can also write a different but equivalent scaling form,
\bea \label{scalingquenched} 
&&C_k^{\rm qu} (t) \simeq 2 T_{\rm eff} \tau \hat g^{\frac{z_s-1}{z_s}} a_s^{-1/z_s} \mathcal{\tilde F}_s^{\rm qu} \left(k/\hat g^{1/z_s}, \gamma t \right) \;, \\
&&\mathcal{\tilde F}_s^{\rm qu} ({\sf x},y) = \int_0^{+\infty} dw \frac{w^{z_s} \left( 1+e^{-4y w^{z_s}} -2e^{-2y(1+w^{z_s})} \right) - (1-e^{-4y w^{z_s}})}{w^{z_s} \left( w^{2z_s} - 1 \right)} 
\cos(2\pi a_s^{-1/z_s} {\sf x} w) \;. \nn
\eea
Note that the two scaling functions are defined with slightly different factors as compared to the annealed case, to make the comparison with Ref.~\cite{SinghChain2020} easier. 

We now study $C_k(t)$ in the joint limit $\hat g \gg 1$, $k\gg 1$ and $t/\tau \gg 1$. There are five regions with different asymptotic behaviours which are displayed in Fig.~\ref{fig:diagram_quenched}
(in the case $0<s<1$). We show how they can be recovered using the two scaling forms introduced above.
\\

%

{\it Small time regime}.
For $y=\gamma t \ll 1$, we can approximate (for any $s>-1$)
\be \label{omega_s}
\mathcal{F}_s^{\rm qu} (x,y\to 0) \simeq \Omega_s(2\pi a_s^{-1/z_s} x) \quad, \quad \Omega_s(\tilde x) = 2^{2-2/z_s} \int_0^{+\infty} dw \left(\frac{1-e^{-\frac{w^{z_s}}{2}}}{w^{z_s}}\right)^2 \cos(\tilde x w) \;.
\ee
In the SR case $z_s=2$, the scaling function $\Omega_s(\tilde x)= \Omega_{\rm SR}(\tilde x)$ has an explicit form given in equation (47) of \cite{SinghChain2020},
with $\Omega_{\rm SR}(\tilde x) = \frac{2\sqrt{\pi}}{3} (2-\sqrt{2}) -\sqrt{\pi} \, (\sqrt{2}-1)\tilde x^2 + O(\tilde x^3) $ at small $\tilde x$ and with exponential decay $\Omega_{\rm SR}(\tilde x) \simeq (\frac{16 \sqrt{\pi}}{\tilde x^4 } + O(\tilde x^{-6})) \,
e^{- \tilde x^2/4}$ at large $\tilde x$.
For the active DBM $s=0$, there is also an explicit expression (we recall that in this case $2\pi a_s^{-1/z_s}=1/\pi$), which reads
\be 
\Omega_0(\tilde x) = \frac{1}{2} \log \left(\frac{3}{4
   \tilde x^2+1}+1\right)-\tilde x \left( \tan ^{-1}(\tilde x)-2 \tan
   ^{-1}(2 \tilde x)+ \frac{\pi}{2}  \right) \;,
\ee 
with $\Omega_0(\tilde x) = \log 2 - \frac{\pi \tilde x}{2} + O(\tilde x^2) $ at small $\tilde x$ and a power law decay $\Omega_0(\tilde x) = \frac{1}{8 \tilde x^2 } + O(\tilde x^{-4})$ 
at large $\tilde x$. More generally, for $- \frac{1}{2} < s<1$ the function $\Omega_s(\tilde x)$ decays as {$\Omega_s(\tilde x) \simeq 
2^{- \frac{3+s}{1+s}} 
\Gamma(2+s)\cos(\frac{\pi s}{2})/\tilde x^{2+s}$} at large $\tilde x$ and is finite at $\tilde x=0$ with a non-analyticity $O(\tilde x^{1+2s})$ at small $\tilde x$ (which also exists for the SR case, setting $s=1$): $\Omega_s(\tilde x) = \omega_0 - \tilde \omega \tilde x^{1+2s} - \omega_2 \tilde x^2 + o(\tilde x^2,\tilde x^{1+2s})$ with $\omega_0 = \frac{2^{\frac{s}{1+s}}(2^{\frac{s}{1+s}}-1)}{1+s}\Gamma(-\frac{1+2s}{1+s})$. Inserting the expression for $\omega_0$ into \eqref{scaling1} recovers the result \eqref{C0_quenched_rtp_superdiff} for $C_0^{\rm qu}(t)$.


The function $\Omega_s(\tilde x)$ describes the crossover between a short distance regime $k \ll (t/\tau)^{1/z_s}$ (regime IV in Fig. \ref{fig:diagram_quenched}) and a large distance regime $k \gg (t/\tau)^{1/z_s}$ (regime III in Fig. \ref{fig:diagram_quenched}). 
At short distances $k \ll (t/\tau)^{1/z_s}$ we find for $s>1/2$, $C^{\rm qu}_0(t)-C^{\rm qu}_k(t) \propto t^{\frac{2s-1}{1+s}} k^2$ (resp. $\propto \sqrt{t} \, k^2$ for $s>1$), while for $-1/2<s<1/2$ we have $C^{\rm qu}_0(t)-C^{\rm qu}_k(t) \propto k^{1+2s}$. The latter result is independent of time at leading order, 
consistent with the system having reached stationarity locally, i.e. on distances smaller than the dynamical length $(t/\tau)^{1/z_s}$.
At large distance $k \gg (t/\tau)^{1/z_s}$ the correlation decays to zero, as
\be \label{Ck_quenched_t3}
C_k^{\rm qu} (t) \simeq \begin{dcases} \frac{(1+s)v_0^2 t^3}{\tau k^{2+s}}\;,  \hspace{6.95cm} \text{for } 0<s<1 \;, \\ 64 \sqrt{\frac{2}{\pi}} ((s+1)\zeta(s))^{3/2} \frac{v_0^2 t^{7/2}}{\tau^{3/2} k^4} \exp\left(-\frac{k^2}{8(s+1)\zeta(s)t/\tau}\right)\;, \quad \text{for } s>1 \;.\end{dcases}
\ee
This decay occurs over the dynamical length $k \sim (t/\tau)^{1/z_s}$. By contrast the annealed correlation $C_k^{\rm ann} (t)$ decays to zero on the static length
$k \sim \hat g^{1/z_s}$, which is much larger (since we recall that we are studying here the time regime $\tau \ll t \ll 1/\gamma$).
\\

{\it Large time, large distance}. In the large time regime $\gamma t \gg 1$, we have
\be
\mathcal{F}_s^{\rm qu} (x,y\to \infty) \simeq 2^{2-2/z_s} y^{-1} \int_0^{+\infty} dw \frac{1-e^{-w^{z_s}}}{w^{z_s}} \cos(\tilde x w) \;, \quad {\rm with}  \quad \tilde x=
2 \pi a_s^{-1/z_s} k/(2t/\tau)^{1/z_s}
\ee
and we recover the quenched Brownian result, see \eqref{Ck_quenched_brownian}, with for the LR case $C^{\rm qu}_0(t)-C^{\rm qu}_k(t) \propto k^s$ ($C^{\rm qu}_0(t) \propto t^{\frac{s}{1+s}}$) for $k \ll (t/\tau)^{1/z_s}$ 
(regime Ib in Fig. \ref{fig:diagram_quenched}) and $C^{\rm qu}_k(t)\simeq 2(s+1) T_{\rm eff} g\rho^{s+2} \frac{t^2}{k^{2+s}}$ for $k \gg (t/\tau)^{1/z_s}$ (regime II in Fig. \ref{fig:diagram_quenched}), as in the annealed case up to some prefactors (equation \eqref{Ck_brownian_largek} with $t\to 2t$ and $T\to T/2$). For the SR case this corresponds to $C^{\rm qu}_0(t)-C^{\rm qu}_k(t) \propto k$ ($C^{\rm qu}_0(t) \propto \sqrt{t}$) for $k \ll (t/\tau)^{1/2}$ 
(regime Ib in Fig. \ref{fig:diagram_quenched}) and $C^{\rm qu}_k(t) \simeq 4\sqrt{\frac{2(s+1)\zeta(s)}{\pi\tau}} T_{\rm eff} \frac{t^{3/2}}{k^2} \exp(-\frac{k^2}{8(s+1)\zeta(s)t/\tau})$ for $k \gg \sqrt{t/\tau}$. 
To summarize, the crossover from region Ib to region II in Fig. \ref{fig:diagram_quenched}) is described by the same scaling function $F_s(x)$
as for the Brownian. 
\\

{\it Large time, fixed distance}. We now discuss the crossover between regions Ia and Ib in Fig. \ref{fig:diagram_quenched}). The region 
Ia was not considered in \cite{SinghChain2020}, hence this crossover was not discussed there. 
For this crossover and the ones below we use the second scaling form \eqref{scalingquenched}. In the limit $y=\gamma t \gg 1$, we obtain
\be \label{crossq} 
\mathcal{\tilde F}_s^{\rm qu} (0,y\to\infty) - \mathcal{\tilde F}_s^{\rm qu} ({\sf x},y\to\infty) \simeq 2 \int_0^{+\infty} dw \frac{\sin^2(\pi a_s^{-1/z_s} {\sf x} w)}{w^{z_s} \left( 1 + w^{z_s} \right)} = \frac{a_s^{1/z_s}}{4} {\sf x}^{z_s-1} {\sf G}_s({\sf x}) \;,
\ee
where ${\sf G}_s({\sf x})$ was defined in \eqref{eqFgap}. This result 
corresponds to $C^{\rm qu}_0(t)-C^{\rm qu}_k(t) = D_k(0)/2$, where $D_k(0)$ is the variance of the gaps in the stationary state 
(it is this $1/2$ of the annealed result). As in the annealed case, there is thus a crossover \eqref{crossq} described by the same scaling function ${\sf G}_s({\sf x})$
of the parameter ${\sf x}=k/\hat g^{1/z_s}$,
between regimes Ia and Ib, see Fig. \ref{fig:diagram_quenched}, which exhibit different behaviors as a function of $k$. 
\\

{\it Large distance, fixed time}. 
There is a new regime, region II in Fig. \ref{fig:diagram_quenched}, as compared to the annealed case, for ${\sf x} = k/\hat g^{1/z_s} \gg 1$ and $y=\gamma t$ fixed. 
For $-1/2<s<1$ consider Eq. \eqref{scalingquenched} in the limit of large ${\sf x}$ with fixed $y$. One finds
\be 
\mathcal{\tilde F}_s^{\rm qu} ({\sf x},y) \simeq  4 (e^{-2 y} -1 + 2 y) y \frac{c_s}{{\sf x}^{2 + s}} \quad, \quad {\sf x} \gg 1 \;,
\ee 
with 
$c_s=\Gamma(2+s) \cos(\frac{\pi s}{2})/ (2\pi a_s^{-1/(1+s)})^{2+s}$. This leads to (for $k\gg \hat g^{1/(1+s)}$)
\be 
C^{\rm qu}_k(t) \simeq \frac{(1+s)T_{\rm eff}}{\gamma^2 \tau k^{2+s}} \Xi_s(y) \quad , \quad \Xi_s(y) = y (e^{-2 y} - 1 + 2 y) \;.
\ee 
One has $\Xi_s(y) \simeq 2y^3$ for $y \ll 1$, compatible with regime III in Fig.~\ref{fig:diagram_quenched} (see \eqref{Ck_quenched_t3}) and $\Xi_s(y) \simeq 2 y^2$ for $y \gg 1$, recovering regime II. Thus the function $\Xi_s(y)$ correctly describes the crossover between these two regions.

The crossover in the SR case $s>1$ occurs in a qualitatively different fashion. To see this, 
let us define $X = 2\pi a_s^{-1/z_s} {\sf x} $ and $y= r X$, and perform the limit of large $X$ at fixed $r$. We obtain
\bea 
\mathcal{\tilde F}_s^{\rm qu} ({\sf x},y) 
&=& \frac{1}{2} \int_{-\infty}^{+\infty} dw \frac{w^{2} \left( 1+e^{-4r X w^{2}} -2e^{-2r X(1+w^{2})} \right) - (1-e^{-4r X w^{2}})}{w^{2} \left( w^{4} - 1 \right)} \cos(X w) \\
&\simeq& \frac{1}{2} {\rm Re} \int_{-\infty}^{+\infty} dw e^{-4r X w^{2} + i X w}  \frac{1}{w^2 (w^2-1)} \simeq \frac{1}{4} \sqrt{\frac{\pi}{r X}} e^{-X/(16 r) } \frac{ (8 r)^4}{1 + (8 r)^2} \;.
\eea 
For the last step we have performed a saddle point approximation around $w=i/(8 r)$. The crossover thus 
occurs along rays $k \propto t$ (at fixed $r$), more precisely for $k\sim (t/\tau)/\sqrt{\hat g} \ll t/\tau$. For $r \ll 1$ one recovers the behavior in the 
regime III, 
see \eqref{Ck_quenched_t3}, while for $r \gg 1$, one recovers the regime II.
\\

{\it Small distance, fixed time}. 
We can also look at the crossover between regions Ia and IV of Fig.~\ref{fig:diagram_quenched}, corresponding to ${\sf x} = k/\hat g^{1/z_s}\ll 1$. 
The leading order in ${\sf x}$ is given by $C_0^{\rm qu}(t)$ and exhibits a crossover from the superdiffusive regime for $\gamma t \ll 1$ to the subdiffusive regime for $\gamma t \gg 1$, see Fig.~\ref{fig:diagram_quenched_C0}. When considering the subleading order in ${\sf x}$, we need to distinguish between $s>1/2$ and $s<1/2$. 
For $s>1/2$, for ${\sf x} \ll 1$, one can expand the sine function to obtain
\be
\mathcal{\tilde F}_s^{\rm qu} (0,y) - \mathcal{\tilde F}_s^{\rm qu} ({\sf x},y) \simeq {\sf x}^2 \psi_s^{\rm qu}(y) \ , \ \psi_s^{\rm qu}(y) = \frac{2\pi^2}{a_s^{2/z_s}} \int_0^{+\infty} dw \, w^{2-z_s}  \frac{w^{z_s} \left( 1+e^{-4y w^{z_s}} -2e^{-2y(1+w^{z_s})} \right) - (1-e^{-4y w^{z_s}})}{w^{2z_s} - 1 } \;,
\ee
where we have defined $\psi_s^{\rm qu}(y)$ which describes the crossover between regions Ia and IV of Fig.~\ref{fig:diagram_quenched}. The asymptotic
behaviors of this function are as follows.
For $y \ll 1$, one finds
$\psi_s^{\rm qu}(y) \simeq
%
(2y)^{\frac{2z_s-3}{z_s}} 2\pi^2 a_s^{-2/z_s} \frac{(2^{2-\frac{3}{z_s}}-2) \Gamma(\frac{3}{z_s}-2)}{z_s}$,
which leads to $C_0^{\rm qu}(t)-C_k^{\rm qu}(t) \propto t^{\frac{2s-1}{1+s}} k^2$, recovering the region IV in Fig.~\ref{fig:diagram_quenched}.
For $y \gg 1$, 
$\psi_s^{\rm qu}(y) \simeq \frac{2\pi^3}{z_s \sin(\frac{(3-z_s)\pi}{z_s}) a_s^{2/z_s}}$, 
and we recover $C_k^{\rm qu}(0)-C_k^{\rm qu}(t) \simeq D_k(0)/2$ in the regime $k \ll \hat g^{-1/(1+s)}$ and $\gamma t \gg 1$.

For $-1/2<s<1/2$, the integral is dominated by small arguments and we have instead
\be
\mathcal{\tilde F}_s^{\rm qu} (0,y) - \mathcal{\tilde F}_s^{\rm qu} ({\sf x},y) \simeq {\sf x}^{1+2s} c'_s \quad , c'_s = 2 \int_0^{+\infty} dv \, \frac{\sin^2(\pi a_s^{-\frac{1}{1+s}}v)}{v^{2(1+s)}} = a_s^{\frac{1}{1+s}-2} \frac{\pi^{2s+\frac{3}{2}} \Gamma(\frac{1}{2}-s)}{(2s+1)\Gamma(1+s)} \;.
\ee
In this case, phase Ia and IV are identical, and the difference 
\be
C_k^{\rm qu}(0)-C_k^{\rm qu}(t) \simeq \frac{2 v_0^2}{(a_s g \rho^{s +2})^2} \frac{\pi^{2s+\frac{3}{2}}}{2s+1} \frac{\Gamma(\frac{1}{2}-s)}{\Gamma(1+s)} k^{1+2s}
\ee
is independent of time and equal to $D_k(0)/2$ in the regime $k \ll \hat g^{-1/(1+s)}$ (see \eqref{gap_rtp_half}).
A local stationary regime has thus been reached. 
}


\subsection{Extremely slow tumbling}

Until now we have always assumed the tumbling rate $\gamma$ to be independent of $N$. Here we briefly consider the case
where $\gamma$ is very small, e.g. scaled as a power of $1/N$. 

\subsubsection{Limit $\gamma \to 0$}

Let us start with the limit where we take $\gamma \to 0$ before the large $N$ limit. This limit
was studied for the active DBM case $s=0$ in \cite{ADBM2}. In that 
limit the system has enough time to converge to one the fixed points of the dynamics, parametrized by fixed $\sigma_i$,
solutions of $0 = v_0\sigma_i - \sum_{j(\neq i)} W'(x_i-x_j)$, before a tumbling event occurs.
Assuming again small relative displacements at these fixed points, one can obtain their
covariance by linearizing the
fixed point equation, and averaging over the $\sigma_i=\pm 1$ with uniform distribution.
Performing this calculation, 
we recover the same result as when setting $\gamma=0^+$ in 
\eqref{cov_rtp}
\be \label{cov_gamma0}
\langle \delta x_i \delta x_j \rangle \underset{\gamma \to 0}{\simeq} 
\frac{2 v_0^2\tau^2}{N} \sum_{q=1}^{(N-1)/2} \frac{\cos(2\pi \frac{q}{N} (i-j))}{f_s(\frac{q}{N})^2} \;,
\ee
where $\tau=1/(g \rho^{s+2})$. This limit is reached when $\gamma \ll 1/(\tau N^{z_s})$, i.e. the slowest  
relaxation rate in the system\footnote{The rate of change of the vector $(\sigma_1,...,\sigma_N)$ which determines the fixed point of the dynamics, is in fact $N\gamma$, but for the quantities we consider the change of a single arbitrary $\sigma_i$ has a negligible effect, so that $1/\gamma$ is indeed the relevant time-scale.}.

To test the validity of the linear approximation in the $\gamma \to 0$ limit (as we did in Section \ref{sec:Dk0_rtp} for finite $\gamma$) 
we now consider the variance of the gaps.
It reads
\be \label{gapssmallgamma}
D_k(0) = \langle (\delta x_i - \delta x_{i+k})^2 \rangle = \frac{8v_0^2\tau^2}{N} \sum_{q=1}^{(N-1)/2} \frac{\sin^2(\pi \frac{q}{N} k)}{f_s(\frac{q}{N})^2} \;.
\ee
In the large $N$ limit there are two cases. 

For $-1<s<1/2$, we obtain for any $k \ll N$
\be \label{Dk0_smallgammahalf}
D_k(0) \simeq 8 v_0^2\tau^2 \int_0^{+\infty} du \frac{\sin^2(\pi k u)}{f_s(u)^2} \;,
\ee
which remains finite as $N \to +\infty$.
It coincides with the limit $\hat g \to \infty$ of \eqref{gaprtp}, which is
independent of $\gamma$, using $T_{\rm eff}/\hat g= v_0^2 \tau$. 
For $k=1$ it shows that the domain of validity is also independent of $\gamma$
and given by $v_0^2 \tau^2 \rho^2 \ll B_s$ where $B_s$
is a constant as obtained in \eqref{validity_rtp}. The large $k$ behavior
of the gap variance, for $1 \ll k \ll N$ and for $-1/2<s<1/2$ is given in
\eqref{gap_rtp_half}, while for $s<-1/2$ it converges to a constant $2\langle \delta x_i^2 \rangle$ on a scale $k\ll N$.


For $s>1/2$ the sum in \eqref{gapssmallgamma} is dominated by the first few terms and in the limit of large $N$, with $k\ll N$, it becomes
\be \label{Dk0_gamma0}
D_k(0) \simeq N^{2z_s-1} \frac{8 v_0^2\tau^2}{a_s^2} \sum_{q=1}^{\infty} \frac{\sin^2(\pi \frac{q}{N} k)}{q^{2z_s}} \simeq 
\begin{cases} N^{2s-1} \frac{8 \pi^2 v_0^2\tau^2}{a_s^2} \zeta(2s) k^2 \quad \text{for } 1/2<s<1 \;, \\ N \frac{v_0^2\tau^2}{12 (s+1)^2 \zeta(s)^2} k^2 \hspace{1.08cm} \text{for } s>1 \;. \end{cases}
\ee
Thus for $s>1/2$, $D_k(0)$ diverges as $N\to+\infty$, and the linear approximation breaks down, unless $v_0\tau$ scales as $N^{-(2s-1)}$ (resp. $N^{-1}$ for $s>1$) or smaller. 

From \eqref{cov_gamma0}, we can also compute the variance for $\gamma=0^+$. One finds 
\be \label{varsmallgamma}
\langle \delta x_i^2 \rangle \simeq N^{2z_s-1} \frac{2 v_0^2\tau^2}{a_s^2} \zeta(2z_s) \;.
\ee
This formula is valid under the same conditions as given above (i.e. for $\rho v_0 \tau$ of order unity for $s<1/2$
and for $\rho v_0 \tau \lesssim N^{-(2s-1)}$ for $s>1/2$ (resp. $N^{-1}$ for $s>1$).


\subsubsection{Crossover for $\gamma \sim N^{-z_s}$: static and dynamic} 

For some observables there is a crossover when $\gamma \sim N^{-z_s}/\tau$ (equivalently $\hat g =1/(2 \gamma \tau) \sim N^{z_s}$) 
between the regime where $\hat g$ is of
order unity but large, and the regime discussed above $\gamma=0^+$. To study this crossover we thus define
\be \label{gammatilde}
\gamma = \tilde \gamma N^{-z_s} \quad , \quad \tilde g = \frac{1}{2\tilde \gamma \tau} = N^{-z_s}\hat g 
\ee 
and consider $\tilde \gamma$ of order unity. 

For the variance of the displacements we obtain from \eqref{var_RTP2} 
the crossover form 
\be \label{var_RTP_gamma0}
\langle \delta x_i^2 \rangle \simeq 
N^{2z_s-1} \frac{2 v_0^2 \tau^2}{a_s} \sum_{q=1}^{\infty} \frac{1}{q^{z_s} (a_s q^{z_s} + \tilde g^{-1})} \;.
\ee
For $\tilde g \gg 1$ it recovers \eqref{varsmallgamma}. As for all observables here and below
the validity criterion discussed above apply. 

For the variance of the gaps in the case $s<1/2$ there is no crossover since as we noted above the result for $1 \ll \hat g \ll N^{z_s}$
is independent of $\gamma$. For $s>1/2$ one finds the following crossover form
\be \label{crossoverDk0smallgamma}
D_k(0) \simeq 
N^{2z_s-3}\frac{8\pi^2v_0^2\tau^2}{a_s} k^2 \sum_{q=1}^{\infty} \frac{q^{2-z_s}}{a_s q^{z_s}+\tilde g^{-1}} \;,
\ee
which for $\tilde g \to \infty$ recovers the estimates in the last equation in \eqref{Dk0_gamma0}. 
\\

Finally, let us consider a dynamical quantity, the displacement with time $C_0(t)$ for $s>0$. 
We first note that, with the present scaling of $\gamma$ in
\eqref{gammatilde},
$C_0(t)$ remains ballistic for any $t \ll 1/\gamma \sim N^{z_s}$ and $t \ll N^{z_s} \tau$. Thus we also rescale the time as $t=N^{z_s} \tilde t$.
We use the equation \eqref{disp_RTP} where now the sum 
is dominated by small values of $q$, and we obtain the crossover form
\be \label{C0_smallgamma}
C_0(t) \simeq 
N^{2z_s-1} 4 v_0^2 \tau^2 \sum_{q=1}^{\infty} \frac{a_s q^{z_s}(1-e^{-2\tilde \gamma \tilde t}) - \tilde g^{-1} (1-e^{-a_sq^{z_s} \tilde t/\tau})}{a_s q^{z_s}(a_s^2 q^{2z_s} - \tilde g^{-2})} \;.
\ee 
When $\tilde g \ll 1$, we can approximate the sum in \eqref{C0_smallgamma} by an integral with integration variable $v=(a_s \tilde g)^{1/z_s} q$. We recover the expression \eqref{C0scaling}, which was obtained for $\hat g$ large but independent of $N$. 

Consider now \eqref{C0_smallgamma} for $\tilde g$ of order unity. At small $\tilde t \ll 1$ (i.e. $t \ll N^{z_s} \tau$ and $t \ll 1/\gamma$) by expanding both exponentials we recover a ballistic behaviour
\be \label{C0_smallgamma_ballistic}
C_0(t) \simeq N^{2z_s-1} 2v_0^2 \tilde t^2 \sum_{q=1}^{\infty} \frac{1}{1+a_s\tilde g q^{z_s}} 
= \frac{2v_0^2 t^2}{N} \sum_{q=1}^{\infty} \frac{1}{1+a_s\tilde g q^{z_s}} = \tilde v_R(\tilde g)^2 t^2  \;.
\ee
One can check, using again the change of variable $v= (a_s \tilde g)^{1/z_s} q$, that the 
renormalized velocity $v_R$ matches in the limit $\tilde g \ll 1$ the behavior obtained for $\hat g \gg 1$ 
in \eqref{vR_large_g}. 
On the other hand, for $\tilde t \gg 1$ we recover the limiting value $2\langle \delta x_i^2 \rangle$ as given by \eqref{var_RTP_gamma0}. 

Finally, Fig.~\ref{fig:time_regimes_rtp} bottom panel corresponds to the regime where the time scales are well separated, i.e. $1/\gamma \gg N^{z_s} \tau$, i.e. $\tilde g \gg 1$. In this case, $C_0(t)$ is still ballistic, given by \eqref{C0_smallgamma_ballistic} for $t\ll N^{z_s} \tau$. For $t \gg N^{z_s} \tau$, the second term in the numerator becomes negligible and one can approximate $C_0(t) \simeq 2\langle \delta x_i^2 \rangle (1-e^{-2\gamma t})$, i.e. it is linear in time for $t \ll 1/\gamma$
and 
converges exponentially fast to its large time limit for $t \sim 1/\gamma$. 

\section{Active Dyson Brownian motion and active Calogero-Moser model in a harmonic trap} \label{sec:harmonic_trap}

\subsection{Active Dyson Brownian motion}

Until now we have considered particles on a circle. However, the method presented here can also be used for confined particles on the real axis. In particular it can be used to extend the results of \cite{ADBM2} concerning the active Dyson Brownian motion (which was introduced in \cite{ADBM1}). This model corresponds to the log-gas ($s=0$), on the real axis, confined inside a harmonic potential $V(x)=\lambda x^2/2$. 
It follows the dynamical equation
\bea \label{ADBM_def} 
 \dot x_i(t) &=& - \lambda x_i(t) +  \frac{2\,g }{N} \sum_{j \neq i} 
\frac{1}{x_i(t)-x_j(t)} +  v_0 \sigma_i(t) \quad {\rm for} \ i=1,2, \cdots, N,
\eea
where the $\sigma_i(t)$ are again independent telegraphic noises with rate $\gamma$ (note that in this case the interaction constant is scaled as $1/N$ so that the support of the density remains finite as $N\to+\infty$).
Since the particles cannot cross, the ordering of the particles is preserved. We choose $x_1(t) > x_2(t) >...>x_N(t)$. In the absence of noise, the dynamics of the system converge to a ground state which is unique for a given initial ordering of the particles, where the position of the particles are given by $x_{\rm eq,1} > x_{\rm eq,2} >...>x_{{\rm eq},N}$. A remarkable property 
is that the scaled equilibrium positions $y_i = \sqrt{\frac{\lambda N}{2g}} x_{\rm eq,i}$ are the zeros of the Hermite polynomial of degree $N$, i.e., the roots of $H_N(y_i) = 0$.
At large $N$ the particles thus form a "crystal", and from the properties of the Hermite polynomials one can show that
the mean density of particles at equilibrium (i.e. for $v_0=0$) in the quadratic well is a Wigner semi-circle with edges at $\pm 2 \sqrt{g/\lambda}$, as is also the case for the DBM \cite{Forrester_book,Mehta_book}.

As in the periodic case, we consider the limit of weak noise, where
the particles remain close to their equilibrium positions $x_{\rm eq,i}$,
so that one can write
\be \label{eq_adbm}
x_i = x_{\rm eq,i} + \delta x_i   \quad , \quad x_{\rm eq,i} = \sqrt{\frac{2g}{\lambda \, N}}\, y_i \quad , \quad H_N(y_i) = 0\;,
\ee 
where now the $\delta x_i$'s are the small deviations from equilibrium which vanish as $v_0 \to 0$. This limit was already studied in \cite{ADBM2}, where the variance and covariance of particle displacements were computed (at fixed time, in the stationary state), but only in the limit of large persistence time $\gamma \to 0^+$. The present method enables us to extend the results to any value of $\gamma >0$. Although we will not expand too much on this aspect, it also makes it possible to compute time-dependent observables, as in the periodic case. The domain of validity of the weak noise approximation was estimated in \cite{ADBM2}, where it was found to be valid when $v_0^2/(g \lambda) \ll 1$. As we will see, the variance and covariance of the particle displacements are decreasing functions of $\gamma$ (as it was conjectured in \cite{ADBM2}), so that the results for $\gamma \to 0^+$ give upper bounds for $\gamma>0$. Therefore the criterion $v_0^2/(g \lambda) \ll 1$ should remain sufficient for the derivation presented below.

We start by linearizing the dynamics \eqref{ADBM_def},
\be \label{small_dx_adbm}
\frac{d}{dt} \delta x_i(t) = - \sum_{j=1}^N H^{DBM}_{ij} \, \delta x_j(t) + v_0 \sigma_i(t) \;,
\ee
where the Hessian matrix reads
\be
H^{DBM}_{ij} = \lambda \mathcal{H}_{ij} = \lambda\left[\delta_{ij}\left(1+ \sum_{k\neq i} \frac{1}{(y_i-y_k)^2}\right) - (1-\delta_{ij}) \frac{1}{(y_i-y_j)^2}\right]\; .
\label{eqHessian_adbm}
\ee
The matrix $\mathcal{H}$ is only a function of the Hermite roots $y_i$, independent of the model parameters. Contrary to the circular case, we cannot use plane waves to diagonalize this matrix. However, in this particular case, the Hessian matrix can be diagonalized exactly, as proved in \cite{eigenvectors}. The eigenvalues of ${\cal H}$ are the $N$ first strictly positive integers $k=1,2,\cdots, N$, and the corresponding normalized eigenvectors read
\begin{equation}
    (\psi_k)_i = \frac{u_k(y_i)}{\sqrt{\sum_{j=1}^N u_k(y_j)^2}} \quad , \quad u_k(y) = \frac{H_N^{(k)}(y)}{H_N'(y)} 
    = 2^{k-1} \frac{(N-1)!}{(N-k)!} \frac{H_{N-k}(y)}{H_{N-1}(y)} \;.
    \label{Hermite_eigenvectors}
\end{equation}
This result, which is crucial for the following computations, is specific to the log-gas, which is why we have to restrict to $s=0$ in this section. The only other known case for which such a result exists is the Calogero-Moser model $s=2$, which will be discussed below.
Note that in Ref \cite{ADBM2} the same approach has been used to study the dynamics of the passive version of both models.

The linearized dynamics \eqref{small_dx_adbm} take the same form as in the circular case \eqref{Eq_delta_x_RTP}, except that here we do not need to subtract the motion of the center of mass since the particles are confined. Assuming from now that the system is in the stationary state, we can thus perform the same derivation (i.e. in the annealed case) to compute the two-time covariance, leading to the equivalent of \eqref{rtp_corr_cm}
\be
\langle \delta x_i(t) \delta x_j(t') \rangle = v_0^2 \int \frac{d\omega}{2 \pi} \frac{4\gamma \, e^{i \omega (t-t')}}{\omega^2+4\gamma^2} [\omega^2 \mathbb{1}_N + \lambda^2 \mathcal{H}^2]^{-1}_{ij} \;.
\label{adbm_corr}
\ee
We can now use the eigendecomposition of $\mathcal{H}$ given in \eqref{Hermite_eigenvectors} to obtain
\begin{eqnarray}
\langle \delta x_i(t) \delta x_j(t') \rangle &=& v_0^2 \sum_{k=1}^N (\psi_k)_i (\psi_k)_j \int \frac{d\omega}{2 \pi} \frac{4\gamma}{\omega^2+4\gamma^2} \frac{e^{i \omega (t-t')}}{\omega^2+(\lambda k)^2} \nonumber \\
&=& \frac{v_0^2}{\lambda^2} \sum_{k=1}^N \frac{u_k(y_i)u_k(y_j)}{\sum_{l=1}^N u_k(y_l)^2} \frac{k e^{-2\gamma|t-t'|}-2\frac{\gamma}{\lambda} e^{-\lambda k|t-t'|}}{k(k^2-4\left(\frac{\gamma}{\lambda}\right)^2)} \;.
\label{corr_ADBM}
\end{eqnarray}
For the equal time correlation this gives
\be
\langle \delta x_i \delta x_j \rangle = \frac{v_0^2}{\lambda^2} \sum_{k=1}^N \frac{u_k(y_i)u_k(y_j)}{\sum_{l=1}^N u_k(y_l)^2} \frac{1}{k(k+2\frac{\gamma}{\lambda})} \;.
\label{corr_ADBM_stat}
\ee
Note that this expression has a well-defined finite limit when $\gamma \to 0^+$, which correctly recovers the result of \cite{ADBM2}. The expressions \eqref{corr_ADBM} and \eqref{corr_ADBM_stat} can be simplified further in the limit of large $N$, using results from \cite{ADBM2}. However this requires to distinguish two regimes, which have a different scaling with $N$: a bulk regime, where $i, N-i \gg 1$, and an edge regime where  $i, N-i = O(1)$.

\subsubsection{Bulk regime}

For bulk particles in the large $N$ limit, due to the fast convergence of the series, one only needs to focus on the terms $k \ll N$. An approximate expression of $u_k(y_i)$ for $k \ll N$ was obtained in \cite{ADBM2} as a function of the Chebyshev polynomials of the second kind $U_k(r)$,
\begin{equation}
    u_k(y_i) \simeq (2N)^{\frac{k-1}{2}} U_{k-1} \left( \frac{y_i}{\sqrt{2N}} \right)  \quad , \quad  U_{k-1}(r) = \frac{\sin(k \arccos(r))}{\sqrt{1-r^2}} \;.
    \label{uk_chebyshev}
\end{equation}
The denominator can then be simplified using the orthonormality of the $U_k(r)$ with respect to the Wigner semicircle measure, 
\begin{equation}
    \sum_{l=1}^N U_{k-1} \left( \frac{y_l}{\sqrt{2N}} \right)^2 \simeq N \int_{-1}^1 dr \frac{2\sqrt{1-r^2}}{\pi} U_{k-1}(r)^2 = N \;.
    \label{normalization}
\end{equation}

\begin{figure}
    \centering
    \includegraphics[width=0.32\linewidth]{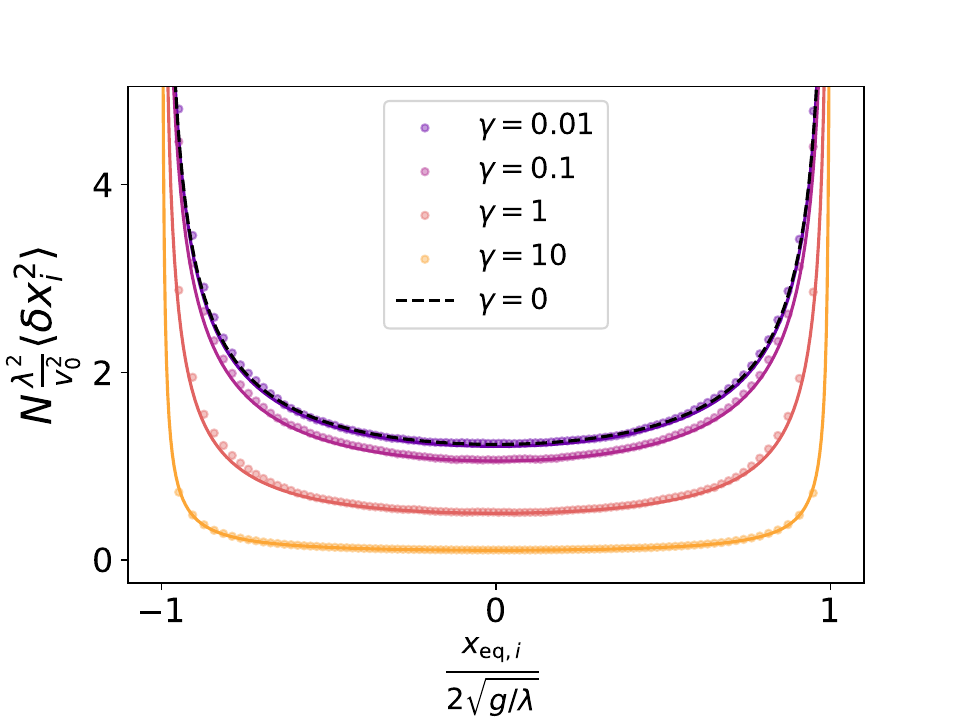}
    \includegraphics[width=0.32\linewidth]{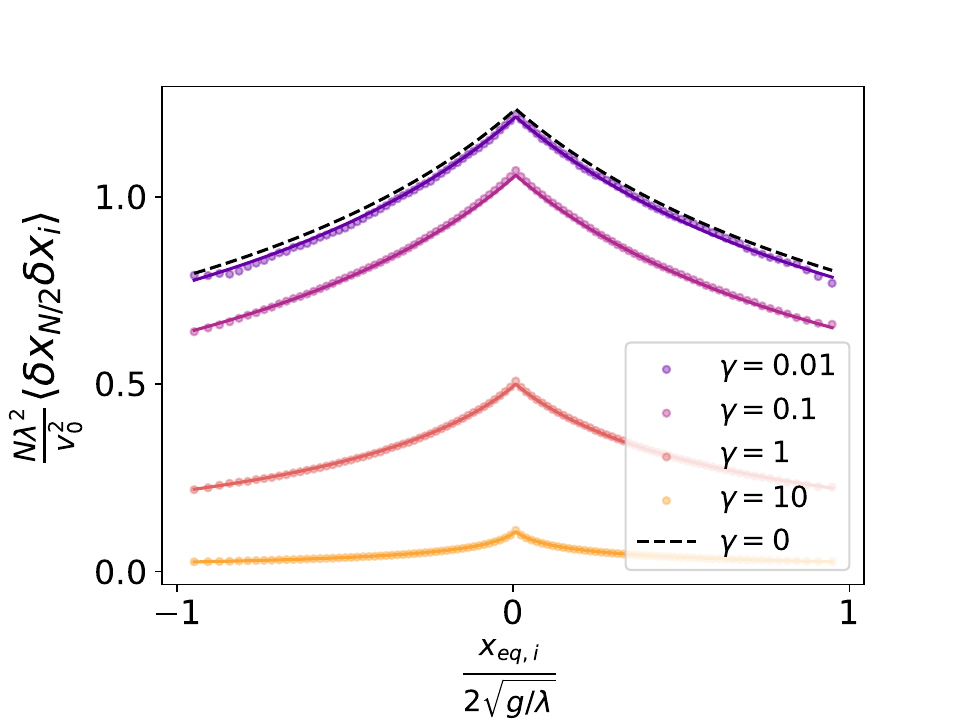}
    \includegraphics[width=0.32\linewidth]{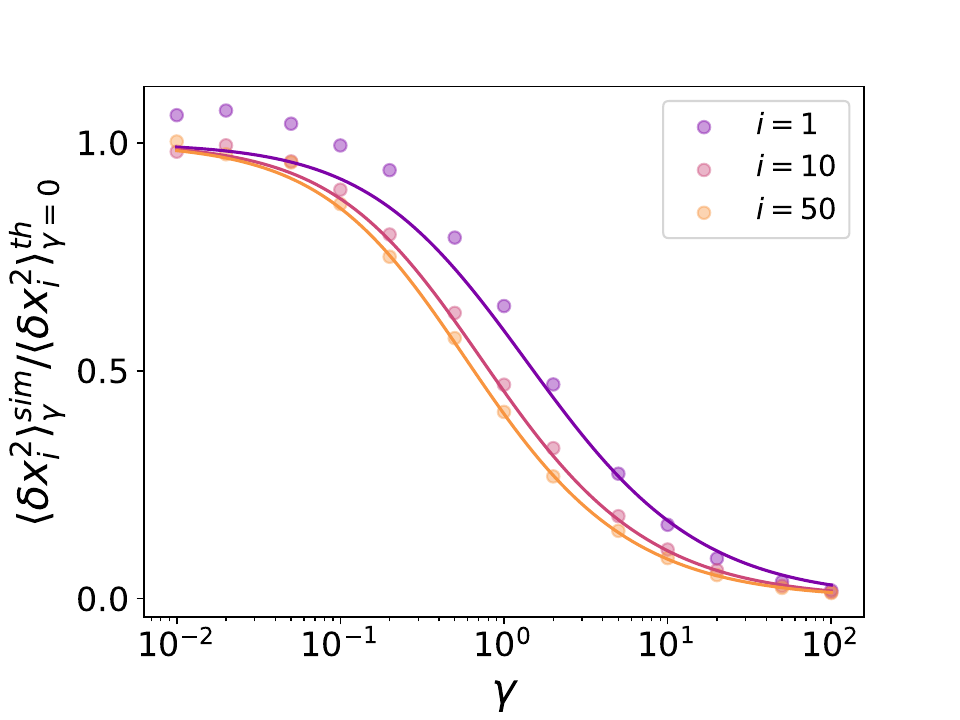}
    \caption{Left: Variance of the displacement of a single particle $\langle\delta x_i^2 \rangle$ in the stationary state, plotted as a function of the equilibrium position $x_{{\rm eq},i}$, for $N=100$, $\lambda=1$, $g=1$, $v_0=0.1$ and different values of $\gamma$. The results of numerical simulations of the dynamics (dots) are compared with the the large $N$ prediction \eqref{varADBM_largeN} (lines) for each value of $\gamma$, showing a very good agreement. The dashed black line corresponds to the $\gamma=0^+$ prediction from \cite{ADBM2}. Center: Same plot for the covariance between particle $i$ and a central particle $i=N/2$, plotted as a function of $x_{{\rm eq},i}$. The full lines correspond to the large $N$ prediction \eqref{covADBM_largeN}. Right: Ratio of the variance obtained from simulations with the theoretical value for $\gamma=0$, as a function of $\gamma$, for different values of $i$, plotted in log-linear scale. The plot shows that the
    variance decreases as $\gamma$ increases.
    The full lines again correspond to the theoretical prediction \eqref{varADBM_largeN}. The variance decreases linearly for $\gamma/\lambda \ll 1$ (this can be seen by performing a small $\gamma$ expansion of \eqref{varADBM_largeN}), and decays as $1/\gamma$ when $\gamma/\lambda \gg 1$. The decrease becomes slower as one gets closer to the edges. Note that in our simulations, the number $N$ of particles is too small to observe a true distinct edge regime.}
    \label{figADBM1}
\end{figure}

Using these approximations, we thus obtain the two-time correlations for the displacements of bulk particles in the stationary state, for $N \gg 1$ (with $\tilde \gamma=\gamma/\lambda$ fixed),
\be
\langle \delta x_i(t) \delta x_j(t') \rangle \simeq \frac{v_0^2}{\lambda^2 N} \mathcal{C}_b^{\gamma/\lambda}\left( \frac{x_{{\rm eq},i}}{2\sqrt{g/\lambda}}, \frac{x_{{\rm eq},j}}{2\sqrt{g/\lambda}}, \lambda |t-t'| \right) \ , \ 
\mathcal{C}_b^{\tilde \gamma}(x,y,\tau) = \sum_{k=1}^\infty \frac{k e^{-2\tilde \gamma \tau}-2\tilde \gamma e^{-k\tau}}{k(k^2-4\tilde \gamma^2)} U_{k-1}(x) U_{k-1}(y) \;,
\label{covADBM_largeN_time}
\ee
and for the equal time correlations
\begin{equation}
\langle \delta x_i \delta x_j \rangle \simeq \frac{v_0^2}{\lambda^2 N} \mathcal{C}_b^{\gamma/\lambda}\left( \frac{x_{{\rm eq},i}}{2\sqrt{g/\lambda}}, \frac{x_{{\rm eq},j}}{2\sqrt{g/\lambda}} \right) \quad , \quad 
\mathcal{C}_b^{\tilde \gamma}(x,y) = \sum_{k=1}^\infty \frac{U_{k-1}(x) U_{k-1}(y)}{k(k+2\tilde \gamma)}  \;,
\label{covADBM_largeN}
\end{equation}
up to a relative error of order $O(\frac{1}{N})$ (as long as $\gamma/\lambda \ll N$). Let us also write the variance of the displacement for a single particle
\begin{equation}
\langle \delta x_i^2 \rangle \simeq \frac{v_0^2}{\lambda^2 N} \mathcal{V}_b^{\gamma/\lambda}\left( \frac{x_{{\rm eq},i}}{2\sqrt{g/\lambda}}\right) \quad , \quad \mathcal{V}_b^{\tilde \gamma}(x) = \mathcal{C}_b^{\tilde\gamma}(x,x) = \sum_{k=1}^\infty \frac{U_{k-1}(x)^2}{k(k+2\tilde \gamma)} \;.
\label{varADBM_largeN}
\end{equation}
We recall that $x_{{\rm eq},i} = \sqrt{\frac{2g}{\lambda \, N}}\, y_i \in (-2\sqrt{g/\lambda}, 2\sqrt{g/\lambda})$ and therefore $\mathcal{C}_b(x,y)$ and $\mathcal{V}_b(x)$ are defined on $(-1,1)^2$ and $(-1,1)$ respectively. For $\gamma=0^+$, the infinite sums can be computed explicitly, leading to simple explicit expressions for these two functions (see \cite{ADBM2}). Note that $\mathcal{C}_b(x,y)$ and $\mathcal{V}_b(x)$ are independent of $N$, and thus the covariance and variance scale as $1/N$. As was anticipated in \cite{ADBM2} from numerical results, both the variance and covariance are decreasing functions of $\gamma$. We have tested numerically the validity of these predictions by comparing with numerical simulations in Fig.~\ref{figADBM1}. For small enough ratio $\frac{v_0}{\sqrt{g\lambda}}$ (equal to $0.1$ in the figure), we find a perfect agreement for any value of $\gamma$.

Finally, as detailed in \cite{ADBM2}, the expression of the covariance can be used to obtain the variance of the distance between two particles, i.e. of a gap of size $k$. Indeed
\begin{equation} \label{corr1} 
    \langle (\delta x_i - \delta x_{i+n})^2 \rangle = \langle \delta x_i^2 \rangle + \langle \delta x_{i+n}^2 \rangle - 2 \langle \delta x_i \delta x_{i+n} \rangle \;.
\end{equation}
For $1 \ll n \ll N$, one can use the above results to estimate \eqref{corr1}. 
This leads to 
\be
\langle (\delta x_i - \delta x_{i+n})^2 \rangle \simeq \frac{v_0^2}{\lambda^2 N} \mathcal{D}_b^{\gamma/\lambda} \left(\frac{x_{{\rm eq},i}}{2\sqrt{g/\lambda}}, \frac{x_{{\rm eq},i+n}}{2\sqrt{g/\lambda}}\right) \ , \ \mathcal{D}_b^{\tilde \gamma} (x,y) = \mathcal{V}_b^{\tilde \gamma}(x) + \mathcal{V}_b^{\tilde \gamma}(y) - 2 \mathcal{C}_b^{\tilde \gamma}(x,y)
= \sum_{k=1}^\infty \frac{\left(U_{k-1}(x)-U_{k-1}(y)\right)^2}{k(k+2\frac{\gamma}{\lambda})} 
\label{var_chn} \;.
\ee
Due to the approximations used, this expression is however valid only for $k=\kappa N$ with $\kappa = O(1)$, i.e. on macroscopic scales in the bulk.

\subsubsection{Edge regime}

The expressions above are valid in the bulk, i.e. for $i$ and $N-i$ of order $N$. Indeed, for $x$ close to 1, one can show that the scaling function $\mathcal{V}_b(x)$ diverges as $\sim 1/\sqrt{1-x}$. 
Writing $\epsilon=1-x$, we have
\be \label{Vb_edge}
\mathcal{V}_b^{\tilde \gamma}(1-\epsilon) \simeq \sum_{k=1}^{\infty} \frac{\sin^2(k\sqrt{2\epsilon})}{2\epsilon k (k+2\tilde \gamma)} \simeq \int_0^\infty \frac{du}{\sqrt{2\epsilon}} \frac{\sin^2(u)}{u(u+2\tilde \gamma \sqrt{2\epsilon})} = \frac{\pi}{2\sqrt{2\epsilon}} +O(1)
\ee
(note that the leading order is independent of $\tilde \gamma=\gamma/\lambda$). This divergence suggests a different scaling at the edge. Indeed at the edge, i.e. for $i$ of order 1, the Hermite polynomials in the definition of $u_k(y_i)$ should instead be approximated using the Airy function $\Ai(x)$. Following the derivation of \cite{ADBM2}, we find
\begin{eqnarray}
\langle \delta x_i \delta x_j \rangle &\simeq& \frac{v_0^2}{\lambda^2 N^{1/3}} \frac{1}{\Ai'(a_i) \Ai'(a_j)} \sum_{k=1}^{\infty} \frac{\Ai(a_i + kN^{-1/3})\Ai(a_j + kN^{-1/3})}{k(k+2\frac{\gamma}{\lambda})} \\
&\simeq& \frac{v_0^2}{\lambda^2 N^{2/3}} \frac{1}{\Ai'(a_i) \Ai'(a_j)} \int_0^{+\infty} dx \ \frac{\Ai(a_i + x)\Ai(a_j + x)}{x(x+2 \hat \gamma)} \quad , \quad \hat \gamma=\frac{\gamma}{\lambda}N^{-1/3} \;,
\label{covADBM_edge}
\end{eqnarray}
where $a_i$ denotes the $i^{th}$ zero of the Airy function. Note the $N^{-2/3}$ overall scaling instead of $N^{-1}$. 
If one keeps $\gamma/\lambda$ fixed, one sees that the 
variance and covariance near the edge are independent of $\gamma$ at large $N$, 
i.e we recover the same result as for $\gamma=0^+$. This should be expected since, as noted in \cite{ADBM2} Section VII for the passive
case, the relaxation time scale is much faster at the edge as compared to the bulk. Hence it is natural to scale time as $N^{-1/3}$,
and $\gamma \sim N^{1/3}$ so that $\hat \gamma$ is fixed. Indeed, for 
the two-times correlations, we obtain in the same way the following scaling form
\bea
\langle \delta x_i(t) \delta x_j(t') \rangle &\simeq& \frac{v_0^2}{\lambda N^{2/3}} \mathcal{C}^{N^{-1/3}\gamma/\lambda}_e(a_i,a_j, N^{1/3} \lambda |t-t'|) \, , \nn \\
{\cal C}_e^{\hat \gamma}(a_i,a_j, \hat \tau) &=& \frac{1}{\Ai'(a_i) \Ai'(a_j)} \int_0^{+\infty} dx \frac{x e^{-2\hat \gamma \hat \tau}-2\hat \gamma e^{-\hat \tau x}}{x(x^2-4\hat \gamma^2)} \Ai(a_i + x)\Ai(a_j + x)   \;,
\label{cov_edge_integral_CM_time_result}
\eea
which extends to the active case the expression Eq. (126) obtained in \cite{ADBM2} for the passive DBM. 

Finally, as explained in \cite{ADBM2} for $\gamma=0^+$, the covariance between a bulk particle and an edge particle (i.e. $i=O(1)$ and $j\gg 1$) is still given by the bulk expression \eqref{covADBM_largeN}.


\subsection{Active Calogero-Moser model}

Similar results can be obtained for the active Calogero-Moser model, which corresponds to the active Riesz gas for $s=2$ on the real axis, confined inside a harmonic potential $V(x)=\lambda x^2/2$. This model is introduced and studied in a companion paper \cite{activeCM}. Its dynamics can be written
\begin{equation}
 \dot x_i(t) = - \lambda x_i(t) +  \frac{8\, \tilde g^2}{N^2} \sum_{j \neq i} 
\frac{1}{(x_i(t)-x_j(t))^3} + v_0 \sigma_i(t) \;.
\label{Calogero}
\end{equation}
To be consistent with the active DBM case, we have scaled the interaction with $N$ so that the density has a finite support when $N\to+\infty$. The mapping to the notations of \cite{activeCM} will be given at the end of this section. In \cite{activeCM}, this model was studied using a combination of numerical and analytical arguments, and it was shown to display a behaviour very similar to the active DBM. In particular, both models display the same crossover between three regimes depending on the amplitude of the noise: a ``crystal'' regime at very weak noise, where the particles are strongly localized and the particle density takes a semi-circular shape, a ``liquid'' regime at intermediate noise with stronger fluctuations but where the Wigner semi-circle still holds, and finally a regime of very strong noise, where the semi-circle breaks down and the density takes a bell shape. The difference between the two models is however more visible when looking at the fluctuations, in particular for the edge particles, due to the short-range nature of the interaction in the CM model. In \cite{activeCM}, we use the method introduced in \cite{ADBM2} to compute the covariance of the particle displacements in the limit of weak noise and large persistent time $\gamma \to 0^+$. The extension to finite $\gamma$ is also given without derivation. We explain here how to obtain these results for $\gamma >0$. The results presented here also generalize the ones for the passive CM, which was studied in \cite{ADBM2}.

Quite surprisingly, for the (active) CM model, the equilibrium positions of the particles are also the rescaled roots of the Hermite polynomial $H_N(y)$ \cite{Calogero75, Moser76, Agarwal2019}. With the present notations one has 
\be \label{eq_CM}
x_{\rm eq,i}  = \frac{1}{\lambda^{1/4}}\sqrt{\frac{2 \tilde g}{ \, N}}\, y_i \quad , \quad H_N(y_i)=0 \;.
\ee 
Once again, we consider the weak noise limit, which in this case can be shown to hold when $\frac{v_0}{\lambda^{3/4}\tilde g^{1/2}} \ll N^{1/2}$ (see below), 
and we linearize the equation of motion \eqref{Calogero} as 
\be \label{small_dx_acm}
\frac{d}{dt} \delta x_i(t) = - \sum_{j=1}^N H^{CM}_{ij} \, \delta x_j(t) + v_0 \sigma_i(t) \;,
\ee
where again $\delta x_i=x_i-x_{{\rm eq},i}$. From here, the correlations $\langle \delta x_i(t) \delta x_j(t') \rangle$ can be directly deduced from the active DBM result, thanks to a non-trivial relation between the Hessian matrices of the two models (see e.g. \cite{Agarwal2019})
\be
H^{CM} = \lambda \mathcal{H}^2 \;.
\ee
Note that we have used similar methods to study the passive version of the CM model in the weak noise limit in Ref. \cite{ADBM2}. 

Thus, one can directly apply the result \eqref{corr_ADBM}, replacing the eigenvalues $k=1,...,N$ by their square,
\be
\langle \delta x_i(t) \delta x_j(t') \rangle = \frac{v_0^2}{\lambda^2} \sum_{k=1}^N \frac{u_k(y_i)u_k(y_j)}{\sum_{l=1}^N u_k(y_l)^2} \frac{k^2 e^{-2\gamma|t-t'|}-2\frac{\gamma}{\lambda} e^{-\lambda k^2|t-t'|}}{k^2(k^4-4\left(\frac{\gamma}{\lambda}\right)^2)} \;,
\label{corr_CM}
\ee
and for the equal time correlation,
\be
\langle \delta x_i \delta x_j \rangle = \frac{v_0^2}{\lambda^2} \sum_{k=1}^N \frac{u_k(y_i)u_k(y_j)}{\sum_{l=1}^N u_k(y_l)^2} \frac{1}{k^2(k^2+2\frac{\gamma}{\lambda})} \;.
\label{corr_CM_stat}
\ee

An important difference with the active DBM is that, in the limit of large $N$, there is no distinct scaling for the edge particles (see below).
Thus the approximations used above for the bulk regime of the active DBM are valid for all particles in the case of the active CM. We thus have, for $N\gg 1$, for any particles $i$ and $j$,
\be
\langle \delta x_i(t) \delta x_j(t') \rangle \simeq \frac{v_0^2}{\lambda^2 N} \mathcal{\widetilde C}_b^{\gamma/\lambda}\left( \frac{x_{{\rm eq},i}}{2\sqrt{\tilde g}/\lambda^{1/4}}, \frac{x_{{\rm eq},j}}{2\sqrt{\tilde g}/\lambda^{1/4}}, \lambda |t-t'| \right) \ , \ 
\mathcal{\widetilde C}_b^{\tilde \gamma}(x,y,\tau) = \sum_{k=1}^\infty \frac{k^2 e^{-2\tilde \gamma \tau}-2\tilde \gamma e^{-k^2\tau}}{k^2(k^4-4\tilde \gamma^2)} U_{k-1}(x) U_{k-1}(y) \;,
\label{covCM_largeN_time}
\ee
and for the equal time correlations
\begin{equation}
\langle \delta x_i \delta x_j \rangle \simeq \frac{v_0^2}{\lambda^2 N} \mathcal{\widetilde C}_b^{\gamma/\lambda}\left( \frac{x_{{\rm eq},i}}{2\sqrt{\tilde g}/\lambda^{1/4}}, \frac{x_{{\rm eq},j}}{2\sqrt{\tilde g}/\lambda^{1/4}} \right) \quad , \quad 
\mathcal{\widetilde C}_b^{\tilde \gamma}(x,y) = \sum_{k=1}^\infty \frac{U_{k-1}(x) U_{k-1}(y)}{k^2(k^2+2\tilde \gamma)}  \;,
\label{covCM_largeN}
\end{equation}
up to a relative error of order $O(\frac{1}{N})$ (as long as $\gamma/\lambda \ll N$). Let us also write the variance of the displacement for a single particle
\begin{equation}
\langle \delta x_i^2 \rangle \simeq \frac{v_0^2}{\lambda^2 N} \mathcal{\widetilde V}_b^{\gamma/\lambda}\left( \frac{x_{{\rm eq},i}}{2\sqrt{\tilde g}/\lambda^{1/4}}\right) \quad , \quad \mathcal{\widetilde V}_b^{\tilde \gamma}(x) = \mathcal{\widetilde C}_b^{\tilde\gamma}(x,x) = \sum_{k=1}^\infty \frac{U_{k-1}(x)^2}{k^2(k^2+2\tilde \gamma)} \;.
\label{varCM_largeN}
\end{equation}
We recall the expression of the Chebyshev polynomials of the second kind,
\begin{equation}
    U_{k-1}(r) = \frac{\sin(k \arccos(r))}{\sqrt{1-r^2}} \;.
    \label{uk_chebyshev2}
\end{equation}
We recall that $x_{{\rm eq},i} \in (-2\sqrt{\tilde g}/\lambda^{1/4}, 2\sqrt{\tilde g}/\lambda^{1/4})$ so that once again $\mathcal{\widetilde C}_b(x,y)$ and $\mathcal{\widetilde V}_b(x)$ are defined on $(-1,1)^2$ and $(-1,1)$ respectively. For $\gamma=0^+$, the infinite sums can again be computed explicitly, leading to the expressions given in \cite{activeCM}. Note also that for $\gamma \gg 1$, for the bulk particles, we recover the results for the passive Calogero-Moser model derived in \cite{ADBM2}, with an effective temperature $T_{\rm eff} = v_0^2/(2\gamma)$.

Contrary to the active DBM case, the expressions \eqref{covCM_largeN_time}, \eqref{covCM_largeN} and \eqref{varCM_largeN} have finite, $\tilde \gamma$-dependent limits for $x,y\to\pm 1$. For instance, one has for $x\to 1^-$, $U_{k-1}(x) \to k$, and thus
\be
\mathcal{\widetilde V}_b^{\tilde \gamma}(x) \xrightarrow[x\to 1^-]{} \sum_{k=1}^{\infty} \frac{1}{k^2+2\tilde \gamma} = \frac{\pi}{2} \frac{\coth(\pi \sqrt{2\tilde \gamma})}{\sqrt{2\tilde \gamma}} - \frac{1}{4\tilde \gamma} \simeq \begin{cases} \frac{\pi^2}{6} \text{ for } \tilde \gamma \ll 1 \\ \frac{\pi}{2\sqrt{2 \tilde \gamma}} \text{ for } \tilde \gamma \gg 1 \end{cases} \;.
\ee
As anticipated, this means that the $1/N$ scaling of the variance remains valid even for edge particles.

As in the active DBM case, the present results also enable us to compute the variance of the gaps. It reads,
\be \label{var_chn_gamma}
\langle (\delta x_i - \delta x_{i+n})^2 \rangle = \frac{v_0^2}{\lambda^2 N} \mathcal{\widetilde D}_b^{\gamma/\lambda} \left(\frac{x_{{\rm eq},i}}{2\sqrt{\tilde g}/\lambda^{1/4}}, \frac{x_{{\rm eq},i+n}}{2\sqrt{\tilde g}/\lambda^{1/4}}\right) \quad , \quad \mathcal{\widetilde D}_b^{\tilde \gamma} (x,y) 
= \sum_{k=1}^{\infty} \frac{\left(U_{k-1}(x)-U_{k-1}(y) \right)^2}{k^2(k^2+2\tilde \gamma)}
\;.
\ee
Contrary to the active DBM case, this result holds even at the microscopic scale $n=O(1)$ (see \cite{activeCM} for a comparison with numerics). This is due to the faster convergence of the series. To compute the behavior of the variance of the gaps for $n=O(1)$ we can thus expand the scaling function $\mathcal{D}_b^{\tilde \gamma}(x,y)$
for $y$ close to $x$. One finds, writing $x=\cos \theta$ for $\theta \in [-\pi/2,\pi/2]$ that it behaves as $\sim (x-y)^2$,
\be 
\mathcal{D}_b^{\tilde \gamma}(x,y) \simeq B(\tilde \gamma,\theta) (x-y)^2 \quad , \quad 
B(\tilde \gamma,\theta) = \frac{1}{\sin^4 \theta} \sum_{k=1}^{\infty} \frac{(k \cos(k \theta) - \cot \theta \sin(k \theta))^2}{k^2( k^2+2\tilde \gamma) } \;.
\ee 

Using that the equilibrium density is a semi-circle at large $N$, one has $\frac{x_{{\rm eq},i}}{2\sqrt{\tilde g}/\lambda^{1/4}} = \frac{y_{{\rm eq},i}}{\sqrt{2N}} \simeq G^{-1}(i/N)$ where $G(x)=\int_{-1}^x du \rho_{sc}(u)$ is the cumulative distribution of the semi-circle density $\rho_{sc}(u) = \frac{2\sqrt{1-u^2}}{\pi}$.
This implies 
\be 
\frac{n}{N} \simeq  \rho_{sc}( \frac{x_{{\rm eq},i}}{2\sqrt{\tilde g}/\lambda^{1/4}}) \frac{(x_{{\rm eq},i}-x_{{\rm eq},i+n})}{2\sqrt{\tilde g}/\lambda^{1/4}} \;,
\ee 
which leads to the variance of the gaps for any $n \ll N$, 
\be 
\langle (\delta x_i - \delta x_{i+n})^2 \rangle \simeq C^{\gamma/\lambda}\left( \frac{x_{{\rm eq},i}}{2\sqrt{\tilde g}/\lambda^{1/4}} \right) \,   n^2 \quad , \quad 
C^{\tilde \gamma}(x) = \frac{v_0^2}{\lambda^2 N^3} \frac{\pi^2}{4} 
\frac{B(\tilde \gamma,\theta) }{\sin^2 \theta}  \quad , \quad x = \cos \theta \;. \label{gapvariance_largeN_gamma}
\ee 
At the center of the trap one has $\theta=\pi/2$ and the series in the amplitude can be explicitly evaluated,
\be 
B(\tilde \gamma,\theta=\pi/2) = \frac{1}{4} \left( \frac{\pi}{\sqrt{2\tilde \gamma}} 
\coth \big( \pi \sqrt{\frac{\tilde \gamma}{2}} \big) -\frac{1}{\tilde \gamma} \right) \;.
\ee 

Thus the growth of the variance of the gaps is $\sim n^2$. This implies so-called giant number fluctuations 
\cite{DasGiant2012}.
Indeed the number ${\cal N}_{[a,b]}$ of particles inside an interval $[a,b]$ can be estimated as, for sufficiently large intervals $1 \ll b-a \ll N$,
\be 
{\rm Var} \, {\cal N}_{[a,b]} \simeq \rho(x_{{\rm eq},i})^2 \langle (\delta x_i-\delta x_{i+n})^2 \rangle \simeq \frac{v_0^2}{4\tilde g\lambda^{3/2} N^3} B(\tilde \gamma,\theta) \, n^2  \quad , \quad x = \cos \theta \;,
\ee 
where $b=x_{{\rm eq},i}$ and $a=x_{{\rm eq},i+n}$.
\\
The criterion for the validity of the weak noise approximation can be obtained from \eqref{gapvariance_largeN_gamma}. Indeed for the approximation to hold one should have
\be
\frac{\sqrt{\langle (\delta x_i - \delta x_{i+1})^2 \rangle}}{x_{{\rm eq},i}-x_{{\rm eq},i+1}} \sim \frac{v_0/(\lambda N^{3/2})}{\sqrt{\tilde g}/(\lambda^{1/4} N)} = \frac{v_0}{\lambda^{3/4} \sqrt{\tilde g N}} \ll 1 \;.
\ee

As we discussed above, these results are also mentioned and compared with numerics in \cite{activeCM}. Since 
the two papers use different notations, let us specify that the results of \cite{activeCM} can be recovered by taking $\lambda \to 1$ and $\tilde g \to N/2$. With these notations, the equilibrium positions of the particles are now directly given by the roots of $H_N(y)$, $x_{{\rm eq},i} \to y_i$, and the edges of the support by $\pm 2 \sqrt{\tilde g}/\lambda^{1/4} \to \pm \sqrt{2N}$.

\section{Conclusion and outlook} \label{sec:conclusion}

In this work, we have performed a weak noise/low temperature analysis of the space time correlations of particle positions in the Riesz gas, with interactions $\propto 1/|x_i - x_j|^s$ with $s>-1$, 
both for passive (Brownian) and 
active (RTP) systems. The case $s>1$ corresponds to short range (SR) interactions, and $s<1$ to long-range (LR) interactions.
The special cases $s=0$ and $s=2$ correspond respectively to the (passive or active) Dyson Brownian motion/log-gas and the Calogero-Moser model. 
Even in the passive case we have found original and non-trivial results, including for the most studied case of the log-gas \cite{Spohn1}.
We have obtained the detailed time dependence of the particle displacements which exhibits the standard single-file diffusion 
for the SR case, and an anomalous subdiffusion for the LR case with an $s$ dependent exponent. We have also obtained detailed results 
for the gap statistics, which are currently studied in the mathematics literature \cite{BoursierCLT,BoursierCorrelations}, and extended our results to 
their time evolution which was an open problem. 
We estimated the validity of the method, and found that our formulas should hold in a wide range of temperature. In particular in the case $s<0$ we found that the system
may undergo a melting transition, as suggested recently by numerical simulations  \cite{Lelotte2023}, and 
obtained a Lindemann estimate for the conjectured melting transition temperature. This leaves open the question
of formulating a more predictive theory for this transition. 

In the active case, i.e., for $N$ run-and-tumble particles, we showed that the behavior of the correlations is similar to the Brownian particles
at large distance and large time, while it differs in the other cases, i.e for short distance or large time, in the
strongly active regime $\hat g = 1/(2 \gamma \tau) \gg 1$. We have performed a detailed analysis of the various dynamical
regimes and crossovers of these correlations. We have shown the existence of a new correlation length associated to the stationary regime 
given by $\hat g^{1/z_s}$, where $z_s=\min(1+s,2)$ is the dynamical exponent. We have made connections
with recent studies of the harmonic chains~\cite{HarmonicChainRevABP,SinghChain2020,PutBerxVanderzande2019,HarmonicChainRTPDhar},
which, in the large $N$ limit, becomes a special case of our study with SR interactions. 

The method introduced in this paper also allowed to obtain new results concerning the fluctuations in the active DBM ($s=0$) and active Calogero-Moder model ($s=2$) in a harmonic trap, generalizing to finite tumbling rates $\gamma>0$ the results of \cite{ADBM2}. The results derived here for the active Calogero-Moder model ($s=2$) will be compared to numerical results in a companion paper focusing on this model \cite{activeCM}.

A tantalizing question is the connection between the weak noise expansion starting from a crystal developed in the present work and the more hydrodynamic 
point of view of macroscopic fluctuation theory (MFT). Indeed, 
remarkably, we found that our formula for the anomalous single-file diffusion regime in the Brownian case coincides
exactly, including prefactors in the LR case, with the recent study \cite{DFRiesz23} using MFT. 
It would be interesting to see whether our systematic weak noise expansion could be pushed to
obtain the higher order correlation functions and their temperature dependence. Another natural extension of the present work would be to extend MFT to the study of the active Riesz gas.
This should give another approach to compute higher order correlation functions as well as other interesting
observables such as the current fluctuations and associated large deviations. 



To conclude, a notable achievement of our method lies in providing a detailed description of fluctuations in a one-dimensional passive or active Riesz gas at microscopic scales. Extending this approach to long-range interacting active systems in higher dimensions would be an exciting direction for future research, particularly to address situations involving topological defects, such as dislocations, in two-dimensional systems 
\cite{ActiveVortexMeltingNature2021,Leticia2022,Chate2023,ActiveMeltingNature2024}. 

\acknowledgments
We thank R. Lelotte and M. Lewin for discussions about Ref. \cite{Lelotte2023} and A. Dhar and S. Santra for sharing with us their preliminary 
numerical data about the active Calogero-Moser model. We acknowledge support from ANR Grant No. ANR- 23-CE30-0020-01 EDIPS.

\appendix

\section{Mapping to other active particle models} \label{app:other_active_models}

In this paper we focus on Brownian particles and run-and-tumble particles. However, since we only compute quantities based on the second order moments, the results can easily be extended to other models of active particles. Let us consider again $N$ interacting particles on a circle of perimeter $L$, but now driven by generic i.i.d. additive noise~$\zeta_i(t)$,
\be \label{Eq_def_gen}
\dot x_i(t) = -\sum_{j(\neq i)} W'(x_i(t)-x_j(t)) + \zeta_i(t) \;.
\ee
The dynamical equations for small particle displacements read
\be 
\frac{d}{dt} \delta x_i(t) = - \sum_{j=1}^N H_{ij} \, \delta x_j(t) + \zeta_i(t) \;,
\ee 
where we recall that the $\delta x_i(t)$ are defined in \eqref{def_delta_x}, and $H$ is the Hessian matrix defined in \eqref{defHessian}.
Since we only consider the moments up to second order, only the first two moments of the noise matter. We consider centered noise, i.e. $\langle \zeta_i(t) \rangle=0$. For RTPs, the second moment is 
\be
\langle \zeta_i(t) \zeta_j(t') \rangle = v_0^2 e^{-2\gamma|t-t'|}\delta_{ij} \;.
\ee
This means that the computations of the previous section are also valid for other types of driving noises $\zeta_i(t)$, as long as they are exponentially correlated in time. The active Ornstein-Uhlenbeck particle (AOUP) corresponds to a driving noise
\be
\dot \zeta_i(t) = -\frac{\zeta_i(t)}{\tau_{ou}} + \sqrt{2D} \, \eta_i(t) \;,
\ee
where the $\eta_i(t)$ are standard Gaussian white noises. Thus the stationary correlations of the driving noise are given by
\be
\langle \zeta_i(t) \zeta_j(t') \rangle = D\tau_{ou} e^{-\frac{|t-t'|}{\tau_{ou}}} \delta_{ij} \;.
\ee
The results for the RTPs can thus be applied to AOUPs by taking $v_0^2 \to D\tau_{ou}$ and $2\gamma \to 1/\tau_{ou}$.
Similarly, for active Brownian particles (ABPs) one has
\be
\zeta_i(t) = v_0 \cos(\theta_i(t)) \quad , \quad \dot \theta_i(t) = \sqrt{2D} \, \eta_i(t) \;,
\ee
which leads to
\be
\langle \zeta_i(t) \zeta_j(t') \rangle = \frac{v_0^2}{2} e^{-D|t-t'|} \delta_{ij} \;.
\ee
We have used that 
\bea
\langle \zeta_i(t) \zeta_i(t') \rangle &=& v_0^2 \langle \cos \theta_i(t) \cos \theta_i(t') \rangle = \frac{v_0^2}{2} (\langle \cos (\theta_i(t) - \theta_i(t')) \rangle + \langle \cos (\theta_i(t) + \theta_i(t')) \rangle)  \nn \\
&=& \frac{v_0^2}{2} \langle \cos (\theta_i(t) - \theta_i(t')) \rangle = \frac{v_0^2}{2} e^{-\frac{1}{2} \langle (\theta_i(t) - \theta_i(t'))^2 \rangle} \;,
\eea
where we assume the initial state to be stationary ($\theta_i(0)$ modulo $2 \pi$ is uniform on the circle).
Therefore the mapping with RTPs requires taking $v_0^2 \to \frac{v_0^2}{2}$ and $2\gamma \to D$. 

\section{Origin of the validity condition \eqref{cond_approx}} \label{app:validity_condition}

Let us justify the validity condition \eqref{cond_approx} by taking the Taylor expansion of the equation of motion \eqref{Eq_def_intro} to the next order. We have:
\be
\delta \dot{x}_i = - \sum_j \frac{\partial^2 E}{\partial x_i \partial x_j}(\{x_i^0\}) \delta x_j - \frac{1}{2} \sum_{j,k} \frac{\partial^3 E}{\partial x_i \partial x_j \partial x_k}(\{x_i^0\}) \delta x_j \delta x_k + O(\delta x^3) + \zeta_i(t) \;,
\ee
where $\zeta_i(t)$ is the driving noise and $x_i^0= (i - \frac{N+1}{2})L$ denote the positions in the ground state. Let us recall that
\be
H_{ij} = \frac{\partial^2 E}{\partial x_i \partial x_j}(\{ x_i^0 \}) = \begin{cases} \sum_{k(\neq i)} W''(x_i^0-x_k^0) \quad {\rm for} \ i=j \;, \\ -W''(x_i^0-x_j^0) \hspace{1.25cm} {\rm for} \ i\neq j \;. \end{cases}
\ee
In addition
\be
\frac{\partial^3 E}{\partial x_i \partial x_j \partial x_k}(\{ x_i^0 \}) = \begin{cases} \sum_{l(\neq i)} W'''(x_i^0-x_l^0) \quad {\rm for} \ i=j=k \;, \\ -W'''(x_i^0-x_k^0) \hspace{1.14cm} {\rm for} \ i=j\neq k \;, \\ -W'''(x_i^0-x_j^0) \hspace{1.14cm} {\rm for} \ i=k\neq j \;, \\ W'''(x_i^0-x_j^0) \hspace{1.42cm} {\rm for} \ i\neq j= k \;, \\ 0 \hspace{3.3cm} {\rm for} \ i= j= k \;. \end{cases}
\ee
This leads to
\be
\sum_j \frac{\partial^2 E}{\partial x_i \partial x_j}(\{x_i^0\}) \delta x_j = \sum_{j(\neq i)} W''(x_i^0-x_j^0) (\delta x_i -\delta x_j) \;,
\ee
and
\be
\sum_{j,k} \frac{\partial^3 E}{\partial x_i \partial x_j \partial x_k}(\{x_i^0\}) \delta x_j \delta x_k = \sum_{j(\neq i)} W'''(x_i^0-x_j^0) (\delta x_i -\delta x_j)^2 \;.
\ee
Thus, the second order term can be neglected if
\be
\forall j\neq i, \ \delta x_i - \delta x_j \ll 2\frac{W''(x_i^0-x_j^0)}{W'''(x_i^0-x_j^0)} \;,
\ee
which is the condition \eqref{cond_approx}.

\section{Next order in the expansion} \label{app:avg_nextorder}

Let us expand the equation of motion for the Brownian case \eqref{Eq_def_brownian} to the next order in $T$,
\be \label{Eq_delta_x_order2}
\delta \dot{x}_i(t) = - \sum_{j=1}^N H_{ij} \delta x_j(t) - \sum_{j,k=1}^N K_{ijk} \delta x_j(t) \delta x_k(t) + \sqrt{2T} \xi_i(t)  - \frac{\sqrt{2T}}{N} \sum_{j=1}^N \xi_j(t) +o(T) \quad , \quad K_{ijk} = \frac{\partial^3 E}{\partial x_i \partial x_j \partial x_k}(\{ x_i^0 \}) \;.
\ee 
Taking the Fourier transform in time (and using \eqref{identity_hessian_fourier} for the last term) we obtain
\be \label{eqfourier_order2} 
\delta \hat x_j(\omega) = \sum_{k,l,m=1}^N  [i \omega \mathbb{1}_N + H]^{-1}_{jk} K_{klm} (\delta \hat x_l * \delta \hat x_m)(\omega) + \sqrt{2T} \sum_{k=1}^N  [i \omega \mathbb{1}_N + H]^{-1}_{jk} \hat \xi_k(\omega) - \frac{\sqrt{2T}}{N} \frac{1}{i\omega} \sum_{k=1}^N \hat \xi_k(\omega) +o(T) \;.
\ee 
Taking the average over realisations of the noise and going back to real space we obtain
\bea
\langle \delta x_j(t) \rangle &=& \int_{-\infty}^{+\infty} \frac{d\omega}{2 \pi} e^{i \omega t}  \sum_{k,l,m=1}^N  [i \omega \mathbb{1}_N + H]^{-1}_{jk} K_{klm} \langle (\delta \hat x_l * \delta \hat x_m)(\omega) \rangle +o(T) \nn \\
&=& \int_{-\infty}^{+\infty} \frac{d\omega}{2 \pi} e^{i \omega t}  \sum_{k,l,m=1}^N  [i \omega \mathbb{1}_N + H]^{-1}_{jk} K_{klm} \int_{-\infty}^{+\infty} dt' e^{-i\omega t'} \langle \delta x_l(t') \delta x_m(t') \rangle +o(T) \nn \\
&=& \frac{2T}{N} \int_{-\infty}^{+\infty} \frac{d\omega}{2 \pi} e^{i \omega t}  \sum_{k,l,m=1}^N  [i \omega \mathbb{1}_N + H]^{-1}_{jk} K_{klm} \int_{-\infty}^{+\infty} dt' e^{-i\omega t'} \sum_{q=1}^{(N-1)/2} \frac{\cos\left(2\pi \frac{q}{N} (l-m)\right)}{\mu_q} +o(T) \nn \\
&=& \frac{2T}{N} \sum_{q=1}^{(N-1)/2} \sum_{k,l,m=1}^N \frac{\cos\left(2\pi \frac{q}{N} (l-m)\right)}{\mu_q} \int_{-\infty}^{+\infty} \frac{d\omega}{2 \pi} e^{i \omega t}   [i \omega \mathbb{1}_N + H]^{-1}_{jk} K_{klm} \delta(\omega)  +o(T) \nn \\
&=& \frac{2T}{N} \sum_{q=1}^{(N-1)/2} \sum_{k,l,m=1}^N \frac{\cos\left(2\pi \frac{q}{N} (l-m)\right)}{\mu_q} [H^{-1}]_{jk} K_{klm} +o(T)
\;.
\eea
Thus the average of $\delta x_i(t)$ is of order $T$, and it is thus subleading compared to the covariance. A similar computation in the RTP case leads to $\delta x_i(t) \propto v_0^2$.

\section{Other choice of scaling} \label{app:other_scaling}

In most of this paper, for the Riesz gas on the circle, we have chosen to keep the density $\rho=N/L$ fixed when taking the limit $N \to +\infty$, i.e. we have chosen $L \propto N$. Another possibity would be to keep the system size $L$ fixed. This would mean that the interparticle distance, and thus the space variable $x$, would scale as $N^{-1}$. Thus we would need to scale the temperature as $T \sim N^{-2}$. In addition, since testing the validity of the linear approximation as in \eqref{validity_brownian} requires comparing $T$ to the product $g\rho^s$, one should also have $g\rho^s \sim N^{-2}$, i.e. since $\rho \sim N$, $g \sim N^{-(s+2)}$. In this way, the interaction time scale $\tau=g \rho^{s+2}$ remains the same as in the paper, i.e. it scales as $N^0$, and thus we can keep the same scaling for the time $t$. For the RTP, this scaling implies that we should scale $\gamma \sim N^0$ and $v_0 \sim N^{-1}$.


Note that for the active DBM and active CM model in a harmonic trap we have made a different choice of scaling. In this case, we chose $g\sim N^{-1}$ for the active DBM and $g \sim N^{-2}$ for the active CM, in order for the support of the density to be finite in the large $N$ limit (note that for the active CM we also did $g\to \tilde g^2$ to have expressions closer to the active DBM). Contrary to the circular case, the choice of scaling for $g$ does not have any effect on the time scale, which, remarkably, in this case 
is fixed only by the strength of the confining potential $\lambda$. This can be seen as follows. 
The support of the density being $\sim \sqrt{gN/\lambda}$ for the active DBM, one has
$\rho \sim \sqrt{\lambda N/g}$ and thus $\tau = 1/(g\rho^2) \sim 1/(\lambda N)$. 
In the active CM case the support is of size $\sim (gN^2/\lambda)^{1/4}$, which leads to $\rho \sim (\lambda N^2/g)^{1/4}$ and thus $\tau = 1/(g\rho)^4 \sim 1/(\lambda N^2)$. In both cases the largest time scale is $N^{z_s} \tau \sim 1/\lambda$. This can also be seen more rigorously by noting that the Hessian is proportional to $\lambda$ and independent of $g$, see e.g. \cite{ADBM2}.

\section{Correspondence between the definitions of the Riesz potential}
\label{app:RieszCorresp}

Let us recall the Poisson summation formula
\be
\sum_{n=-\infty}^\infty f(x + n a) = \frac{1}{a} \sum_{m=-\infty}^\infty \hat f(\frac{2 \pi}{a} m) e^{ i \frac{2 \pi}{a} m x} \quad {\rm where} \quad \hat f(y) = \int_{-\infty}^{+\infty} dt f(t) e^{- i y t} dt \;.
\ee 
Choosing $f(t)={\rm sgn}(t)/|t|^{s+1}$, one finds for $-1<s<1$
\be 
\hat f(y) = 2 i \, {\rm sgn}(y) |y|^s \Gamma(-s) \sin (\frac{\pi s}{2}) = -\frac{\sqrt{\pi}}{2^s} i \, {\rm sgn}(y) |y|^s \frac{\Gamma(\frac{1-s}{2})}{\Gamma(1+\frac{s}{2})},
\ee 
which leads to the identity between the two expressions of $W'(x)$ introduced in Eq. \eqref{defRiesz} and \eqref{defRieszFourier} respectively,
\be
W'(x) = - g \sum_{m=-\infty}^\infty \frac{{\rm sgn}(x+m L)}{|x+mL|^{s+1}} = i \pi^{s+\frac{1}{2}} \frac{\Gamma(\frac{1-s}{2})}{\Gamma(1+\frac{s}{2})} \frac{g}{L^{s+1}} \sum_{p \neq 0} |p|^{s} \, {\rm sgn} (p) \, e^{2 i \pi p \frac{x}{L}} \;.
\ee

Note that with the definition in Fourier space, the second derivative reads
\be
W''(x) = -2 \pi^{s+\frac{3}{2}} \frac{\Gamma(\frac{1-s}{2})}{\Gamma(1+\frac{s}{2})} 
\frac{g}{L^{s+2}} \sum_{p \neq 0} |p|^{s+1}  \, e^{2 i \pi p \frac{x}{L}}
= -4 \pi^{s+\frac{3}{2}} \frac{\Gamma(\frac{1-s}{2})}{\Gamma(1+\frac{s}{2})} \frac{g}{L^{s+2}} {\rm Re} \, {\rm Li}_{-1-s}(e^{2 i \pi \frac{x}{L}})
\ee
which, using formula \eqref{eigenvals}, leads to (with $q=1,\dots,N-1$ and $u=q/N$), for $s>-1$,
\bea  
\mu_q &=& -2 \pi^{s+\frac{3}{2}} \frac{\Gamma(\frac{1-s}{2})}{\Gamma(1+\frac{s}{2})} 
\frac{g}{L^{s+2}} \sum_{p=-\infty}^\infty |p|^{s+1}  \, \sum_{\ell=1}^{N-1} e^{2 i \pi p \frac{\ell}{N}} (1 - \cos(\frac{2 \pi q \ell}{N})) \\
&=& -2 \pi^{s+\frac{3}{2}} \frac{\Gamma(\frac{1-s}{2})}{\Gamma(1+\frac{s}{2})} g \rho^{s+2} \sum_{n=-\infty}^\infty ( |n|^{s+1} - \frac{1}{2} |n+u|^{s+1} - \frac{1}{2} |n-u|^{s+1})
\eea  
where the last sum comes from the terms such that $p,p+q,p-q$ are multiples of $N$ (the other terms vanish).
The sum converges for $-1<s<0$ and we have checked that it agrees with the result \eqref{mu_Riesz} for $\mu_q$ for these values.

\section{Useful integrals and identities for the gamma function} \label{app:integrals}

For any $1<\alpha\leq 2$,
\be
\int_0^{+\infty} du \frac{\sin^2(\pi u)}{u^{\alpha}} = \frac{\pi^{\alpha-\frac{1}{2}}}{2(\alpha-1)} \frac{\Gamma(\frac{3-\alpha}{2})}{\Gamma(\frac{\alpha}{2})} = -2^{\alpha-2} \pi ^{\alpha-1} \sin \left(\frac{\pi  \alpha}{2}\right) \Gamma (1-\alpha) = -\frac{2^{\alpha-3} \pi^\alpha}{\cos(\frac{\pi\alpha}{2})\Gamma(\alpha)} \;.
\ee
This leads to, for $0<s<1$,
\be
a_s = 2\pi^{s+\frac{3}{2}} \frac{\Gamma(\frac{1-s}{2})}{\Gamma(1+\frac{s}{2})} = -2^{2+s}\pi^{s+1} \sin(\frac{\pi s}{2}) \Gamma(-s) = \frac{2^{1+s} \pi^{2+s}}{\cos(\frac{\pi s}{2})\Gamma(1+s)} \;.
\ee
We also have, for $\alpha>1$,
\be
\int_0^{+\infty} dv \frac{1-e^{-v^{\alpha}}}{v^{\alpha}} = \frac{\Gamma(1/\alpha)}{\alpha-1}
\quad \text{and} \quad \int_0^{+\infty} dv \left(\frac{1 - e^{-v^{\alpha}}}{v^{\alpha}}\right)^2 = \frac{2(2^{\frac{\alpha-1}{\alpha}}-1)}{\alpha} \Gamma\left(-\frac{2\alpha-1}{\alpha}\right) \;,
\ee
as well as
\be
\int_0^{+\infty} \frac{dv}{1+v^{\alpha}} =\Gamma(\frac{\alpha-1}{\alpha}) \Gamma(\frac{\alpha+1}{\alpha}) = \frac{\pi}{\alpha \sin(\frac{\pi}{\alpha})} \;.
\ee

\section{An argument for the large time limit of $C_0(t)$} \label{app:Ckrtp_argument}

Consider the large time limit of $C_0(t)$ in the RTP case
\be \label{C0rtp_app}
C_0(t) \simeq
 4 T_{\rm eff} \tau \int_0^{1/2} du \frac{(1-e^{- f_s(u) t/\tau})-\hat g f_s(u) (1-e^{-2\gamma t})}{f_s(u)(1-\hat g^2 f_s(u)^2)} \;.
\ee
We focus on the case $s>0$. In this appendix, we consider two different limits of this integral to justify some of the statements made in section \ref{sec:C0(t)}. We first consider the limit $t \gg \tau$ at fixed $\hat g$, for which we argue that the main contribution to the integral comes from $u\sim (t/\tau)^{-1/z_s}$. We then consider $\hat g \gg 1$, in which case we find that it is dominated by $u \sim \hat g^{-1/z_s}$. Both results also hold for $C_k(t)$, which justifies the scaling forms \eqref{Ck_scaling1} and \eqref{Ck_scaling2}. The large time result also serves as an argument for the replacement $f_s(u) \simeq a_s u^{z_s}$ in the Brownian case. 

We start with the limit $t\gg \tau$, with $\hat g$ fixed (of order 1 or smaller). This implies that $\gamma t = \frac{t/\tau}{2\hat g} \gg 1$. Let us consider $u\sim (t/\tau)^{-\alpha}$ with $\alpha\geq 0$. The first term in the numerator of \eqref{C0rtp_app} is of order 1 if $\alpha\leq 1/z_s$ and of order $\sim (t/\tau)^{1-z_s\alpha}$ if $\alpha>1/z_s$. Since $\gamma t \gg 1$, the second term in the numerator is of order $\sim (t/\tau)^{-z_s\alpha}$ and can be neglected compared to the first term. The second term in the denominator can also be discarded, the remaining part of the denominator being $\sim (t/\tau)^{-z_s\alpha}$. To evaluate the contribution to the integral, the result should be weighted by a factor $(t/\tau)^{-\alpha}$. This leads to a total contribution $\sim (t/\tau)^{f(\alpha)}$ where $f(\alpha)=\alpha(z_s-1)$ for $\alpha \leq 1/z_s$ and $f(\alpha) =1-\alpha$ for $\alpha > 1/z_s$. The function $f(\alpha)$ is  maximal at $\alpha=1/z_s$ (with $1/2\leq 1/z_s<1$). The integral is thus indeed dominated by small $u \sim (t/\tau)^{-1/z_s} \ll 1$. This allows us to use the asymptotic expression of $f_s(u)$ \eqref{fasympt} and to neglect the second term in the numerator and in the denominator. Note however that this is not enough to allow us to expand the first exponential, its argument being of order 1 for such values of $u$. We thus have in this regime ($t/\tau \gg 1$ and $\gamma t \gg 1$),
\be \label{C0_large_t_app}
C_0(t) \simeq
 4 T_{\rm eff} \tau \int_0^{+\infty} du \frac{1-e^{- a_s u^{z_s} t/\tau}}{a_s u^{z_s}} = \begin{dcases} U_s T_{\rm eff} \tau^{\frac{1}{s+1}} \, t^{\frac{s}{s+1}} \;,\quad
\hspace{1cm}\text{for } 0<s<1 \;, \\ 
2T_{\rm eff} \sqrt{\frac{\tau \, t}{\pi (s+1) \zeta(s)}}\;, \quad
\quad \text{for } s>1 \;, \end{dcases}
\ee
which is the same as the limit $t\gg \tau$ of the Brownian regime \eqref{displacement_Riesz}.

We now consider the limit $\hat g \gg 1$, keeping $t/\tau$ fixed. This implies that $\gamma t = \frac{t/\tau}{2\hat g} \ll 1$. In this case, we can make a similar reasoning as above, by assuming now that $u\sim \hat g^{-\alpha}$. Writing $\gamma t = \frac{t/\tau}{2\hat g}$ to eliminate $\gamma t$, we see that both terms in the numerator are of order $\sim \hat g^{-z_s \alpha}$, while the denominator is of order $\sim \hat g^{2-3z_s \alpha}$ if $\alpha\leq 1/z_s$ and of order $\sim \hat g^{-z_s\alpha}$ if $\alpha > 1/z_s$. Adding a weight $\hat g^{-\alpha}$ we obtain a total contribution $\sim \hat g^{f(\alpha)}$ with $f(\alpha)=(2z_s-1)\alpha-2$ for $\alpha \leq 1/z_s$ and $f(\alpha)=-\alpha$ for $\alpha>1/z_s$. For any $z_s>1/2$ (i.e. $s>-1/2$), $f(\alpha)$ is again maximal for $\alpha=1/z_s$, which means that the integral is dominated by $u\sim \hat g^{-1/z_s}$. This means that we can expand both exponentials (since $(t/\tau) f_s(u) \sim (t/\tau)/\hat g \sim 2\gamma t$) in \eqref{C0rtp_app}, leading to
\be \label{C0_large_g_app}
C_0(t) \simeq
 2 v_0^2 t^2 \tau \int_0^{1/2} \frac{du}{1+\hat g u^{z_s}} = v_R(\hat g)^2 t^2  \simeq \begin{cases} \frac{2\pi}{(s+1)\sin(\frac{\pi}{s+1})} \frac{v_0^2 t^2}{(a_s \hat g)^{\frac{1}{s+1}}}\;, \quad  \text{ for } 0<s<1 \;, \\ \frac{\pi v_0^2 t^2}{\sqrt{a_s \hat g}}\;, \quad  \hspace{2.38cm} \text{ for } s>1 \;, \end{cases}
\ee
and thus we recover the ballistic regime $\gamma t \ll 1$ for $\hat g \gg 1$, as in Eq.~\eqref{vR_large_g}.

What happens if both $t/\tau$ and $\hat g$ are large ? In this case, writing everything in terms of $\hat g$, $t$, $\tau$ and $v_0$, we find that \eqref{C0_large_t_app} gives $C_0(t) \sim v_0^2 \hat g \tau^2 (t/\tau)^{1-1/z_s}$ and \eqref{C0_large_g_app} gives $C_0(t) \sim v_0^2 t^2 \hat g^{-1/z_s}$. Thus the two expressions become of the same order when $t/\tau \sim \hat g$, i.e. when $\gamma t \sim 1$. This confirms that, when $\hat g \gg 1$, the ballistic expression \eqref{C0_large_g_app} holds whenever $\gamma t \ll 1$, while the large time expression \eqref{C0_large_t_app} holds for $\gamma t \gg 1$.

\section{Large $k$ limit of $C_k(t)$ for the RTPs} \label{app:computation_details}

We show here how to recover the large separation ballistic regime \eqref{Ck_ballistic3} for $C_k(t)$ for the RTPs, for $0<s<1$. The computation is similar to the one presented in Sec.~\ref{sec:spacetime_brownian}. Using the scaling form \eqref{Ck_scaling2}, this corresponds to the limit ${\sf x}\gg 1$ of $\mathcal{F}({\sf x},y)$. We first perform two integrations by parts and a change of variable $w={\sf x}v$ to obtain 
\be
\mathcal{F}({\sf x},y) = - \frac{a_s^{\frac{1}{s+1}}}{\pi^2 {\sf x}^3} \int_0^{+\infty} dw \cos(2\pi a_s^{-\frac{1}{s+1}} w) h_y''(\frac{w}{{\sf x}}) \quad , \quad h_y''(u) \underset{u\to 0}{\simeq} -2y^2 s(s+1) u^{s-1} +O(u^{2s}) \;,
\ee
which leads to
\be
\mathcal{F}({\sf x},y) \simeq 2y^2 \frac{s(s+1)a_s^{\frac{1}{s+1}}}{\pi^2 {\sf x}^{2+s}} \int_0^{+\infty} dw \frac{\cos(2\pi a_s^{-\frac{1}{s+1}} w)}{w^{1-s}} = 8y^2 \frac{a_s\cos(\frac{\pi s}{2})\Gamma(2+s)}{(2\pi {\sf x})^{2+s}} = \frac{4(s+1)y^2}{{\sf x}^{2+s}} \;.
\ee
This gives, for $k\gg\hat g^{1/z_s}$ and $0<s<1$,
\be
C_k(t) \simeq \frac{4(s+1)T_{\rm eff}}{g\rho^{s+2}} (\gamma t)^2 \frac{\hat g^2}{k^{2+s}} = \frac{(s+1)\hat g}{k^{2+s}} v_0^2 t^2 = \frac{(s+1)T_{\rm eff} t^2}{\tau k^{2+s}} \;,
\ee
which coincides with the result obtained for the Brownian particles with the change $T\to T_{\rm eff}$.

This result is in theory valid for finite $y$, but it is actually also valid in the limit $\gamma t \ll 1$. We now show how to recover this result using the scaling form of the ballistic regime \eqref{Ck_ballistic2}. Let us define $\kappa = \frac{2\pi k}{(a_s \hat g)^{1/z_s}}$. Performing two successive integration by parts in \eqref{Ck_ballistic2}, we find (the boundary terms vanish)
\bea
\int_0^{+\infty} dv \frac{\cos (\kappa v)}{1+v^{1+s}} &=& \int_0^{+\infty} \frac{dv}{\kappa^2} \frac{\cos (\kappa v)}{(1+v^{1+s})^2} \left( s(s+1) v^{s-1} - \frac{2(s+1)^2 v^{2s}}{1+v^{1+s}} \right) \nn \\
&=& \int_0^{+\infty} \frac{dw}{\kappa^3} \frac{\cos(w)}{(1+(\frac{w}{\kappa})^{1+s})^2} \left( s(s+1) (\frac{w}{\kappa})^{s-1} - \frac{2(s+1)^2 (\frac{w}{\kappa})^{2s}}{1+(\frac{w}{\kappa})^{1+s}} \right) \nn \\
&\underset{\kappa\gg 1}{\simeq}& \frac{s(s+1)}{\kappa^{2+s}} \int_0^{+\infty} dw \frac{\cos(w)}{w^{1-s}} = \frac{\cos(\frac{\pi s}{2}) \Gamma(2+s)}{\kappa^{2+s}} \;.
\eea
Putting everything together and using the expression for $a_s$ given in \eqref{fasympt} we obtain, for $k \gg \hat g^{1/z_s} \gg 1$ and $0<s<1$, $\phi_s \simeq (s+1) x^{-(2+s)}$, i.e.,
\be
C_k(t) \simeq \frac{(s+1) \hat g}{k^{2+s}} v_0^2 t^2 \;.
\ee


\begin{thebibliography}{100}
\bibitem{footnoteBoursier}
We were not able to find agreement however concerning the prefactor. 
The prefactor given in formula 1.9 Thm 5 of 
\cite{BoursierCorrelations} seems $4$ times bigger than our prefactor setting $g=2s$. 
The prefactor given in Corrolary 1.1 of Thm 3 of \cite{BoursierCLT} does not seem consistent
with either result.
\end{thebibliography}
\end{document}